\documentclass[11pt]{report}         
\usepackage{utdiss2}              
\usepackage[dvips]{graphicx}
\usepackage{amsmath,amssymb,amsthm,amsfonts,amscd}
\usepackage{eucal}              
\usepackage{verbatim}           
\usepackage{makeidx}            
\usepackage{psfig}              
\usepackage{epsfig}             
\usepackage{citesort}           %
\usepackage{float}           %
\usepackage{url}                
\newcommand{\beq}[1]{\begin{equation} #1 \end{equation}}
\newcommand{\beqa}[1]{\begin{eqnarray} #1 \end{eqnarray}}
\newcommand{\greekbf}[1]{\mbox{\boldmath$#1$}}

\newcommand{\T}{\hspace*{0.5cm}}
\newcommand{\makeline}[1]{\underline{\hspace*{#1}}}

\newcommand{\mathbig}[1]{\mbox{\large$#1$}}
\newcommand{\mathBig}[1]{\mbox{\Large$#1$}}

\newcommand{\mathsmallest}[1]{\mbox{\scriptsize$#1$}}

\newcommand{\switchfonts}{\usefont{OT1}{cmr}{m}{it}}
\newcommand{\lie}{\switchfonts \mbox{\symbol{36}} \normalfont}
\newcommand{\deriv}[2]{\frac{ d #1 }{ d #2 }}
\newcommand{\pderiv}[2]{\frac{ \partial #1 }{ \partial #2 }}
\newcommand{\bfbar}[1]{\mathbf{\bar{#1}}}
\newcommand{\qbar}{\bfbar{q}}
\newcommand{\ibanez}{J. $\mathrm{M}^{\underline{a}}$ Ib\'{a}\~{n}ez}
\newcommand{\marti}{J. $\mathrm{M}^{\underline{a}}$ Mart\'{i}}

\author{Scott Charles Noble}       

\address{8933 S.W. Chevy Circle \\
Stuart, FL  34997
}

\title{A Numerical Study of Relativistic Fluid Collapse}    

\supervisor
        [Philip J. Morrison]
        {Matthew W. Choptuik}

\committeemembers
        [Richard Fitzpatrick]
        [Richard A. Matzner]
        {Paul R. Shapiro}

\previousdegrees{B.S.}
\graduationmonth{December}      
\graduationyear{2003}   
\oneandonehalfspacing

\oneandonehalfspacequote

\newcommand{\latexe}{{\LaTeX\kern.125em2%
                      \lower.5ex\hbox{$\varepsilon$}}}

\chardef\bslash=`\\     

\makeatletter           
\def\square{\RIfM@\bgroup\else$\bgroup\aftergroup$\fi
  \vcenter{\hrule\hbox{\vrule\@height.6em\kern.6em\vrule}%
                                              \hrule}\egroup}
\makeatother            

\makeindex    

\begin{document}

\copyrightpage          

%
%
\commcertpage           

\titlepage              

%
\begin{dedication}
\index{Dedication@\emph{Dedication}}%
To my mother... 
\end{dedication}

\begin{acknowledgments}         
\index{Acknowledgments@\emph{Acknowledgments}}%
\vspace{-1cm}
First, I thank my supervisor, Matt Choptuik, for all that he has taught me, his refreshingly helpful 
manner, and the great patience he exhibited in working on this dissertation with me.  It was truly 
a pleasure to work with him.  In addition, I thank all my colleagues at UBC and the Center for Relativity 
at UT-Austin.  I have learned much from the innumerable discussions with them and respect them all greatly.  
Especially, I wish to show my appreciation for the many helpful conversations I shared with I\~{n}aki 
Olabarrieta, who served as a sounding board for many of my ideas.  

I owe a great deal of gratitude to my mother, who  always  encouraged me to follow my passions in 
life and to achieve great things.  Without her continuous support, I fear I would not have been able to 
come this far along in my studies.  For all she has provided, I give my deepest love and appreciation to her.  
I also wish to thank my father for never ceasing to be 
interested in my endeavors, for fostering my interest in science, and for always being proud of his son.  

For the past year I have been blessed with the love and emotional support from my dearest Somayeh.
She has made this the happiest time of my life, and made writing this dissertation a much more delightful
experience than it would have been without her.   

Finally, I acknowledge financial support from the National Science Foundation grant PHY9722068 and 
and from the Natural Sciences and Engineering Research Council of Canada (NSERC).   All the calculations 
referred to in this work were performed on the \textrm{vn.physics.ubc.ca} cluster at UBC
funded by the Canadian Foundation for Innovation, NSERC and the Canadian Institute for 
Advanced Research. 

\end{acknowledgments}

%
\utabstract
\index{Abstract}%
\indent

We investigate the dynamics of self-gravitating, spherically-symmetric
distributions of fluid through numerical means.  In particular, systems
involving neutron star models driven far from equilibrium in the strong-field
regime of general relativity are studied.  Hydrostatic solutions of
Einstein's equations using a stiff, polytropic equation of state are
used for the stellar models.  Even though the assumption of spherical symmetry
simplifies Einstein's equations a great deal, the hydrodynamic equations
of motion coupled to the time-dependent geometry still represent a set of
highly-coupled, nonlinear partial differential equations that can only be
solved with computational methods.  Further, many of the scenarios we
examine involve highly-relativistic flows that require improvements upon
previously published methods to simulate.
Most importantly, with techniques such as those used and developed in this
thesis, there is still considerable physics to be extracted from
simulations of perfect fluid collapse, even in spherical symmetry.
Here our particular focus is on the physical behavior of the coupled
fluid-gravitational system at the threshold of black hole
formation---so-called black hole critical phenomena.

To investigate such phenomena starting from conditions representing
stable stars, we must drive the star far from its initial stable configuration. We
use one of two different mechanisms to do this: setting the initial velocity
profile of the star to be in-going, or collapsing a shell of massless scalar
field onto the star.
Both of these approaches give rise to a large range of dynamical scenarios
that the star may follow.  These scenarios have been extensively surveyed
by using different initial star solutions, and by varying either the magnitude
of the velocity profile or the amplitude of the scalar field pulse.
In addition to illuminating the critical
phenomena associated with the fluid collapse,
the resulting phase diagram of possible outcomes provides
an approximate picture of the stability of neutron stars to large, external
perturbations that may occur in nature.

Black hole threshold, or critical, solutions, occur in
in two varieties:  Type~I and Type~II.  Generically, a Type~I solution is
either static or periodic and exhibits a finite black hole mass at threshold,
whereas a Type~II solution is generally either discretely or continuously
self-similar and characterized by infinitesimal black hole mass at
threshold.  We find both types of critical behavior in our space of star
solutions.  The Type~I critical solutions we find are perturbed
equilibrium solutions with masses slightly larger than their progenitors.
In contrast, the Type~II solutions are continuously self-similar solutions
that strongly resemble those
found previously in ultra-relativistic perfect fluids.  The boundary between
these two types of critical solutions is also discussed.

\tableofcontents   

\listoftables      
\listoffigures     

\chapter{Introduction}
\label{chap:introduction}

The dynamics of compact gravitating objects out of equilibrium has always been a topic of much interest
in astrophysics.  Physical systems that fall under this subject
include supernovae, failed supernovae such as hypernovae or collapsars, gamma-ray 
burst (GRB) progenitors, coalescing  neutron star binary systems, accreting compact stars, and 
neutron stars that undergo sudden phase transitions, to only name a few.  In many 
of these cases, 
a compact star is in such an excited state that it must catastrophically collapse and/or explode. 

For those involving supernovae, the star has reached a non-equilibrium state either through accretion 
from a companion star (Type~Ia), or---if sufficiently massive---by reaching the ultimate end in the 
thermonuclear cycles when fusion is no longer exothermic (Type~II,Ib,Ic). 
In the former case, the unstable star is a white dwarf that has accreted past its Chandrasekhar
limit, and consequently its electron degeneracy pressure is no longer sufficient 
to support it from gravitational collapse.  The latter case, on the other hand, involves a star that has burned
through successive elemental cores until an iron core has developed
and can no longer support the star through thermonuclear processes \cite{carroll-ostlie}.  Instead the degeneracy 
pressure of the relativistic electrons holds the star together until the outer layers produce 
enough iron to overwhelm the supporting electron pressure.  In both cases, the onset of instability 
brings about a sudden homologous collapse that is ultimately halted by the matter stiffening
from the increased presence of neutrons in the core.  
As the star collapses upon itself, the outer layers of the stellar core typically form a shock and 
recoil from the dense inner core once a maximum central density and pressure are reached. 
For core collapse supernovae, the shock propagates outward, heating the matter and leaving 
a convective region in its wake.  It has been found in many detailed 
simulations---e.g. \cite{muller-janka} and references therein---that the hydrodynamic bounce 
scenario eventually stalls as the shock becomes thin to neutrinos and thermal photons are 
absorbed by the dissociation of Fe nuclei into $\alpha$ particles.  The explosion is re-energized
by a ``hot-bubble'' region heated by neutrinos from the core \cite{bethe-wilson} that forms 
between the core and the stalling shock front. 
Once the radiation-dominated bubble has been heated, convection drives  a dynamic overturn of 
the neutrino-heated matter and the cold matter located behind the shock.  The transport of 
the hot matter to the shock front re-energizes the supernova explosion.
Even though the purely gravitational
hydrodynamic bounce and shock scenario is not solely responsible for the ultimate 
explosion associated with Type~II/Ib/Ic supernovae, it still plays an important role in 
determining whether the progenitor object is a neutron star or a black hole.  In addition, matter 
can fall back upon the nascent neutron star and initiate a new collapse scenario.  

The increase in neutron density in the core results from inverse
$\beta$-decay,  which becomes an energetically-favorable process as the electrons become more 
relativistic, and from neutrons that ``drip'' off of neutron-rich nuclei that is 
caused by the core's extremely high pressure.  The neutrons in turn form a condensed fermionic gas whose 
degeneracy pressure may be able to support the continuing collapse of the star; if the star 
does not collapse to a black hole, the neutron gas will form a hot neutron star that will cool
very quickly---going from tens of MeV to less than $1$~MeV\cite{glendenning} in a matter of 
seconds.  Since a 
neutron's mass is significantly larger than an electron's, the neutrons in the cooled neutron 
star are non-relativistic at these temperatures and---consequently---can be described by stiff equations of state. 
In fact, these temperatures are far below the Fermi temperature of the condensed neutron gas and 
can therefore be neglected in most cases.

Once the neutron star is formed, it may undergo additional evolution.  If it is born out of a Type~II 
supernova, the outwardly-moving shock wave of matter may stall and collapse onto the nascent neutron 
core \cite{zampieri-etal-1998}.  In contrast, if the neutron star is in a binary system with a less 
compact companion star, accretion from the companion  may push the neutron star over its 
Chandrasekhar limit.  In either of these cases, the resultant non-equilibrium system will most likely undergo  
a hydrodynamic implosion that will often result in black hole formation. 

Because all these examples involve the often complex dynamics of compact objects, it is essential 
to be able to model the systems of interest in great detail \emph{and} breadth.  
Making detailed models of these systems requires the inclusion of a plethora of 
physical effects---such as radiation transport, multi-species
flows, general relativistic gravitation and magnetohydrodynamics.  Simulating objects with all these attributes
will be impossible in the near future given the current rate at which computational power is 
increasing.  Hence, the systems must be simplified in some fashion for their simulations 
to be tractable.  In this study, we wish to consider hydrodynamical systems in the strong-field regime of 
gravity, where compact stars are set far from  equilibrium  and follow highly-relativistic evolutions.
A specific topic we wish to cover is how such stars 
collapse to black holes, which are regions of spacetime that are so greatly curved that 
nothing---not even light---can escape.  Consequently, we will restrict our investigation to 
the most relativistic, compact stellar objects known: neutron stars (other objects consisting of more exotic
matter may exist, such as a so-called quark star comprised of free quarks \cite{glendenning}). 

Being able to examine compact objects on the verge of black hole formation also allows us 
to investigate the critical phenomena that will likely arise.  Critical phenomena in general relativity 
involves the study of the solutions---called \emph{critical} solutions---that lie at the boundary between 
black hole-forming and black hole-lacking spacetimes \cite{choptuik-1998,gundlach,gundlach-rev2}.  Because 
of critical solutions' intriguing 
characteristics, critical phenomena are one of the most exciting new topics in general relativity 
over the past few decades.  Not only are the critical solutions exotic, they represent a new class 
of solutions that are universal in some sense, independent---to a degree---of the initial data 
from which they evolved. 

The first critical solutions to be discovered were Type~II critical solutions \cite{choptuik-1993,evans-coleman}, 
named after the analogous behavior observed in statistical mechanics.  
Across this so-to-speak gravitational phase transition, the mass of the resulting black 
hole---$M_\mathrm{BH}$---can be thought of as the order parameter.  Hence,
Type~II critical behavior is such that the transition from black hole-lacking solutions to black 
hole-forming solutions is continuous in the black hole mass.  That is,
as one adjusts, or \emph{tunes}, the initial data toward the critical solution, arbitrarily small 
black holes can be formed.  In addition, the critical solution generically contains a massless
curvature singularity that is not shrouded by an event horizon.
Furthermore, solutions at a Type~II threshold typically display 
continuous self-similarity (CSS) or discrete self-similarity (DSS) in which 
the solutions' dynamical scales shrink as they in-fall toward the origin. 
This type of critical solution 
is particularly intriguing to the study of the cosmic censorship conjecture, which suggests that nature 
tries to hide---or censor---singularities from the rest of the universe by shrouding them in a black hole.
With the singularity in a black hole, it
cannot be probed in any way, because any signals traveling into the hole are forever trapped.  However, 
if the singularity is \emph{naked}---e.g. without a surrounding event horizon---then it exists in the 
causal structure of the universe and consequently is observable.  However, since this singularity 
represents an ``infinity'' in spacetime, it fails to be describable by our laws of physics as we now 
know them.  Hence, in a sense, the cosmic censorship conjecture asserts that the universe ``protects'' 
observers from seeing something they cannot describe.  Even though critical phenomena has lead to 
interesting consequences in the cosmic censorship conjecture, we will not discuss cosmic censorship 
any further in this thesis. 

In addition to the peculiar Type~II solutions, Type~I solutions have also been observed in 
a variety of matter models.  By continuing the analogy from statistical mechanics, these 
solutions are discontinuous in their order parameter, $M_\mathrm{BH}$.  Hence, the critical 
solutions have finite mass, and they are typically
static or oscillatory in nature.  In contrast to the Type~II case, the Type~I critical solution is not
singular,
but is typically a meta-stable distribution of matter with compact support.  Such critical solutions 
are usually observed in models that have known bound states.  

In this work, we investigate both types of critical behavior using a perfect fluid model. 
The initial conditions, which we adjust, entail a compact star and some sort of ``perturbing agent.''
The compact star solutions which we use are the spherically-symmetric hydrostatic solutions to the coupled 
Einstein-fluid equations, the so-called Tolman-Oppenheimer-Volkoff (TOV) solutions 
\cite{oppenheimer-volkoff,tolman-book,tolman-paper}.
To approximate the stiff flows commonly thought to exist in the cores of neutron stars, we use the 
stiffest, causal polytropic equation of state.  The methods by which we drive a star to a non-equilibrium 
state involve giving the star an initially in-going velocity profile, and collapsing 
a spherical shell of scalar field onto it.  Both methods can hardly be considered as perturbative 
since they can often drive the star to total obliteration, or prompt collapse to a black hole, but 
we use this term sometimes since a better one is lacking. 

Since the perfect fluid equations of motion have an intrinsic length scale and are known to have bound
states, we are able to study both types of critical phenomena within the same model.  
Chapters~\ref{chap:theoretical-basis}~and~\ref{chap:numerical-techniques}, respectively, provide an introduction to the 
theory describing our systems and the numerical methods we use to simulate them.
In Chapter~\ref{chap:veloc-induc-neutr},
we begin our study of stellar collapse by extensively covering the parameter space of initial 
conditions for velocity-perturbed stars.  The results from this chapter provide a broad view of the 
range of dynamical scenarios one can expect in the catastrophic collapse of neutron stars.  We then 
employ this knowledge in our examination of the solutions on the verge of black hole formation.  Both Type~I 
and Type~II solutions are found and studied.  In Chapter~\ref{chap:type-ii-critical}, we analyze the 
the observed Type~II behavior and compare it to recent work in the field.  The stars' critical behavior 
is further explored in Chapter~\ref{chap:type-i-critical}, where we extend the scope to Type~I
phenomena.  The nearly critical solutions we calculate from the Type~I study are then compared to 
perturbed unstable TOV solutions.  The boundary between the two types of phenomena is discussed along the 
way.  Finally, we conclude in Chapter~\ref{chap:concl-future-work} with some closing remarks and 
notes on anticipated future work. 

\section{Notation, Conventions and Units}
\label{sec:notat-conv-units}

In the following work, so-called geometrized units are used and are such that $G = c = 1$. 
Abstract index notation, which is a way of referring to a tensor's components in a covariant manner, is 
 is used with the first few Latin letters (i.e. $a, b, c, d, ...$) \cite{wald}. 
Greek indices will always refer to \emph{all} spacetime 
components (i.e. $\mu, \nu, \ldots \in \{0, 1, 2, 3\}$), and
$i, j, \ldots, n \in \{1, 2, 3\}$.  Also, we follow \cite{wald} in tensor definitions, 
definitions of the Christoffel symbols,  and sign conventions. 
The Einstein summation convention is always used (but only in 
regards to repeated indices that are not ``$t, x, y, z, r, \theta, \phi$''), e.g. 
$g_{\mu \nu} n^\mu \equiv \sum_{\mu = 0}^{3} g_{\mu \nu} n^\mu$ but $g_{t t}$ just represents a 
metric component. 

When referring to discretized quantities, subscripts $i, j, k, \ldots$ typically refer 
to locations on a discrete grid 
of coordinates, while the superscript $n$ represents the  index of the
quantity's discrete time step.  Quantities in bold-face, e.g. $\mathbf{q}, \mathbf{f}$, 
are generally state vectors. 

In addition, when referring to the Tolman-Oppenheimer-Volkoff (TOV) solutions, the fluid's 
equation of state sets a scale in the system.  This scale is typically set to $1$ in order 
to remove unit-dependence from the system of equations.   Restoring quantities mentioned 
herein to physical units is discussed in Appendix~\ref{app:unit-conversion}. 

Two acronyms will occur quite frequently in this dissertation, so we will define them here:
DSS = Discretely Self-Similar (or Discrete Self-Similarity), CSS = Continuously Self-Similar 
(or Continuous Self-Similarity). 

Finally, we will use a star, $\star$, in the superscript position to denote that a quantity 
pertains to a critical solution.  On the other hand, a quantity with an asterisk in the subscript 
position should suggest that it refers to a star solution.

\chapter{Theoretical Basis}
\label{chap:theoretical-basis}

\section{Introduction to General Relativity and the ADM Formalism}
\label{sec:gen-rel}

The General Theory of Relativity is a geometric description of gravity.  
A central idea of this theory, the equivalence principle, suggests that 
motion commonly attributed to a gravitational force field is to be interpreted as free-fall motion
in a curved spacetime.  Since the motion is due to the spacetime's curvature, all objects in the 
spacetime  are affected equivalently.  This spacetime curvature may be thought of as the geometry's 
deviation from \emph{flat} Minkowski spacetime.  In the language of differential geometry, 
spacetime can be described as a $4$-dimensional, real, differentiable manifold---$\mathcal{M}$ on which a 
metric, $g_{a b}$, with Lorentzian signature is defined.  As its name suggests, $g_{a b}$ allows 
one to measure spacetime separations in a coordinate-independent manner.  It is 
the fundamental tensor field that describes gravity since all measurable properties of 
the spacetime can be derived from it.
Because of the metric's Lorentzian signature, $g_{a b}$---defined at a non-singular 
event---asymptotes to the flat spacetime metric as the spacetime interval about this event tends to zero.  
Hence, in the flat space limit, all equations reduce to those of  special relativity.  

The final key feature of relativity is that matter and energy in the spacetime make it curved, while 
the spacetime's curvature dictates how the matter and energy propagate.  This intuitively explains 
the nonlinear interplay between geometry and energy in Einstein's equation:
\beq{
G_{a b} = 8 \pi T_{a b} \quad . \label{einsteins-equations} 
}
Heuristically, the left-hand side is a measure of spacetime's local curvature, while the 
right-hand side contains the stress-energy tensor---$T_{a b}$---that characterizes the matter-energy 
content of the spacetime. ( For an example of a stress-energy tensor, please refer to that of  
a perfect fluid in Equation~(\ref{stress}).)  In any given coordinate system, this tensor equation 
represents a set of second-order, coupled partial differential equations for the metric components 
$g_{a b}$, and---generically---require numerical solution due to their complexity.  

To aid in the numerical solution of Einstein's equation, one often enlists the help of the so-called 
``3+1'', or ADM (Arnowitt-Deser-Misner), formalism  that decomposes the 
$4$-dimensional manifold structure of spacetime into a space-plus-time framework \cite{adm}.  
It is a constrained
Hamiltonian formalism which arranges Einstein's equations into a Cauchy, or initial-value, problem.  
Our explanation of the formalism is based primarily on York's reformulation \cite{york-1979}, 
a concise summary of which was written by Choptuik \cite{choptuik-adm-1998}.  

Because $g_{a b}$ is Lorentzian, we may foliate spacetime in a series of space-like
hypersurfaces---$\Sigma_t$---that are level surfaces of a scalar field, $t$.   
Note that the progress of time is \emph{relative} in general relativity---i.e. there is no 
global definition of time---then $t$ here is only to be interpreted here as a parameter.  
Orthogonal to the hypersurfaces lie local time-like, dual vectors, $n_a$, defined by 
\beq{
n_a = - \alpha \nabla_a t \label{adm-time-norm}
}
where the lapse function $\alpha$ is such that it normalizes $n_a$: $n_a n^a = -1$, and $\nabla_a$ 
is the covariant derivative operator that is associated with our metric, $\nabla_a \, g_{b c} = 0$. 
Since $n^a$ are
orthogonal to the foliations, or slices, they naturally allow for the creation of \emph{projection operators}, 
${\gamma^a}_b$,  that project $4$-dimensional spacetime tensors onto the space-like hypersurface:
\beq{
{\gamma^a}_b = {\delta^a}_b + n^a n_b  \quad ,  \label{adm-projector}
}
where ${\delta^a}_b$ is the $\delta$-function.  If we apply the projection operator to the spacetime 
metric, which is equivalent to lowering the contravariant index of the projection operator, 
we obtain the \emph{spatial metric} 
\beq{
\gamma_{a b} = g_{a b} + n_a n_b \quad . \label{spatial-metric}
}
induced on the hypersurfaces.  The projection of a tensor onto the hypersurface is called a \emph{spatial}
tensor.  Let $\bot$ represent the operator that projects an arbitrary $4$-dimensional tensor, 
${T^{a_1 \ldots a_n}}_{b_1 \ldots b_n}$, onto the hypersurface.  Finding the spatial version of this tensor
entails applying the projection operator on every index:
\beq{
\bot \ {T^{a_1 \ldots a_n}}_{b_1 \ldots b_n} = {T^{c_1 \ldots c_n}}_{d_1 \ldots d_n}
\prod_{i=1}^n \ \prod_{j=1}^n {\gamma^{a_i}}_{c_i} \ {\gamma^{d_j}}_{b_j} 
  \label{general-projection}
}
Indices of spatial tensors can be raised and lowered with the spatial metric, e.g. if 
$s_a$ is a spatial one-form then $g^{a b} s_b = \gamma^{a b} s_b$.

While $\gamma_{a b}$ contains the complete geometric information that an observer can gather
from measurements constrained to $\Sigma_t$, any particular $3$-dimensional slice could be embedded into a
$4$-dimensional spacetime in an infinite number of ways.  The manner in which a slice is embedded 
can be described by the \emph{extrinsic curvature}, $K_{a b}$, which describes 
how the spatial projection of the gradient of the surface normal, $n^a$, varies over the slice:
\beq{
K_{a b} \equiv - \bot \nabla_a n_b  \quad . \label{extrinsic-curvature}
}
By using properties of $n^a$ and Lie derivatives, it can be shown that this definition is equivalent to 
\beq{
K_{a b} = - \frac{1}{2} \lie_n \gamma_{a b} \quad . \label{extrinsic-momentum}
}
where $\lie_n$ is the Lie derivative with respect to the vector $n^a$.  
This new definition suggests how $K_{a b}$ can be thought of as 
the ``conjugate momentum'' or ``velocity'' to the ``generalized coordinates'' $\gamma_{a b}$  
in this Hamiltonian formulation. 

In order to demonstrate how Einstein's equations are expressed in this formulation, let us first define
the Einstein tensor:
\beq{
G_{a b} \equiv R_{a b} - \frac{1}{2} R g_{a b}  \label{einstein-tensor}
}
where $R_{a c}={R_{a b c}}^b$ is the Ricci tensor defined from the Riemann tensor, ${R_{a b c}}^d$, and 
$R \equiv {R^a}_a$ is the Ricci scalar.  
The Riemann tensor is related to the failure of a vector, or equivalently a one-form, $p_c$, to remain 
unchanged after parallel transport around a small closed curve:
\beq{
\nabla_a \nabla_b \, p_c - \nabla_b \nabla_a \, p_c =  {R_{a b c}}^d \, p_d  \quad.  \label{riemann}
}
It can be expressed in these coordinates from the connection:
\beq{
{R_{a b c}}^d = \partial_b {\Gamma^d}_{a c} - \partial_a {\Gamma^d}_{b c} + 
{\Gamma^{e}}_{a c} {\Gamma^d}_{e b}  - {\Gamma^e}_{b c} {\Gamma^d}_{e a}  \quad ,
\label{riemann-components}
}
where the Christoffel symbols, ${\Gamma^c}_{a b}$,   are calculated from the metric
\beq{
{\Gamma^a}_{b c} = \frac{1}{2} g^{a d} \left( \partial_b g_{c d} + \partial_c g_{b d} 
- \partial_d g_{b c} \right) \quad . \label{connection-defintion}
}
The deviation of the covariant derivative from the ordinary derivatives in a specific coordinate system 
can also be written in terms of the connection:
\beq{
\nabla_a p_b  = \partial_a p_b - {\Gamma^c}_{a b} p_c  \quad , \quad
\nabla_a p^b  = \partial_a p^b - {\Gamma^b}_{a c} p_c  \quad . \label{affine-1}
}

To describe the intrinsic curvature of the hypersurface, we need to define a spatial 
covariant derivative:
\beq{
D_a \equiv \bot \nabla_a  = {\gamma^b}_a \nabla_b  \quad , \label{spatial-derivative}
}
which leads to a natural way of calculating the spatial Riemann tensor associated with the $\gamma_{a b}$:
\beq{
D_a D_b \, p_c - D_b D_a \, p_c  =  {}^{(3)}{R_{a b c}}^d \, p_d  \quad .
\label{spatial-riemann}
}
In deriving the form of Einstein's equations on these hypersurfaces, it is essential to 
know the spatial projection of the $4$-dimensional Riemann tensor on them.  We will not derive the 
resultant equations, but give the reader a sense of how they would be derived.  First, the Gauss-Codazzi
equations express the spatial projection of the Riemann tensor in terms of both the intrinsic and the extrinsic 
curvature (see \cite{choptuik-adm-1998} for a lucid derivation of these equations):
\beq{
\bot R_{a b c d} = {}^{(3)}{R_{a b c d}} + K_{a c} K_{b d} - K_{a d} K_{b c} 
\quad , \quad 
\bot \left(R_{a b c d} n^a \right) = D_{d} K_{c b} - D_{c} K_{d b} \quad . \label{gauss-codazzi}
}
We also need a description of the matter content defined with respect to an observer moving orthogonal 
to the slices. This is easily found by projecting different components of the stress-energy tensor 
onto the hypersurface, yielding the energy density, momentum density and 
spatial stress tensor---respectively---that such observers would measure:
\beq{
\varrho \ = \ n^a n^b \, T_{ a b } \label{adm-density}
}
\beq{
j^i \ = \ \bot \, n_a T^{ a b } \label{adm-momentum}
}
\beq{
S_{a b} \ = \ \bot T^{ a b } \ = \ {\gamma^c}_a \, {\gamma^d}_b \, T_{c d}  \label{adm-stress}
}
Using (\ref{gauss-codazzi}-\ref{adm-stress}), it can be shown that the contraction of Einstein's equations
along the direction of $n^a$, 
\beq{
G_{a b} \, n^a n^b = 8 \pi T_{a b} \, n^a n^b \label{pre-ham-const}
}
can be expressed in the following form, called the \emph{Hamiltonian constraint}:
\beq{
{}^{(3)}R + K^2 - {K^a}_b {K^b}_a \ = \ 16 \pi \varrho  \quad . \label{hamiltonian-constraint}
}
Here, ${}^{(3)}R$ is the spatial Ricci scalar derived from the spatial Riemann tensor, 
${}^{(3)}{R_{a b c}}^d$, and $K={K^a}_a$ is the trace of the extrinsic curvature.
Similarly, if only one index is contracted with $n_a$ while the other is projected onto the hypersurface, 
\beq{
\bot \,  G^{a b} \, n_a = 8 \pi \bot \, T^{a b} \, n_a \label{pre-mom-const}
}
the \emph{momentum constraint} is obtained:
\beq{
D_b \, K^{a b} - D^a K \ = \ 8 \pi j^a   \quad . \label{momentum-constraint}
}
The two equations (\ref{hamiltonian-constraint}),(\ref{momentum-constraint}) only involve spatial
quantities, and in particular, do not contain any terms involving second time derivatives of the metric.
Hence, they can be thought of as constraint equations that must be satisfied on every 
slice, including the initial slice at $t=0$.

\begin{figure}[htb]
\centerline{\includegraphics*[scale=0.85]{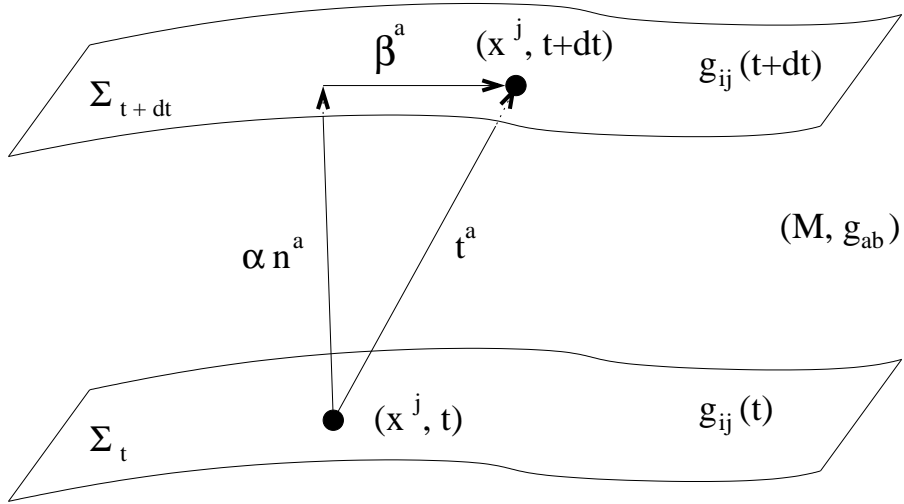}}
\caption[Foliation of spacetime, $(\mathcal{M},g_{a b})$,  into space-like hypersurfaces.]{Foliation of 
spacetime, $(\mathcal{M},g_{a b})$,  into space-like hypersurfaces.  Only two hypersurfaces,
$\Sigma_t$ and $\Sigma_{t+dt}$, are shown here.  The time-direction, $t^a$, 
can be decomposed into a part orthogonal to the slice, $\alpha n^a$, and a part tangent 
to it, $\beta^a$. 
\label{fig:adm}}
\end{figure}

Once the initial data is known, evolution equations are required to describe how the spatial metric and 
curvature vary slice to slice.  It is useful to consider time differentiation---specifically Lie differentiation
with respect to a vector field $t^a = \left(\pderiv{}{t}\right)^a$---using a $t^a$ which is more general than $n^a$.  
In particular, we take
\beq{
t^a = \alpha n^a + \beta^a  \quad ,  \label{time-vector}
}
where $\alpha$ is the lapse function defined previously, and $\beta^a$ is a spatial vector known as the 
shift vector.  
The vector field $t^a$ can be thought of as the tangent vectors to the world lines of coordinate-stationary 
observers.  If we choose the coordinate basis $\{x^\mu\} = \{t,x^j\}$ 
(where $\mu \in \{0,1,2,3\}$ and $j \in \{1,2,3\}$, see Section~\ref{sec:notat-conv-units} for a reminder of 
what values different indices can represent), then
the metric of the ADM formulation can be written as:
\beq{
ds^2 = \left( -\alpha^2 + \beta^j \beta_j \right) dt^2 + 2 \beta_j dx^j dt 
+ g_{i j} dx^i dx^j 
\label{adm-metric}
}
where we have used $g_{i j} = \gamma_{i j}$ to represent the spatial part of the metric.  
All of the quantities in (\ref{adm-metric}) are illustrated in Figure~\ref{fig:adm}.  
The decomposition of $t^a$ into parts 
tangent and orthogonal to the hypersurface is clearly seen.  Note that the coordinates remain 
the same along $t^a$, not along $n^a$. 

In this coordinate basis, the normal vector and its dual have components given by
\beq{
n^\mu = \left[\frac{1}{\alpha} , \, -\frac{\beta^j}{\alpha} \right]^T \quad , \quad 
n_\mu = \left[-\alpha, 0, 0, 0 \right]  \label{adm-norm}
}

Using $t^a$ as the ``time-direction'', the equations of motion for the spatial metric follow from the 
definition of the extrinsic curvature (\ref{extrinsic-momentum}):
\beq{
\lie_t \gamma_{a b} \ = \ \lie_N \gamma_{a b} + \lie_\beta \gamma_{a b} 
\ = \ \alpha \lie_n \gamma_{a b} + \lie_\beta \gamma_{a b} 
\ = \ -2 \alpha K_{a b} + \lie_\beta \gamma_{a b} \label{metric-evolution}
}
where $\lie_N$ is the Lie derivative along vector $N^a = \alpha n^a$.  The equations of motion 
for the spatial metric's conjugate momentum, $K_{a b}$, are found from 
the spatial projection of Einstein's equation
\beq{
\bot \,  G^{a b} = 8 \pi \bot \, T^{a b} \quad . \label{pre-ext-curv-evol}
}
After massaging this equation a great deal, the final form of the evolution equations for 
${K^a}_b$ can be found to be:
\beqa{
\lie_t {K^a}_b \ = &&\lie_\beta {K^a}_b \, - \, D^a D_b \, \alpha  \nonumber \\
&& + \alpha \left\{ {}^{(3)}{R^a}_b + K {K^a}_b +
8 \pi \left[ \frac{1}{2} \, {\gamma^a}_b \left({S^c}_c - \varrho \right) - {S^a}_b \right] \right\} 
\quad . 
\label{extrinsic-curvature-evolution}
}

Since the $4$-dimensional metric, $g_{a b}$, is symmetric, one might naively expect the Hamiltonian 
description to represent a system of 10 degrees of freedom.  However, the choice of the kinematic variables, 
$\alpha$ and $\beta^j$, are coordinate, or ``gauge'', conditions that 
can be made arbitrarily.  Since $\alpha$ determines how the hypersurfaces are embedded in the 
$4$-dimensional manifold, the equation that specifies it over each hypersurface is called the 
\emph{slicing condition}.  The equations that determine $\beta^j$ describe how the spatial
coordinates vary with respect to $n^a$ as well as how they vary from slice to slice.   Using these 
$4$ coordinate conditions with the 
$4$ constraint conditions (\ref{hamiltonian-constraint},\ref{extrinsic-curvature-evolution}) 
leaves 2 degrees of freedom left for the evolution equations 
(\ref{metric-evolution},\ref{extrinsic-curvature-evolution}).  
Or, rather, since each continuous gauge symmetry eliminates $2$ degrees of freedom, then 
the $4$ gauge symmetries  in general relativistic gravity eliminate $8$ of the $10$ degrees of freedom.  
These $2$ remaining degrees of freedom are
the dynamical degrees of freedom inherent to gravity and, at least in certain limits, describe 
the gravitational wave content of the theory.  Such waves can be thought of disturbances in 
the metric that travel at the speed of light, transversally-deforming the spacetime through which 
it travels.  
After describing how these $2$ degrees of freedom evolve, there are still $4$ out of the $6$ 
sets of evolution equations, $\{\dot{\gamma}_{i j}, \dot{K}_{i j}\}$, left.  This redundancy, however,
allows the numericist to choose how to go about solving them.  One need not use all the constraint 
equations, but can instead---at least naively---use all the evolution equations and only use the 
constraints to determine the initial data.  

Following this general overview we now proceed to a discussion of the specific set of Einstein equations 
that is used in this work. 

\subsection{Polar-Areal Coordinates}
\label{sec:geometry}

Hereafter, we restrict attention to spherically-symmetric spacetimes and adopt topologically spherical-polar
coordinates $\left(t, r, \theta, \phi \right)$.  The most general, time-dependent spherically-symmetric 
metric can be written as the following in this coordinate basis \cite{choptuik-phd}:
\beq{ 
ds^2 = \left(- \alpha + a^2 \beta^2 \right) dt^2 + 2 a^2 \beta \, dt \, dr + a^2 dr^2 
+ r^2 b^2 d\Omega^2  \quad , \label{ss-metric}
}
where $d\Omega^2 \equiv d\theta + \sin^2 \theta \, d\phi^2$ is the metric on the unit $2$-sphere.  
Here, $\alpha$, $\beta$, $a$, and $b$ are functions of $r$ and $t$, and $\beta$ is the single non-trivial 
component of the shift vector $\beta^i = \left[ \beta , 0 , 0 \right]$.
Since the coordinate conditions and the constraints are enough to specify all the metric
functions in spherical symmetry, gravity is no longer dynamical in that case.  This means that we will not be able 
to produce gravitational radiation in our simulations.  

Instead of the most general metric (\ref{metric}), we use the polar-areal form named after the gauge conditions 
used: the areal condition and the polar slicing condition.  The areal condition sets 
$r$ to be the areal coordinate so 
that $2\pi r$ is the proper area of a sphere; this requires $b(r,t)=1$.  The polar slicing 
condition---$K = {K^i}_i = {K^r}_r$---requires $K_{\theta \theta} = K_{\phi \phi} = 0$ for 
all $t$.  The consequence of these two conditions 
is $\beta=0$, inferred from the evolution equation for $g_{\theta \theta}$.  This leads to a far simpler
metric,
\beq{ 
ds^2 = - \alpha\left(r,t\right)^2 dt^2 + a\left(r,t\right)^2 dr^2 
+ r^2 d\Omega^2  \label{metric}
}
that now only depends on $2$ metric functions, $\alpha$ and $a$.  

For completeness, we tabulate the non-vanishing Christoffel symbols associated  with (\ref{metric}):
\beq{\begin{array}{lll} 
{\Gamma^t}_{r r} = a \dot{a} / \alpha^2 , 
  &{\Gamma^t}_{t t} = \dot{\alpha} / \alpha , 
  &{\Gamma^t}_{r t} = \alpha '/ \alpha \\[0.2cm]
{\Gamma^\theta}_{\theta r} = {\Gamma^\phi}_{\phi r} = 1/r , 
  &{\Gamma^\theta}_{\phi \phi} = - \sin\theta \cos\theta , 
  &{\Gamma^\phi}_{\phi \theta} = \cot \theta      \\[0.2cm]
{\Gamma^r}_{\phi \phi} = - r \sin^2\theta / a^2 , 
  &{\Gamma^r}_{\theta \theta} = - r / a^2 , 
  &{\Gamma^r}_{t t} = \alpha \alpha '/ a^2  \\[0.2cm]
{\Gamma^r}_{r r} = a'/ a   , 
  &{\Gamma^r}_{r t} = \dot{a} / a  & 
\end{array}   \label{connection}
}
These can be used to calculate the non-zero components of the extrinsic curvature (\ref{extrinsic-curvature}):
\beq{ 
K_{r r} = - \frac{a \dot{a}}{\alpha} \quad
\Rightarrow \quad K \equiv {K^i}_i = {K^r}_r = - \frac{\dot{a}}{a \alpha} 
\label{K}
}
The non-zero spatial Ricci tensor components and spatial Ricci scalar are
\beq{ 
{}^{(3)}{R^r}_r = \frac{2 a'}{r a^3} \qquad
{}^{(3)}{R^\theta}_\theta = {}^{(3)}{R^\phi}_\phi = 
\frac{1}{r^2 a^3} \left( r a' + a^3 - a \right) \label{spatial-riccitensor}
}
\beq{ 
{}^{(3)}{R} \equiv {}^{(3)}{R^i}_i = \frac{2}{r^2 a^3} 
\left( 2 r a' + a^3 - a \right) \label{spatial-ricciscalar}
}
For completeness, the $4$-dimensional Ricci scalar for our metric (\ref{metric}) is 
\beqa{
R \ = \ && \frac{ 2 }{ \alpha^3 a^3 r^2 } \left[ 
r^2 \left( \ddot{a} \alpha a^2  - \alpha^{\prime \prime} \alpha^2 a 
           - \dot{\alpha} \dot{a} a^2 + \alpha^{\prime} a^{\prime} \alpha^2 \right) \right. \nonumber\\
	 &&  + \left. 2 r \alpha^2 \left( a^\prime \alpha - \alpha^\prime a  \right) 
	   + \alpha^3 a \left( a^2 - 1 \right) \right]  \quad . 
\label{full-ricciscalar}
}
Or, using Einstein's equation, we may write the Ricci scalar in terms of the 
fluid variables (see (\ref{stress}) for the stress-energy tensor of a perfect fluid):
\beq{
R \ = \ - 8 \pi T \ = \ 8 \pi \left( \rho - 3 P \right)   \quad ,
\label{ricciscalar-stressscalar}
}
where $P$ is the pressure and $\rho$ is the total energy density of the perfect fluid. 
By comparing $g_{r r}$  components of the polar-areal metric to the Schwarzschild metric 
in our coordinates, 
\beq{
ds^2 = -\left( 1 - \frac{2M}{r} \right) dt^2  + \left( 1 - \frac{2M}{r} \right)^{-1} dr^2 + r^2 d\Omega^2
\label{schwarzschild-metric}
}
we can define the \emph{mass aspect function}, 
\beq{  
m(r,t) \equiv \frac{r}{2} \left( 1 - 1 / a^2 \right) \quad . \label{massaspect}
}
In the polar-areal metric, the Hamiltonian constraint reduces to a first-order differential equation
for $a$:
\beq{
\frac{a^\prime}{a} = a^2 \left[ 4 \pi r \varrho - \frac{m}{r^2} \right]  \quad .
\label{polar-areal-hamiltonian-const}
}
Here and subsequently, the primes indicate differentiating with respect to $r$, and a dot indicates
differentiation with respect to $t$. 
The evolution equation for $a$ is found from the definition of the extrinsic curvature 
(\ref{extrinsic-curvature}) 
and the fact that $K_{r r}$ is algebraically constrained by the momentum constraint, yielding
\beq{
\dot{a} = - 4 \pi r \alpha a j_i  \quad . \label{polar-areal-momentum-const}
}
Finally, the slicing equation for $\alpha$ is derived from the evolution equation of 
$K_{\theta \theta}$, or equivalently from that of $K_{\phi \phi}$.  In particular, from 
$K_{\theta \theta}(r,t) = \dot{K}_{\theta \theta}(r,t)=0$, we derive the following homogeneous, linear
differential equation for $\alpha$:
\beq{
\frac{\alpha^\prime}{\alpha} = \frac{a^\prime}{a}  + \frac{1}{r} \left(a^2 - 1 \right) 
- \frac{8 \pi a^2}{r} \left[ T_{\theta \theta} - \frac{r^2}{2} \left( {T^i}_i - \varrho \right) \right] 
\quad \label{polar-areal-slicing-condition}
}

\section{Critical Phenomena in General Relativity}
\label{sec:crit-phen-gener}

Published work in  general relativistic critical phenomena  began just over a decade ago with the 
seminal paper by Choptuik \cite{choptuik-1993}.  The work numerically investigated the dynamics of
the spherically-symmetric Einstein-massless-Klein-Gordon (EMKG) field, which is a model for 
a scalar field---$\phi(r,t)$---coupled to gravity.  To specify initial conditions, Choptuik needed 
only to provide the form of $\phi(r,0)$,  which was set to a distribution dependent only on a 
single parameter, $p$, as well as $\dot{\phi}(r,0)$ which was generally chosen so that the scalar field was initially
in-going.  For example, one of the distributions he used was a Gaussian:
\beq{
\phi(r,0) = p \, r^3 \, e^{-\left[\left(r-r_0\right)/\delta\right]^2}
\label{choptuik-data}
}
He found that the ensuing dynamics would result in a black hole for large $p$---$p=p_\mathrm{high}$---as 
the scalar field collapsed to the origin, but for small $p$---$p=p_\mathrm{low}$---the scalar field would completely
disperse to infinity.  
By repeatedly bisecting between these limits, he was able to tune towards a solution that lay right 
at the threshold of black hole formation.  On the black hole-forming side of this threshold, he 
found that the mass of the black holes decreased as $p$ approached the threshold value. 
Specifically, the black hole mass dependence on $p$ was found to follow the scaling law
\beq{
M_\mathrm{BH} \propto \left|p - p^\star\right|^\gamma  \quad .
\label{mass-scaling}
}
remarkably well.  

Further, Choptuik found---for solutions of $p\simeq p^\star$---that the spacetime and matter 
distributions followed a discretely self-similar symmetry (DSS) as the matter distribution accumulated 
toward the origin.  
A snapshot of the solution at a given time resembled itself---on a smaller spatial scale---after a certain, 
ever-decreasing period of time. 
If $Z(r,\tilde{T}_0)$ represents any field that exhibited 
DSS, then Choptuik found that---as $p\rightarrow p^\star$---the field would asymptote to a solution that 
was precisely DSS:
\beq{
Z( r,\tilde{T}_0 ) = 
Z( \, e^{\pm n \Delta} r , \, e^{\pm n \Delta } \tilde{T}_0 \, )  
\quad , \quad n \in \mathbb{Z}^+
\label{dss-rt}
}
Here, we have adopted a new time coordinate, $\tilde{T}_0$, 
\beq{
\tilde{T}_0 \equiv T^\star_0 - T_0 \quad , 
\label{T-rescaled}
}
where $T_0$ is the elapsed central proper time:
\beq{
T_0(t) \equiv \int_0^t \alpha(0,t^\prime) \, dt^\prime \quad . 
\label{central-propertime}
}
and $T_0^\star$ is the central proper time at which the self-similar solution ``accumulates'' at the origin.   
This suggested that the \emph{critical solution}---the solution obtained by setting $p=p^\star$ 
exactly---was this precisely DSS solution.  

Choptuik also found that the critical solutions were \emph{universal} by using different 
$1$-parameter families of initial data.  Indeed, the critical solutions and the scaling 
exponents---$\gamma$---obtained from tuning the distinct families of initial data all matched.  

This first study in critical phenomena touched upon the three fundamental aspects of 
the critical behavior: 1) universality and 2) scale invariance of the critical solution with 3) power-law 
behavior in its vicinity.   All three have also been seen in a multitude of matter models, such as 
perfect fluids, $\mathrm{SU}(2)$ Yang-Mills model, and collision-less matter to name a few.  
A tabulation of all the matter models in which critical phenomena has been found is given in 
\cite{gundlach,gundlach-rev2}, which reviews the field in general as well.  Another excellent introduction
to general relativistic critical phenomena is given by Choptuik \cite{choptuik-1998}. 

Through all these investigations, different types of critical behavior have been  illuminated: Type~I 
and Type~II behavior.  
Type~II behavior entails critical solutions that are either continuously self-similar (CSS)
or DSS.  Super-critical solutions---those that form black holes---give rise to black holes with 
masses that scale as a power-law (\ref{mass-scaling}), implying that arbitrarily small black holes can be 
formed.  Since $M_\mathrm{BH}(p)$ is continuous across
$p=p^\star$, this type of critical behavior was named ``Type~II'' since it parallels 
Type~II (continuous) phase transitions of statistical mechanics.  

As in the statistical mechanical case, 
there is a Type~I behavior, where the black hole mass ``turns on'' at a finite value.  Also, 
Type~I critical solutions are quite different from their Type~II counterparts, tending to be 
meta-stable star-like solutions that are static or periodic.  Hence, the critical solutions 
are described by a continuous or discrete symmetry in time, analogous to Type~II's 
CSS and DSS solutions.  Unlike the Type~II behavior, however, the black hole masses of super-critical 
solutions do not follow a power-law scaling.  Instead, the span of time---as measured by an 
observer at the origin---that a given solution resembles the critical solution scales with the 
solution's deviation in parameter space from the critical one:
\beq{
\Delta T_0(p) \propto - \sigma \ln \left| p - p^\star \right|  \quad . \label{type-i-scaling-0}
}
The longer a solution emulates the critical solution, the closer it has been tuned. 

\subsection{Type~II Scaling Behavior}
\label{sec:type-ii-scaling}

The accepted model that describes the scaling behavior near the critical solution 
was suggested by Evans and Coleman \cite{evans-coleman}.  They found that the critical solution 
of a radiation fluid---$P=\rho/3$---obtained dynamically was the same as a precisely CSS solution
of the fields equations.  By assuming the solution is CSS, the field equations reduce to 
a set of ODE's, which is further an eigenvalue problem that can be solved with standard shooting methods. 
They also suggested that the scaling behavior 
could be explained through dimensional arguments by examining linear perturbations about the 
CSS critical solution.  This was finally done by Koike et al. \cite{koike-etal-1995} for the 
radiation fluid, who showed that the scaling exponent, $\gamma$, was the inverse of the Lyapunov exponent of
the critical solution's single, unstable eigenmode.  Later, it was found that the scaling exponent, $\gamma$, 
was not a universal constant of general relativity, but was dependent on the critical solution's matter model. 
The first evidence for this non-universality in scaling behavior was given in concurrent works by 
Maison \cite{maison-1996} and Hara et al. \cite{hara-etal-1996}, who first found that 
$\gamma$ was dependent on the adiabatic index, $\Gamma$, in an ``ultra-relativistic'' fluid's equation
of state (\ref{ultra-eos}) by similar means as \cite{koike-etal-1995,koike-etal-1999}. 

Below, we will review the heuristic explanation for scaling in critical solutions, taking 
Type~II CSS solutions as our specific case.  We follow the description given in \cite{gundlach-rev2}. 
We first adopt coordinates tailored to the CSS symmetry
\beq{
\mathcal{X} = \ln \left( \frac{r}{\tilde{T}_0} \right)
\label{X1-ss}
}
\beq{
\mathcal{T} \equiv \ln \left(\tilde{T}_0 \right) \quad . 
\label{T-ss}
}
In particular, general relativistic continuous self-similarity corresponds to a symmetry with respect 
to a homothetic Killing vector field, $\xi$, \cite{cahill-taub}:
\beq{
\lie_\xi g_{a b} = 2 g_{a b}  \quad . \label{hometheticity}
}
In the CSS coordinates, $\xi = \partial / \partial \mathcal{T}$.  
The CSS nature of the critical solution, $Z^\star$, is then independent of the time coordinate
in this system:
\beq{
Z^\star = Z^\star(\mathcal{X}) \quad . \label{css-critical-solution}
}

Let $Z(\mathcal{X}, \mathcal{T})$, represent a solution that is near the critical solution. 
The solution $Z(\mathcal{X}, \mathcal{T})$ only resembles the critical solution in the so-called
``intermediate attractor regime'' where the solution has evolved past initial transients, but 
before the solution begins to disperse or collapse to a black hole.  
In this regime, we assume that the deviation of $Z(\mathcal{X}, \mathcal{T})$ from $Z^\star(\mathcal{X})$ can be 
expanded in terms of discrete modes:
\beq{
\delta Z(\mathcal{X}, \mathcal{T}) \equiv Z(\mathcal{X}, \mathcal{T}) - Z^\star(\mathcal{X}) 
\simeq  \sum_n C_n(p) \,  Z_n(\mathcal{X}) \, e^{ - \omega_n \mathcal{T}}  \quad , 
\label{critical-expansion}
}
where $\omega_n$ are the eigenvalues, and $Z_n(\mathcal{X})$ are the associated eigenmodes. 
Since the system is governed by a Cauchy problem, the solution's evolution is a function of the 
initial data.  Hence, the coefficients $C_n(p)$ in the expansion can be thought of as complicated 
functions of all the parameters that define the initial data even though we only highlight its 
dependence on the tuning parameter.  

The $\omega_n$ are, in general, complex.  The presence of sharp scaling behavior depends 
on the existence of only one unstable mode \cite{koike-etal-1995}, which we will assume is the first mode 
of this expansion.  Since we have defined $\mathcal{T}$ in such a way that it tends to $-\infty$ 
as $T_0\rightarrow T_0^\star$, then the growing mode has $\omega_0>0$, while all other modes are 
damped or oscillate in time: $\omega_{n\ne0} < 0$ or $\Re \omega_{n\ne0} = 0$.  
Neglecting the possibility of oscillating modes for the sake of simplicity, $\delta Z(\mathcal{X}, \mathcal{T})$
will then asymptote to only depend on this growing mode:
\beq{
\lim_{\mathcal{T}\rightarrow -\infty}  \delta Z(\mathcal{X}, \mathcal{T}) = 
C_0(p) \,  Z_0(\mathcal{X}) \, e^{ - \omega_0 \mathcal{T}}  \quad . 
\label{large-T-crit-deviation}
}
This illustrates how the one unstable mode is responsible for the ultimate departure of the solution 
from the intermediate linear regime.  Since $Z(\mathcal{X}, \mathcal{T}) = Z^\star(\mathcal{X})$ for 
$p=p^\star$, then $C_0(p^\star)=0$.  This suggests that we perform an expansion of 
(\ref{large-T-crit-deviation}) in terms of the length-scale set by the deviation in the 
parameter---$\left(p - p^\star\right)$---and keep only the linear term since we are assuming that 
$Z(\mathcal{X}, \mathcal{T}) \simeq Z^\star(\mathcal{X})$:
\beq{
\lim_{\mathcal{T}\rightarrow -\infty}  Z(\mathcal{X}, \mathcal{T}) \simeq
Z^\star(\mathcal{X}) +  \, \left( p - p^\star\right) \left. \deriv{C_0(p)}{p}\right|_{p=p^\star} 
Z_0(\mathcal{X}) \, e^{ - \omega_0 \mathcal{T}}   \quad . 
\label{linear-crit-deviation}
}
As $p$ is tuned closer to the critical value, we can see from this expression how 
the resulting solution's resemblance to the critical solution increases.  However, the growing 
mode ultimately drives the dynamics away from the critical solution.

Let $\mathcal{T}(p)$ be the \emph{departure time}---or the time at which 
$Z(\mathcal{X}, \mathcal{T})$ begins to leave the intermediate linear regime.  We do not 
wish to differentiate between the deviation of supercritical and subcritical solutions, but 
the sign of the deviation depends on whether $p$ is greater or less than $p^\star$.  Hence, we 
measure the solution's departure time---independent of the fact that its supercritical or 
subcritical---when its deviation reaches a specific value:
\beq{
\varepsilon   \equiv \left| p - p^\star\right| \left. \deriv{C_0(p)}{p}\right|_{p=p^\star} 
 \, e^{ - \omega_0 \mathcal{T}(p)} 
\label{departure-deviation}
}
Solving for $\mathcal{T}(p)$, we obtain:
\beq{
\mathcal{T}(p) \propto  \frac{1}{\omega_0} \ln \left|p - p^\star \right| 
\label{generic-lifetime-scaling}
}
This relationship represents the scaling behavior intrinsic to solutions near the critical one. 
If we substitute $\varepsilon$ into (\ref{linear-crit-deviation}), the near-critical solution takes
the form 
\beq{
Z(\mathcal{X}, \mathcal{T}(p)) \simeq Z^\star(\mathcal{X}) \pm \varepsilon Z_0(\mathcal{X}) 
\quad . 
\label{linear-crit-deviation-departure}
}
Here, the ``plus-minus'' represents the fact that $\left(p-p^\star\right)$ can take both signs, which
was ignored in (\ref{departure-deviation}).  Since $\varepsilon$ is chosen to represent the value 
at which the solution deviates from the linear regime---i.e. when the mode grows to approximately
the same magnitude as the critical solution---then $\varepsilon \sim O(1)$ as measured in the
$\mathcal{X}$ coordinates.  If $p$ is supercritical, then the growing term will form a black hole
whose size, $\mathcal{X}_\mathrm{BH}$ is comparable to the term's size as the solution leaves the
critical solution, implying that $\mathcal{X}_\mathrm{BH} \sim O(1)$ \cite{koike-etal-1995}.  
The black hole formation is also characterized by the ``time'' $\mathcal{T}_\mathrm{BH}=\mathcal{T}(p)$.  
These two scales of the black hole formation in the $(\mathcal{X}, \mathcal{T})$ coordinates 
determine the extent of the black hole in normal radial coordinates, using (\ref{X1-ss},\ref{T-ss}):
\beqa{
r_\mathrm{BH} & \equiv & r(\mathcal{X}_\mathrm{BH},\mathcal{T}_\mathrm{BH}) 
= \ln\left(\mathcal{X}_\mathrm{BH}\right) +  \ln\left(\mathcal{T}_\mathrm{BH}\right) 
= \ln\left(1\right) +  \ln\left(\mathcal{T}_\mathrm{BH}\right) \\
& \propto & \left| p - p^\star \right|^{1/\omega_0}   \label{mass-scaling-exponent}
}
Since $r_\mathrm{BH} \propto M_\mathrm{BH}$, we get the final black hole mass scaling 
relationship:
\beq{
M_\mathrm{BH} \propto \left| p - p^\star \right|^{1/\omega_0}  \label{mass-scaling-lyapunov}
}
Comparing this relation to the ``empirically'' determined one (\ref{mass-scaling}), we find
that the scaling exponent, $\gamma$, is just the inverse of the Lyapunov exponent of the one, unstable
mode:
\beq{
\gamma = \frac{1}{\omega_{Ly}}   \label{exponent-mode}
}
where $\omega_{Ly} \equiv \omega_0$.    

The subcritical counterpart to $r_\mathrm{BH}$---i.e. $r_\mathrm{DIS}$---can describe a scaling behavior
of those solutions near the critical one that do not form a black hole.  For instance, 
Garfinkle and Duncan \cite{garfinkle2} found that measuring the global maximum, $R_{\max}$, over $r$ and $t$ of 
the Ricci scalar yields the scaling behavior
\beq{
R_{\max} \propto \left|p - p^\star\right|^{-2\gamma} \label{ricci-scaling}
} 
since $R$ has units of $(\mathrm{Length})^{-2}$.  By contracting the Einstein equation, we can obtain 
the same scaling relation for the trace of the stress-energy tensor:
\beq{
T_{\max} \propto \left|p - p^\star\right|^{-2\gamma} \quad. 
\label{stress-scaling}
}
Since $T=3P - \rho$ is much easier to calculate than $R$ (\ref{full-ricciscalar}), we generally use 
(\ref{stress-scaling}) to calculate $\gamma$ for our perfect fluid computations.  

When analyzing numerical solutions that follow CSS behavior, it is helpful to transform into some sort
of coordinates adapted to the CSS symmetry so that we can readily see the solutions' departures from 
self-similarity.  The particular form of $\mathcal{X}$ that we will use is
\beq{
\mathcal{X} = \ln \left( \frac{r}{r_s} \right) \quad .
\label{X-ss}
}
The sonic point, $r_s$, is defined as the point at which $v=c_s$.  This choice of $\mathcal{X}$
is sufficient to track the self-similar behavior since---from past studies---we anticipate $r_s$
to represent a natural co-moving length scale of nearly critical solutions.  For fluids that follow the 
ideal gas equation of state (\ref{ideal-eos}), the determination of the sonic point is not very accurate. 
Instead of using $r_s$ to specify the solution's length scale, we sometimes use 
$r_{a_{\max}}$:
\beq{
\mathcal{X}_a = \ln \left( \frac{r}{r_{a_{\max}}} \right) \quad ,
\label{X-ss2}
}
where $r_{a_{\max}}$ is the position of the local maximum of $a(r)$ closest to $r=0$.

\subsection{Type~I Scaling Behavior}
\label{sec:type-i-scaling-explanation}

The  analysis performed in the previous section also sheds light on the scaling behavior in a 
solution's lifetime time, $\Delta T_0$, observed in Type~I behavior (\ref{type-i-scaling-0}).  
In this case, the critical solution is---let us say---static, so that it takes the form 
\beq{
Z^\star(r,t) = Z^\star(r) \label{type-i-crit-soln}
}
A solution that has been tuned near this critical solution enters an intermediate linear
regime just as in the Type~II case.  Hence, we can follow the same logical steps as in 
Section~\ref{sec:type-ii-scaling} except that we need to use $(r,T_0)$ coordinates
instead of $(\mathcal{X}, \mathcal{T})$.  Also, since $T_0$ is future-directed, then exponents in the 
perturbative expansion about the critical solution have the opposite sign than in (\ref{critical-expansion}):
\beq{
\delta Z(r, T_0) \equiv Z(r, T_0) - Z^\star(r) 
\simeq  \sum_n C_n(p) \,  Z_n(r) \, e^{ \omega_n T_0} 
\label{type-i-critical-expansion}
}
Assuming that the first mode is the only growing mode, then for late times and $p\simeq p^\star$ this
deviation can be expanded to first-order in $\left(p - p^\star\right)$:
\beq{
\lim_{T_0\rightarrow T_0^\star} \delta Z(r,T_0) \simeq  \left( p - p^\star \right) 
\left. \deriv{C_0(p)}{p}\right|_{p=p^\star} 
Z_0(r) \, e^{ \omega_0 T_0}  \quad.  \label{type-i-linear-expansion}
}
Using similar arguments, the lifetime time---$\Delta T_0$---is defined as the time when the mode grows 
to approximately the same order as the critical solution:
\beq{
\varepsilon = \left| p - p^\star \right|
\left. \deriv{C_0(p)}{p}\right|_{p=p^\star}  \, e^{ \omega_0 \Delta T_0}  \label{departure-time-def-type-i}
}
which finally gives 
\beq{
\Delta T_0 \propto  - \frac{1}{\omega_0} \ln \left| p - p^\star \right|  \quad , \label{type-i-scaling-lyapunov}
}
which suggests from (\ref{type-i-scaling-0}) that the Type~I scaling exponent is equal 
to the inverse of the Lyapunov exponent, $\omega_{Ly} = \omega_0$ of the one unstable mode associated with the 
critical solution.  

\section{Relativistic Perfect Fluids}
\label{sec:relat-perf-fluids}

As is the case for most material objects in nature, neutron stars consist of an assortment of hadrons, leptons, 
and photons.  
Since we are primarily interested in the star's interaction with gravity, we will neglect all but the 
heaviest particles and assume that we only have a large distribution of baryons of identical mass, $m_B$.  
Further, to reduce the number of degrees of freedom in this large assembly of particles, we use 
the hydrodynamic approximation and study bulk characteristics of the particles within volumes---called
fluid elements---whose lengths are large compared to the mean free path of their collisions.  
Thus, the particles in each fluid element are assumed to be in local thermodynamic equilibrium, 
and have velocities that are isotropic---randomly distributed in space---in the frame where
the average velocity vanishes.  The isotropic velocity distribution then implies that the pressure the
particles exert on the sides of the fluid element are also isotropic.  

In order to calculate the stress-energy tensor in a covariant form, let us first describe 
what the isotropic stress tensor should look like.  Assuming that the 
fluid element is small enough compared to the macroscopic curvatures of spacetime, 
the metric in the rest frame of the fluid element should be close to that of the Minkowski spacetime.  
In this frame, the $T_{0 0}$ component of the stress-energy tensor, consequently, represents the total 
energy density in the fluid element, while the average value of the particles 
momentum density is given by $T_{0i}$.  However, $T_{0i}=0$ since the average flow of the particles
vanishes.  Hence, the stress-energy tensor takes a diagonal form in the rest frame\cite{mtw}:
\beq{
T_{\mu \nu} = \left[ \begin{array}{cccc} \rho &&& \\ &P&& \\ &&P& \\ &&&P \end{array} \right]
\label{rest-frame-stress}
}
Here, $\rho$ and $P$ are---respectively---the total energy density and 
the pressure as measured in the local rest frame of the fluid element.  To get a covariant version of
the stress tensor, we note that the $4$-velocity of this frame is $u^\alpha=(1,0,0,0)$ and separate the 
``temporal'' and ``spatial'' parts of the tensor using the space-like projection operator that the 
$4$-velocity defines: ${\delta^a}_b + u^a u_b$.  Performing the separation, we then obtain
\beq{
T_{a b} = \rho u_a u_b + P \left( \eta_{a b} + u_a u_b \right) \label{flat-stress-tensor}
}
where $\eta_{a b}=\mathrm{diag}(-1,1,1,1)$ is the Minkowski metric.  A covariant form is finally obtained 
by taking $\eta_{a b}\rightarrow g_{a b}$, and rearranging terms so that the expression takes the more 
traditional form:
\begin{equation}
T_{a b} = \left( \rho + P \right) u_a u_b + P g_{a b} \label{stress}
\end{equation}
Isotropic fluids described by such stress tensors are often called perfect fluids since they are free 
of heat conduction and viscous effects. 
The presence of any of these would result in a stress-energy tensor with non-diagonal terms 
or different values along the spatial part of the diagonal \cite{mtw}.  

This description of the fluid has, so far, neglected the microscopic nature of the fluid.  As we mentioned
at the very beginning of the section, the particles are assumed to be baryons each of mass 
$m_B$.  The rest mass energy density of fluid, as measured in its local rest mass frame, is then
\begin{eqnarray}
\rho_{\circ} & = & m_B n \label{restmass}
\end{eqnarray}
where the $n$ is the number density  or number of baryons per fluid element.  The total energy density 
of the fluid also contains contributions from the particles' internal degrees of freedom, 
called the \emph{internal energy} of the fluid:
\beq{
\rho = \left( 1 + \epsilon \right) \rho_{\circ} \quad ,  \label{rho}\\
}
where $\epsilon$ is the internal energy per unit rest-mass, or \emph{specific} internal energy. 
The internal energy includes, for example, the particles' thermal energy, inter-particle energies, and intra-particle
(binding) energies. 
Further, the specific enthalpy of the fluid is defined as 
\beq{ 
h = 1 + \epsilon + \frac{P}{\rho_\circ} \quad . \label{h}
}
It is important to remember that the set of thermodynamic quantities, 
$\{\rho_\circ, \epsilon, P\}$ are all measured in the rest frame, or \emph{Lagrangian}, frame 
of the fluid element.  However, we wish to take a \emph{Eulerian} perspective and choose coordinates
not necessarily tied to the flow. Therefore, we will need the $4$-velocity of the fluid element, $u^a$, 
to describe how the fluid flows with respect to the Eulerian coordinates.  The $4$-velocity has the usual 
normalization
\beq{
u^a u_a = -1  \quad . \label{vel-norm}
}

To describe the fluid's dynamics, two conservation laws are used:
the \emph{local conservation of energy}
\beq{
\nabla_a {T^a}_b = 0  \quad , \label{energycons}
}
and the \emph{local conservation of baryon number}
\beq{
\nabla_a  J^a  = 0  \quad . \label{currentcons}
}
Here, $J^a$ is the conserved current of the flow, 
\beq{ 
J^a \equiv \rho_{\circ} u^a  \quad .   \label{current}
} 
An  important feature of perfect fluids is that they are naturally adiabatic along the direction 
of the fluid's $4$-velocity.  This can be proven using the above conservation laws and the 
First Law of Thermodynamics, which states that while the fluid is
in thermodynamic equilibrium, 
\beq{
d\epsilon \ = \ T \, ds + \frac{ P }{ \rho_\circ^2 } d\rho_\circ \quad ,
\label{first-thermo-law}
}
where $s$ is the specific entropy as measured in the fluid's rest frame.   Projecting $\nabla^a {T^a}_b$
along the fluid's $4$-velocity, we obtain:
\beqa{
0 & = & u^b \nabla^a {T^a}_b = u^b \left[ \nabla_a \left( \rho_\circ h u^a u_b \right) 
+ \nabla_a\left(P {\delta^a}_b \right) \right] \label{energy-cons1} \\
& = & -u^a \nabla_a \rho - \rho_\circ h \nabla u^a \label{energy-cons2}
}
where we have used the fact that $u^b u^a \nabla_a u_b = \frac{1}{2} u^a \nabla_a \left( u^b u_b \right) = 0$. 
Associated with the energy conservation equation of (\ref{energy-cons2}) is Euler's equation, 
which is obtained from taking the projection perpendicular to the flow 
$\left({\delta^a}_b + u^a u_b\right) \nabla_c {T^c}_a$.

The First Law of Thermodynamics (\ref{first-thermo-law}) implies that 
\beq{
u^a \nabla_a \epsilon = T u^a \nabla_a s + \frac{P}{\rho_\circ^2} u^a \nabla_a \rho_\circ 
\quad . \label{diff-first-thermo-law}
}
Using this form of the law and the identity that we get by expanding the continuity equation, 
we obtain
\beq{
\rho_\circ T u^a \nabla_a s  = 0 \label{adiabatic}
}
which means that entropy is conserved along flow lines, assuming that the fluid has non-vanishing 
rest-mass density and temperature.  Hence, from the definition of the 
perfect fluid stress-energy tensor (\ref{stress}), the first law of thermodynamics (\ref{first-thermo-law})
and the fluid's conservation equations (\ref{energycons}-\ref{currentcons}), we have proven the 
lower bound of the Second Law of Thermodynamics:
\beq{
u^a \nabla_a s \ge 0 \quad . \label{second-thermo-law}
}
The second law is satisfied throughout the fluid.  The ``greater-than'' part of the inequality---e.g.
an increase in entropy---happens when the fluid is not in thermal equilibrium and is not necessarily governed
by the first law. This enables the entropy to momentarily increase before the fluid finally settles to thermal 
equilibrium, 
which occurs---for example---when shocks arise.  Shocks always border fluid states with different 
entropies, hence 
the adiabatic condition is satisfied only outside regions with shocks.  
Specifically, shocks always increase entropy in the fluid into which they travel. 
Hence, a shock travels from high-entropy to low-entropy regions \cite{thorne}.
In fact, a distribution of perfect fluid will always remain isentropic---$\nabla_\mu s = 0$---if it is initially
and never produces a shock.  The increase in entropy due to shocks is associated with the 
transfer of energy 
of bulk motion into internal energy, or heat.  We will encounter this phenomena repeatedly in our simulations. 

Another useful quantity to calculate from the fluid's properties  and the laws of thermodynamics is the speed of 
sound, $c_s$.  The speed of sound is the speed of the characteristics of the wave equations one obtains 
from linearizing the equations of motion.  
After the linearization and a few simplifications, one obtains $c_s$ ~\cite{landau-lifshitz}
\beq{
c_s \ = \  \left[ \left( \frac{ \partial P }{ \partial \rho } 
\right)_s \right]^{1/2} \quad . \label{sound-speed-definition}
}
Since this form of $c_s$ cannot be readily calculated from the equations of state that we use, we must seek 
an alternative form. By employing the first law of thermodynamics with the 
the Maxwell relation (see \cite{huang} or most any other text on thermodynamics)
\beq{
dh = T ds + \frac{ dP }{ \rho_\circ}  \quad . \label{maxwell-relation2}
}
Then, we have
\beq{
d \left( \rho_\circ h \right) \ = \ h \, d\rho_\circ \, + \, \rho_\circ \, dh 
\ = \ h \, d\rho_\circ \, + \, \rho_\circ \, T \, ds + dP  
\label{max-temp1}
}
and, from the definition of $\rho$, we have
\beq{
d\left( \rho_\circ h \right) \ = \ d \left( \rho + P \right) \ = \ d\rho \, + \, dP 
\quad . \label{max-temp2}
}
Equating (\ref{max-temp1}) and (\ref{max-temp2}) and simplifying, we get
\beq{
d\rho \ = \ h \, d\rho_\circ \, + \, \rho_\circ \, T \, ds \quad . 
\label{maxwell-relation3}
}
\beq{
\Longrightarrow \quad 
\left( \frac{ \partial \rho_\circ }{ \partial \rho } \right)_s  \quad . 
\ = \ \frac{ 1 }{ h } 
\label{drho0-drho}
}
From the first law of thermodynamics (\ref{first-thermo-law}),  we get
\beq{
\left( \frac{ \partial \epsilon }{ \partial \rho_\circ } \right)_s
\ = \ \frac{ P }{ \rho_\circ^2 }  \quad . 
\label{depsilon-drho0}
}
Since  $\rho - ( 1 + \epsilon ) \rho_\circ = 0$, and by a
partial derivative identity \cite[pg. 20]{huang}, we find that 
\beq{
\left( \frac{ \partial \epsilon }{ \partial \rho } \right)_s
\ = \ \left( \frac{ \partial \epsilon }{ \partial \rho_\circ } \right)_s  \,
\left( \frac{ \partial \rho_\circ }{ \partial \rho } \right)_s
\ = \  \frac{ P }{ h \rho_\circ^2 }  
\quad . \label{depsilon-drho}
}
Thus, (\ref{sound-speed-definition}) can be put into a form we can immediately calculate by
using (\ref{drho0-drho}), (\ref{depsilon-drho}) and the fact that 
$P = P( \rho_\circ , \epsilon )$  :
\beqa{
\left( \frac{ \partial P }{ \partial \rho } \right)_s
 & = & \left( \frac{ \partial \rho_\circ }{ \partial \rho } \right)_s  
\left( \frac{ \partial P }{ \partial \rho_\circ } \right)_\epsilon  
 +  \ \left( \frac{ \partial \epsilon }{ \partial \rho } \right)_s  
\left( \frac{ \partial P }{ \partial \epsilon } \right)_{\rho_\circ} \nonumber \\[0.3cm]
& = & \frac{ 1 }{ h } 
\left( \frac{ \partial P }{ \partial \rho_\circ } \right)_\epsilon  
 +  \frac{ P }{ h \rho_\circ^2 } 
\left( \frac{ \partial P }{ \partial \epsilon } \right)_{\rho_\circ} \quad . 
\label{sound-speed2} 
}
Finally, we obtain the final form of the speed of sound:
\beq{
c_s^2 \ = \ \frac{ 1 }{ h } \left( \chi + \frac{ P }{ \rho_\circ^2 } 
\mathBig{\kappa} \right) 
\quad , \label{sound-speed}
}
where $\chi$ and $\kappa$ are defined as 
\beq{
\chi \equiv \left( \frac{ \partial P }{ \partial \rho_\circ} \right)_\epsilon
\quad , \quad 
\kappa \equiv \left( \frac{ \partial P }{ \partial \epsilon } \right)_{\rho_\circ}
\quad . \label{chi-kappa}
}
We will see later that $\chi$ and $\kappa$ are easily found from the closed-form state equations we use.

Next, we will discuss the hyperbolicity of fluid's equations of motion.  This topic will be important 
to the particular numerical methods we use to evolve the fluid in time.  
First, it can be shown that the equations of motion (\ref{energycons}-\ref{currentcons}) 
of the fluid take the form of a system of $N$ quasi-linear (see Courant and Hilbert
\cite{courant-hilbert} for discussions regarding quasi-linear PDE's) first-order partial 
differential equations:
\beq{
\mathbf{B}^\mu(\mathbf{w}) \nabla_\mu \mathbf{w} = \mathbf{c}(\mathbf{w})
\label{general-conservation-eq}
}
where $\mathbf{w}$ is the $N$-dimensional vector of primitive variables 
for the fluid, $\mathbf{c}(\mathbf{w})$ is a differentiable $N$-dimensional 
vector function and $\mathbf{B}^\mu$ are real $N \times N$ matrices.   The primitive variables 
typically include independent fluid variables of the fluid's rest frame (e.g. ($\{P,\rho_\circ\}$),
and the fluid's velocity---$v^j$---with respect to the space-like hypersurface.

A system of the type (\ref{general-conservation-eq}) is said 
to be in \emph{conservation form} \cite{anile} if there exist 
real vector functions $\greekbf{\mathcal{F}}^\mu$ such that $\mathbf{B}^\mu$ are the
Jacobian matrices of $\greekbf{\mathcal{F}}^\mu$, i.e. that 
\beq{
{{\mathrm{B}^\mu}^{(n)}}_{(m)} = 
\frac{\partial {\greekbf{\mathcal{F}}^\mu}^{(n)} }
{\partial w^{(m)}} 
\label{conservation-form-jacobian}
}
where ${{\mathrm{B}^\mu}^{(n)}}_{(m)}$ are the 
components of the matrix $\mathbf{B}^\mu(\mathbf{w})$.  

In order for $\mathbf{w}$ to be a solution to the Cauchy problem, 
the equations in (\ref{general-conservation-eq}) must maintain their 
hyperbolicity \cite{courant-hilbert} as defined in the following \cite{anile}:

Let $n^a$ be a differentiable time-like unit-norm vector that lies in an 
open subset $\mathcal{W}$ of our 4-dimensional manifold $\mathcal{M}$,  
$\mathcal{W} \subseteq \mathcal{M}$.  Equations 
(\ref{general-conservation-eq}) are said to be \emph{hyperbolic}
along $n^a$ (the \emph{time direction}) if they obey the following 
two conditions:
	\begin{enumerate}
	\item $\det{\left( \mathbf{B}^\mu n_\mu \right)} \ne 0$

	\item The eigenvalue problem 
	\beq{
	\mathbf{B}^\mu \left( \xi_\mu - \lambda n_\mu \right) 
		\greekbf{\eta} = 0
	\label{eigenvalue-problem}
	}
	has $\tilde{N}$ distinct real eigenvalues 
	$\{ \lambda_p \} \ \left( p = 1, \cdots, \tilde{N} \right)$  and $N$ 
	linearly independent $N$-dimensional eigenfunctions $\greekbf{\eta}$
	for any space-like vector $\xi^a$ in $\mathcal{W}$.  
	\end{enumerate}
The system is considered \emph{strictly hyperbolic} if
all the eigenvalues are distinct, i.e. $\tilde{N} = N$. 

Banyuls et al. \cite{banyuls}
have presented a formulation of the equations of motion for a general, $3$-dimensional fluid with the 
ADM metric (\ref{adm-metric}).  They were able to find a system of flux functions such that
\beq{
\partial_\mu \greekbf{\mathcal{F}}^\mu\left(\mathbf{w}\right) = 
\greekbf{\psi}\left(\mathbf{w},g_{a b}\right)  \quad , 
\label{flux-conservation-eq}
}
where some terms that include the metric functions and derivatives of the metric functions have been moved into
the source function $\greekbf{\psi}\left(\mathbf{w}, g_{a b}\right)$ and others have been
absorbed into the flux functions.  Also, no derivatives of the fluid variables $\mathbf{w}$ 
appear in $\greekbf{\psi}\left(\mathbf{w}, g_{a b}\right)$, which is required for the equations to 
remain hyperbolic.  
These flux functions are such that they form an eigensystem
\beq{
\left( \mathbf{B}^j - \lambda \mathbf{B}^0 \right) \greekbf{\eta} = 0 \quad .
\label{eigenvalue-problem-example}
}
Typically, the following identification is made
\beq{
\mathbf{q} = \greekbf{\mathcal{F}}^0 
\left( \mathbf{w}\left(\mathbf{q}\right) \right)
\label{q-eq-F0}
}
\beq{
\mathbf{f}^j(\mathbf{q}) = 
\greekbf{\mathcal{F}}^j\left( \mathbf{w}\left(\mathbf{q}\right) \right)
\quad , \label{fj-eq-Fj}
}
where $\mathbf{q}(\mathbf{w})$ is the $N$-dimensional vector of 
\emph{conservative variables} and $\mathbf{f}^j\left(\mathbf{q}\right)$ is 
an $N$-dimensional function of $\mathbf{w}$ alone.
Then, it can be clearly seen that 
the system (\ref{flux-conservation-eq})  becomes
\beq{
\partial_t \, \mathbf{q} + \partial_j \, \mathbf{f}^j(\mathbf{q}) = 
\greekbf{\psi}(\mathbf{q}) 
\quad . \label{simple-conservation-eq}
}
This last formulation is the one that will be used in our simulations. 

In order to numerically solve the system of equations using the particular methods we employ, it must
first be put into quasi-linear form:
\beq{
\partial_t \, \mathbf{q} + \mathbf{A}^j \, \partial_j \, \mathbf{q} = 
\greekbf{\psi}(\mathbf{q})  \quad , 
\label{quasi-linear-form}
}
where, 
\beq{
{{\mathbf{A}^j}^{\left(a\right)}}_{\left(b\right)} \equiv
\frac{\partial \, {\mathrm{f}^j}^{\left(a\right)} }
{ \partial \, q^{\left(b\right)}} 
\label{A-matrix-def}
}
Calculating $\mathbf{A}^j$ is difficult to do in general since $\mathbf{f}^j$ 
is usually expressed in terms of $\mathbf{q}$ and $\mathbf{w}$ (see
(\ref{ideal-piphi-state-vectors}) for an example) 
and $\mathbf{w} = \mathbf{w}(\mathbf{q})$ is not known in closed form, generally.

Using (\ref{conservation-form-jacobian}), (\ref{q-eq-F0}), (\ref{fj-eq-Fj}), 
and (\ref{A-matrix-def}), we can transform $\mathbf{A}^j$ into a more 
convenient form
\beq{
{{\mathrm{A}^j}^{\left(a\right)}}_{\left(b\right)} \equiv
\frac{\partial \, {\mathrm{f}^j}^{\left(a\right)} }
{ \partial \, \mathrm{q}^{\left(b\right)}} 
\ = \   
\frac{\partial \, {\mathrm{F}^j}^{\left(a\right)} }
{ \partial \, {\mathrm{F}^0}^{\left(b\right)}} 
\ = \ 
\frac{\partial \, {\mathrm{F}^j}^{\left(a\right)} }{\partial \, \mathrm{w}^{(c)}}  
\ 
\frac{\partial \, \mathrm{w}^{(c)}}{ \partial \, {\mathrm{F}^0}^{\left(b\right)}} 
\ = \ 
{{\mathrm{B}^j}^{(a)}}_{(c)} \ 
{\left[ \left( \mathbf{B}^0 \right)^{-1} \right]^{(c)}}_{(b)}
\label{B-to-A-derivation}
}
\beq{
\Longrightarrow \quad \mathbf{A}^j \ = \ 
\mathbf{B}^j \left(\mathbf{B}^0\right)^{-1}
\label{B-to-A}
}

Thus, in order to find $\mathbf{A}^j$, we need to know $\mathbf{B}^\mu$.  
It is somewhat interesting to note that the eigenvectors and eigenvalues 
for $\mathbf{A}^j$ are related to those for $\mathbf{B}^\mu$ \cite{font-etal1}, as we now discuss.
Let $\{ \greekbf{\eta}^j_m \}$ and $\{ \lambda^j_m \}$ $(m = 1, \cdots, N)$
be, respectively, the eigenvectors and eigenvalues for $\mathbf{A}^j$, 
and let $\{ \tilde{\greekbf{\eta}}^j_m \}$ and 
$\{ \tilde{\lambda}^j_m \}$ $(m = 1, \cdots, N)$
be, respectively, the eigenvectors and eigenvalues for the system 
(\ref{eigenvalue-problem-example}).  Note that the superscript $j$ is not a tensor index but
only specifies that the corresponding quantity is associated with the matrix $\mathbf{A}^j$. 
 By inspection, it is obvious that the 
eigenvalue problem for $\mathbf{A}^j$
\beq{
\left( \mathbf{A}^j - \lambda \mathbf{I} \right) \mathbf{\eta}^j = 0
\quad \mbox{or} \quad 
\left[ \mathbf{B}^j \left(\mathbf{B}^0\right)^{-1} - 
\lambda \mathbf{I} \right] \mathbf{\eta}^j = 0
\label{A-eigenvalue-problem}
}
is the same as that for the $\mathbf{B}^\mu$ system:
\beq{
\left( \mathbf{B}^j - \tilde{\lambda} \mathbf{B}^0 \right) \tilde{\mathbf{\eta}}^j = 0
\quad \mbox{or} \quad 
\left[ \mathbf{B}^j \left(\mathbf{B}^0\right)^{-1} - 
\tilde{\lambda} \mathbf{I} \right] \mathbf{B}^0 \tilde{\mathbf{\eta}}^j = 0
\label{B-eigenvalue-problem}
}
where $\mathbf{I}$ is the identity matrix.  Specifically, the eigenvalues and eigenvectors
of the two problems are related by:
\beq{
\left\{ \lambda^j_m \right\} = \left\{ \tilde{\lambda}^j_m \right\}  
\label{eigenvalue-comparison}
}
\beq{
\left\{ \greekbf{\eta}^j_m \right\} 
= \mathbf{B}^0 \left\{ \tilde{\greekbf{\eta}}^j_m \right\}
\quad . 
\label{eigenvector-comparison}
}

Explicit calculations of  $\{\greekbf{\eta}^j_m\}$ and $\{ \lambda^j_m \}$ 
for the case of current interest  can be found in Section~\ref{sec:spher-symm-perf}.

\subsection{Equations of State}
\label{sec:equations-state}

In general, there are $6$ fluid quantities  that describe the fluid: $\rho_\circ$, $\epsilon$, $P$, 
and $v^i$---the latter being the $3$-velocity of the fluid as measured by coordinate stationary observers. 
However, there are only $5$ equations of motion (EOM) (\ref{energycons},\ref{currentcons}), requiring a 
$6^\mathrm{th}$ equation to close the system.   This relation is called the equation of state (EOS) and 
provides a connection
between the microscopic properties of the particles and the thermodynamic quantities with which they are 
associated.  In practice, the equation of state is an equation that describes how the pressure 
in the matter varies with two independent quantities, such as $\rho_\circ$, $T$ or $\epsilon$.  
In this sense the equation of the state gives a measure of how the matter responds
when in a particular thermodynamic state.  

Since we wish to perform large parameter space surveys consisting of hundreds, if not thousands, 
of runs and are primarily interested in the hydrodynamical processes of stellar collapse, we wish 
to use simple equations of state that can be given in closed form.   This is in contrast to 
what is commonly done when studying core collapse supernovae or detailed simulations of neutron 
star dynamics, where tabulated data representing the state equation are used.  Such tables are calculated
from sophisticated nuclear physics models of cold, degenerate matter above nuclear densities.  The 
characteristics of this kind of matter are not well known primarily for  two reasons.  First, nuclear
physics experiments are unable to form \emph{cold} degenerate matter above nuclear densities because 
the only current way to produce such matter is to collide heavy nuclei together, and this always results in 
very \emph{hot} nuclear states.  Second, Quantum Chromodynamic (QCD) theory, which describes the 
nature of the strong force and its effect on hadrons, is not 
not completely understood at these densities.  Even with a complete QCD theory, calculating a resultant
state equation at specific fluid states would most likely be quite laborious and require the numerical astrophysicist to 
calculate a tabulated state equation  beforehand  in order to efficiently simulate systems of interest. 
These tabulated equations of state have additional error due to its finite resolution 
that closed-form state equations do not.  Hence, we will only use closed-form equations of state
for this initial study, but may eventually study the effect more realistic equations of state have on the behavior
seen here. 

A common, closed-form equation of state is called the polytropic equation of state, which---in 
general---is any equation
that depends on more than one field.  One that describes \emph{ideal}, or non-interacting, 
degenerate matter takes the form 
\beq{
P = K(s) \rho_\circ^\Gamma  \label{polytrope-eos}
}
where $K(s)$ is a function of entropy and $\Gamma$ is known as the adiabatic index.  
For example, this state equation can describe \emph{relativistic} fermion ideal gases for 
$\Gamma=4/3$---such as found in white dwarfs that are supported by degenerate relativistic electrons. 
For $\Gamma-5/3$, this EOS describes nonrelativistic degenerate fermi gases, 
such as the gas of neutrons found in neutron stars.  

Since neutrons stars are typically at temperatures far below their fermi energy, they are effectively 
at $T=0$. Hence, the degenerate neutrons in a static configuration  can be  well modeled by 
adiabatic flow, i.e. with $K(s)=\mathrm{const.}=K$.   Integrating the first law of thermodynamics with the 
adiabatic assumption (\ref{depsilon-drho0}) and using (\ref{polytrope-eos}) for the pressure yields 
the following relationship between the internal energy and the rest-mass density for cold degenerate matter:
\beq{
\epsilon = \frac{K \rho_\circ^{\Gamma-1}}{\Gamma-1} = \frac{P}{\rho_\circ \left(\Gamma-1\right)} 
\label{ideal-eos-derivation}
}
this then yields the \emph{relativistic ideal gas} law:
\beq{
P = \left( \Gamma - 1 \right) \rho_{\circ} \epsilon \quad . \label{ideal-eos}
}
This equation was used by Synge \cite{synge} to model a monatomic, nondegenerate, noninteracting relativistic gas
and serves as a relativistic version 
of Boyle's Law: 
\beq{
P = \frac{k_B}{m} \rho_\circ T \quad , \label{boyles-law}
}
and is thus often known as the ``ideal gas'' EOS.  With the adiabatic assumption, the equations 
(\ref{polytrope-eos},\ref{ideal-eos}) together represent a \emph{barotropic} EOS, 
which is defined as one in which the pressure is a function of the density alone.

The adiabatic index, $\Gamma$,  is not a constant
in general but a function of $\rho_\circ$ and $\epsilon$, with a range of physically-acceptable values 
$\Gamma\in[4/3,5/3]$ \cite{anile}.  Its determination in arbitrary dynamical systems typically requires
the use of tabulated equations of state.  However, the equation of state can be used as a model to describe
stiffer fluids of $\Gamma>5/3$ that result in the most compact stellar configurations. 
For example, $\Gamma=2$ is the 
maximum value allowed for fluid to remain causal---i.e. $c_s < c$---and was 
found to correspond to the equation of state that describes  baryons interacting through a meson
vector field (see Zel'dovich \cite{zeldovich} as referred to in Tooper \cite{tooper}).
Also, Salgado et al. \cite{salgado1,salgado2}  compared equilibrium solutions of rotating, relativistic
fluid systems generated by different equations of state.  They found that the equation of state represented
by equations (\ref{polytrope-eos},\ref{ideal-eos}) and $\Gamma=2$ lead to neutron star models 
that qualitatively resemble those with realistic state equations.  However, this is not too surprising
since it is commonly known that global features of the spherically-symmetric hydrostatic solutions in 
general relativity are independent---to a degree---of the EOS \cite{harrison-etal}. 

Since (\ref{polytrope-eos},\ref{ideal-eos}) with $\Gamma=2$ seems to be the best closed-form equation 
of state for neutron star matter, we shall use it to determine our initial neutron star 
models.  If we were to use both equations after the initial time, however, it would effectively 
constrain the internal energy of the flow to remain barotropic and never increase if and when 
shocks arise.  This consequence is because the equation, (\ref{polytrope-eos}), eliminates
the equation of motion for $\epsilon$.   An example of what happens when both state
equations are used throughout the fluids evolution is shown in \cite{font-etal2}, which examines the effect 
the state equation has on simulating dynamic stellar oscillations.  Thus, we use both 
(\ref{polytrope-eos},\ref{ideal-eos}) at $t=0$ to calculate the star solution, and only use 
(\ref{ideal-eos}) for $t>0$. 

Previous critical phenomena studies of perfect fluids have focussed on those governed by
 the so-called ``ultra-relativistic'' EOS:
\beq{
P = \left( \Gamma - 1 \right) \rho \label{ultra-eos}
}
This can be thought of as an ultra-relativistic limit of (\ref{ideal-eos}) 
wherein the fluid's internal energy becomes much greater than its rest mass density:
\beq{ 
\rho_{\circ} \epsilon \gg \rho_{\circ}
\quad \Rightarrow \quad \rho \simeq \rho_{\circ} \epsilon  \quad \quad .
\label{ultrarelativistic-limit}
}

In the following section, we will give the equations for both the general, spherically-symmetric
perfect fluid as well as the special case of an ultra-relativistic fluid.

\subsection{Spherically-Symmetric Perfect Fluids}
\label{sec:spher-symm-perf}

We first describe the equations governing a perfect fluid that is described by a general equation of 
state $P=P(\rho_\circ,\epsilon)$.  In some places, however, we use the ideal gas EOS (\ref{ideal-eos})
to simplify expressions and we indicate such specialization accordingly.  We use the formulation of Romero et al. 
\cite{romero}, which was the first implementation of high-resolution shock-capturing schemes for fluids 
coupled to a time-dependent geometry, primarily since their methods seemed to be quite successful.  
In the following development, we will assume that the metric takes the polar-areal form
(\ref{metric}).  

We begin by defining a few quantities that characterize the fluid.  Instead of the 
$4$-velocity of the fluid, a more useful quantity is the radial component of the Eulerian---or physical---velocity 
of the fluid as measured by a Eulerian observer:
\beq{
v = \frac{ a u^t }{\alpha u^r} 
}
where $u^\mu = \left[ u^t, u^r, 0, 0\right]$ (recall that we are working in spherical symmetry).  
The associated ``Lorentz gamma function'' is defined by 
\beq{ 
W = \alpha u^t  \quad . 
\label{w}
}
Given the fact that the 4-velocity is time-like and unit-normalized, i.e. $u^\mu u_\mu=-1$, 
$v$ and $W$ are related by
\beq{
W^2 = \frac{1}{1 - v^2} \quad . \label{vwrelation}
}
We will shortly
see that the equations of motion in spherical symmetry can take a conservation-law form, with 
\emph{conservative} variables defined by
\begin{eqnarray}
D & = & a \rho_{\circ} W \label{D}  \\ 
S & = & \left( \rho + P \right) W^2 v \  = \ \rho_\circ h W^2 v \label{S}\\
E & = & \left( \rho + P \right) W^2  - P \  = \ \rho_\circ h W^2 - P \label{E} \\
\tau & = & E - D  \ = \ \rho_\circ h W^2 - P - a \rho_{\circ} W    \quad . \label{tau}
\end{eqnarray}
The above variables can be thought of as the rest-mass density, momentum density, 
total energy density, and internal energy density as measured in a Eulerian-frame defined by the ADM 
slicing, respectively. 

In order to perform a few simplifications in the source terms of the equations of 
motion, the geometric constraints and evolution equation will be used.  
The ADM local energy density and ADM momentum density for a perfect fluid can be easily calculated:
\beq{
\varrho \equiv  n_\mu n_\nu T^{\mu \nu} = \tau + D  \quad . \label{ideal-adm-rho}
}
\beq{ 
j_i \equiv - n_\mu {T^\mu}_i  = \alpha {T^t}_i = \left[ a S, 0, 0 \right]
\label{momentum}
}
The Hamiltonian constraint can then be shown to take the form:
\beq{ 
\frac{a '}{a} = a^2 \left[ 4 \pi r \left( \tau + D \right) - \frac{m}{r^2} \right]
\quad , \label{hamconstraint}
}
while the slicing condition  and the momentum constraint are respectively:
\beq{
\frac{\alpha '}{\alpha} 
= a^2 \left[ 4 \pi r \left( S v + P \right) + \frac{m}{r^2} \right] 
\label{slicingcondition}
}
\beq{
\dot{a} = - 4 \pi r \alpha a^2 S   \quad . \label{momentumconstraint}
}

As we saw previously, the equations of motion for the perfect fluid can be cast into conservation 
form.  Deriving them from (\ref{energycons}-\ref{currentcons}) is fairly straightforward, especially 
in spherical symmetry.  The continuity equation yields 
\beqa{
0 = \nabla_\mu J^\mu \ & = & \ 
\frac{1}{\sqrt{\left| g \right|}} 
\partial_\mu \left( \sqrt{\left| g \right|} J^\mu \right) 
\label{current-eom1} \\[0.5cm]
& = & \ \frac{1}{ \alpha a } \left[ 
\partial_t \left(\alpha a \ \frac{ D }{ \alpha a } \right)
+ \frac{1}{r^2} \partial_r \left( \alpha a r^2 \ \frac{ D v }{ a^2 } \right) 
 \right] \label{current-eom2}
}
\beq{ 
\Longrightarrow \quad \dot{D} + \frac{1}{r^2} \left( r^2 X D v \right)^{ \prime}
\ = \ 0  
\label{ideal-D-eom}
} 
where (\ref{current-eom1}) used a well-known identity (see \cite[pg.49]{wald}) and
\beq{ 
g \equiv \det{\left( g_{a b} \right)} = - \alpha^2 a^2 r^4 \sin^2{\theta}
\quad . \label{det-g}
}
The other two equations of motion follow from the two components of the equation of local energy conservation. 
From  $\nabla_\mu {T^\mu}_t = 0$ we have
\beqa{
0 & = & \nabla_\mu {T^\mu}_t  =  \partial_\mu {T^\mu}_t 
+ {\Gamma^\mu}_{\mu \nu} {T^\nu}_t - {\Gamma^\nu}_{\mu t} {T^\mu}_{\nu}
\nonumber \\
& = & - \dot{E} - \frac{1}{r^2} \left( r^2 X S \right)'
-  \frac{\dot{a}}{a} \left( E + S v + P \right) 
- X S \left( \frac{a '}{a} + \frac{\alpha '}{\alpha} \right) 
\label{eom1step1}
}
\beq{ 
\Longrightarrow \qquad \dot{E} + \frac{1}{r^2} \left( r^2 X S \right)' = 0
\label{E-eom}
}
where $X \equiv \alpha / a$, 
and in proceeding from (\ref{eom1step1}) to (\ref{E-eom}) we used the 
Hamiltonian constraint (\ref{hamconstraint}),  
slicing condition (\ref{slicingcondition}), and momentum constraint
(\ref{momentumconstraint}). 
Similarly, from $\nabla_\mu {T^\mu}_r = 0$ we have
\beqa{
0 & = & \nabla_\mu {T^\mu}_r = \partial_\mu {T^\mu}_r 
+ {\Gamma^\mu}_{\mu \nu} {T^\nu}_r - {\Gamma^\nu}_{\mu r} {T^\mu}_{\nu}
\nonumber \\[0.25cm]
& = & \frac{\dot{S}}{X} + \frac{2 \dot{a}}{a} \frac{S}{X}  
+ \left( S v + P \right)' - \frac{2 P}{r} 
+ \frac{2}{r} \left( S v + P \right)
+ \frac{\alpha '}{\alpha} \left( S v + P + E \right)
\label{eom2step1}
}
\beq{ 
\Longrightarrow \qquad \dot{S} 
+ \frac{1}{r^2} \left[ r^2 X \left( S v + P \right) \right]' = \Sigma
\label{S-eom}
}
where
\beq{ 
\Sigma \equiv \Theta + \frac{2 P X}{r} \label{Sigma}
}
\beq{ 
\Theta \equiv \alpha a \left[ 
\left( S v - \tau - D \right) \left(8 \pi r P + \frac{m}{r^2} \right) 
+ P \frac{m}{r^2} \right]  \label{Theta}
}
Again, in going from (\ref{eom2step1}) to (\ref{S-eom}) we have used the 
the Hamiltonian constraint (\ref{hamconstraint}), the slicing condition 
(\ref{slicingcondition}), and the momentum constraint (\ref{momentumconstraint}). 

The variable $\tau$ is often evolved in place of $E$ in order to separate the rest-mass and internal
energy densities, which can often take values that differ by orders of magnitude.  For instance, if 
$D \ll \tau$, then the numerical error involved in calculating $E$ will be on the order of $D$, and this 
feature has been found to cause inaccuracies in the entire numerical scheme \cite{romero}.  To find the evolution equation 
for $\tau$, (\ref{ideal-D-eom}) is subtracted from (\ref{E-eom}), yielding:
\beq{ 
\dot{\tau}  
+ \frac{1}{r^2}\left[ r^2 X v \left( \tau + P \right) \right]^{\prime}
= \ 0
\label{ideal-tau-eom}
}
where the following identity was used:
\beq{ 
S - v D = v \left( \tau + P \right)  \quad or \quad S = v \left( E + P \right)
\quad . \label{S-E-identity}
}

To date, the above formulation is the one that most researchers have used to study spherically-symmetric fluids in 
conservation form \cite{brady_etal,neilsen-crit,neilsen,novak}.  We can clearly see 
that (\ref{ideal-D-eom},\ref{S-eom},\ref{ideal-tau-eom}) 
form a set partial differential equations in conservation form.  However, we found that for extremely relativistic 
flows near the threshold of black hole formation, this formulation was not very stable.  
In an attempt to stabilize the evolution during such collapse scenarios, we use a different formulation 
motivated by work of Neilsen and Choptuik \cite{neilsen} who studied  fluid collapse with the ultra-relativistic EOS. 
As the fluid becomes extremely relativistic, $\tau$ and $S$ become similar in magnitude, and 
Neilsen and Choptuik found that evolving $\tau \pm S$ allowed for a 
more precise calculation.  

The new variables for a general perfect fluid take the form
\beq{
\Pi  \ \equiv \  \tau + S  \ = \ 
     \frac{ 1 }{ 1 - v } \left[ \rho_\circ 
         \, + \, P \left( \frac{1}{\tilde{\kappa}} + v \right) \right]
     \, - \,  a \rho_\circ W 
\label{Pi-ideal}
}

\beq{
\Phi  \ \equiv  \ \tau - S  \ = \ 
     \frac{ 1 }{ 1 + v } \left[ \rho_\circ  
         \, + \, P \left( \frac{1}{\tilde{\kappa}} - v \right) \right]
     \, - \, a \rho_\circ W 
\label{Phi-ideal}
}
where $\tilde{\kappa} \equiv \Gamma - 1$.  Since $\tau$ and $S$ are conservative variables, any linear combination of them 
are also conservative variables, and, hence, the equations of motion for $\Pi$ and $\Phi$ are also conservation
laws.  These equations  can be easily found by following similar procedures as that used for the $\tau$ EOM.  
The new EOM for $\Pi$ and $\Phi$ with the EOM for $D$ then form the set of $3$ conservation 
equations that we will use hereafter:
\beq{ 
\partial_t \mathbf{q} 
+ \frac{1}{r^2} \partial_r \left( r^2 X \mathbf{f} \right) = \greekbf{\psi} 
\quad . \label{conservationeq}
}
where the state vectors take the form
\beq{ 
\mathbf{q} =  \left[ \begin{array}{c} D \\ \Pi \\ \Phi \end{array}\right]  
\ , \
\mathbf{f} = \left[ \begin{array}{c} D v \\ v \left( \Pi + P \right) + P
        \\ v \left( \Phi + P \right) - P \end{array} \right]
\ , \
\greekbf{\psi} = \left[ \begin{array}{c} 0 \\ \Sigma \\ -\Sigma \end{array} \right]
\ , \
\mathbf{w} = \left[ \begin{array}{c} P \\ v \\ \rho_\circ \end{array} \right]
\label{ideal-piphi-state-vectors}
}
These are the equations that we will use for simulating the fluid without any other matter models present.  
Note, that we have 
also defined $\mathbf{w}$ which represents the vector of primitive variables that will be used. 
We also note that flat space equations of motion are obtained by setting $\Theta = 0$ and $X=1$. 

We use high-resolution shock-capturing (HRSC) techniques for solving the above conservative system
of partial differential equations.  These methods often utilize the characteristic structure of the 
differential equations in order to elucidate how the various waves of the solution move from one 
grid cell to the next.  Let us provide the equations that determine the characteristic structure here. 
In order to find the characteristics, we need to put the conservative equation (\ref{conservationeq})
into quasi-linear form
\beq{
\partial_t \mathbf{q} 
+ \frac{1}{r^2} \mathbf{A} \partial_r \left( r^2 X \mathbf{q} \right) = \greekbf{\psi} 
\quad . \label{quasilinear-conservationeq}
}
In our case, and in general, the system of partial differential equations are highly-coupled and 
so result in a non-diagonal characteristic matrix, $\mathbf{A}$, which is just the Jacobian matrix
defined in (\ref{A-matrix-def}).  Since $\mathbf{f}$ is a function of 
$\mathbf{w}$ and $\mathbf{q}$, we cannot directly calculate $\mathbf{A}$ from its definition (\ref{A-matrix-def}).  
Instead, we typically use (\ref{B-to-A}) and explicitly calculate $\mathbf{B}^r$ and $\mathbf{B}^t$:
\beq{
\mathbf{B}^t \equiv \pderiv{\mathbf{q}}{\mathbf{w}}  \quad , \quad 
\mathbf{B}^r \equiv \pderiv{\mathbf{f}}{\mathbf{w}}  \quad .
\label{sub-jacobians}
}
The explicit forms of the matrix elements are not important and are quite complicated, so they will not be shown here. 
All that is needed from $\mathbf{A}$ is its eigenvalues and eigenvectors, which we have determined 
using the mathematical software \texttt{Maple}.  As far as the author knows, no one else 
has ever used this particular formulation of the perfect fluid equations, and---consequently---the 
characteristic structure is given here for the first time.  Since the transformation 
from $\{D, S, \tau\}$ to $\{D, \Pi, \Phi\}$ is linear, we expect the two sets of eigenvalues 
to be the same for the corresponding two sets of equations.  We have verified this fact with our \texttt{Maple} routine,
and find
\beq{
\lambda_1 = v 
\qquad , \qquad  
\lambda_{2 \atop 3} \ = \ \lambda_\pm \ = \ \frac{ v \pm c_s }{ 1 \pm v c_s }  \quad .
\label{ideal-piphi-evalues}
} 
However, the right eigenvectors $\eta_m$, defined in (\ref{A-eigenvalue-problem}), take very 
different forms for the two sets of equations.  Using the typical normalization for the
eigenvectors ($\greekbf{\eta}_m^{(2)} = \lambda_m$ ), leads to a 
very complicated set of eigenvectors.  Hence, we used the following normalizations:
\beq{
\greekbf{\eta}_m^{(1)} = 1   \quad \quad  \forall \ m  \quad ,
\label{ideal-piphi-evect-normalization}
}
which leads to significant simplification.  
With this normalization the right eigenvectors associated with (\ref{conservationeq}) become:
\beq{
\greekbf{\eta}_1 \ = \ \left[ \begin{array}{c} 1 
\\[0.5cm]  \frac{W \left(1+v\right)}{a} - 1 
\\[0.5cm]  \frac{W \left(1-v\right)}{a} - 1 \end{array}\right]   \qquad , \qquad
\greekbf{\eta}_{2 \atop 3} \ = \greekbf{\eta}_\pm \ = \ \left[ \begin{array}{c} 1 
\\[0.5cm] \frac{ W  \left( 1 + v \right) }{ a }  h \left( 1 \pm c_s \right) - 1
\\[0.5cm] \frac{ W  \left( 1 - v \right) }{ a }  h \left( 1 \mp c_s \right) - 1
       \end{array}\right] 
\label{ideal-piphi-right-evects}
}
The left eigenvectors are also useful. If we define a matrix whose column are the right eigenvectors, 
\beq{
\greekbf{\mathcal{N}} \equiv \left[ \greekbf{\eta}_1 \  \greekbf{\eta}_2  \ \greekbf{\eta}_3 \right]
\quad , \quad 
\greekbf{\eta}_m = \left[ 
\begin{array}{c} {\eta_m}^{(1)} \\ \vdots \\ {\eta_m}^{(N)} \end{array}
\right]
\label{column-matrix}
}
then the left eigenvectors can be defined from the rows of the inverse of $\greekbf{\mathcal{N}}$:
\beq{
\greekbf{\mathcal{N}}^{-1} = \left[ \begin{array}{c} \mathbf{l}_1 \\ \mathbf{l}_2 \\ \mathbf{l}_3 \end{array} \right]
\quad , \quad 
\mathbf{l}_m = \left[ {l_m}^{(1)} \  {l_m}^{(2)} \ {l_m}^{(3)} \right]
\label{left-eigenmatrix}
}
Using \texttt{Maple}, we found these to be:
\beqa{
\mathbf{l}_1 \ & = & \ \left[ \begin{array}{c} 
  1  + \frac{\mathbig{\tilde{\kappa}}}{h \,c_s^2} \left( 1 - a W \right)
\\[0.5cm]  -\frac{a}{2 h \,c_s^2} \, \tilde{\kappa} \, W \left( 1 - v \right)
\\[0.5cm]  -\frac{a}{2 h \,c_s^2} \, \tilde{\kappa} \, W \left( 1 + v \right)
    \end{array}\right]^T  \label{ideal-piphi-left-evects1} \\[0.3cm]
\mathbf{l}_{2 \atop 3} \ = \ \mathbf{l}_\pm \ = & \mathBig{\frac{1}{2 h \, c_s^2}} &
\left[ \begin{array}{c} 
a W \left( \tilde{\kappa} \mp v c_s \right) - \tilde{\kappa} 
\\[0.5cm]  \frac{1}{2} a W \left(1 - v \right) \left( \tilde{\kappa} \pm c_s \right)
\\[0.5cm]  \frac{1}{2} a W \left(1 + v \right) \left( \tilde{\kappa} \mp c_s \right)
    \end{array}\right]^T
\label{ideal-piphi-left-evects2}
}
Note that in calculating the eigenvalues and eigenvectors, we have now explicitly used the ideal gas equation of state 
(\ref{ideal-eos}).  The speed of sound was assumed to be the one associated with this EOS:
\beq{
c_s^2 \, = \, \frac{ \left( \Gamma - 1 \right) \Gamma P }
                   { \left( \Gamma - 1 \right) \rho_\circ + \Gamma P }
\label{ideal-cs}
}
and we also have
\beq{
\tilde{\kappa} \equiv \kappa/\rho_\circ  = \Gamma - 1 \label{kappa-tilde} \quad . 
}
In addition, when calculating the eigensystem we used the 
following identity, which is derived from the ideal gas EOS (\ref{ideal-eos}):
\beq{
h {c_s}^2 =  \frac{ \Gamma P }{ \rho_\circ }  \quad . \label{h-cs-identity}
}

In our simulations of  self-gravitating, ideal-gas fluids, the fluid is integrated in time with 
equations (\ref{conservationeq}-\ref{ideal-piphi-state-vectors}), while the geometry is simultaneously
calculated using the Hamiltonian constraint (\ref{hamconstraint} and the slicing condition 
(\ref{slicingcondition}).  The specific methods we used to numerically integrate these equations are 
explained in Chapter~\ref{chap:numerical-techniques}. 

\subsection{The Ultra-relativistic Fluid}
\label{sec:ultra-relat-fluid}

The ultra-relativistic fluid is a perfect fluid in which microscopic particles,
which constitute the fluid, move at extremely relativistic speeds.  Thus, the thermal energy of such a fluid 
is much greater than the rest-mass density, and the flow can be well described by the ultra-relativistic limit 
(\ref{ultrarelativistic-limit}).  Since
$\rho_\circ$ is irrelevant in ultra-relativistic flows, we can easily see that $D$ is similarly irrelevant.  
Let us define the ultra-relativistic fluid to be the limiting case where 
$\rho_\circ \epsilon = \rho$, $D=\rho_\circ=0$, and $\rho_\circ h = \rho + P$.  This reduces the 
set of $3$ fluid EOM to $2$, and 
simplifies the numerical procedure significantly.  For example, in order to calculate the 
flux vectors $\mathbf{f}$, we need to find the primitive variables $\mathbf{w}$ 
from the conservative variables $\mathbf{q}$.  Even though the solution $\mathbf{q}=\mathbf{q}(\mathbf{w})$
is straightforward---via the definitions (\ref{D}-\ref{tau}), the inverse transformation 
$\mathbf{w}=\mathbf{w}(\mathbf{q})$ is rather difficult to determine
when using the more general ideal gas equations since no known closed-form solution is known.  Hence, 
we need to rely on approximate, numerical solutions, which are sometimes imprecise and whose 
determination represents a large part of the code's runtime. 
However, with the ultra-relativistic system, the calculation of $\mathbf{w}=\mathbf{w}(\mathbf{q})$ 
reduces to simple algebraic expressions that can be calculated in closed-form.  Also, the 
ultra-relativistic system is intrinsically
scale-free, making it ideal for investigating self-similar flows such as those found in Type~II critical 
behavior.  In fact, the methods we use to simulate ultra-relativistic flows are based entirely on 
those used to study critical phenomena of ultra-relativistic fluids \cite{neilsen-crit,neilsen}. 

In this section, we will give the equations that describe ultra-relativistic flows in spherical 
symmetry.  To derive them, we may start with the system described in the previous section, set $\tau=E$, 
and then remove $D$'s EOM from the system.  The equations still have the same conservative form 
(\ref{conservationeq}), but the state vectors are now defined as:
\beq{ 
\mathbf{q} =  \left[ \begin{array}{c} \Pi \\ \Phi \end{array}\right]  
\ , \
\mathbf{f} = \left[ \begin{array}{c}  v \left( \Pi + P \right) + P
        \\ v \left( \Phi + P \right) - P \end{array} \right]
\ , \ 
\greekbf{\psi} = \left[ \begin{array}{c} \Sigma_u \\ -\Sigma_u \end{array} \right]
\ , \
\mathbf{w} = \left[ \begin{array}{c} P \\ v \end{array} \right]
\label{newstatevectors}
}
where $\Sigma_u$ and $\Theta_u$ are essentially the same as previously except that we now have $D=0$:
\beq{
\Sigma_u \equiv \Theta_u + \frac{2 P X}{r} \quad , \quad
\Theta_u \equiv \alpha a \left[ 
\left( S v - \tau \right) \left(8 \pi r P + \frac{m}{r^2} \right) 
+ P \frac{m}{r^2} \right]  \label{sigma-theta-ultra}
}
Here, the ultra-relativistic versions of  $\Pi$ and $\Phi$ are defined as 
\beq{
\Pi = W^2 \left( \rho + P \right) \left( 1 + v \right) - P  \quad , \quad
\Phi = W^2 \left( \rho + P \right) \left( 1 - v \right) - P  \label{ultra-piphi}
}

Notice that the number of primitive variables is reduced to just two---$P$ and $v$---since $\rho_\circ=0$. 
The total energy density, $\rho$, is calculated from the ultra-relativistic equation of state 
(\ref{ultra-eos}).  The velocity can be determined from the ultra-relativistic version of 
(\ref{S-E-identity}):
\beq{
v = \frac{ S }{ \tau + P }  \label{ultra-v-calc}
}
and $P$ can be calculated from $v$ and the definitions of $\Pi$ and $\Phi$ (\ref{ultra-piphi}):
\beq{ 
P = - \beta \left( \Pi + \Phi \right) 
+ \left[ \beta^2 \left( \Pi + \Phi \right)^2 
+ \left( \Gamma - 1 \right) \Pi \Phi \right]^{1/2} 
\label{ultra-P-calc}
}
where $\beta \equiv ( 2 - \Gamma ) / 4 $. 
For large values of $W$, equation (\ref{ultra-v-calc}) leads to 
unphysical velocities ( $|v| > 1$ ) because of round-off errors in its numerical evaluation. 
Hence, we use an equation which is more accurate in this regime:
\beq{
v = \frac{1}{2 \Lambda} \left( \sqrt{ 1 + 4 \Lambda^2 } - 1 \right) 
\label{numvelocity}
}
where 
\beq{
\Lambda \  \equiv \ W^2 v \ = \ \frac{ ( \Gamma - 1 ) S }{\Gamma P } \quad . 
\label{Lambda}
}
Equation (\ref{numvelocity}) is merely an identity derived from 
the definition of $W$ (\ref{vwrelation}), and (\ref{Lambda}) follows from the 
definition of $S$, (\ref{S}), and the equation of state (\ref{ultra-eos}). 
However, when $\Lambda > 10^{-4}$,
equation (\ref{ultra-v-calc}) is used to calculate $v$.  

The geometrical variables in the ultra-relativistic case are calculated using equations  
(\ref{hamconstraint}-\ref{momentumconstraint}), where, in the Hamiltonian constraint (\ref{hamconstraint}), 
we set $D=0$.

We also need the characteristic structure of the ultra-relativistic fluid in order to use HRSC 
methods.  Since there are now only two PDE's, the linear system is only two-dimensional. 
The Jacobian matrix from the quasi-linear form of the equations of motion is
\beq{
\mathbf{A} = \left[ \begin{array} {c c} 
{\mathrm{A}^1}_1 & {\mathrm{A}^1}_2 \\ 
{\mathrm{A}^2}_1 & {\mathrm{A}^2}_2 \end{array} \right]  \label{A-matrix}
}
\beq{
\begin{array}{ll}
{\mathrm{A}^1}_1 & = \frac{1}{2} \left( 1 + 2 v - v^2 \right) 
+ \left( 1 - v^2 \right) \frac{\partial P}{\partial \Pi}  \\[0.3cm]
{\mathrm{A}^1}_2 & = - \frac{1}{2} \left( 1 + v \right)^2 
+ \left( 1 - v^2 \right) \frac{\partial P}{\partial \Phi} \\[0.3cm]
{\mathrm{A}^2}_1 & = \frac{1}{2} \left( 1 - v \right)^2 
+ \left( v^2 - 1 \right) \frac{\partial P}{\partial \Pi} \\[0.3cm]
{\mathrm{A}^2}_2 & = \frac{1}{2} \left( v^2 + 2 v - 1 \right)
+ \left( v^2 - 1 \right) \frac{\partial P}{\partial \Phi}
\end{array}
\label{new-A-elements}
}
where 
\beqa{
\frac{\partial P}{\partial \Pi} & = & - \beta \ 
+ \ \frac{ 2 \beta^2 \left( \Pi + \Phi \right) 
+ \left( \Gamma - 1 \right) \Phi}{\, 2 
\left[ \beta^2 \left( \Pi + \Phi \right)^2 + \left( \Gamma - 1 \right) \Pi \Phi 
\right]^{1/2} \, } \label{dpdpi} \\[0.5cm]
\frac{\partial P}{\partial \Phi} & = & - \beta \
+ \ \frac{ 2 \beta^2 \left( \Pi + \Phi \right) 
+ \left( \Gamma - 1 \right) \Pi}{\, 2 
\left[ \beta^2 \left( \Pi + \Phi \right)^2 + \left( \Gamma - 1 \right) \Pi \Phi 
\right]^{1/2} \, } \label{dpdphi} 
}
For completeness, we note that the following was used in deriving (\ref{new-A-elements}):
\beqa{ 
\frac{\partial v}{\partial \Pi} & = & 
\frac{v}{ \Pi - \Phi } 
\left( 1 - v - 2 v \frac{\partial P}{\partial \Pi} \right) 
\label{dvdpi}  \\[0.5cm]
\frac{\partial v}{\partial \Phi} & = & 
- \frac{v}{ \Pi - \Phi } 
\left( 1 + v + 2 v \frac{\partial P}{\partial \Phi} \right) 
\label{dvdphi}
}
The right eigenvectors associated with this matrix are then:
\beq{
\greekbf{\eta}_\pm = \left[ 1 \atop {Y_\pm} \right]
\qquad , \quad Y_\pm \equiv \frac{ \lambda_\pm - {\mathrm{A}^1}_1 }{ {\mathrm{A}^1}_2 } 
\label{ultra-eigenvectors}
}
with eigenvalues
\beq{
\lambda_\pm  =  \frac{1}{2} \left[ {\mathrm{A}^1}_1 + {\mathrm{A}^2}_2 
\pm \sqrt{ \left( {\mathrm{A}^1}_1 - {\mathrm{A}^2}_2 \right)^2 
+ 4\, {\mathrm{A}^1}_2 \, {\mathrm{A}^2}_1  } \ 
\right]
\label{ultra-eigenvalues}
}

\subsection{Minimally-Coupled Scalar Field}
\label{sec:minim-coupl-scal}

The evolution of a scalar field minimally-coupled to a perfect fluid is an 
interesting problem since it is still uncertain whether the 
collapse of a perfect fluid (scalar field) in a scalar field (fluid) 
background would lead to the same critical phenomenon as with no scalar 
field (fluid).  Also, we use the gravitational interaction between the scalar field 
and the fluid to dynamically drive equilibrium star solutions to collapse. 
In this section, we give the evolution and constraint equations for 
a scalar field and perfect fluid system.  
We assume that the two fields are not directly coupled but only interact 
by how each one affects the local spacetime geometry.  Since there is no explicit 
interaction between the fluid and scalar field the total stress-energy
tensor of the system is given by
\beq{
T_{a b} = {\tilde{T}}_{a b} + {\hat{T}}_{a b}  \label{totalstress}
}
where ${\tilde{T}}_{a b}$ is the scalar's stress-energy and ${\hat{T}}_{a b}$
is that of the fluid.  The stress-energy for scalar field, $\phi$, is given by
\beq{
{\tilde{T}}_{a b} = \nabla_a \phi \nabla_b \phi - \frac{1}{2} g_{a b} 
\left(\, \nabla_c \phi \nabla^c \phi + 2 V(\phi) \ \right)  
\label{scalarstress}
}
where $V(\phi)$ is the scalar's potential; in the following equations, 
We will assume
that $V(\phi)$ is non-zero however in subsequent calculations we will take $V(\phi) = 0$.   
Since the 
two fields are not directly interacting, then the local conservation of 
energy equation holds \textit{separately} for each stress-energy, specifically:
\beq{
\nabla^a T_{a b} = \nabla^a {\tilde{T}}_{a b} = \nabla^a {\hat{T}}_{a b} = 0 \ .
\label{bothenergyconservation}
}
This equation yields the usual 
equation of motion for the scalar field:
\beq{
\Box \phi \equiv \nabla^a \nabla_a \phi = \partial_{\phi} V(\phi)  \label{generalscalareom} .
}
Given the metric (\ref{metric}), the scalar's EOM simplifies to 
\beq{
\frac{1}{r^2} \partial_r \left( X r^2 \phi ' \right) 
- \partial_t \left( \frac{a}{\alpha} \dot{\phi} \right) 
= \alpha a \partial_\phi V \ .
\label{simplescalareom}
}
We can convert this to a system of first-order (in time) PDE's by making the substitution
\beq{ 
\Xi \equiv \phi '  \quad , \quad \Upsilon \equiv \frac{a}{\alpha} \dot{\phi} \quad . 
\label{Xi-Upsilon}
}
With these definitions the ``new'' EOM's are 
\beq{
\dot{\Xi} = \left( X \Upsilon \right) '  \label{scalareom1}
}
\beq{
\dot{\Upsilon} = \frac{1}{r^2} 
\left( r^2 X \Xi \right) ' - \alpha a \partial_\phi V 
\label{scalareom2}
}
where $X \equiv \alpha / a$ as before. The equation (\ref{scalareom1}) follows from the 
definitions of $\Xi$ and $\Upsilon$ and the fact that $\partial_t$ and $\partial_r$ 
commute, while the second 
EOM (\ref{scalareom2}) is merely (\ref{simplescalareom})
with the definitions (\ref{Xi-Upsilon}).  For completeness, 
we note that the non-zero components of the scalar field's stress tensor are:
\beq{
{\tilde{T^t}}_t = - \frac{1}{2 a^2} \left( \Xi^2 + \Upsilon^2 \right) 
- V(\phi) 
\quad , \quad 
{\tilde{T^t}}_r = - \frac{ \Xi \Upsilon }{ \alpha a }
\quad , \quad 
{\tilde{T^r}}_t = \frac{1}{a^2} X \Xi \Upsilon  \label{scalar-stress1} 
}
\beq{
{\tilde{T^r}}_r = \frac{1}{2 a^2} \left( \Xi^2 + \Upsilon^2 \right) - V(\phi)
\quad , \quad 
{\tilde{T^\theta}}_\theta = {\tilde{T^\phi}}_\phi = 
\frac{1}{2 a^2} \left( \Upsilon^2 - \Xi^2 \right) - V(\phi)  
\label{scalar-stress2}
}

In order to state the equations for the 
geometry without specifying the fluid type, we need only replace $E$ in the following
with the appropriate quantity for that model as follows:
\beq{
\begin{array}{l l}
E \, = \, \tau  \quad  \quad &\mbox{Ultra-relativistic Fluid} \\[0.1cm]
E = \tau + D   &\mbox{Ideal Gas}  \quad.
\end{array}
\label{E-limits} 
}

The ADM energy density is 
\beq{
\varrho = E + V(\phi) + \frac{1}{2 a^2} \left( \Xi^2 + \Upsilon^2 \right)
\quad . \label{adm-rho-both}
}
We can clearly see that the total energy density is composed of a fluid
part and a scalar part:
\beq{
\varrho = \varrho_{\,_{\mathsmallest{\mathrm{fluid}}}} + 
\varrho_{\,_{\mathsmallest{\mathrm{scalar}}}} 
\label{adm-rho-both-parts}
}
where 
\beq{
\varrho_{\,_{\mathsmallest{\mathrm{fluid}}}} \, \equiv \, E 
\label{adm-rho-fluid}
}

\beq{
\varrho_{\,_{\mathsmallest{\mathrm{scalar}}}}  
\ \equiv \ \frac{1}{2 a^2} \left( \Xi^2 + \Upsilon^2 \right)
+ V(\phi)
\label{adm-rho-scalar}
}
Since we know that 
\beq{
\pderiv{ m }{ r } \ = \ 4 \pi r^2 \varrho 
\label{dmdr}
}
from the Hamiltonian constraint and definition of the mass aspect function $m(r)$, 
we can also define relations for the mass functions associated with each matter 
part:
\beq{
\pderiv{ m }{ r } \ = \ \pderiv{ m_{\mathrm{fluid}} }{ r } 
\ + \ \pderiv{ m_{\mathrm{scalar}} }{ r } \label{dmdr-total}
}
where 
\beq{
\pderiv{ m_{\mathrm{fluid}} }{ r } \ = \ 4 \pi r^2 
\varrho_{\,_{\mathsmallest{\mathrm{fluid}}}} 
\ = \ 4 \pi r^2 E 
\label{dmdr-fluid}
}
\beq{
\pderiv{ m_{\mathrm{scalar}} }{ r } \ = \ 
4 \pi r^2 \varrho_{\,_{\mathsmallest{\mathrm{scalar}}}}  
\ = \  4 \pi r^2 \left[ \frac{1}{2 a^2} \left( \Xi^2 + \Upsilon^2 \right)
+ V(\phi) \right] \quad .
\label{dmdr-scalar}
}
However, the two mass contributions can only be unambiguously differentiated in regions 
of non-overlapping support, since---for instance---$\partial m_\mathrm{scalar} / \partial r$ depends on metric
quantities which in turn depends on the local energy content of all present matter distributions of the spacetime.   

The Hamiltonian constraint takes the form:
\beq{
\frac{ a^\prime }{a} = a^2 \left[ 4 \pi r \left( E + V(\phi) \right) - 
\frac{m}{r^2} \right]  + 2 \pi r \left( \Xi^2 + \Upsilon^2 \right) 
\nonumber
}
or
\beq{
\frac{ a^\prime }{a} = a^2 \left[ 4 \pi r \left( E + V(\phi) \right) - 
\frac{1}{2 r} \right] + \frac{1}{2 r} + 2 \pi r \left( \Xi^2 + \Upsilon^2 \right) 
\label{hamiltonian-both}
}
The ADM momentum density is :
\beq{
j_i = \left[ a S - \frac{ \Xi \Upsilon }{a} \, , \, 0 \, , \, 0 \, \right]
\quad . \label{adm-momentum-both}
}
The momentum constraint is: 
\beq{
\dot{a} = 4 \pi r \alpha \left( \Xi \Upsilon - a^2 S \right)
\quad . 
\label{momentum-constraint-both}
}
The slicing condition becomes:
\beq{
\frac{\alpha^\prime}{\alpha} = a^2 \left[ 4 \pi r \left( S v + P - V(\phi) \right)
+ \frac{m}{r^2} \right] + 2 \pi r \left( \Xi^2 + \Upsilon^2 \right) 
\nonumber
}
or equivalently
\beq{
\frac{\alpha^\prime}{\alpha} = a^2 \left[ 4 \pi r \left( S v + P - V(\phi) \right)
+ \frac{1}{2 r} \right] - \frac{1}{2 r} + 2 \pi r \left( \Xi^2 + \Upsilon^2 \right) 
\quad .
\label{slicing-both}
}

As a weak check of the derivation of the geometry equations, we can see that 
the geometry equations for the fluid-only case are obtained when the scalar field 
variables ($\Xi$ and $\Upsilon$) are set to zero (and vice versa).  

Since the fluid EOM's involve the geometry equations, they are now different than for the case
without the scalar field.  In the following two subsections we will present 
the equations for the ideal gas 
(Section~\ref{sec:spher-symm-perf}) and the ultra-relativistic fluid
(Section~\ref{sec:ultra-relat-fluid}), with the addition of a scalar field in each case.

\subsubsection{Ideal Gas EOM for ``Scalar+Fluid'' System}

From (\ref{eom2step1}), the evolution equation for $S$, before the use of 
the geometry equations, takes the form:
\beq{
\dot{S} + \frac{1}{r^2} \left[ r^2 X \left( S v + P \right) \right]^{ \prime} 
= \Sigma 
\label{S-eom-both}
}
here $\Sigma = \Theta + 2 P X / r$ and
\beq{
\Theta = \, - \frac{2 \dot{a}}{a} S \, - \, 
\frac{a^\prime}{a} X \left( S v + P \right)
\, - \, \frac{\alpha^\prime}{\alpha} X E   
\label{Theta-general-both}
}
Using the geometry equations (\ref{hamiltonian-both})-(\ref{slicing-both}), 
$\Theta$ becomes
\beqa{
\Theta & = & \alpha a \left\{  \left( S v - E \right) \left[ 
4 \pi r \left( 2 P - V(\phi) \right) + \frac{m}{r^2} \right]
+ P \left( \frac{m}{r^2} - 4 \pi r V(\phi) \right) \right\} \nonumber \\[0.1cm]
& & - \, 2 \pi r X \left[ 4\, \Xi \Upsilon S \, + \, 
\left( \Xi^2 + \Upsilon^2 \right)
\left( S v + P + E \right) \right] 
\label{Theta-both}
}
where, for the ideal gas code, 
we always replace $E$ with  $\left(\tau + D \right)$ in the equations.
Notice that (\ref{Theta-both}) reduces to (\ref{Theta}) when 
$\Xi = \Upsilon = V(\phi) = 0$.

The EOM for $D$ is independent of the geometry equations, so 
it remains the same as before. 
However, the EOM for $E$ does depend on the constraint/evolution equations 
for the geometry.  From (\ref{eom1step1})  we see that :
\beq{
\dot{E} + \frac{1}{r^2} \left( r^2 X S \right)^{ \prime} = \Psi_E
\label{E-eom-both}
}
where
\beq{
\Psi_E \equiv
- \, \frac{ \dot{a} }{ a } \left( S v + P + E \right) 
\ - \ X S \left( \frac{ a^{ \prime} }{ a } + \frac{ \alpha^{ \prime} }{ \alpha }
\right)  
\quad .
\label{E-source-general-both}
}
Using the geometry equations (\ref{hamiltonian-both})-(\ref{slicing-both}), 
this simplifies to
\beq{
\Psi_E =
- 4 \pi r X \left[ \, S \left( \Xi^2 + \Upsilon^2 \right) 
\ + \ \Xi \Upsilon \left( S v + P + E \right) \right] 
\quad .
\label{E-source-both}
}
As a check, it is clear that $\Psi_E = 0$ when $\Xi = \Upsilon = 0$ 
as it is in (\ref{E-eom}). 
Using (\ref{E-eom-both}) and (\ref{ideal-D-eom}) with the definition of $\tau$, 
$\tau = E - D$, we get the EOM for $\tau$:
\beq{
\dot{\tau} + \frac{1}{r^2} \left[ r^2 X v \left( \tau + P \right) \right]^{ \prime}
= \Psi_E 
\label{tau-eom-ideal-both}
}
where $\Psi_E$ is given in (\ref{E-source-general-both}), (\ref{E-source-both}).

In state vector notation, the EOM's for $( D, \Pi, \Phi )$ obey a conservation law 
(\ref{conservationeq}), where the state vectors, except the source $\greekbf{\psi}$,  remain the same:
\beq{ 
\mathbf{q} =  \left[ \begin{array}{c} D \\ \Pi \\ \Phi \end{array}\right]  
\quad , \quad
\mathbf{f} = \left[ \begin{array}{c} D v \\ v \left( \Pi + P \right) + P
        \\ v \left( \Phi + P \right) - P\end{array} \right]
\quad , \quad 
\greekbf{\psi} = \left[ \begin{array}{c} 0 \\ \Psi_\Pi \\ \Psi_\Phi \end{array} \right]
\label{ideal-piphi-state-vectors-both}
}
The new sources are given by
\beqa{
\Psi_\Pi \ &= \ &\Sigma_\circ \ - \ 
     2 \pi r X  \, 
	\frac{ \rho_\circ \, h \left( 1 + v \right) }{ \left( 1 - v \right) }
	\left( \Xi + \Upsilon \right)^2 
\label{psi-pi} \\[0.3cm]
\Psi_\Phi \ &= \  &-\Sigma_\circ  \ + \ 
     2 \pi r X  \,
	\frac{ \rho_\circ \, h \left( 1 - v \right) }{ \left( 1 + v \right) }
	\left( \Xi - \Upsilon \right)^2 
\label{psi-phi}
}
where 
\beq{
\Sigma_\circ = \Theta_\circ + \frac{2PX}{r}  \label{Theta-0}
}
and $\Theta_\circ$ is the first term on the right-hand side of (\ref{Theta-both}):
\beq{
\Theta_\circ \ \equiv \alpha a \left\{  \left( S v - E \right) \left[ 
4 \pi r \left( 2 P - V(\phi) \right) + \frac{m}{r^2} \right]
+ P \left( \frac{m}{r^2} - 4 \pi r V(\phi) \right) \right\} 
\label{theta-o}
} 

\subsubsection{Ultra-relativistic Fluid EOM for ``Scalar+Fluid'' System}

The ultra-relativistic fluid shares the same EOM's as the ideal gas, except that $E = \tau$ 
and $D$ drops out of the system.  Hence, it can be described by the last two EOM of the ideal gas, 
which take the form (\ref{conservationeq}) with the original state vectors (\ref{newstatevectors}).
However, the source vector is now
\beq{
\greekbf{\psi} = \left[ \begin{array}{c} \Psi_\Pi \\ \Psi_\Phi \end{array} \right]
\label{newstatevectors-both}
}
where $\Psi_\Pi$ and $\Psi_\Phi$ are given by (\ref{psi-pi}),(\ref{psi-phi})---respectively---and
$E$ is replaced by $\tau$. 

\section{Initial Star Solutions}
\label{sec:init-star-solut}

In this thesis, we model neutron stars as spherically-symmetric, static solutions to Einstein's
equations with a stiff equation of state.  The equations describing spherical, hydrostatic  solutions 
in general relativity were first derived---to the best of our knowledge---in 1934 by Tolman 
\cite{tolman-book}.  The equations he found 
are similar to those that we 
use, which are:
\beqa{
\frac{d m}{d r} \ & = & \  4 \pi r^2 \rho \label{TOV-dmdr}  \\
\frac{d P}{d r} \ & = & \  
- \, \frac{ \left( \rho + P \right) \left( m + 4 \pi r^3 P \right) }
{ r \left( r - 2 m \right) }  \label{TOV-dPdr} \\[0.35cm]
\frac{d \varphi }{d r} \ & = & - \, \frac{ 1 }{ \, \rho + P \, } \,  \frac{ d P }{ d r } 
\label{TOV-dphidr}
}
where 
\beq{
\varphi \equiv \ln{ \alpha } \quad ,
\label{tov-polar-areal}
}
These equations are derived from the Einstein-fluid equations under the assumption that
the fluid and geometry are both spherically-symmetric and static.  

Tolman found closed-form solutions---both new and previously known---to (\ref{TOV-dmdr}-\ref{TOV-dphidr}) 
by making explicit 
assumptions about the metric functions \cite{tolman-paper}.  In the preceding article of the same journal volume, 
Oppenheimer and Volkoff \cite{oppenheimer-volkoff} used Tolman's methods and equations to 
calculate models for neutron stars.  Similar to white dwarfs, neutron stars are thought to be 
supported by the degeneracy pressure of a fermionic gas.  In the case of the neutron star, the 
neutrons form a degenerate gas, which can be considered to be at a negligible temperature since its
Fermi energy is well above the neutrons' anticipated thermal energies.  The momenta of the 
neutrons in the Fermi levels then results in an effective pressure that counters the inward pull of gravity.  
Earlier that decade, Chandrasekhar \cite{chandra-1931,chandra-1935} studied the equation of 
state of non-relativistic and relativistic degenerate electron gases in order to describe the 
state of matter at the core of white dwarfs.  Since degenerate neutron and electron gases
are fundamentally the same---i.e. they consist of particles obeying Fermi statistics---Oppenheimer 
and Volkoff were able to adopt Chandrasekhar's equation of state to their study of neutron star 
solutions.  Through numerical means, they solved the system of equations for a series of central 
densities (Oppenheimer and Volkoff actually use another parameter, but this 
parameter---in turn---monotonically parameterizes the central density).  
In their investigation, they found evidence suggesting there was a maximum stable mass for 
these equilibrium solutions.  A similar mass limit was found first for white dwarfs 
in 1931 by Chandrasekhar \cite{chandra-1931} using Newtonian gravity.  Since the two matter models 
are fundamentally the same, both mass limits are named after Chandrasekhar and are called the 
\emph{Chandrasekhar mass limits} for neutron stars and white dwarfs.  In addition, since Oppenheimer 
and Volkoff were the first to solve Tolman's equation with a realistic equation of state, the system
of equations (\ref{TOV-dmdr}-\ref{TOV-dphidr}) are named after the trio as the 
Tolman-Oppenheimer-Volkoff (TOV) equations. 

In order to numerically solve the TOV equations, we must close the system with a state equation.  
We use the ideal gas law (\ref{ideal-eos}), (\ref{rho}), and the polytropic equation of state,
\beq{
P  \ = \ K \rho_\circ^\Gamma  \quad .
\label{polytropic-eos}
}
In geometrized units ($G=c=1$)
the constant $K$ sets the length scale of the system, so we have set $K=1$ for all 
cases given here in order to make all quantities dimensionless \cite{cook-shap-teuk-1992}.
The transformation from these units to more common, astrophysical units is discussed in Appendix~\ref{app:unit-conversion}.

Since the only freedom in the TOV equations is the central value of the pressure and 
the EOS governing the fluid, we may parameterize our TOV solutions 
with the value of the rest-mass density at the origin---$\rho_c$---and the adiabatic
index---$\Gamma$.  We use $\Gamma = 2$ for all results shown in this 
thesis, so the solutions only depend on $\rho_c$.  
Once $\rho_c$ has been specified, the TOV equations are solved using a similar 
method to one described in \cite{shapiro-and-teuk}, first described by Tolman \cite{tolman-book}.  
However, in contrast to \cite{shapiro-and-teuk},
we do not integrate the equations until $P \ge 0$ is no longer satisfied, but rather we
pick a non-zero, positive threshold for $P$ that determines when we stop the outwards integration.  Specifically, 
we integrate the equations
from the origin out to the radius of the star, $R_\star$,  which we define as 
the smallest radius to satisfy $P(r) \le P_{\mathrm{floor}}$ with $P_{\mathrm{floor}}$ 
being a small constant that is usually $10^{-13} P(r=0)$.  This allows us to continuously  match the 
star solution to a constant atmosphere---or \emph{floor} (see Section~(\ref{sec:floor}) for a 
description of what the floor is and why it is used)---outside of the star so that 
$P(r>R_\star) = P_{\mathrm{floor}}$.  The metric functions are continued past the star's radius by 
matching to the Schwarzschild solution: 
\beq{
m(r>R_\star) = m(r=R_\star) \equiv M_\star \quad , \quad 
\varphi(r>R_\star) =  \ \frac{1}{2} \log{ \left( 1 - \frac{ 2 M_\star }{ r } \right)}   
\label{schwarzschild-matching}
}
Finally, in order to recover the metric functions $\alpha$ and $a$ we use the inverse of the relations 
(\ref{massaspect}) and (\ref{tov-polar-areal}).  

We will call a set $\{\alpha, a, P\}$ of functions calculated in the previously prescribed manner
a ``TOV solution.''  Note, however, that the such a solution will not strictly be completely static since
the energy density of the fluid in the atmosphere region outside of the star is not a
solution of the TOV equations.  Also, the interior---or star-like---part of the solution will 
not be perfectly static as it is perturbed slightly by inherit inaccuracies in the discretization of 
the TOV equations and the equations of motion, and by accretion of the atmosphere onto the star. 
The atmosphere's total mass is typically below $0.01 M_\star$, it has been observed to have little effect on the 
dynamics of the collapsing stars.  

\begin{figure}[htb]
\centerline{\includegraphics*[bb=0.3in 2.2in 8.2in 9.8in, scale=0.4]{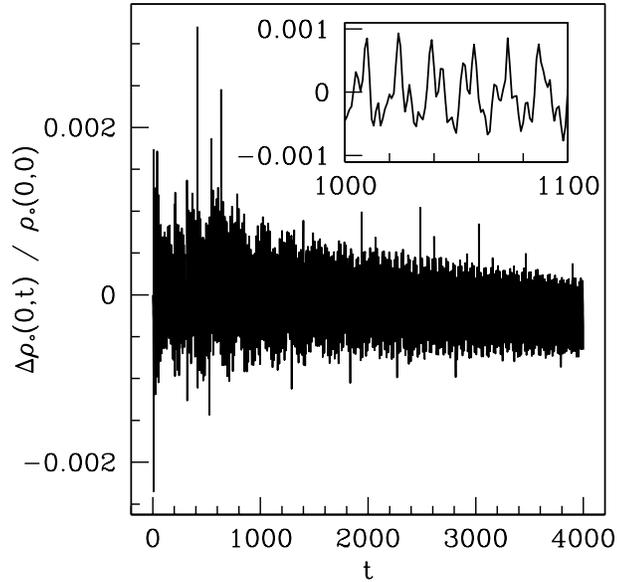}}
\caption[The relative change in the central density of a TOV solution is shown as a function of 
time measured by an observer at space-like infinity. ]{The relative change in the central 
density of a TOV solution is shown as a function of time measured by an observer at space-like infinity.  
The oscillations are due to truncation error in finding the 
numerical representation of the TOV solution, and from interactions with the artificial atmosphere resulting 
from the floor imposed on the pressure. 
The dissipation inherent in the numerical methods and 
the star's transfer of energy to the atmosphere causes the average value of $\rho_\circ(0,t)$
to decay over time.   A closer 
view of the oscillation over a few fundamental periods is given in the inset plot in the 
upper-right corner.  The fundamental period can be measured from time separation of the largest
peaks, which is approximate $t_0\simeq14.5$ for this solution.  Hence, the larger plot shows 
approximately $275$ fundamental oscillations.   The particular TOV solution used 
for the initial data has $\rho_c=0.05$ and $\Gamma=2$.
\label{rhoc-tov}}
\end{figure}

\clearpage

\begin{figure}[htb]
\centerline{\includegraphics*[bb=0.3in 2.2in 8.2in 9.8in, scale=0.4]{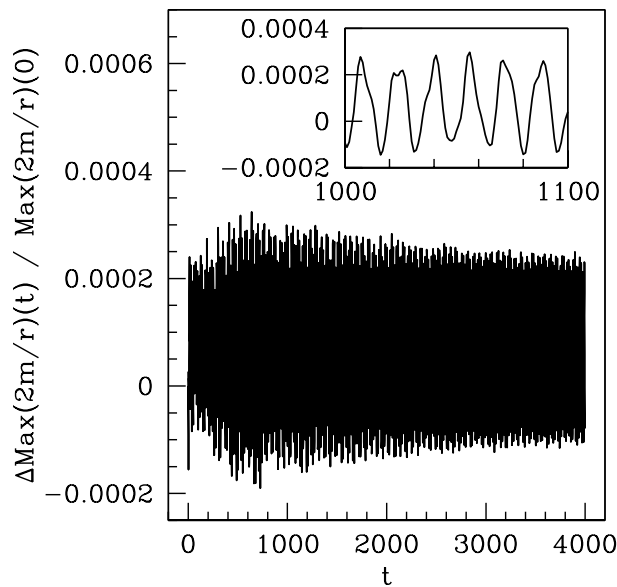}}
\caption[The relative change in ${\max}(2m/r)$ of a TOV solution is shown as a function of 
time measured by an observer at space-like infinity. ]{The relative change in ${\max}(2m/r)$ 
of a TOV solution is shown versus the time measured by an observer at space-like infinity. 
The oscillations are explained in the caption of Figure~\ref{rhoc-tov}.  The inset shows a detailed
view of a few periods.  We set $\rho_c=0.05$ and $\Gamma=2$ for the initial 
TOV solution shown here. 
\label{maxtmr-tov}}
\end{figure}

\begin{figure}[htb]
\centerline{\includegraphics*[bb=0.3in 2.2in 8.1in 9.8in, scale=0.55]{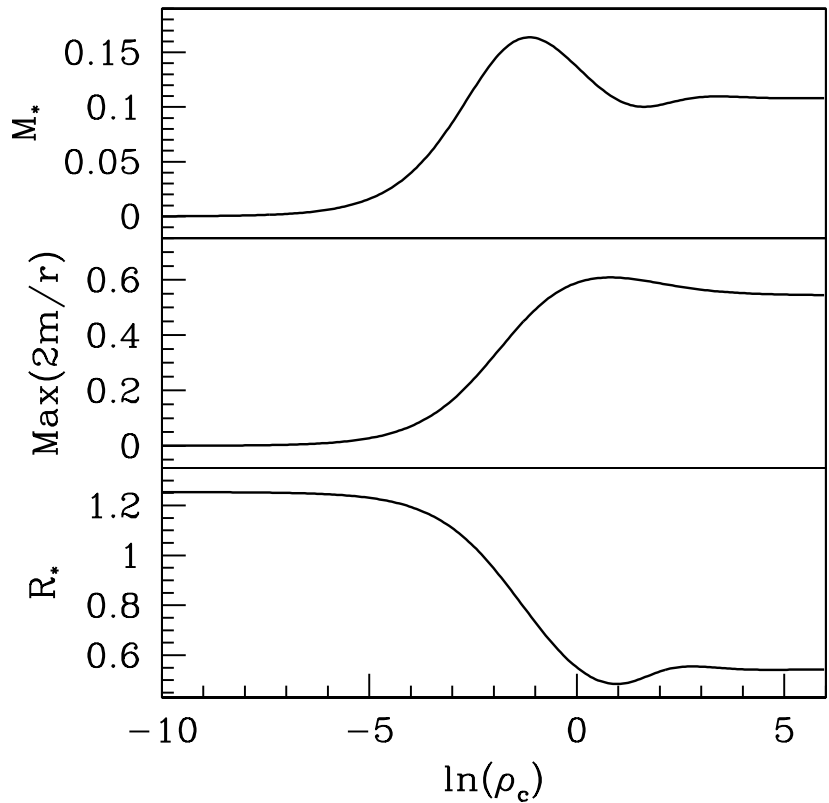}}
\caption[The mass, ${\max}(2m/r)$, and radius of TOV solutions as a function of central density.]{The mass, 
${\max}(2m/r)$, and radius of TOV solutions as a function of central density. These solutions were 
found using the polytropic EOS (\ref{polytropic-eos}) with $\Gamma=2$ and $K=1$. 
\label{tov-soln} }
\end{figure}

Since the initial work summarized above, the TOV solutions have been studied a great deal.  An excellent historical
account of these analyses was written by Harrison et al. \cite{harrison-etal}, but since that 
reference is a little out of date, we will defer to the description given in Shapiro and Teukolsky 
\cite{shapiro-and-teuk}.  As suggested above,  the TOV solutions may be uniquely 
parameterized by their central densities, $\rho_c$. 
A solution can be further characterized  by its mass ($M_\star$), its radius
($R_\star$) and the maximum value that  $2m/r$ takes within the star (${\max}(2m/r)$).  Even though
each solution has a unique $\rho_c$, $M_\star$, $R_\star$, and ${\max}(2m/r)$
are not necessarily one-to-one with respect to $\rho_c$.
To illustrate this, we have shown these three quantities versus $\rho_c$ in Figure~\ref{tov-soln}.  
From these distributions, we can clearly identify that there exists a global maximum mass that these
solutions can have, which is the previously mentioned Chandrasekhar mass for neutron stars. 
Also, the solutions are all finite and non-zero in extent, with compactification 
factors---${\max}(2m/r)$---less than $\simeq 0.61$ for the particular equation of 
state used here. 
Even though these distributions all represent hydrostatic  solutions to Einstein's equations in 
spherical symmetry, the question of stability must still be considered.  By calculating the normal, radial 
modes of oscillation of these static solutions, we can determine which solutions are stable or 
unstable---i.e. which perturbations are oscillatory and which are exponentially growing.  
If we define $\rho_c^{\max}$ as the central density of the maximum mass solution, 
it can be shown that stable TOV solutions are those for which $\rho_c < \rho_c^{\max}$. 

\begin{figure}[htb]
\centerline{\includegraphics*[bb=0.3in 2.2in 8.1in 9.8in, scale=0.55]{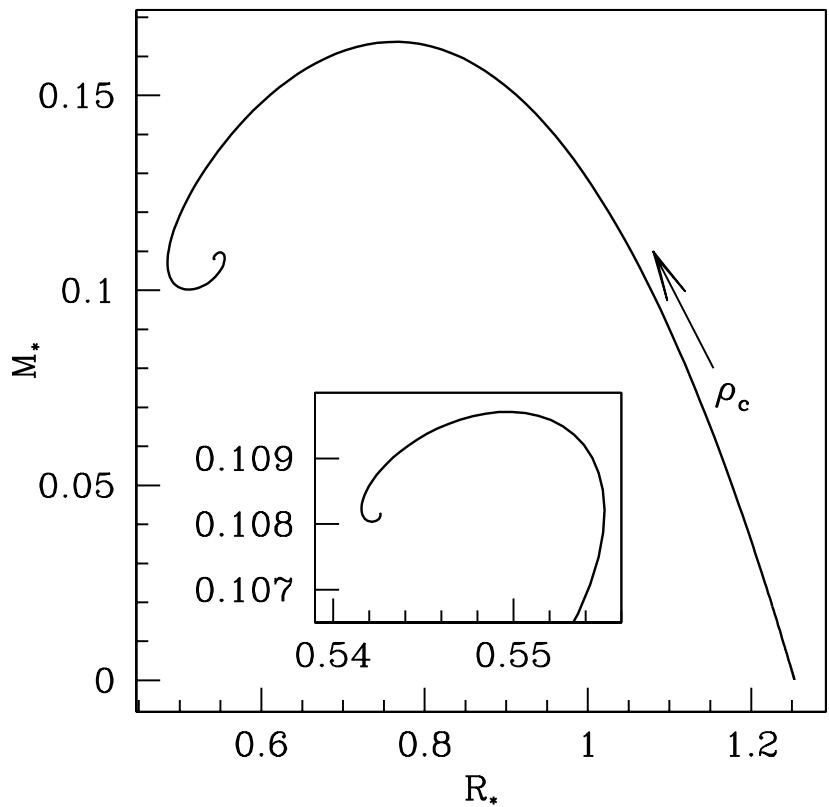}}
\caption[Mass versus radius of TOV solutions using $\Gamma=2$ and $K=1$ with the polytropic EOS 
(\ref{polytropic-eos}). ]{Mass versus radius of TOV solutions using $\Gamma=2$ and $K=1$ with the 
polytropic EOS (\ref{polytropic-eos}). In the inset, we show a detailed view of the spiraling behavior. 
The arrow along the right side of the curve indicates the direction of increasing central density. 
\label{tov-m-v-r}}
\end{figure}

The stability properties of the solutions can be further illustrated by looking at the distribution of 
$M_\star$ versus $R_\star$, Figure~\ref{tov-m-v-r}.  Here, we see that $M_\star(R_\star)$ winds-up with increasing
central density.  At the global maximum of $M_\star(R_\star)$ the fundamental, or lowest, mode 
becomes unstable.  After each subsequent local extremum of $M_\star(R_\star)$ in the direction of increasing
$\rho_c$, the next lowest mode becomes unstable.  For instance, there are four local extrema of
$M_\star(R_\star)$ shown in Figure~\ref{tov-m-v-r}, so those solutions with the largest $\rho_c$ will 
have their four lowest modes exponentially grow.  

As discussed previously, black hole critical solutions are typically characterized by a single growing mode.
Hence, the Type~I behavior associated with ``perturbed'' TOV solutions can be immediately anticipated to entail 
those TOV solutions that lie between the first and second extrema of $M_\star(\rho_c)$.  

After the initial, star-like solution is calculated, an in-going velocity profile
is sometimes added to drive the star to collapse.  In order to do this, we follow
the prescription used in \cite{gourg2} and \cite{novak}.  The method described therein entails specifying the 
coordinate velocity 
\beq{
  U \equiv \frac{ d r }{ d t } = \frac{ u^r }{ u^t }   \quad .
  \label{radial-coord-velocity}
}
of the star, and then finding the physical velocity, $v$, once the geometry has been calculated.  
In general, the profile takes the algebraic form:
\beq{ 
U_g(x) = A_0 \left( x^3 - B_0 x \right) \quad . \label{v-profile-general}
} 
The two profiles that were used in \cite{novak} are 
\beqa{
U_1(x) & = &  \frac{U_\mathrm{amp}}{2} \left( x^3 - 3 x \right)  \label{v-profile-1} \\[0.4cm]
U_2(x) & = &  \frac{27 \, U_\mathrm{amp}}{10 \sqrt{5}} \left( x^3 - \frac{5 x}{3} \right)  
\label{v-profile-2} \\[0.4cm]
x & \equiv & \frac{r}{R_\star} \label{x-def}
} 
Unless stated otherwise, $U_1$ profile will be used for all the results herein. 

Specifying the coordinate velocity instead of the \emph{Eulerian} velocity, 
$v= a U / \alpha$, 
couples the Hamiltonian constraint (\ref{hamconstraint}) and the slicing condition 
(\ref{slicingcondition}) by introducing $\alpha$ and $a$ into the right-hand sides of 
them.  In order to explicitly show how the right-hand side  changes, the conservative variables
must be expressed in terms of the coordinate velocity and primitive variables 
via (\ref{D}-\ref{tau}):
\beq{  
\frac{a '}{a} = a^2 \left\{ 4 \pi r \left[ 
\frac{\rho_\circ h}{1 - \left(\frac{a U}{\alpha}\right)^2} - P \right] - \frac{1}{2 r^2} \right\} 
+ \frac{1}{2 r^2}
\quad , \label{vp-hamconstraint}
}
 
\beq{
\frac{\alpha '}{\alpha}  
= a^2 \left\{ 4 \pi r \left[ \rho_\circ h 
\frac{ \left(a U / \alpha\right)^2}{1 - \left(\frac{a U}{\alpha}\right)^2}
  + P \right] + \frac{1}{2 r^2} \right\}  - \frac{1}{2 r^2}
\label{vp-slicing-condition} 
}

The coupling of these equations makes their numerical solution more involved, and 
the following is the prescription used to solve them:
\begin{enumerate}
\item $\left\{P(r), \rho_\circ(r), a(r), \alpha(r)\right\}_{\mathrm{TOV}}$ are calculated 
  using (\ref{tov-polar-areal})~-~(\ref{polytropic-eos}) with the usual
  regularity conditions (see Section~\ref{sec:geom-bound-cond}
  for a more thorough discussion of the regularity conditions imposed on the geometric fields) 
  at the origin, and with a match to the Schwarzschild metric at the star's boundary via 
  reparameterization of $\alpha$ such that $\alpha a |_{r=r_{\max}} = 1$;
\item Given $U_\mathrm{amp}$, $U(r)$ is specified via (\ref{v-profile-1}) or (\ref{v-profile-2}), and
$\left\{\alpha(r) , a(r)\right\}_\mathrm{VP}$ are calculated via a 2-dimensional Newton-Raphson 
method which solves (\ref{vp-hamconstraint})-(\ref{vp-slicing-condition}) at each grid point.  The 
integration starts at the origin with
\[\left\{\alpha(r=0) , a(r=0)\right\}_\mathrm{VP}
= \left\{\alpha(r=0) , a(r=0)\right\}_\mathrm{TOV}\] and continues 
outwards to $r_{\max}$.  The Eulerian velocity, $v$, is then calculated by 
$v = U a_\mathrm{VP} / \alpha_\mathrm{VP}$. 

\item Since the parameterization for $\alpha_\mathrm{VP}$ was chosen at the origin, the 
  outer boundary condition, $\alpha a |_{r=r_{\max}} = 1$, will not necessarily be satisfied.  
  In order to impose this outer boundary condition on the solution and to calculate the 
  final values of $\alpha$ and $a$, the uncoupled 
  Hamiltonian (\ref{hamconstraint}) and the slicing condition (\ref{slicingcondition}) 
  are solved using the $v$ calculated in the previous step.  
\end{enumerate}

The process of recalculating $a$ and $\alpha$ from the uncoupled equations 
(\ref{vp-hamconstraint}-\ref{vp-slicing-condition}) and using 
$v = U a_\mathrm{VP} / \alpha_\mathrm{VP}$ in the source terms of those equations means that 
$v$ will no longer be consistent with the initial coordinate velocity profile, $U(r)$, 
since---in general---$\left\{ \alpha , a \right\} \ne \left\{ \alpha , a \right\}_\mathrm{VP}$. 
If we define $U_\mathrm{final} = \left(v \alpha / a\right)$  to be the coordinate velocity at the 
end of the procedure outlined above, then $U_\mathrm{amp}$ parameterizes a family of 
functions $U_\mathrm{final}$ just as it parameterizes the final Eulerian velocity function $v$. 
We have found through an extensive numerical search, that for any $\rho_c$ we tried, 
the minimum of $U_\mathrm{final}(r)$ as a function of $U_\mathrm{amp}$ had at least one extremum 
suggesting that every star has degenerate values of the minimum of $U_\mathrm{final}$. 
Let $\tilde{U}_\mathrm{amp}(\rho_c)$ be the value of $U_\mathrm{amp}$ at which the first 
extremum is located for a given star with central density $\rho_c$.  Then, we find that 
only for $U_\mathrm{amp} < \tilde{U}_\mathrm{amp}(\rho_c)$, are $U_\mathrm{final}$ 
and $U(r)$ proportional to within truncation error with the constant of 
proportionality equal to $\left(\alpha_\mathrm{VP} a_\mathrm{VP}\right)|_{r=r_{\max}}$.  
In other words, it seems that $\alpha$ can still be freely reparameterized when 
$U_\mathrm{amp} < \tilde{U}_\mathrm{amp}(\rho_c)$ even though the  
coupled set of equations (\ref{vp-hamconstraint}-\ref{vp-slicing-condition}) are inhomogeneous 
in $\alpha$.  
For $U_\mathrm{amp} > \tilde{U}_\mathrm{amp}(\rho_c)$, the solution we obtain is made consistent 
with the outer boundary conditions because of the last step of the procedure; this very step, however, 
makes $U_\mathrm{final}(r)$ not proportional to its intended form, $U_1(x)$ (\ref{v-profile-1}).  
Hence, we will term those cases with 
$U_\mathrm{amp} > \tilde{U}_\mathrm{amp}(\rho_c)$ as \emph{not} being a solution to our 
procedure for calculating the initial data for velocity-perturbed TOV stars.  Fortunately, most of 
the phenomena we are interested in lies within this region $U_\mathrm{amp} < \tilde{U}_\mathrm{amp}(\rho_c)$.  
Also, it seems that Novak 
\cite{novak} was unable to find solutions at all above $\tilde{U}_\mathrm{amp}(\rho_c)$;  comparing 
our values for $\tilde{U}_\mathrm{amp}(\rho_c)$, seen as the line dividing the two uppermost regions
in Figure~\ref{fig:pspace}, to his we find fair agreement.  


\chapter{Numerical Techniques}
\label{chap:numerical-techniques}

In this section we describe the numerical techniques used 
to simulate the highly-relativistic flows encountered in the driven
collapse of neutron stars.  The simulations entail solution of a system of coupled, partial and ordinary 
differential equations that describe how the fluid, scalar field, and gravitational field evolve in time.  
The following sections contain explanations and a few numerical tests of the procedures we employ.  
Most of the discussion regards those methods used to treat the hydrodynamical flow since they are the 
most complicated and innovative.  Without the fluid methods we developed for this
work,  a large portion of the results would have been unattainable.  A description of the problems 
encountered and their solution is given along the way. 

To handle check-pointing, input/output, and memory management, 
we use the Rapid Numerical Prototyping Language (RNPL) written by Marsa and Choptuik \cite{marsa-choptuik}. 
RNPL is a high-level language that frees the user 
from having to write procedures common to most finite difference programs.  RNPL's language 
and infrastructure requires the user to specify the grid functions and run-time parameters, a 
list of all the finite difference equations to solve, and calls to external routines if any other 
calculations need to be done which cannot be performed within the RNPL environment.  During compilation, RNPL 
generates all the code needed to solve the finite difference equations. 
Even though RNPL is straightforward to use for finite difference algorithms, we use 
various finite volume techniques
that cannot be implemented in RNPL's framework.  Thus, we use secondary, external  routines 
that are called by the primary RNPL procedure in order to update all grid functions.  RNPL, then,
is used only to \emph{drive} the time-stepping process.

\section{Finite Differencing}
\label{sec:finite-differencing}

Finite difference (FD) algorithms are computational techniques used to solve 
partial differential equations (PDEs) by approximating them as systems of discrete algebraic equations.  They are 
typically used to solve equations with no known closed-form solutions, allowing the user to find solutions 
within some degree of accuracy, depending on the particular implementation used.  Even though they 
have existed for  hundreds of years, it was not until the invention of the computer that they 
became prevalent \cite{kopal}.  The computer allows scientists to perform the tedious, repetitive 
calculations necessary to solve FD equations (FDEs).  As FDE solutions became easier to calculate, methods 
grew more complex in order to improve solution accuracy and/or stability.  Now, the subject
of finite difference approximations is fundamental to numerical analysis.  
In this section, we will provide a brief introduction to techniques used for solving PDEs with FDEs,
and for ensuring that the numerical solution is a good approximation to the continuum solution.
For notation and guidance, we will use an introduction to the finite difference solution of PDEs written by Choptuik 
\cite{choptuik-taller-1999}.   

Let us consider a differential equation of the form 
\beq{
L u = f    \label{differential-operator}
}
where $L$ represents a differential operator, $u$ is the continuum function which we are trying to calculate,
and $f$ is a source term.  For simplicity, let us consider a system that is only dependent on  a 
time-coordinate, $t$, and a space-coordinate, $x$;  however, our discussion on FD methods is 
valid for vector equations---where $u$ and $f$ are vectors---and where $u$ and $f$ depend on an 
arbitrary set of coordinates.  We thus have $u=u(x,t)$ and $f=f(u,x,t)$, $f$ may  explicitly
depend on $u$.   In these coordinates, the differential operator must take the form 
$L=\partial_t$, $L=\partial_x$, or $L=\partial_{tt}  - v^2(x,t) \partial_{xx}$, for example, where
the last case describes a wave equation with characteristic velocities $\pm v(x,t)$.  
In order to make a FD approximation to this differential equation, a discrete domain of points
must be introduced on which the solution will be defined.  The spacing between each adjacent 
pair of grid points, $h$, can---in general---be a function of $x$ and $t$, but we will only consider
grids with constant $h$ for our introductory discussion.  Also, any function defined on this \emph{grid}
of points will be called a \emph{grid function}.
Then, the discrete version of (\ref{differential-operator}) would be 
\beq{
L^h u^h = f^h  \quad , \label{discrete-differential-operator}
}
where $u^h$ is the grid function representing the FDE solution, $f^h$ is the  discrete version of the
source, and $L^h$  is now an operator acting on discretized quantities.  As we will see, $L^h$---called
the difference operator---can be defined in a number of ways, and the accuracy of the resulting solution 
will depend on details of its construction.

FD operators are often found by approximating $u(x,t)$ by Taylor series 
expansions that are truncated to some order in order to obtain a discrete, or finite difference, 
approximation to the continuum one.  The quantity that represents the error in curtailing the series 
is called the \emph{truncation error}:
\beq{
\tau^h =  L^h u - f^h  \quad . \label{truncation-error}
}
In order for the FD approximation to be \emph{consistent} with the original PDE in the continuum limit, 
the truncation error must vanish:
\beq{
\lim_{h\rightarrow 0} \tau^h  = 0 \label{fd-consistency}
}

The consistency of the FDEs does not necessarily guarantee that the FD solution tends to the 
continuum solution.  For that, the \emph{convergence} of the numerical solution must be examined, which 
is done by considering the \emph{solution error}:
\beq{  
e^h \equiv u - u^h  \quad . \label{solution-error}
}
Specifically, a FD approximation is said to be convergent if 
\beq{
e^h \rightarrow  0   \quad \mathrm{as} \quad h \rightarrow 0  \quad .\label{convergence}
}
Hence, convergence measures how well $u^h$ approximates $u$, while consistency is how well 
the FDE approximates the PDE.  A connection between the two can be made with \emph{Richardson's expansion}, 
which predicts that the finite difference solution  deviates from the continuum solution can be expressed as an 
asymptotic series in terms of the grid spacing $h$:
\beq{
e^h = u - u^h = \sum_{n=1}^{\infty} h^n e_n \label{richardson-expansion-general}
}
where $e_n$ are functions of $(x,t)$ but not the grid spacing.  The expansion can be proven 
in some cases, but requires that the solution remain smooth.  This last fact is critical in understanding
the convergence properties of fluid flows with shocks.  We will define the 
order of the FD approximation, $O(h^l)$, to be the first non-zero order 
in (\ref{richardson-expansion-general}).  For instance,  Richardson's expansion for a so-called centered difference
approximation is one with only even-order terms:
\beq{
e^h = \sum_{n = 1}^{\infty} h^{2n} e_{2n} \label{richardson-expansion-2nd-order}
}
so the order of such a scheme would be $O(h^2)$ or \emph{$2^\mathrm{nd}$-order}.  
By default, all FD approximations we use for this work are  $2^\mathrm{nd}$-order or better, except in the 
vicinity of shocks (see Section~\ref{sec:reconstr-at-cell} for a discussion about the 
accuracy of finite volume methods near shocks).  However, the numerical solution is not expected 
to follow Richardson's expansion in such cases since the solution is inherently discontinuous.  

From the previous definitions, 
the truncation error can be shown to be related to the solution error:
\beq{ 
\tau^h = L^h u - f^h =  L^h \left( u^h + e^h \right)  - f^h  =  L^h e^h  - f^h   
\label{constinency-convergence}
}
where we have used (\ref{richardson-expansion-general}) in the second equality and 
(\ref{discrete-differential-operator}) in the third.   Even though the above expression (\ref{constinency-convergence})
assumed that $L^h$ is a linear operator, a similar asymptotic behavior can be gleaned from the general
case by linearizing the nonlinear equation about the solution, $u$.
Hence, the solution error
should have the same leading-order dependence on $h$ than the truncation error assuming that 
Richardson's expansion is valid.  

In order to determine the order at which a certain code is converging, the form of Richardson's expansion
can be exploited.  For example, if two numerical solutions  $u^{2h}$ and $u^h$ are calculated 
at resolutions $2h$ and $h$---respectively---with $O(h^l)$ methods, then their 
difference can be given in terms of the Richardson's expansion:
\beqa{
u^{2h} - u^h & = & \left( u - \sum_{n=l}^{\infty} (2h)^n e_n \right)  
- \left( u - \sum_{n=l}^{\infty} h^n e_n \right)  
\nonumber \\[0.2cm]
& = & \sum_{n=l}^{\infty} \left(2^n-1\right)  h^n e_n 
 = \left(2^l-1\right) h^l e_l + O(h^{l+1})   \quad . 
\label{convergence-levels}
}
Repeating this process for $u^{4h} - u^{2h}$ yields 
\beq{
u^{4h} - u^{2h} = 2^l \left(2^l-1\right) h^l e_l + O(h^{l+1})   \quad . 
\label{convergence-levels2}
}
To leading order then, we can relate $u^{2h} - u^h$ and $u^{4h} - u^{2h}$ to each other by the 
so-called convergence factor, $Q_\mathrm{cf}$, defined by the following relationship:
\beq{
Q_\mathrm{cf}(x,t)  \equiv  \frac{u^{4h} - u^{2h}}{u^{2h} - u^{h}}   \quad . 
\label{convergence-factor}
}
If we assume that the FD approximate employed is precisely $O(h^l)$ for all $x$ and $t$, 
then---by (\ref{convergence-levels}-\ref{convergence-factor})---the convergence factor should be a 
constant to leading order:
\beq{
Q_\mathrm{cf}(x,t)  \simeq 2^l  \quad.  \label{affective-convergence-factor}
}
Since quantities such as  $u^{4h} - u^{2h}$ may sometime vanish at certain points, we often simultaneously
plot $u^{4h} - u^{2h}$ and $2^l\left(u^{2h} - u^{h}\right)$; any regions where the curves 
do not overlap signifies a departure from the anticipated Richardson expansion.  If the deviation is 
fairly small, then the FD approximation follows a Richardson's expansion---thereby suggesting that the scheme is 
convergent.

However, even if $Q_\mathrm{cf}(x,t)$ is calculated to be the expected value from Richardson's expansion, 
the FDE may be approximating the wrong PDE.   For instance, a particular FD approximation can be
$O(h^l)$ accurate if its FDEs are incorrectly derived from the PDEs in such a way as to 
approximate another set of PDEs to $O(h^l)$ accuracy.   A trivial example of such an 
error would be to add an erroneous constant to $f$ when making the FD approximation, so that 
$f \rightarrow f^h + \mathrm{const.}$. 
To test for such errors, \emph{independent residual operators}
are used.  The key idea here is that a given $L$ can be approximated by many 
finite difference operators that each approximate $L$ to some order.  Let $u^h$ be the 
FD solution resulting from the use of the $O(h^l)$ difference operator, $L^h$, and let $\tilde{L}^h$ 
be a distinct $O(h^l)$ operator that also approximates $L$.  We also note that FD operators generally can be 
formally expanded in terms of $L$ and additional differential operators, $E_n$, by the definition of $L^h$:
\beq{
\tilde{L}^h = L + \sum_{n=l}^{\infty}  h^n E_n  \label{difference-richardson}
}
where the summation starts at $n=l$ since $\tilde{L}^h$ is $O(h^l)$.  
Then, we have
\beqa{
\tilde{L}^h u^h  & = & 
\left( L + \sum_{n=l}^{\infty}  h^n E_n \right) \left( u + \sum_{n=l}^{\infty}  h^n e_n \right) 
\nonumber \\[0.2cm]
& = & L u  + h^l \left( E_n u + L e_n \right) + O(h^{l+1}) \label{indep-resid-operator}
}
Since $E_n u$ and $L e_n$ are $O(1)$ quantities, $\tilde{L}^h u^h$ converges at the same 
order as the individual FDEs so that computation of $\tilde{L}^h u^h$ at resolutions $h,2h, \ldots$
can be used to validate the convergence of $u^h$. 
If $L^h$ and $\tilde{L}^h$ do not approximate the same $L$, then the expansion in (\ref{indep-resid-operator})
can be made and $\tilde{L}^h u^h$ will not converge 
as $O(h^{l})$.   In general, any inconsistency between $L^h$ and $L$ would lead to 
a $O(1)$ error in $u^h$, making $\tilde{L}^h u^h$  $O(1)$ accurate as well. 

Typically, the difference operator, $L^h$, is such that it yields a set of  
algebraic equations for $u^h$ whose solution can be found explicitly or implicitly.  Implicit FD 
approximations are often solved through iterative methods that solve (\ref{discrete-differential-operator}) 
to a preset precision.  Let $u^h_{(n)}$ represent an estimate for $u^h$ found after $n$ iterations. 
Since $u^h_{(n)}$ approximates $u^h$ to some precision, then $u^h_{(n)}$ will not satisfy 
(\ref{discrete-differential-operator}) exactly:
\beq{
r^h_{(n)} \equiv L^h u^h_{(n)} - f^h  \label{residual}
}
This deviation, $r^h_{(n)}$, is defined as the \emph{residual} of the difference equation after 
$n$ iterations.  The  goal of the implicit scheme is to then provide a  value of $u^h_{(n)}$ that 
reduces $r^h$ below some maximum allowed tolerance.   This tolerance is usually 
set to a value small enough so that the error in $u^h$ due to the implicit 
scheme's inability to drive $r^h$ precisely to zero is much smaller than the actual 
solution error, $e^h$.  Hence, this iteration error can usually be assumed to play an insignificant role in the 
error analysis described above.  

For completeness, we now give an example on the derivation of a FD
operator.  The particular operator that we will derive approximates  $\partial_x$.  In order to 
derive FD operators, we must define the discretization used.  Let the $x$-domain be discretized by 
width $\Delta x = h$, while the $t$-domain is discretized by $\Delta t = \lambda \Delta x = \lambda h$. 
The discrete coordinates can then be defined by $x_j = x_0 + j \Delta r$ and $t^n = t^0 + n \Delta t$.  
Then, we can define a grid function,  $u^n_j$, to be the FD approximation to the continuum 
value $u(x_j,t^n)$.  With these definitions and assuming that $h$ is small, then 
$u(x_{j+k},t^n)=u^n_{j+k}$ can be approximated by Taylor series expansion about point $x_j$:
\beq{
u(x_{j+k},t^n) = u^n_j + k h \, u'(x_j,t^n) 
+ \frac{(kh)^2}{2} u''(x_j,t^n) + \frac{(kh)^3}{3!} u'''(x_j,t^n)
+ O(h^4) \label{taylor-exp-1}
}
In general, (\ref{taylor-exp-1}) are solved for a set of $k$ about $k=0$---so that the derivative 
operator is ``centered.''  For a given order of accuracy, the act of centering the operator usually leads 
to a difference operator that requires the finite difference stencil of minimum width.
In order to calculate an $O(h^2)$  estimate for $\partial_x$, we need only calculate
(\ref{taylor-exp-1}) for $k=-1,1$ and solve for $u'(x_j,t^n)$:
\beq{
u'(x_j,t^n) = \frac{u^n_{j+1} - u^n_{j-1}}{2 h} - \frac{h^2}{3!} u'''(x_j,t^n) \label{d-dx}
}
We then take, 
\beq{
D_x u^n_j = \frac{u^n_{j+1} - u^n_{j-1}}{2 h} \label{discrete-dx}
}
as the $O(h^2)$ accurate difference operator for $\partial_x$ acting on $u^n_j$.

\begin{figure}[htb]
\centerline{\includegraphics*[bb=0.7in 2.2in 8in 9.9in, scale=0.4]{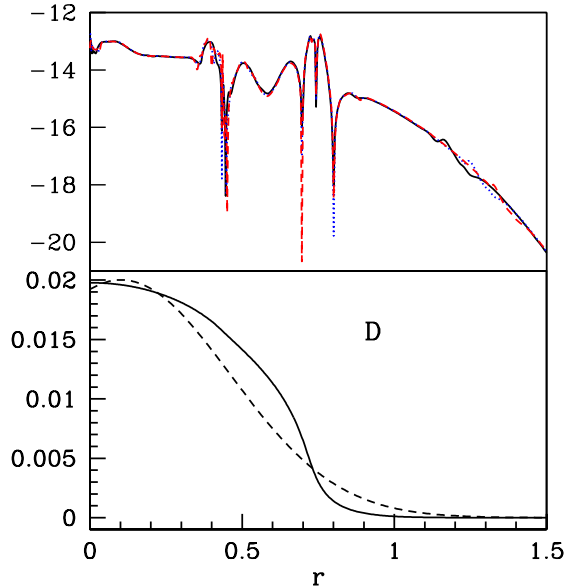}}
\caption[Convergence test for the fluid variable $D$.]{Convergence test for the fluid variable $D$.  
The top panel shows 
$\ln(3\tau^{8h}/4)\equiv D^{8h} - D^{4h}$ 
(black line), $\ln(3\tau^{4h})\equiv 4\left(D^{4h} - D^{2h}\right)$ (blue dots), 
$\ln(12\tau^{2h})\equiv 16\left(D^{2h} - D^{h}\right)$ (red dashes) which have been scaled such 
that they will look identical if our solutions are well-described by a Richardson expansion.  The bottom 
panel shows  $D(r,0)$ (black dashes) and $D(r,t)$, where $t$ is the time at which we performed the convergence test.
The initial data consisted of a self-gravitating fluid specified by a Gaussian function for 
$\rho_\circ$ centered at $r=0.1$ with an initial,
linear velocity profile. The initial grid used for the coarsest solution shown is defined by the parameters 
$\{N_a,N_b,N_c,\Delta r_a\}=\{ 200, 300, 20, 0.005 \}$;  please see Section~\ref{sec:grid} for definitions of 
these variables. 
\label{fig:D-converge}}
\end{figure}

\begin{figure}[htb]
\centerline{\includegraphics*[bb=0.5in 2.2in 8in 9.9in, scale=0.38]{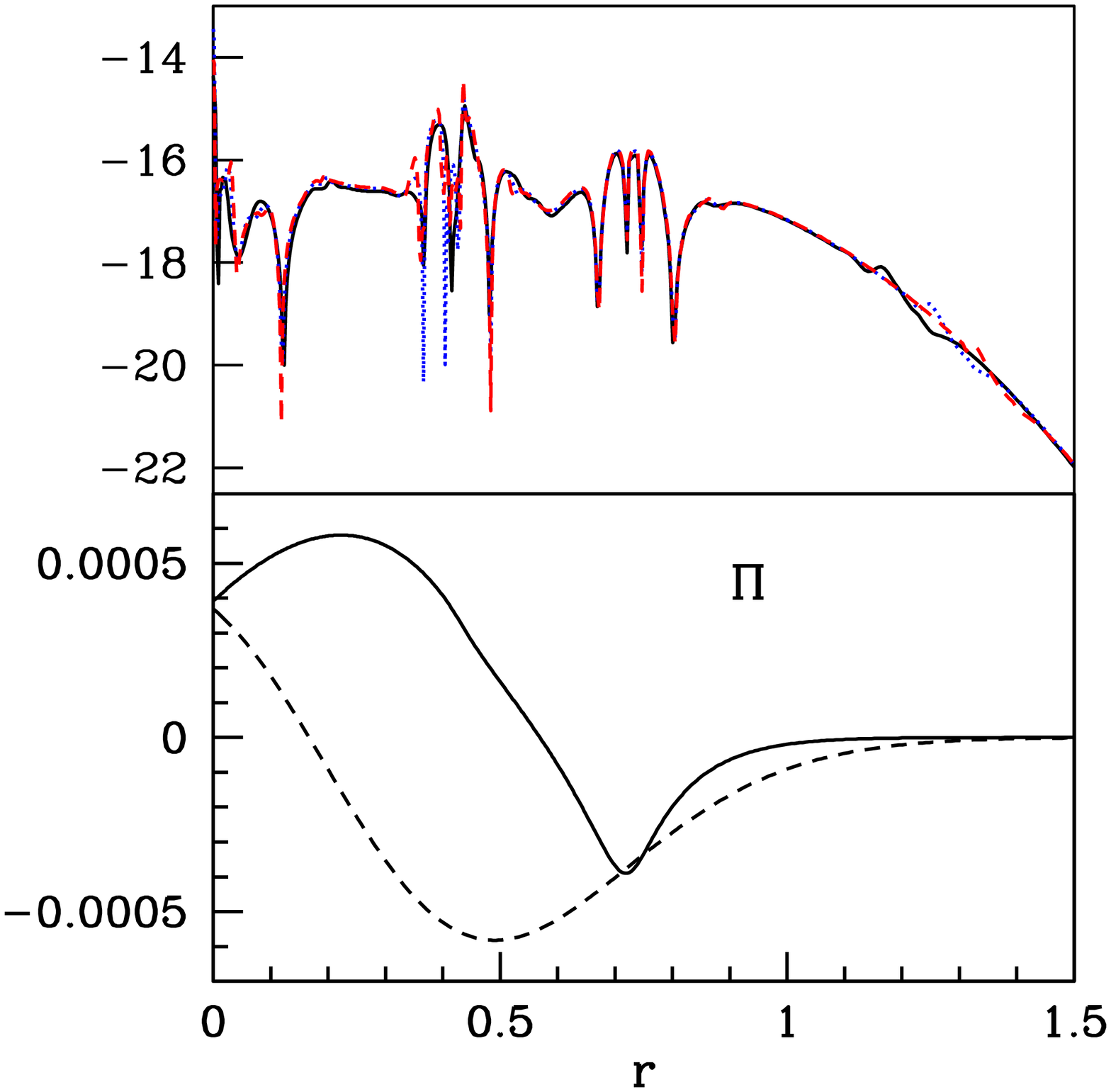}
\includegraphics*[bb=0.5in 2.2in 8in 9.9in, scale=0.38]{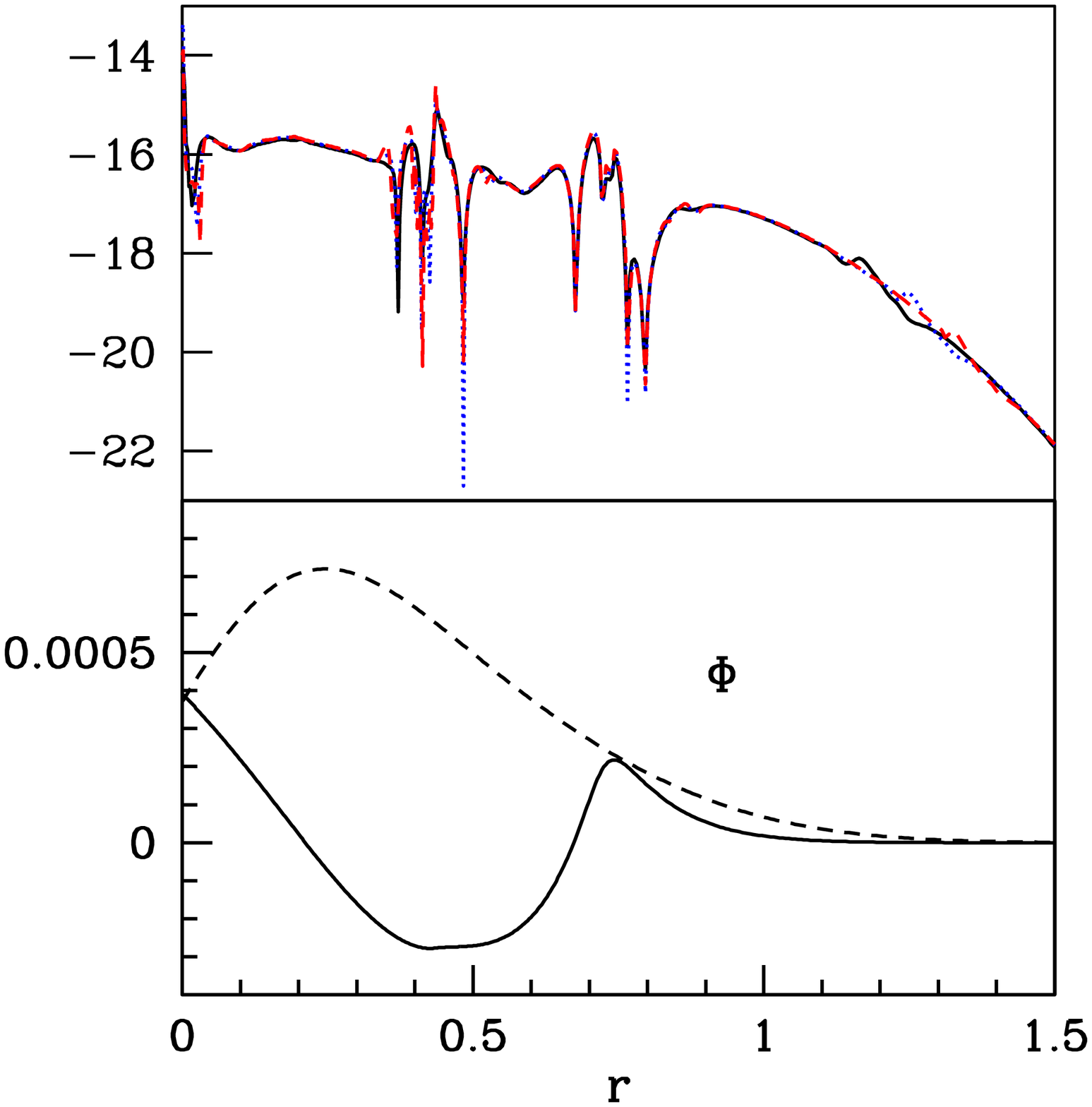}}
\caption[Convergence test for the fluid variable $\Pi$ (left) and $\Phi$ (right).  ]{Convergence 
test for the fluid variable $\Pi$ (left) and $\Phi$ (right).  The top panel 
of each figure show the scaled truncation error estimates for the respective fluid variable
as described for the grid function $D$ in the caption to Figure~\ref{fig:D-converge}.  The bottom
panels show $\Pi(r,0)$ and $\Phi(r,0)$ (dashed), and $\Pi(r,t)$ and $\Phi(r,t)$ (solid) where $t$ is the time
at which convergence is tested.
\label{fig:Pi-Phi-converge}}
\end{figure}

To illustrate convergence properties of finite difference approximations, we show the 
convergence of results from our hydrodynamic code in Figures~\ref{fig:D-converge}-~\ref{fig:Pi-Phi-converge}.  
Even though \emph{finite volume} methods (finite volume methods will be discussed in Section~\ref{sec:fluid-methods})
are based upon the idea of approximating integral equations instead of differential equations, 
the above analysis still holds for finite volume solutions \cite{leveque1}.  Shown in 
Figures~\ref{fig:D-converge}-~\ref{fig:Pi-Phi-converge} are---respectively---$D$ (\ref{D}), 
$\Pi$ (\ref{Pi-ideal}), and $\Phi$ (\ref{Phi-ideal}).  The scaled truncation error estimates shown in the 
top panels of each figure demonstrate how the code exhibits the expected dependence on $h$ 
from Richardson's expansion (\ref{convergence-levels}-\ref{convergence-levels2}) for each dynamic
fluid variable.  The data taken from a time step before any discontinuities were observed in the solution
to ensure that Richardson's expansion would remain valid.  In order to test the convergence of 
our regridding procedure---as described in Section~\ref{sec:refinement-procedure}---we calculated
the truncation error estimates at a time after the grid was refined once.  

\section{Introduction to Conservative Methods}
\label{sec:fluid-methods}

We employ High-Resolution Shock Capturing (HRSC) algorithms to solve the equations of 
motion for the fluid (\ref{conservationeq}).  Such methods have become increasingly popular 
in the field of relativistic hydrodynamics since they 
ensure: 1) conservation of the variables $\mathbf{q}$,  and 2) discontinuities---e.g. shocks---are 
well resolved and travel at correct speeds.   A key ingredient to these schemes is their use of solvers 
for the Riemann problem (see Section~\ref{sec:riemann-solvers} for a discussion of the Riemann problem)  at 
every cell interface.  This is crucial for the conservative nature of 
these schemes since the solution to the Riemann problem is 
always a weak solution of the hyperbolic conservation laws.  
The ``high-resolution'' aspect of the algorithms denotes that in regions where the grid 
functions are smooth, 
the integration procedure is at least $O(\Delta r^2)$ accurate.
The HRSC methods used in this work have been used in several previous works such as \cite{romero}, 
\cite{novak}, and \cite{neilsen} to name only a few relevant papers.  Also, many excellent references on 
conservative methods have been written by LeVeque \cite{leveque1,leveque2}; most of the development discussed here
has been gleaned from these texts.  

Conservation laws typically take the form of a differential equation, for example 
(\ref{flux-conservation-eq}) or specifically (\ref{conservationeq}).
However, these ``differential formulations'' of the conservation laws do not 
\emph{directly} follow from the original physical concepts involved and require that the dynamical variables be 
differentiable.  Recall that our fluid fields are really thermodynamics 
quantities and, therefore, averages over finite fluid elements---which we will call \emph{cells}
in numerical contexts.  Thus, the conservation laws result more naturally from \emph{integral} equations. 

In order to show the connection between the integral and differential formulations of conservation laws, let us 
consider the general case where ${ x^k }$ is an $N$-dimensional orthogonal, spatial, coordinate 
system, and let $V_i$ and $S_i$ represent the volume and surface---respectively---of cell 
$\mathcal{C}_i$.  A more covariant approach to conservative methods is given in \cite{pons-font-etal}.
So, in general, the \emph{differential} form of the conservation 
law that we wish to consider is:
\beq{
\partial_t \mathbf{q}\left(\vec{x},t\right) \ = \ 
- \nabla \cdot \vec{\mathbf{f}}\left(\mathbf{q}\right) 
\ + \ \greekbf{\psi}(\mathbf{q})
\label{general-diff-conserv-law}
}
where $\vec{\mathbf{f}}$ is the flux density vector with components $\{ \mathbf{f}^k \}$ 
in the basis of space-like coordinates $\{ x^k \}$, and $\greekbf{\psi}$ is a source term that 
involves no derivatives of the conserved variables $\mathbf{q}$.  A variable in 
boldface represents a vector or set of quantities 
(as in equation (\ref{newstatevectors-both})).  Such a differential conservation law
can be defined from the more general \emph{integral} equation:
\beq{
\frac{\partial}{\partial t} \int_{V_i} \mathbf{q}\left(\vec{x},t\right) dV 
\ = \ - \oint_{S_i} \vec{\mathbf{f}} \cdot d\vec{S} \ + \ \int_{V_i} \greekbf{\psi} \, dV 
\label{general-int-conserv-law}
}
where---for example---we have assumed that the volume and surface that are being 
integrated over are that of the cell $\mathcal{C}_i$, but any arbitrary volume can be used
in general.  This equation implies that any change over time in the ``amount'' of $\mathbf{q}$
in volume $V_i$ is due to its flux at the surface of $V_i$, and from its source or sink within $V_i$.  
Integrating this equation with respect to time, we get:
\beq{
\int_{V_i} \mathbf{q}\left( \vec{x}, t_2 \right) dV \, - \, 
\int_{V_i} \mathbf{q}\left( \vec{x}, t_1 \right) dV  
\ = \ - \int^{t_2}_{t_1} \oint_{S_i} \vec{\mathbf{f}} \cdot d\vec{S} \, dt \ + \ 
\int^{t_2}_{t_1} \int_{V_i} \greekbf{\psi} \, dV dt 
\label{general-int-conserv-law2}
}
The differential form (\ref{general-diff-conserv-law}) of the conservation law is 
derived from (\ref{general-int-conserv-law2}) by using Gauss' Theorem: 
\beq{
\oint_{S_i} \vec{\mathbf{f}} \cdot d\vec{S} = \int_{V_i} \nabla \cdot \vec{\mathbf{f}} dV
\quad . 
\label{gauss-theorem}
}
Since Gauss' theorem assumes that the functions are differentiable, 
the differential form only holds valid for systems that can be described by 
differentiable functions. 

In order to arrive at a discretized form of (\ref{general-int-conserv-law2}), we
must first define a few quantities.  The average value of the conserved variable over
the cell volume, $V_i$, is given by 
\beq{
\bar{\mathbf{q}}_i (t)  \ = \ \frac{1}{V_i} \int_{V_i} 
\mathbf{q}\left( \vec{x}, t \right) dV  
\quad . 
\label{general-cell-average}
}
If the cell $\mathcal{C}_i$ centered at $\vec{x}_i = (x^1_i , \ldots ,x^N_i)$ 
has boundaries 
$[x^1_{i-1/2} , \, x^1_{i+1/2}] \times \ldots \times [x^N_{i-1/2} , \, x^N_{i+1/2}]$---where  
$x^k_i \equiv x_\mathrm{min}^k + (i - 1) \Delta x^k$---then 
the flux integral in (\ref{general-int-conserv-law}) and 
(\ref{general-int-conserv-law2}) can be written as:
\beq{
\oint_{S_i} \vec{\mathbf{f}} \cdot d\vec{S} \  =  \  
\sum_{k=1}^{N} \left( \int_{S_{i+1/2}^k} \mathbf{f}^k \, d S  \quad - \quad
\int_{S_{i-1/2}^k} \mathbf{f}^k \, d S  \right) \label{general-new-flux1} \\
}
where the surface $S^k_{i+1/2}$---for instance---is defined as the isosurface
of constant $x^k = x^k_{i+1/2}$.  If we define a 
\emph{generalized numerical flux}
function to be the time average between two time steps of one of these integrals, 
\beq{
\mathbf{\mathcal{F}}^k_{i+1/2} \left(\mathbf{q}(x , t^n)\right) = \frac{1}{\Delta t}
\int^{t^{n+1}}_{t^n} \int_{S_{i+1/2}^k} \mathbf{f}^k \, d S  \, dt
\quad , \label{general-numerical-flux}
}
then we can rewrite (\ref{general-int-conserv-law2})---with (\ref{general-new-flux1}) and 
(\ref{general-cell-average})---as 
\beq{
\bar{\mathbf{q}}_i(t^{n+1}) \, - \, \bar{\mathbf{q}}_i(t^n) \ = \ 
- \frac{\Delta t}{V_i} 
\sum_{k=1}^{N} \left( \mathbf{\mathcal{F}}^k_{i+1/2}  - \mathbf{\mathcal{F}}^k_{i-1/2} 
\right) \ + \ \frac{1}{V_i} \int^{t^{n+1}}_{t^n} \int_{V_i} \greekbf{\psi} \, dV dt
\label{general-discret-eq}
}

\subsection{Example: Spherical Symmetry}

As a specific illustration, we will show how to go about deriving the
spherically-symmetric version of the discretized conservation equation (\ref{general-discret-eq}).

First, note that all functions are independent of $\phi$ and $\theta$.  This means
that the only non-zero flux component is the $r$-component, which we will denote $\mathbf{f}(q)$.  
Thus the numerical flux, $\mathbf{\mathcal{F}}_{i+1/2}$, 
becomes:
\beqa{
\mathbf{\mathcal{F}}_{i+1/2} \left(\mathbf{q}(r , t^n)\right) \ &=& \ \frac{1}{\Delta t}
\int^{t^{n+1}}_{t^n} \int_{S_{i+1/2}} \mathbf{f} \, d S  \, dt
\label{spherical-numerical-flux1} \\[0.5cm]
\ &=& \
4 \, \pi \, r^2_{i+1/2} \, \mathbf{F}_{i+1/2} 
\label{spherical-numerical-flux2}
}
where $\mathbf{F}_{i+1/2}$ is the \emph{numerical flux} defined by 
\beq{
\mathbf{F}^n_i = \frac{1}{\Delta t} \int^{t^{n+1}}_{t^n} 
\mathbf{f}(\mathbf{q}(x_i, t)) \ dt   \quad  .
\label{intnumflux} 
}

With 
\beq{
V_i \ = \ \frac{4 \pi}{3} \left( r^3_{i+1/2} - r^3_{i-1/2} \right)
\label{ss-cell-volume}
}
the discretized equations for a \emph{finite volume} are:
\beq{
\mathbf{\bar{q}}^{n+1}_i \ = \ \mathbf{\bar{q}}^{n}_i 
\, - \, \frac{3 \, \Delta t}{r^3_{i+1/2} - r^3_{i-1/2}} 
\left[ \left( r^2 X \mathbf{F} \right)^n_{i+1/2} 
- \left( r^2 X \mathbf{F} \right)^n_{i-1/2}  \right]
\, + \, \Delta t \, \greekbf{\bar{\psi}}_i 
\label{ss-discreteq} 
}
where 
\beq{
\greekbf{\bar{\psi}}_i \ \equiv \ \frac{1}{\Delta t \left( r^3_{i+1/2} -  r^3_{i-1/2} \right) }
\ \int^{r_{i+1/2}}_{r_{i-1/2}} \int^{t^{n+1}}_{t^n} \greekbf{\psi} \, r^2 \, dr \, dt
\quad . \label{numerical-source}
}
In practice, the average of the source, $\greekbf{\bar{\psi}}_i$, is 
approximated by the ``source of the average'', $\greekbf{\psi}(\mathbf{\bar{q}}_i)$.

The above equation (\ref{ss-discreteq}) provides the basic form of the discretized equation that we solve. 
However, to complete the solution, we need a good way of determining the numerical flux, $\mathbf{F}_{i+1/2}$. 
Computing a good numerical flux constitutes the real art of implementing conservative methods.

\section{The Riemann Problem and Godunov-type Methods}
\label{sec:riemann-solvers}

\begin{figure}[htb]
\centerline{\includegraphics*[scale=0.85]{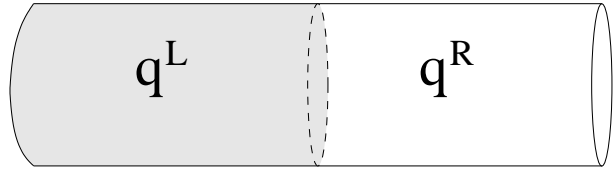}}
\caption{A shock tube representing a Riemann problem.  The tube is filled with fluid in two 
different states, separated by a removable interface.  \label{fig:riemann}}
\end{figure}

A Riemann problem seeks a solution to the equation
\beq{
\partial_t q + \partial_x f = 0  \label{cart-conservation-eq}
}
given piecewise constant initial data about an interface at $x=0$:
\beq{
q(x,0) = \left\{ \begin{array}{ll}  q^L   &x<0 \\ q^R &x>0 \end{array} \right. \quad . \label{riemann-data}
}
A realization of this one-dimensional problem can be represented as a tube with two states of fluid 
that are separated by a removable partition. For example, the two states can have  different 
pressures and/or densities.  If the gas in either side has a non-zero velocity initially, then the 
problem falls under the more general class of \emph{shock tube} problems.  At $t=0$, the interface
is removed and the two fluid components are allowed to mix.  Shock tubes have been 
studied for years in the laboratory to understand how shock waves develop and propagate, and a schematic 
illustration of a shock tube is given in Figure~\ref{fig:riemann}. 

The solution to the one-dimensional (scalar) Riemann problem obviously depends on the nature of the flux 
function $f$, since that function  provides information 
regarding the characteristics of the equation.  The Riemann problem has been studied extensively for many 
different flux functions, and much has been deduced from these investigations.
For example, it is easy to show that a shock wave will develop and move toward $x=\infty$ 
if $q^L > q^R$ and $f(q^L),f(q^R)>0$.  Using, the \emph{Rankine-Hugoniot jump condition}
\beq{
\mathbf{f}(\mathbf{q}^R) - \mathbf{f}(\mathbf{q}^L) = s \left( \mathbf{q}^R - \mathbf{q}^L \right)
\label{rankine-hugionot}
}
we can find the shock speed, $s$.  The solution to the Riemann problem is
then
\beq{
q(x,t) = \left\{ \begin{array}{ll}  q^L   &x<st \\ q^R &x>st \end{array} \right. \quad . 
\label{riemann-shock-solution}
}

If instead we have $q^L < q^R$, $\{f(q^L),f(q^R)\}>0$, and $f''(q)>0$, then the resulting evolution will describe 
a \emph{rarefaction} fan, which is a self-similar solution \cite{leveque1}:
\beq{
q(x,t) = \left\{ \begin{array}{ll}  q^L   &x<f'(q^L)t \\[0.2cm] Z(x/t) & f'(q^L)t < x < f'(q^R)t 
\\[0.2cm] q^R &x>f'(q^R)t \end{array} \right. \quad . 
\label{riemann-rarefaction-solution}
}
where $Z(\mathcal{X})$ is the solution to the characteristic equation $f'(Z(\mathcal{X}))=\mathcal{X}$. 

\begin{figure}[htb]
\centerline{\includegraphics*[bb=0.6in 0in 9in 4.6in, scale=0.65]{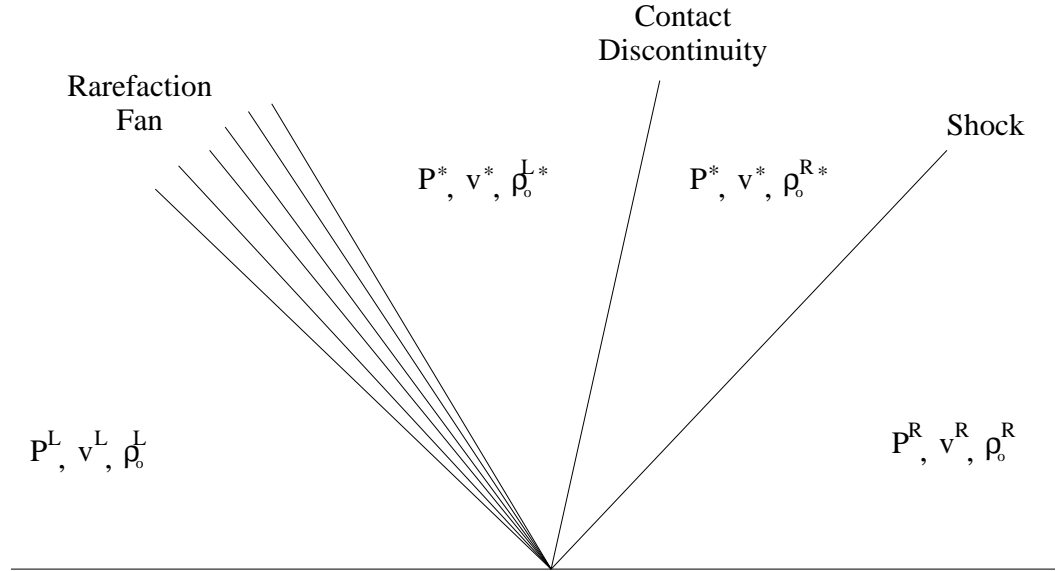}}
\caption[A graphical representation of a generic solution to the vector Riemann problem for the 
Euler equations.  ]{A graphical representation of a generic solution to the vector Riemann problem for the 
Euler equations.  The world lines of the waves are shown as straight lines that separate unique 
states of the fluid.  Hence, the effect the different waves have on the fluid is clearly seen. 
The values of the primitive variables in these distinct states are also shown.  The rarefaction
represents a continuum of states that is represented here by a fan-like ensemble of world lines. 
\label{fig:riemann-solution}}
\end{figure}

However, when the conservation equation consists of many, coupled equations, much of the knowledge 
gleaned from the one-dimensional case cannot be directly used.  However, it is still instructive to study 
the one-dimensional case, and many successful vector Riemann solvers have been based on 
features of the scalar problem.  Since a vector problem can be approximated---to some extent---as a 
linear combination of scalar problems, we expect to find both shocks and rarefaction waves 
coming from a single interface.  In fact, the vector Euler equations---(\ref{cart-conservation-eq}) where
$q$ and $f$ are now vectors---yields three primary wave solutions: shock, rarefaction and contact 
discontinuity.  Let us take  $\mathbf{w} = \left[P,v,\rho_\circ\right]^T$ to be the 
vector of primitive variables, and let us setup a vector Riemann problem with two states,
$\mathbf{q}^L=\mathbf{q}(\mathbf{w}^L)$ and $\mathbf{q}^R=\mathbf{q}(\mathbf{w}^R)$. 
Then an example of a possible solution to this Riemann is given in Figure~\ref{fig:riemann-solution}, 
where we have assumed that the left state is the one of greater pressure and density.  The 
solution can be described by four basic states:
\beq{
\mathbf{w}^L =  \left[ \begin{array}{c} P^L \\ v^L \\ \rho_\circ^L \end{array}\right]  
\ , \
\mathbf{w}^{L*} =  \left[ \begin{array}{c} P^* \\ v^* \\ \rho_\circ^{L*} \end{array}\right]  
\ , \
\mathbf{w}^{R*} =  \left[ \begin{array}{c} P^* \\ v^* \\ \rho_\circ^{R*} \end{array}\right]  
\ , \
\mathbf{w}^R =  \left[ \begin{array}{c} P^R \\ v^R \\ \rho_\circ^R \end{array}\right]  
\ , \
\label{riem-prim-soln}
}
The rarefaction, as usual, travels into the high-density region and represents a continuum of 
states between $\mathbf{w}^L$ and $\mathbf{w}^{L*}$.  The two intermediate states, $\mathbf{w}^{L*}$
and $\mathbf{w}^{R*}$, are separated by the \emph{contact discontinuity} that travels 
with velocity $v^*$ and is discontinuous only in the density, $\rho_\circ$.  The shock represents
a discontinuity in all three fields and has the reverse role of the rarefaction---it travels into the 
less dense region and increases the pressure in its wake.  

Considering the discretized equation (\ref{ss-discreteq}) derived in the previous section, we
see that it describes a solution for averages of $\mathbf{q}$ over cell volumes.  The 
cell averages, $\qbar_i$ and $\qbar_{i+1}$, can be viewed as representing piecewise constant
initial data about the interface at $r=r_{i+1/2}$.  Since  $\qbar_i \ne \qbar_{i+1}$ in general, 
then we can think of the update procedure along this cell border as a Riemann problem.  
Such methods that describe the numerical solution in this way and utilize the Riemann solution 
at each cell border are called \emph{Godunov} methods \cite{leveque1}.  Specifically, a Godunov method is 
one in which  the data are assumed to be piecewise-\emph{constant}, but other methods extend the basic idea by employing
higher-order interpolation schemes that assume that the data to be piecewise-linear,
piecewise-parabolic, etc.  Such  higher-order schemes are thus called Godunov-type methods.  
In Section~\ref{sec:reconstr-at-cell}, we will describe the interpolation routines we used to 
make our solutions $2^\mathrm{nd}$-order away from shocks.   We will also discuss how the Riemann solution is used to 
create a numerical flux function, $\mathcal{F}_j^n$, in the next two sub-sections.  The two specific methods 
used are called \emph{Roe's approximate} solver, and the \emph{Marquina} flux formula.  These two methods 
use approximate solutions to the Riemann problem since finding the exact solution is often inefficient and 
not always necessary.  

\subsection{Roe's Approximate Solver}
\label{sec:roes-appr-solv}

We use a variant of Roe's 
approximate Riemann solver \cite{roe-1981}, which is a Godunov-type method outlined 
in \cite{leveque1}.  It is an approximate Riemann solver since it uses the exact Riemann solution to 
the approximate Riemann problem:
\beq{
\partial_t \, \mathbf{q} \, + \, 
\frac{1}{r^2} \mathbf{A} \cdot \partial_r \left(r^2 \mathbf{q} \right) 
\, = \, \greekbf{\psi}   \quad .
\label{quasilinear-eom} 
}
where $\mathbf{A}$ represents a constant matrix at each step in the Riemann solution. 
The approximation then lies in that conservation equation has been linearized. 
However, this serves as a fair approximation if we choose $\mathbf{A}$ appropriately.  
Since we use Roe's method to solve the Riemann problem at a cell border, the matrix 
$\mathbf{A}$  should only be dependent on the two states:  
$\mathbf{A}=\mathbf{A}(\mathbf{q}^L,\mathbf{q}^R)$.   Determining this dependence lies at the heart 
of the method.  A strict Roe method satisfies the following conditions:
\begin{enumerate}
\item $\mathbf{A}=\mathbf{A}(\mathbf{q}^L,\mathbf{q}^R) \rightarrow \mathbf{f}'(\qbar)$ as 
               $\mathbf{q}^L,\mathbf{q}^R \rightarrow \mathbf{q} $
\item $\mathbf{A}(\mathbf{q}^L,\mathbf{q}^R) \left[ \mathbf{q}^L - \mathbf{q}^R \right]
= \mathbf{f}(\mathbf{q}^L) - \mathbf{f}(\mathbf{q}^L) $ 
\item $\mathbf{A}(\mathbf{q}^L,\mathbf{q}^R)$ has real eigenvalues and is non-singular.
\end{enumerate}
The first condition guarantees that the linear problem will tend to the nonlinear one in smooth regions.
The second criterion ensures that shock speeds are correctly calculated, as the Rankine-Hugoniot 
condition (\ref{rankine-hugionot}) dictates.  The tie between these two equations can be seen by 
diagonalizing the linear system in  criterion 2 and realizing that the 
eigenvalues---the diagonal elements---are the velocities of the shocks or contact discontinuities. 
Finally, the third criterion ensures that the system is hyperbolic. 

Since a matrix meeting all this criteria is not known for the relativistic, spherically-symmetric 
case, we use a further approximation to Roe's approximate Riemann solver.  Specifically, we choose 
(following Romero et al. \cite{romero})
\beq{
\mathbf{A}(\mathbf{q}^L,\mathbf{q}^R) 
= \left. \pderiv{\mathbf{f}}{\mathbf{q}}\right|_{\mathbf{q}=\hat{\mathbf{q}}}  \quad , \quad 
\hat{\mathbf{q}} = \frac{1}{2} \left( \mathbf{q}^L + \mathbf{q}^R \right) \quad . 
\label{approx-roe-matrix}
}
After solving the linear Riemann problem, the numerical flux of the solution can be taken to be
\cite{leveque1}:
\beq{
\mathbf{F}_{k+1/2}(t) = \frac{1}{2} \left[ \mathbf{f}(\mathbf{q}^L_{k+1/2}(t)) + 
\mathbf{f}(\mathbf{q}^R_{k+1/2}(t))  
- \sum_m \left| \lambda_m \right| \omega_m \greekbf{\eta}_m \right]  \quad . 
\label{num-flux}
}
Here, $\lambda_m$ and $\greekbf{\eta}_m$ are the eigenvalues and right eigenvectors, respectively,
of $\mathbf{A}$, $\mathbf{q}^L$ and $\mathbf{q}^R$ are, respectively, the values 
of $\mathbf{q}$ to the left and right of the cell boundary, and $\omega_m$ are the decomposed 
values of the jumps in the space of characteristic values:
\beq{
\mathbf{q}^R - \mathbf{q}^L \ = \ \sum_m \omega_m \greekbf{\eta}_m  \quad . 
\label{decomposed-jumps}
}
In order to calculate all quantities associated with $\mathbf{A}$, such as $\lambda_m$ and 
$\greekbf{\eta}_m$, we use the average of the left and right states, $\hat{\mathbf{q}}$ (\ref{approx-roe-matrix}).

\begin{figure}[htb]
\centerline{\includegraphics*[bb=0.6in 2.2in 9in 9.9in, scale=0.5]{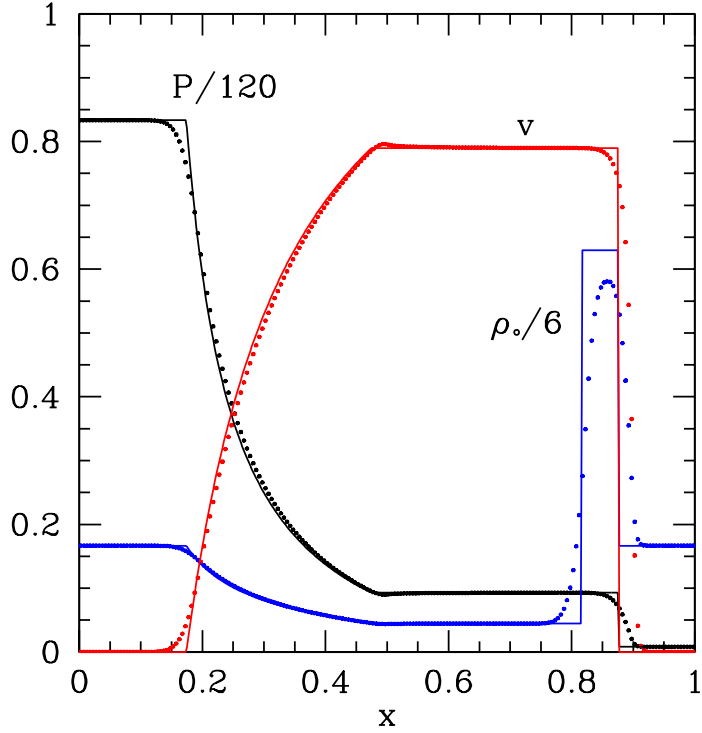}}
\caption[The Riemann solution using the approximate Roe method (dots), with initial data 
$\{P^L,v^L,\rho_\circ^L\}=\{100, 0, 1 \}$, $\{P^L,v^L,\rho_\circ^L\}=\{1, 0, 1 \}$ with $\Gamma=5/3$,
using 200 cells.  ]{The Riemann solution using the approximate Roe method (dots), with initial data 
$\{P^L,v^L,\rho_\circ^L\}=\{100, 0, 1 \}$, $\{P^L,v^L,\rho_\circ^L\}=\{1, 0, 1 \}$ with $\Gamma=5/3$,
using 200 cells.  The $P$ and $\rho_\circ$ have been normalized to 
fit into the same plot.  The solid line is the exact solution.   \label{fig:roe-riemann}}
\end{figure}

We note that this method is only an approximate Roe solver since it does not always satisfy Roe's second 
criterion.  Even though it does not guarantee the Rankine-Hugoniot condition in general, 
the method works well in practice.  In Figure~\ref{fig:roe-riemann}, we show a solution to a Riemann problem
using this approximate Roe solver for each cell, where the Riemann problem was set up in the middle 
of the grid, i.e.  at $x=0.5$.  The solid line is the exact solution of the Riemann problem calculated by 
a routine given in \cite{marti-muller-living}.  The approximate solution compares favorably to the 
exact solution, especially in smooth regions where the Roe solution should be close to the 
exact solution.  The numerical dissipation intrinsic to the method is observable near the shock 
and the edge of the rarefaction fan.  In fact, the density is diffused to such a degree that it does not 
quite reach the exact solution's value between the contact discontinuity and the shock. 

As we will see in Section~\ref{sec:instability}, the Roe solver leads to difficulties in certain situations. 
Since it is based on the solution to the linear Riemann problem, the solution---at the cell-scale---consists
of only shocks and discontinuities.  This leads to a problem when transonic rarefactions arise, i.e. 
when $f(q^L) < 0 < f(q^R)$. 

\subsection{Marquina's Method}
\label{app:marquina-flux-method}

The Marquina Flux equation, as described in \cite{donat-marquina} and
extensively tested in 2-D in \cite{donat-font}, is an 
amalgamation of a Roe flux method and a Lax-Friedrichs method for a 
general system of conservation laws.  The addition of the Lax-Friedrichs-like
method acts as an entropy-fix to the Roe flux.  Hence, Marquina's equation---in many cases---seems 
to effectively add extra dissipation  to the 
system.  An example of this is shown in \cite{donat-font}, where it was found that 
the use of an approximate Roe solver leads to the ``carbuncle phenomenon'' in
front of the bow shock of a supersonic relativistic jet.  The Marquina flux
seems to eliminate the carbuncle and replicate the physics involved with the 
relativistic jet quite well.  This suggests that it may be a useful 
technique in two---and even one---dimensions. 

\begin{table}[htb]
\centerline{\fbox{\parbox{3.5in}{
\begin{list}{}{}
\item For \ $m = 1 , \ldots , N$ \ do:
	 \begin{list}{}{}
	 \item If  \ 
	   $\left( \lambda_m(\tilde{\mathbf{q}}^L) \ 
                   \lambda_m(\tilde{\mathbf{q}}^R) \right)
	   \, \ge \,  0 $  then
\vspace{0.2cm}
		 \begin{list}{}{}
		 \item If \ $\lambda_m(\tilde{\mathbf{q}}^L) > 0$ \ then 
\vspace{0.2cm}
			 \begin{list}{}{}
			 \item $\greekbf{\phi}^{+}_m \ = \ \greekbf{\phi}^L_m$
			 \item $\greekbf{\phi}^{-}_m \ = \ 0$
			 \end{list}
		 \item else
			 \begin{list}{}{}
			 \item $\greekbf{\phi}^{+}_m \ = \ 0$
			 \item $\greekbf{\phi}^{-}_m \ = \ \greekbf{\phi}^R_m$
			 \end{list}
\vspace{0.2cm}
		 \item end if
		 \end{list}
	 \item else
		 \begin{list}{}{}
		 \item $\xi_m = {\max} \left( 
			 \left| \lambda_m(\tilde{\mathbf{q}}^L) \right| , 
			 \left| \lambda_m(\tilde{\mathbf{q}}^R) \right| \right)$
\vspace{0.1cm}
		 \item $\greekbf{\phi}^{+}_m \ = 
			    \frac{1}{2} \left( \greekbf{\phi}^L_m
			      \ + \ \xi_m \omega_m^L \right)$
\vspace{0.2cm}
		 \item $\greekbf{\phi}^{-}_m \ = 
			    \frac{1}{2} \left( \greekbf{\phi}^R_m 
			      \ - \ \xi_m \omega_m^R \right)$
		 \end{list}
	 \item end if
	 \end{list}
\end{list}
\vspace{0.5cm}
\[
\mathbf{F}(\tilde{\mathbf{q}}^L,\tilde{\mathbf{q}}^R) = 
\sum_{m = 1}^N \left( 
\greekbf{\phi}^+_m \greekbf{\eta}_m(\tilde{\mathbf{q}}^L) 
+ \greekbf{\phi}^-_m \greekbf{\eta}_m(\tilde{\mathbf{q}}^R) \right) 
\]
}}}
\caption{\label{table:marquina-method}
The Marquina Flux Calculation.  }
\end{table}

The method utilizes the characteristic variables and fluxes, which are 
spectral expansions of the conservation variables and fluxes, 
in order to determine how Roe-like or Lax-Friedrichs-like the numerical 
flux will be.  The characteristic variables, $\greekbf{\omega}_m$, and fluxes, 
$\greekbf{\phi}_m$,  are defined as :
\beq{\begin{array}{l l}
\greekbf{\omega}_m^L = \mathbf{l}_m ( \mathbf{q}^L ) 
\cdot \mathbf{q}^L 
& \quad 
\greekbf{\omega}_m^R = \mathbf{l}_m ( \mathbf{q}^R ) 
\cdot \mathbf{q}^R \\[0.5cm]
\greekbf{\phi}_m^L = \mathbf{l}_m ( \mathbf{q}^L ) 
\cdot \mathbf{f} ( \mathbf{q}^L )
& \quad 
\greekbf{\phi}_m^R = \mathbf{l}_m ( \mathbf{q}^R ) 
\cdot \mathbf{f} ( \mathbf{q}^R )
\end{array}
\label{characteristic-vars}
}
Here, $\mathbf{l}_m(\mathbf{q})$ and $\greekbf{\eta}_m(\mathbf{q})$ are the 
left and right eigenvectors, respectively, of $\mathbf{A}(\mathbf{q})$, the 
Jacobian matrix appearing in the conservation equation.  Also, $m=1,\ldots, N$ enumerates
the $N$ eigenvectors.  
The algorithm for calculating the Marquina flux is described in Table~\ref{table:marquina-method}, 
where we recall that $\lambda_m(\mathbf{q})$ are the eigenvalues of $\mathbf{A}(\mathbf{q})$.

\begin{figure}[htb]
\centerline{\includegraphics*[bb=0.6in 2.2in 9in 9.9in, scale=0.5]{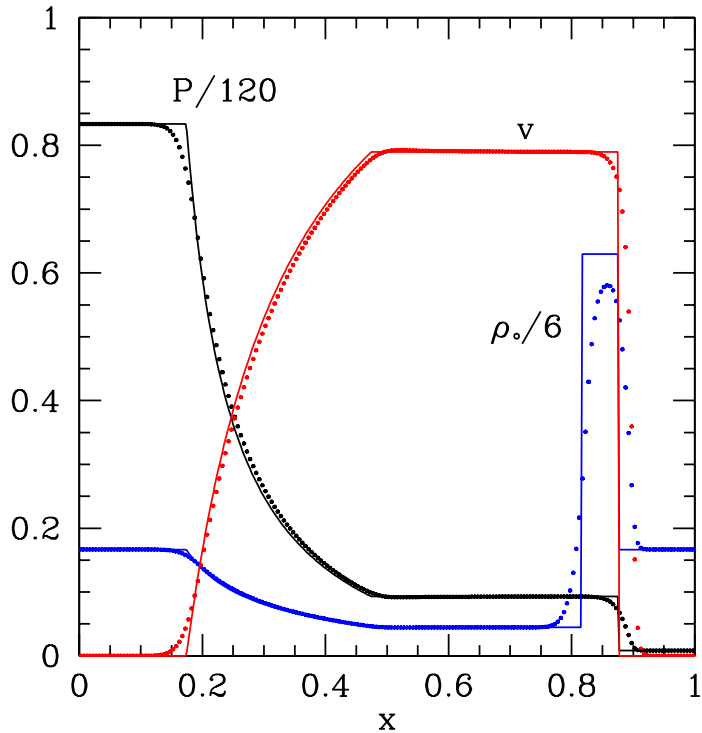}}
\caption[The Riemann solution using Marquina's method (dots), with initial data 
$\{P^L,v^L,\rho_\circ^L\}=\{100, 0, 1 \}$, $\{P^L,v^L,\rho_\circ^L\}=\{1, 0, 1 \}$ with $\Gamma=5/3$,
using 200 cells.  ]{The Riemann solution using Marquina's method (dots), with initial data 
$\{P^L,v^L,\rho_\circ^L\}=\{100, 0, 1 \}$, $\{P^L,v^L,\rho_\circ^L\}=\{1, 0, 1 \}$ with $\Gamma=5/3$,
using 200 cells.  The $P$ and $\rho_\circ$ have been normalized to 
fit into the same plot.  The solid line is the exact solution.   \label{fig:marquina-riemann}}
\end{figure}

Figure~\ref{fig:marquina-riemann} shows a solution 
to the same Riemann problem considered in Figure~\ref{fig:roe-riemann} using the Marquina flux.  For this 
particular Riemann problem, it seems that Roe's method produces a slight Gibbs phenomenon near 
the origin of the rarefaction fan, while Marquina's method results in the functions undershooting there.
However, the two methods produce nearly identical results near the shock and contact discontinuity. 

Even though we primarily use the approximate Roe solver, since it is computationally more efficient, the Marquina 
solver plays an important role in examining the instability near the sonic point in the self-similar 
solutions we obtain near Type~II critical solutions.  See Section~\ref{sec:instability} for 
further discussion of this point. 

\section{Reconstruction at the Cell Borders} 
\label{sec:reconstr-at-cell}

Since the accuracy of the spatial differencing is constrained by the order of interpolation used 
to calculate the cell boundary values, a way to improve upon the $1^\mathrm{st}$-order accuracy of generic 
Godunov schemes 
is to increase the accuracy of the interpolation scheme.  For example, Godunov methods are 
$1^\mathrm{st}$-order accurate since they assume that the data is piecewise-constant, but we can make the
spatial differencing be $2\mathrm{nd}$-order or $3\mathrm{rd}$-order accurate by using piecewise-linear or 
piecewise-parabolic data, respectively.  Even though piecewise-parabolic methods---such as PPM by Colella and 
Woodward \cite{colella-woodward}---have become more popular in recent years \cite{font-review}, 
we only use piecewise-linear methods here since they are straightforward to implement yet still provide 
well-resolved discontinuities. 

Since shocks naturally arise in fluid dynamical evolutions, we require that the interpolation procedure 
capture shocks well so that spurious oscillations---in the form of Gibbs phenomena---do not 
occur.  To minimize such numerical artifacts, we use linear, 
slope-limiting interpolators to calculate the border values $\mathbf{\bar{q}}^L$ and 
$\mathbf{\bar{q}}^R$.  These are found by first interpolating for the primitive variables 
$\mathbf{\bar{w}}^L$, $\mathbf{\bar{w}}^R$ at the border, then setting 
$\qbar^L = \mathbf{q}(\bfbar{w}^L)$ and $\qbar^R = \mathbf{q}(\bfbar{w}^R)$ using the 
definitions of $\mathbf{q}(\mathbf{w})$ (\ref{D}-\ref{tau}).  
We have found that by interpolating $\bfbar{w}$ instead of $\bfbar{q}$ the numerical 
procedure generally leads to smoother solutions and fewer instabilities.  
Specifically, the slope-limiting interpolation is carried out in the following fashion:
\beq{
\bfbar{w}^L_{k+1/2} \ = \ \bfbar{w}_k + \greekbf{\sigma}_k \left( r_{k+1/2} - r_k \right)
\label{w-L}
}
\beq{
\bfbar{w}^R_{k+1/2} \ = \ \bfbar{w}_k + \greekbf{\sigma}_{k+1} \left( r_{k+1/2} - r_{k+1} \right)
\label{w-R}
}
where the $\greekbf{\sigma}_k$ is the slope obtained from the slope-limiting 
function of given slopes 
\beq{
\mathbf{s}_{k+1/2} \equiv \left(\bfbar{w}_{k+1} - \bfbar{w}_k\right)/\left(r_{k+1}-r_k\right)
\label{recon-slopes}
}
and $\mathbf{s}_{k-1/2}$.  For instance, if we use the \textrm{minmod} slope-limiter defined by 
\beq{
\mathrm{minmod}( a , b ) = \left\{ \begin{array} {r@{\qquad}l}
0 & \mathrm{if} \quad a b < 0 \\
a & \mathrm{if} \quad |a| < |b| \quad \mathrm{and} \quad  a b > 0 \\
b & \mathrm{if} \quad |b| < |a| \quad \mathrm{and} \quad  a b > 0
\end{array} \right.   \label{minmod}
}
then 
\beq{
\greekbf{\sigma}_k=\mathrm{minmod}(\mathbf{s}_{k-1/2},\mathbf{s}_{k+1/2}) \quad . \label{sigma-slope-limited}
}

\begin{figure}[htb]
\centerline{\includegraphics*[bb=0.4in 3in 8in 6.8in, scale=0.6]{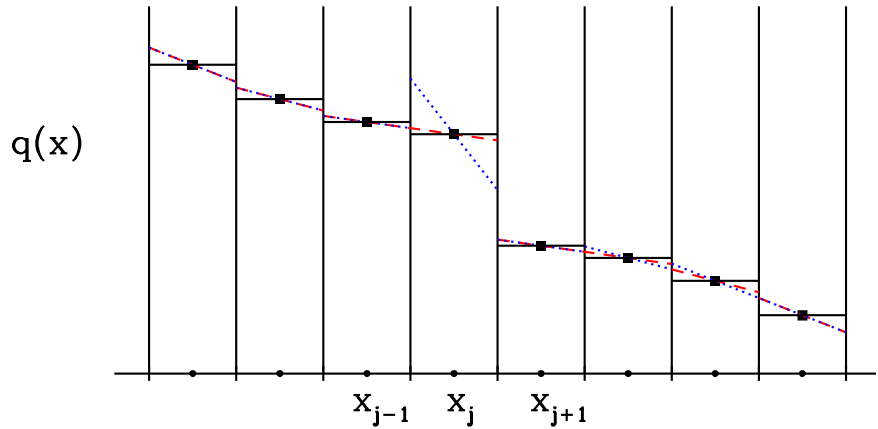}}
\caption[Results of different methods for calculating the slopes used in the 
linear interpolation procedure that estimates $\qbar$ at cell borders. ]{Results of different 
methods for calculating the slopes used in the 
linear interpolation procedure that estimates $\qbar$ at cell borders.  Here, the black 
horizontal lines represent piecewise-constant interpolation,  
the blues dots illustrate second order interpolation without limiting, and the red dashes represent second order
interpolation  with 
limiting. \label{fig:recon}}
\end{figure}

Determining what slopes to use at each border ultimately decides how shocks are resolved.  
In Figure~\ref{fig:recon}, we plot $\greekbf{\sigma}_k$ computed using different schemes. 
The slopes represented by the black line are calculated by setting $\greekbf{\sigma}_k=0$, 
the blue dots from setting $\greekbf{\sigma}_k=\mathbf{s}_{k+1/2}$ always, and the red dashes
are from $\greekbf{\sigma}_k=\mathrm{minmod}(\mathbf{s}_{k-1/2},\mathbf{s}_{k+1/2})$.  Like
most slope-limiters, the \textrm{minmod} function attempts to diffuse numerical oscillations---whose
wavelengths are $2\Delta r$---by setting the slope to $0$ when adjacent slopes change sign.  Also, 
it always uses the less steep slope, so that discontinuities are always well resolved with little 
overshooting or Gibbs phenomena.  As can be seen in 
the figure, use of a non-limited slope  obviously makes the solution  overshoot the shock. 

We have tried other slope-limiters, such as the \textrm{MC} limiter and the 
\textrm{Superbee} limiter, but found the \textrm{minmod} limiter to provide the most stable 
calculations of near-threshold solutions (see \cite{neilsen-thesis} and references therein for descriptions of the
\textrm{MC} and \textrm{Superbee} limiters).  Since the \texttt{MC} and \texttt{Superbee} limiters 
resolve discontinuities more accurately, they lack \textrm{minmod}'s diffusiveness that 
seems to help dampen the instability observed near the sonic point of  near-critical 
solutions (see Section~\ref{sec:instability} for a discussion of the instability mentioned here). 

From the definition of $\mathrm{minmod}$, the limited slopes can be shown to be $0$ 
at extrema of $\qbar$, such as near discontinuities and shocks.  Thus, the interpolation 
procedure at the extrema reduces to a $1^\mathrm{st}$-order scheme, making the numerical solution there 
$1^\mathrm{st}$-order accurate.  Fundamentally, with methods such as those used here, such behavior near shocks
cannot be avoided since shocks in inviscid flow 
are not resolvable in the continuum limit.  Thus, a piecewise-constant representation of the functions in 
a shock's neighborhood is the best that  can be done in any case, unless the position of shocks can be exactly
traced.  However, this makes convergence testing 
difficult since the solution's convergence will be reduced from $2^\mathrm{nd}$-order to $1^\mathrm{st}$-order 
in regions where shocks---or any other extrema of $\qbar$---propagate. 

Non-oscillatory interpolation schemes have even been devised for arbitrary interpolation orders.
For the regridding process described in Section~\ref{sec:refinement-procedure},
we use the so-called \emph{Essentially Non-Oscillatory} (ENO) scheme originally developed
by  Shu \cite{shu1997}.
The algorithm is especially powerful since it can be used to perform an interpolation of arbitrary 
order---only restricted by the number of available grid points from which to sample.  
The particular routine we use was written by Olabarrieta \cite{olabarrieta}, and we have found that 
a $3^\mathrm{rd}$-order ENO interpolation is sufficient for our current work.

\section{Time Integration Procedures} 
\label{sec:method-lines}

In order to make the entire differencing procedure $2^\mathrm{nd}$-order accurate, 
the differencing with respect 
to $t$ needs to also be adjusted since (\ref{ss-discreteq}) is only differenced to $O(\Delta t)$.  
Explicit methods are usually used for performing the time integration in conservative 
schemes since conservative methods usually entail a myriad of other expensive steps.  A simple way 
of making the method 
explicit is to split temporal and spatial difference operators using the  method of lines, which 
entails integrating along each direction----spatial and temporal---separately.  
Since all conservation equations take the form
\beq{
\frac{d \mathbf{q} }{d t} = L(\mathbf{q}) \label{diffeqgeneral} 
}
we can solve this equation, after spatial discretization, as a system of ODEs.  Here, $L$ 
includes the spatial differential operator as well
as the source term.  The discrete version, $\hat{L}$, can easily be inferred from 
the discretized EOM (\ref{ss-discreteq}).  

In predictor-corrector methods, intermediate values---$\qbar^{n*}_j$ or $\qbar^{n+1/2}_j$---are first 
calculated by the \emph{predictor} step and then 
used in the \emph{corrector} step to obtain the final updated values, $\qbar^{n+1}_j$.
For the modified Euler or Huen's method, the following two equations define the predictor 
and corrector steps, respectively:
\beq{ 
\bar{\mathbf{q}}^*_j = \bar{\mathbf{q}}^n_j
+ \Delta t \, \hat{L}(\bar{\mathbf{q}}^n) \label{predictor}
}
\beq{ 
\bar{\mathbf{q}}^{n+1}_j = \frac{1}{2} \left[ \bar{\mathbf{q}}^n_j +
\bar{\mathbf{q}}^*_j + \Delta t \, \hat{L}(\bar{\mathbf{q}}^*) \right] 
\label{corrector}
}
The predicted values, $\qbar^*_j$, can be interpreted as $1^\mathrm{st}$-order approximations to the 
corrected values, $\qbar^{n+1}_j$. 

Another commonly used method is the \emph{half-step} predictor-corrector method, which is equivalent to a
$2^\mathrm{nd}$-order Runge-Kutta technique.  
The half-step update does a $1^\mathrm{st}$-order predictor step integration 
to $t=t^{n+1/2}$, and then  uses the 
slope at the half-step to evolve to $t=t^{n+1}$
\beq{
\mathbf{q}^{n+1/2} = \mathbf{q}^n + \frac{\Delta t}{2} \hat{L}(\mathbf{q}^n)
\label{half-predictor}
}
\beq{ 
\mathbf{q}^{n+1} = \mathbf{q}^n + \Delta t \hat{L}(\mathbf{q}^{n+1/2})
\label{half-corrector}
}
We see little difference  between the two methods in practice, even though Huen's method has 
been shown to be the only $2^\mathrm{nd}$-order predictor-corrector method to be 
Total-Variation-Diminishing (TVD).
TVD analysis is a way to demonstrate whether an algorithm is stable by seeing whether the quantity:
\beq{
\mathrm{TV}\left(\qbar^n\right) \equiv \sum_j \left| \qbar^n_{j+1} - \qbar^n_j \right| \label{TV}
}
monotonically decreases over time \cite{leveque1}.  If it does, then the method is TVD.  Hence, 
we use Huen's method for all of the computations described here.

\begin{figure}[htb]
\centerline{\includegraphics*[bb=1in 4in 8in 9in, scale=0.6]{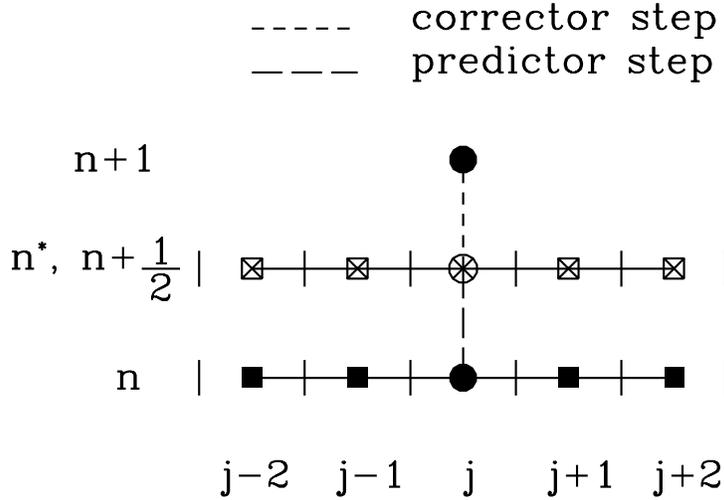}}
\caption[Stencil depicting the update scheme for cell $\mathcal{C}_j$ from 
time step $n$ to time step $n+1$  using piecewise-linear reconstruction.]{Stencil depicting 
the update scheme for cell $\mathcal{C}_j$ from 
time step $n$ to time step $n+1$ using piecewise-linear reconstruction.  The unfilled shapes 
represent the ``predicted'' grid function values, those calculated during the 
predictor step.  The predicted state is labeled $n^*$ or $n+1/2$ depending on whether Huen's
method or the half-step method is used, respectively.  The vertical bars represent cell boundaries.
\label{fig-stencil}}
\end{figure}

The stencil used for a generic predictor-corrector method (such as Huen's method) is shown in Figure~\ref{fig-stencil}.  
The stencil is $5$ cells wide in our case because the slope-limiter uses this many points to determine
the optimal process for reconstructing the values, $\mathbf{q}^L$ and $\mathbf{q}^R$, at each border.  

The Courant-Friedrichs-Lewy (CFL) condition for these schemes essentially reduces to ensuring 
that the physical domain of dependence as determined by the largest characteristic speed $\lambda_{\max}$
is contained in the numerical domain of dependence:
\beq{
 \lambda_\mathrm{CFL}  \ \equiv \ \frac{\Delta t}{\Delta r} 
< \frac{1}{\left|\lambda_{{\max}}\right|} 
\label{cfl}
}
Throughout this work, we use  $\lambda_\mathrm{CFL} = 0.4$.

\section{Primitive Variable Calculation}
\label{sec:prim-var-calc}

Since only the conservative variables are evolved by the HRSC schemes discussed above, the primitive 
variables must be derived
from the conservative variables after each predictor or corrector step in order to 
compute fluxes $\mathbf{f}$ and source functions $\greekbf{\psi}$ for the next evolution step.  
This involves inverting the three equations 
$\mathbf{q}=\mathbf{q}(\mathbf{w})$, which are given by the definitions of the conservative 
variables (\ref{D}-\ref{tau}), for the three unknown primitive variables 
$\mathbf{w}$.
We know of  no closed-form expressions for the inverted equations, and so 
we must solve for $\mathbf{w}(\mathbf{q})$ numerically. 
At each grid point, we use a Newton-Raphson method to find the values of $\mathbf{w}$ 
that minimize the residuals of the conservative variable 
definitions (\ref{D}-\ref{tau}).  Instead of solving the full $3$-by-$3$ 
system at each point, an identity function $\mathcal{I}$---derived 
from (\ref{D}-\ref{tau})---is
used as a residual, making the solution process one-dimensional.  This makes the procedure
much more efficient, especially since it needs to be executed $2N$ times per 
time step, where $N$ is the number of spatial grid points.  

One obvious choice of the identity, which 
is commonly used \cite{romero,neilsen-thesis}, is:
\beq{ 
S - v \left( E + P \right) \  = \ 0\quad . \label{S-E-identity2}
}
We divide by $v$ and express $S$ in terms of $D$ and $\mathbf{w}$ to 
get the final residual of 
\beq{
\mathcal{I}_1(P)  \ = \  D \left[ \frac{W(P)}{a} - 1 \right] 
                        + P \left[ \, \mathcal{G} {W(P)}^2 - 1 \, \right] - \tau   \label{pvar-resid1}
}
where (\ref{ideal-eos}) has been used to eliminate $\epsilon$, and $W(P)$ is given by (\ref{S-E-identity}):
\beq{
W(P) = \sqrt{ \frac{ \left(\tau + D + P \right)^2 }{ \left(\tau + D + P \right)^2 - S^2 } } \quad .  
\label{W-of-P}
}
In practice, this residual leads to inaccurate calculations of $\mathbf{w}$ 
for relativistic flows since the numerical evaluation of (\ref{W-of-P}) as $|v| \rightarrow 1$ becomes 
less precise due to the fact that the numerator---$S$---and the 
denominator---$\left( \tau + D + P \right)$---both grow to infinity in that limit.  
In order to calculate $v$ more accurately in this limit, a different residual, which was first developed in 
\cite{neilsen-thesis}, is used:
\beq{
\mathcal{I}_2(H) \ = \ H W^2 - \tau - D - P   \label{pvar-resid2}
}
where $H$ is the enthalpy, 
\beq{
H \ \equiv \ \rho_\circ h = \rho_\circ \left( 1 + \epsilon + P / \rho_\circ \right)  \quad .
\label{big-enthalpy}
}
Using $H$ as the independent variable instead of $P$ allows us to 
calculate $v$ more precisely in relativistic flows and avoids the calculation of super-luminal velocities.  
Specifically, we compute:
\beq{
v = v( \Lambda ) = \frac{1}{2 \Lambda} \left( \sqrt{ 1 + 4 \Lambda^2 } - 1 \right)
\label{v-of-Lambda}
}
where $\Lambda \equiv S / H $.  We note that equation
(\ref{v-of-Lambda}) is simply an identity based on the definition of $W$ and 
the fact that $\Lambda = W^2 v$.  

The overall method used to find the primitive variables is outlined in 
Table~\ref{table:rel-prim-vars}.  Depending on how \emph{relativistic} or \emph{non-relativistic}
the flow is, different methods for calculating the residual $\mathcal{I}_2$ and its 
``Jacobian'' $\mathcal{I}_2^\prime = \partial \mathcal{I}_2 / \partial H$ are used in order to 
increase the accuracy of $\mathbf{w}$; these methods are 
described in Table~\ref{table:rel-prim-vars-detail}.  The ``non-relativistic'' and 
``intermediate'' methods originated from Neilsen \cite{neilsen-thesis}, where flows in the 
ultra-relativistic limit were also studied.  However, we have found that, in the ultra-relativistic
limit where $\Lambda \rightarrow \infty$, the intermediate method still gives imprecise results 
that are essentially due to the diminishing precision of calculating the deviation
\beq{
1 - v\left(\Lambda\right) = 1 - \sqrt{1+1/4\Lambda^2} + 1/2\left|\Lambda\right|  \quad . 
\label{inter-imprecision}
}

Even though the above methods improved the accuracy of the primitive variable calculation, 
significant errors still remain for highly-relativistic flows ($W > 10^5$) with $P$ 
and $\rho_\circ$ being different by orders of magnitude---.e.g. when 
$P \gg \rho_\circ$  or $\rho_\circ \gg P$.  In these regimes, machine precision limits the 
accuracy of the calculation since terms in $\mathcal{I}_2$ and $\mathcal{I}_2^\prime$ 
become numerically insignificant even though their presence is essential to finding the solution.


\centerline{\fbox{\parbox{4.5in}{$\underline{\mathrm{Given} \ 
\left\{D , S, \tau , a \right\}^\mathrm{new} \ \mathrm{at}  \ t = t^{\mathrm{new}} \ \mathrm{and}
\ \left\{ \rho_\circ , P \right\}^\mathrm{old} \ \mathrm{at} \ t = t^{\mathrm{old}}}$ :
\begin{eqnarray*}
\mathbf{\#1)} && \mathcal{G} \equiv \frac{ \Gamma }{ \Gamma - 1 } 
                 \quad , \quad H^\mathrm{new} = \rho_\circ^\mathrm{old}
                                                      + \mathcal{G} P^\mathrm{old} \\ 
 && \makeline{1in} \\[0.25cm]  
\mathbf{\#3)} && H = H^\mathrm{new} \quad , \quad \Lambda = S \, / \, H \\[0.25cm]
\mathbf{\#4)} && \mathrm{Calculate} \ \left\{ \mathcal{I}_2 \, , \, \mathcal{I}_2^{\,\prime} \, , 
                                               P , v , \rho_\circ \right\}  \\[0.25cm]
\mathbf{\#5)} && \Delta H = - \, \mathcal{I}_2 \, / \, \mathcal{I}_2^{\,\prime}  \quad , \quad 
                   H^\mathrm{new} = H + \Delta H  \\[0.25cm]
\mathbf{\#6)} && \mbox{Repeat Steps \textbf{\#3 -- \#5} \ until } 
                    \ \left(\, |\Delta H / H| < \mathtt{tol} \, \right)\\[0.25cm]
 && \makeline{1in} \\[0.25cm]
\mathbf{\#7)} && P^\mathrm{new} \, = \, P  \quad , \quad v^\mathrm{new} \, = \, v \quad , \quad
	 \rho_\circ^\mathrm{new} \, = \, \rho_\circ
\end{eqnarray*}
}}}
\begin{table}[h] 
\caption[The Point-wise Newton-Raphson method used to construct the primitive variables from the 
conservative variables and geometry.]{\label{table:rel-prim-vars}
The Point-wise Newton-Raphson method used to construct the primitive variables from the 
conservative variables and geometry. The calculation is performed after the conservative 
variables have been integrated to a new time step at $t = t^\mathrm{new}$, and 
$a$ has been found via the Hamiltonian constraint.  A few variables---$P$ and $\rho_\circ$---are 
needed from the previous time step, $t^\mathrm{old}$, as guesses for the iteration process. 
Here, $\mathcal{I}_2$---given by
(\ref{pvar-resid2})---is the residual that is numerically minimized.
}
\end{table}

\clearpage

\centerline{\fbox{\parbox{5.5in}{
\begin{tabbing}
\hspace*{0.5cm}\= \hspace{0.5cm} \= \kill	
\textbf{\texttt{If }} $\left(|\Lambda| > \Lambda_\mathrm{High} \right) 	\ 
                     \mathbf{\mathtt{then}}$ \\[0.25cm]
              \> $b = \mathBig{\frac{1}{ 2 \left|\Lambda\right| }} \quad , \quad B \equiv b^2 \quad , \quad X(b) \equiv 1 / W^2 = 2 \sqrt{B} \left(\sqrt{B+1} - \sqrt{B}\right)$\\[0.25cm]
              \> $\rho_\circ = \mathtt{TAYLOR}_8\left[\mathBig{\frac{D}{a}}\sqrt{X(b)}  \right] \quad , \quad P = \mathtt{TAYLOR}_8\left[ \mathBig{\frac{1}{\mathcal{G}}} \left(H - \mathBig{\frac{D}{a}}\sqrt{X(b)}\right)  \right] $\\[0.25cm]
	      \> $v = \mathtt{sign}(S) \ \mathtt{TAYLOR}_8\left[ \left( \sqrt{B+1} - \sqrt{B} \right) \right] $\\[0.25cm]
	      \> $\mathcal{I}_2 = \mathtt{TAYLOR}_8 \left[ H \mathBig{ \left(\frac{1}{X(b)} - \frac{1}{\mathcal{G}}\right)} - \tau + D \left( \mathBig{\frac{\sqrt{X(b)}}{a \mathcal{G}}} - 1 \right)  \right] $\\[0.25cm]
	      \> $\mathcal{I}_2^\prime = \mathtt{TAYLOR}_8 \left[ \mathBig{\frac{1}{2 \sqrt{B+1}\left(\sqrt{B+1} - \sqrt{B}\right)} - \frac{1}{\mathcal{G}} + \left(\frac{D}{a H \mathcal{G}}\right) \frac{ B^{1/4} \left(\sqrt{B+1} - \sqrt{B}\right)^{3/2}}{\sqrt{2} \sqrt{B+1}}} \right] $\\[0.35cm]
$\mathbf{\mathtt{Else }}$   \\
              \> $\mathbf{\mathtt{If }} \left(  |\Lambda| > \Lambda_\mathrm{Low}\right) 	\ \mathbf{\mathtt{then}}$ \\[0.25cm]
	         \>\> $\quad \mathcal{Y} \ = \ \sqrt{ 1 + 4 \Lambda^2 } \quad , \quad v \ = \ \mathBig{\frac{1}{2 \Lambda}} \left( \mathcal{Y} - 1 \right) \quad , \quad \mathBig{\pderiv{v}{H} \ = \ - \frac{ S }{ H^2 } \left[ \frac{2}{\mathcal{Y}} - \frac{\left( \mathcal{Y} - 1 \right)}{ 2 \Lambda^2 } \right]} $\\[0.25cm]
              \>  $\mathbf{\mathtt{Else }}$\\
                 \>\>  $\quad v \ = \  \left(1+\left(-1+\left(2+\left(-5+\left(14-42\,{\Lambda}^{2}\right){\Lambda}^{2}\right){\Lambda}^{2}\right){\Lambda}^{2}\right){\Lambda}^{2}\right)\Lambda  $\\[0.25cm]
	         \>\> $\quad \mathBig{\pderiv{v}{H} \ =  \ - \frac{ S }{ H^2 } } \ \left[ 1+\left(-3+\left(10+\left (-35+\left (126-462\,{\Lambda}^{2}\right){\Lambda}^{2}\right){\Lambda}^{2}\right){\Lambda}^{2}\right){\Lambda}^{2} \right]  $\\[0.25cm]
              \>  $\mathbf{\mathtt{End}} \, \mathbf{\mathtt{If}}$\\[0.25cm]
          \>  $W = 1 \, / \sqrt{1 - v^2} \quad , \quad P = \mathBig{\frac{1}{\mathcal{G}}} \left( H - \mathBig{\frac{ D }{ a W }} \right) \quad , \quad f(H) = H W^2  \, - \tau - D - P  $ \\[0.25cm]
          \>  $f^{\,\prime}(H) \ = \  W^2 \left( 1 \, + \, 2 H W^2 \, v \, \mathBig{\pderiv{v}{H}} \right) \, - \, \mathBig{\frac{1}{\mathcal{G}} \left( 1 \, + \, \frac{D W v}{ a } \, \pderiv{v}{H} \right) } \quad , \quad \rho_\circ = \mathBig{\frac{D}{a W}} $\\[0.25cm]
$\mathbf{\mathtt{End}} \, \mathbf{\mathtt{If}}$
\end{tabbing}
}}}
\begin{table}[h] 
\caption[Pseudo-code for the calculation of
$\left\{\mathcal{I}_2 , \mathcal{I}_2^{\,\prime} , P , v , \rho_\circ \right\}$ used in the 
primitive variable construction procedure described in Table~\ref{table:rel-prim-vars}. ]{Pseudo-code for the calculation of
$\left\{\mathcal{I}_2 , \mathcal{I}_2^{\,\prime} , P , v , \rho_\circ \right\}$ used in the 
primitive variable construction procedure described in Table~\ref{table:rel-prim-vars}. 
See the next page for this figure's caption. \label{table:rel-prim-vars-detail}}
\end{table}

\begin{table}[h] 
(Caption for Table~\ref{table:rel-prim-vars-detail}) Pseudo-code for the calculation of
$\left\{\mathcal{I}_2 , \mathcal{I}_2^{\,\prime} , P , v , \rho_\circ \right\}$ used in the 
primitive variable construction procedure described in Table~\ref{table:rel-prim-vars}.  
The procedure performs the calculation in three different ways depending in which 
regime the system resides.  In the ultra-relativistic regime, $\left|\Lambda\right|$ becomes
quite large and, subsequently, expanding the nonlinear expressions in powers of 
$b=1/2\left|\Lambda\right|$ becomes numerically more accurate.  In the above table, 
$\mathtt{TAYLOR}_8$ represents the operation of taking the series expansion of its argument 
to $O(b^8)$. 
Also, for the case when the system is 
non-relativistic---e.g. when $\left|\Lambda\right| \ll 1$---we use expansions up 
to $O(\Lambda^9)$.   In practice, the ultra-relativistic regime 
is defined by an adjustable parameter $\Lambda_\mathrm{High}$ and the non-relativistic regime by
$\Lambda_\mathrm{Low}$.  For example, in all the results shown here we used 
$\Lambda_\mathrm{High} = 10^2$  and $\Lambda_\mathrm{Low} = 10^{-4}$; these values ensure 
that the leading-order error terms in the ultra-relativistic and non-relativistic expansions 
are below the round-off error of the machines used.
\end{table}

\section{The Floor}
\label{sec:floor}

Contrary to evolutions in Lagrangian coordinates, flows computed using Eulerian coordinates often 
give rise to evacuated regions where the pressure and/or density 
vanish and near-luminal fluid velocities develop.  Due to the 
finite precision of the calculations and the nature of the 
numerical methods employed, the evacuation often ``overshoots''
the vacuum state generating negative pressures or densities, which in turn leads
to a plethora of unphysical, numerical consequences such as a complex $c_s$ or super-luminal
velocities.  This is one of the more troublesome problems encountered in numerical 
relativistic hydrodynamics and a completely satisfactory resolution is unfortunately still outstanding.  
In order to alleviate the evacuation problem and to avoid such catastrophic consequences, we require the dynamic
fluid quantities---the conservative variables $\mathbf{q}$ 
(\ref{ideal-piphi-state-vectors})---to have values greater than or equal to 
a so-called ``floor'' state.  In order to determine the floor state, we require 
$P,\rho_\circ > 0$ and $|v| < 1$ which implies that 
\beq{
D \ , \ \left( \tau \pm \left|S\right| \right) \  > \  0 
\label{floor-requirements}
}
Using the transformed (``new'') variables $\Pi,\Phi$, we implement this requirement in the following way 
\beqa{
D \ & = & \ {\max}\left(D , \delta \right) \\
\Pi \ & = & \ {\max}\left(\Pi + D , 2 \delta \right) - D \\
\Phi \ & = & \ {\max}\left(\Phi + D , 2 \delta \right) - D 
\label{floor-method}
}
Notice that the $\Pi$ and $\Phi$ need not remain positive since $\tau \le 0$ is physical as 
long as $E>0$.  
Since the floor state involves very little mass-energy, its use does not significantly affect the overall
dynamics of the star.  For example, Figure~\ref{fig:scaling-difffloors} shows how 
the scaling of the global maximum of ${T^a}_a$ as a function of  $\ln \left(p^\star - p\right)$ is independent 
of the floor values.  The most striking indication of the relative insignificance of the floor is the fact
that the computed values for the critical velocity amplitude $p^\star$ are surprisingly in agreement,
to within $4 \times 10^{-5}$. 

\section{Description of the Numerical Grid and Refinement Procedures}
\label{sec:grid}

In this section we describe the basic structure of the numerical domain used 
in the code.  We start by describing the most basic grid---a uniform grid---to 
introduce the cell-centered grids used for finite volume methods.  Then we describe 
the more complicated nonuniform and adaptive grid structure used to study self-similar 
collapse.  

\subsection{Ghost Cells and Uniform Grids} \label{sec:ghost-cells}
The entire numerical grid domain, $\Omega$, consists of two sub-domains: 
the physical domain, $\Omega_o$, and the \textit{ghost} domain, $\Omega_g$.  
The physical domain represents the physical space that we are modeling, 
whereas the ghost domain is used to 
``extend'' the grid so that the same update algorithm can be used on the entire 
physical domain (even for boundary points).  For example, we require 
two ghost cells at each boundary because, as shown in Figure~\ref{fig-stencil}, the update method uses a 5-cell
stencil.
In this thesis, we define $N_g$ to be the number of ghost cells at each 
boundary (so that if $\Omega_o$ has two boundaries, and $N_g=2$, then a total of $4$ ghost cells are used).

In the following, we describe how ghost cells are defined in spherical 
symmetry. 

Let  $\mathcal{C}_i$ denote the $i^{th}$ cell that is centered about 
$r = r_i$, with $i = 1, \ldots, N_r$ and where $N_r$ is the number of cells 
in $\Omega$.  The domains are defined as:
\beq{ 
\Omega_g = \left( \mathcal{C}_1, \ldots , \mathcal{C}_{N_g}, 
	\mathcal{C}_{N_r - N_g + 1}, \ldots , \mathcal{C}_{N_r} \right)
\label{ghost-domain}
}
\beq{
\Omega_o = \left( \mathcal{C}_{N_g+1}, \ldots , \mathcal{C}_{N_r - N_g} \right)
\label{physical-domain}
}
so that the two domains \emph{do not} overlap:
\beq{
\Omega_o \cap \Omega_g = \emptyset  \label{origin-intersection}
}
This way, all the physical cells in $\Omega_o$ are updated using the same 
stencil; however,
ghost cells near the origin are updated differently than those at the 
outer boundary (see the following sections).

The coordinate vector, $r_i$, that we use is defined as
\beq{
r_i \equiv r_{\min} + \left( i - N_g - 1/2 \right) \Delta r  
\quad ,  \ i = 1, \ldots, N_r  \quad ,\label{ghost-coords}
}
 where
\beq{ 
\Delta r \equiv \frac{ r_{\max} - r_{\min} }{ N_r - 2 N_g }  \label{dr-def}
}
and
where $r_{min}$ is the coordinate of the first physical cell's left border and 
$r_{\max}$ is the coordinate of the last physical cell's right border.  
That is, the first physical cell $\mathcal{C}_{N_g + 1}$ is 
located at $r_{N_g + 1} = r_{min} + \Delta r / 2$,  while the last
physical cell $\mathcal{C}_{N_r - N_g}$ is located at 
$r_{N_r - N_g} = r_{\max} - \Delta r / 2$ .

\begin{figure}[htb]
\centerline{\includegraphics*[bb=1in 2.5in 8in 6.5in, scale=0.6]{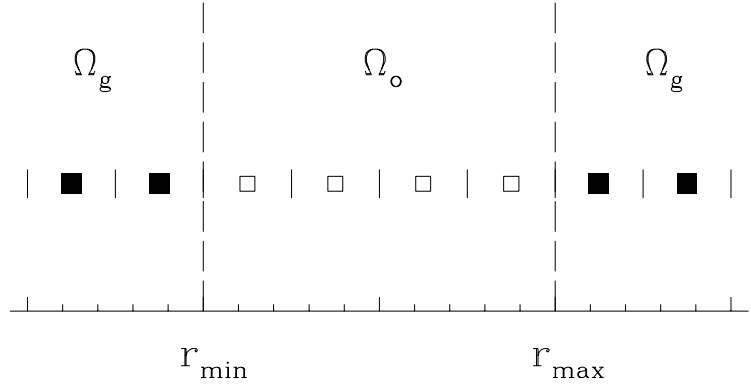}}
\caption[Illustration of the spatial discretization of the solution domain.  This example shows 
a sample discretization with $N_r = 8$  and $N_g = 2$.]{ Illustration of the spatial 
discretization of the solution domain.  This example shows a sample discretization with $N_r = 8$  and $N_g = 2$.
Squares denote the centers of cells, and the short vertical lines denote cell boundaries.  
The dashed vertical lines located at $r=r_{\min}$ and $r=r_{\max}$ separate the ghost cell domain, 
$\Omega_g$, from the physical domain, $\Omega_o$.  The filled squares represent ghost cells, 
while the empty squares represent physical cells.  
\label{fig-grid}}
\end{figure}

In spherical symmetry, one would typically set $r_{min} = 0$.  
From the definition of the grid coordinates  (\ref{ghost-coords}), it can easily be
seen that the first physical cell is offset by $\Delta r / 2$ from the origin; 
this way of discretizing the grid ensures that meshes can be refined in a consistent 
manner.  We will denote  discretized grid functions 
such that $Q_i^n$ denotes the value $Q(r_i, t^n)$. 
Figure~\ref{fig-grid} depicts how the domains are discretized. 

\subsection{The Nonuniform Mesh}
\label{sec:non-uniform-mesh}

In order to track the CSS behavior of near critical solutions, we need to 
numerically resolve the dynamics that take place on continuously-decreasing
spatio-temporal scales.  From previous work in critical phenomena with perfect fluids, 
we know the qualitative behavior of the collapse and so we can tailor our refinement
procedure accordingly.  If this were not the case, we would have to resort to more sophisticated and 
general Adaptive Mesh Refinement (AMR) techniques such as Berger and Colella's 
algorithm \cite{berger-colella} for conservative systems. 
Since the self-similar regime focuses onto the origin 
at some finite proper time, the grid refinement is only  needed near 
the origin.  By using a nonuniform grid that is spaced logarithmically, we are able 
to concentrate computational resources on the most important region.  Also, the logarithmic
grid allows us to extend the outer boundary further than we would be able to with the 
same number of uniformly-spaced points, which subsequently reduces the effect the boundary 
conditions have on the interior solution.  

The prescription for defining the initial grid and the refinement process was inspired
by an algorithm used in \cite{neilsen-thesis}.  However, we believe that there are a few 
details omitted, or not implemented in that work, that improve the method without any additional complexity, 
and so we provide them here. 

As in \cite{neilsen-thesis}, the portion of the grid not containing ghost zones consists of $3$ subdomains:
\beq{
\begin{array}{l@{\ : \ }lll}
\Omega_a & 0 \le r \le r_a , &N_a \ \mathrm{cells}, &\Delta r = \Delta r_a ;\\
\Omega_b & r_a < r < r_b , &N_b \ \mathrm{cells}, &\mathcal{R}_i \equiv \ln(r_i), \ 
             \Delta \mathcal{R} = \mathcal{R}_{i+1/2} - \mathcal{R}_{i-1/2} ;\\
\Omega_c & r_b \le r \le r_c , &N_c \ \mathrm{cells}, &\Delta r = \Delta r_c.\\
\end{array}
\label{subgrids}
}
where $\Delta r_a$, $\Delta \mathcal{R}$, and $\Delta r_c$ are all constant. 
The cell centers are always  defined as the points that lie midway between
two consecutive cell boundaries, so $\mathcal{C}_i$ is located at 
$r = r_i$ with boundaries at $r_{i-1/2}$ and $r_{i+1/2}$.  This motivates the definition of $\Delta \mathcal{R}$ 
in (\ref{subgrids}).  

The logarithmically-spaced grid segment, $\Omega_b$, smoothly (in $\Delta \mathcal{R}$) connects
the higher resolution $\Omega_a$ grid adjoining the origin, to the 
lower resolution $\Omega_c$ grid abutting the outer boundary.  In order for the different
subdomains to connect smoothly, we demand that the grid-defining parameters satisfy the following 
relations:
\beq{
e^{ \mathcal{R}_a + \Delta \mathcal{R}} \, - \, e^{\mathcal{R}_a} \  = \ \Delta r_a  
\label{cont-exp-coords} \\
}
\beq{
e^{\mathcal{R}_b} \, - \, e^{\mathcal{R}_b - \Delta \mathcal{R}} \ = \ \Delta r_c
\label{cont-exp-coords-bc}
}
In addition, we have three more equations that relate the lengths of the discrete subdomains to their resolutions:
\beq{
r_a = N_a \Delta r_a   \quad , \quad \mathcal{R}_b = N_b \Delta  \mathcal{R} + \mathcal{R}_a 
\quad , \quad r_c = N_c \Delta r_c + r_b    \quad . 
\label{uniform-grid-constants}
}
Since we have five equations and 9 unknowns, $\{N_a,\Delta r_a, r_a, N_b,\Delta R, r_b, 
N_c,\Delta r_c, r_c\}$, we need only provide any four parameters to uniquely specify the grid. 
There are many ways of specifying 
such a grid, but we have found that one way in particular ensures that the subdomains match 
smoothly for any choice of parameter values.  First, notice that some parameters are 
integers, $\{N_a, N_b, N_c\}$, and some are floating-point values, $\{\Delta r_a, r_a, \Delta R, r_b, 
\Delta r_c, r_c=r_{\max}\}$.  Specifying the floating-point 
parameter values---in general---will lead to non-integer values of $\{N_a, N_b, N_c\}$ which, in 
turn, would lead to a numerical inconsistency in the matching conditions 
(\ref{cont-exp-coords})~-~(\ref{cont-exp-coords-bc}).  Thus, it is best to specify the 
integer-valued parameters and derive the rest.  Since
there are only three integer-valued parameters, we must specify one floating-point parameter 
as well.  Because we are most interested in the dynamics that takes place within and near domain 
$\Omega_a$, we have found it convenient to specify $\Delta r_a$.  Hence, we specify 
$\{N_a, N_b, N_c, \Delta r_a \}$ and obtain the remaining parameters as follows.

From (\ref{cont-exp-coords}), we obtain an equation for 
$\Delta \mathcal{R}$: 
\beq{
\Delta \mathcal{R} = \ln \left( \frac{ N_a + 1 }{ N_a } \right) 
\label{exp-resolution}
}
Using this with  (\ref{cont-exp-coords-bc}) and (\ref{uniform-grid-constants}), it can easily be seen that
\beq{
\Delta r_c \ = \  \Delta r_a \left( \frac{ N_a + 1 }{ N_a } \right)^{N_b - 1} 
\label{dr-c}
}
Finally, using (\ref{uniform-grid-constants}) and (\ref{dr-c}), we get the remaining two grid parameters
\beq{
r_b \ = \ \Delta r_c \left( N_a + 1 \right)
\label{r-b}
}
\beq{
r_{{\max}} \ = \ r_b + N_c  \Delta r_c  \ = \ \Delta r_c \left( N_a + N_c + 1 \right)
\quad .
\label{rmax-avm}
}

\subsection{The Refinement Procedure}
\label{sec:refinement-procedure}
In order to properly resolve CSS solutions, it is necessary to periodically 
add cells near the origin since this is where the spatial and temporal scales of the solution 
become the smallest.  This is done by reducing $\Delta r_a$ by some fraction, $f_{\mathrm{reg}}$, 
and adding cells to $\Omega_b$ so as to maintain smoothness in $\Delta r(r)$ across the 
two subdomain boundaries.  It can then easily be derived that the following is the transformation 
law of grid parameters during a refinement process:
\beqa{
\Delta r_a \ & \longmapsto \ & \frac{ \Delta r_a } { f_{\mathrm{reg}} }\\[0.2cm]
N_a &\longmapsto& N_a \\
N_b &\longmapsto& N_b + \Delta N_b \\
N_c &\longmapsto& N_c
}
where 
\beq{
\Delta N_b \ \equiv \ \mathrm{NINT}\left[ 
\frac{ \ln (f_{\mathrm{reg}}) }{ \Delta \mathcal{R}} \right] \quad ,
\label{d-Nb}
}
And the  \textrm{NINT()} function returns the nearest integer to its argument.  Since the user-specified
$f_{\mathrm{reg}}$ will not in general be such that $\Delta N_b$ is precisely integer-valued, we need to 
recalculate it from the $\mathrm{NINT}(\Delta N_b)$.  
This is done by initially setting
\beqa{
\Delta N_b & \ \equiv \ & \mathrm{NINT}\left[ 
\frac{ \ln (f_{\mathrm{reg}}^{\prime}) }{ \Delta \mathcal{R}} \right] \quad  
\label{d-Nb2} \\[0.5cm]
f_{\mathrm{reg}} &=& e^{ \Delta N_b \ \Delta \mathcal{R} } \quad , 
\label{ref-factor-transform}
}
where $f_\mathrm{reg}^\prime$ is the user-specified value of $f_\mathrm{reg}$.
After the refined coordinates are calculated, the grid functions are interpolated
onto the new grid from the original grid via $3^{\mathrm{rd}}$-order ($4$-point) interpolation. 
Note, however, that the newly introduced coordinates  exist only in $\Omega_a$ and for the first $\Delta N_b$ cells
of $\Omega_b$.  In particular, all cells in $\Omega_c$ and the original part of $\Omega_b$ 
remain at the same coordinate locations and consequently interpolation is not required there. 

The decision as when to refine the grid is determined by tracking a feature of the solution and ensuring 
that there are a minimum number of cells between it and the origin. 
Since the solutions under study are CSS, this process is an easy one: in the self-similar regime, the solution looks 
functionally the same for all time, but on ever-decreasing scales.
We have chosen to track the local maximum of $2m(r)/r$ that lies nearest to the origin, since it
has empirically been found to always lie near the self-similarity horizon for near-critical solutions.
Tracking ${\max}(2m(r)/r)$ also ensures that the approximate locations of any black hole surfaces that may arise will be 
resolved since ${\max}(2m(r)/r) \rightarrow 1$ as they form in our Schwarzschild-like 
coordinates.   
Hence, we refine the mesh when this maximum 
first passes within $r_a/f_\mathrm{reg}$, thereby requiring there to be between $N_a/f_\mathrm{reg}$ 
and $\sim \! N_a$ cells within the self-similar region.  

In order to perform convergence tests of our nonuniform evolutions, we need a quantitative way for 
refining the mesh locally.  Let \[\{\Delta r_a(l), \Delta \mathcal{R}(l), \Delta r_c(l), N_a(l), N_b(l), N_c(l)\}\] 
represent the grid parameters for a grid at \emph{level of refinement}, $l$.  The $l=0$ grid is called the 
base grid, from which the grid attributes of all other grids of $l \ne 0$ are calculated.  The following is 
how we define the parameters of grids at higher levels of refinement:
\beq{
\begin{array}{lll} 
N_a(l) = 2^l N_a(0) ,  &N_b(l) = 2^l N_b(0) ,  &N_c(l) = 2^l N_c(0) \\[0.2cm]
\Delta r_a(l) = \Delta r_a(0) / 2^l ,  &\Delta \mathcal{R}(l) = \Delta \mathcal{R}(0) / 2^l   ,
&\Delta r_c(l) = \Delta r_c(0) / 2^l   
\end{array}   \quad .  \label{nonuniform-levels}
}

\begin{figure}[htbb]
\centerline{\includegraphics*[bb=0.4in 2.1in 8.0in 9.6in, scale=0.5]{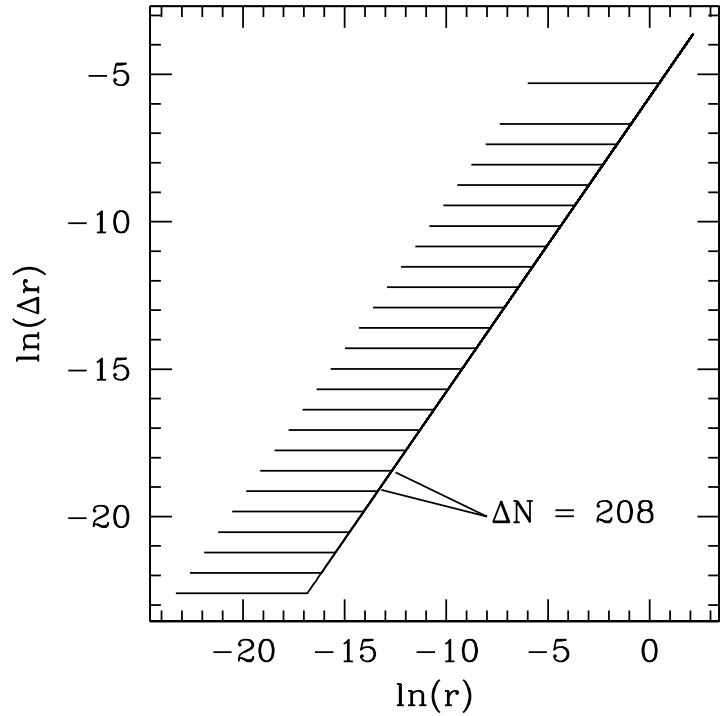}}
\caption[The logarithm of the local resolution, $\Delta r(r)$, is plotted here as a function
of radius.  This particular run was for the nearest subcritical solution we were able to obtain 
for a star with $\rho_c = 0.05$, $\Gamma = 2$, and $K = 1$; all but the second refined grid used in this 
evolution is shown here.]{The logarithm of the local resolution, $\Delta r(r)$, is plotted here as a 
function of radius.  This particular run was for the nearest subcritical solution we were able to obtain 
for a star with $\rho_c = 0.05$, $\Gamma = 2$, and $K = 1$; all but the second refinement is shown here.  
The free parameters that define the grid structure and refinement for this run 
are $N_a=300$, $N_b(t=0)=500$, $N_c=20$, $\Delta r_a = 0.005$.  These particular values are such 
that 208 cells are added to $\Omega_b$ at each refinement.  The final value 
for $\Delta r_a$ is about $1.5 \times 10^{-10}$.
\label{fig:regridding}}
\end{figure}

In all our runs, we have used $f_\mathrm{reg}^\prime = 2$ so as to approximately double 
the resolution in $\Omega_a$ during each refinement.  The other free grid parameters are 
usually chosen as functions of the given star data we start with.  We pick an initial grid such that 
the $r_a$ is slightly larger than the stellar radius, $R_\star$, so that $r_{\max}$ is about $5-10R_\star$, and 
$N_a \simeq 300-600$.  This places the outer boundary sufficiently far from the dynamical region
while providing adequate initial resolution of the star.   For example, the grid parameters used for the 
runs shown in Figure~\ref{fig:regridding} for $l=0$, $\rho_c=0.05$ 
($R_\star(\rho_c=0.05) \simeq 1.11$ in our unit-less coordinates, $K=1$)
are $N_a=300$, $N_b(t=0)=500$, $N_c=20$, $\Delta r_a = 0.005$ which leads to $r_a = 1.5$ and 
$r_{\max} \simeq 8 R_\star$. 

\section{The Numerical Solution of the Metric Functions}
\label{sec:numer-solut-metr}

In this section, we describe the finite difference approximations we use to solve the Hamiltonian 
constraint (\ref{hamconstraint}) and the slicing condition (\ref{slicingcondition}).  As examples, 
we use the form of the Hamiltonian constraint and slicing condition that they take when only the perfect
fluid is present; it should be straightforward to extrapolate the discretization procedure for the geometric equations
to the cases with different matter models.

Since we are using nonuniform grids, it is important to always use discretizations that are centered about the 
cell center.  Also, we need to keep in mind that the metric functions are calculated at the cell borders 
as opposed to the cell centers, where the fluid quantities are calculated.   Because of the particular form 
of the equations, it is best to difference $\ln(a_j)$ and $\ln(\alpha_j)$ instead of $a_j$ and $\alpha_j$ in order
to increase the calculation's precision.  
Thus, we difference the Hamiltonian constraint  in the following way:
\beq{ 
\ln(a_{j+1/2}) - \ln(a_{j-1/2})  =  \left(r_{j+1/2} - r_{j-1/2} \right) \exp\left( \ln(a_{j+1/2}) + \ln(a_{j-1/2})\right) \nonumber
}
\beq{
\quad \times  \left[ 4 \pi r_j \left( \tau_j  + D_j \right) - \frac{1}{2 r_j} \left( 1 - 
4 \exp\left( -\ln(a_{j+1/2}) - \ln(a_{j-1/2})\right) \right) \right]
\quad \quad \quad , \label{discrete-hamconstraint}
}
and difference the slicing condition similarly:
\beq{ 
\ln(\alpha_{j+1/2}) - \ln(\alpha_{j-1/2})  =  \frac{1}{4}  \left(r_{j+1/2} - r_{j-1/2} \right)
\left(a_{j+1/2} + a_{j-1/2} \right)^2 \nonumber 
}
\beq{
\quad \quad \quad  \times  \left[ 4 \pi r_j \left( S_j v_j  + P_j \right) + \frac{1}{2 r_j} \left( 1 - 
\frac{4}{\left( a_{j+1/2} + a_{j-1/2}\right)^2} \right) \right]
\quad . \label{discrete-slicingcondition}
}

Since the slicing condition is a homogeneous ODE  in $\alpha$, we start from $r=r_{\max}$ and solve the 
algebraic equation (\ref{discrete-slicingcondition}) for the unknown neighbor value, continuing 
the process to $r=r_{\min}$.  However, integrating the Hamiltonian constraint is more difficult since it is inhomogeneous
in $a$.  A Newton-Raphson method is used to minimize the residual of (\ref{discrete-hamconstraint}) to 
solve for the unknown neighbor value, which is $a_{j+1/2}$, since we start the integration from $r=r_{\min}$. 

In order to perform an independent residual test on our numerical solutions, we use the following 
distinct discretizations for the above metric equations.  The independent residual for the Hamiltonian 
constraint is
\beqa{
\mathrm{Resid}^\mathrm{HC}_{j} & = & \frac{a_{j+3/2} + a_{j+1/2} - a_{j-1/2} - a_{j-3/2}}{2\left(r_{j+1} - r_{j-1}\right)} \nonumber \\[0.3cm]
                   & - & \frac{1}{4} \left(a_{j+1/2} + a_{j-1/2}\right)^3  \label{indep-discrete-hamconstraint} \\[0.3cm]
& \times & \left[ 4  \pi  r_{j} 
                       \left( \tau_j + D_j \right) - \frac{1}{2 r_j} \left( 1 - 
\frac{4}{\left(a_{j+1/2}+a_{j-1/2}\right)^2} \right) \right]
\quad , 
}
and the independent residual for the slicing condition is
\beqa{
\mathrm{Resid}^\mathrm{SC}_{j} & = & \frac{\alpha_{j+3/2} + \alpha_{j+1/2} - \alpha_{j-1/2} - 
\alpha_{j-3/2}}{2\left(r_{j+1} - r_{j-1}\right)} \nonumber \\[0.3cm]
& - & \frac{1}{8} \left( \alpha_{j+1/2} + \alpha_{j-1/2}\right) \left( a_{j+1/2} + a_{j-1/2}\right)^2 \label{indep-discrete-slicingconstraint}  \\[0.3cm]
& \times &  \left[ 4  \pi  r_{j}  \left( S_j v_j + P_j \right) + \frac{1}{2 r_j} \left( 1 - 
\frac{4}{\left(a_{j+1/2}+a_{j-1/2}\right)^2} \right) \right]
\quad . 
}

\section{Boundary Conditions}
\label{sec:bound-cond}


Since computers only have a finite amount of memory at their reserve, the number of grid cells in the 
domain must, of course, also be finite.  Since the normal update procedures for a given cell require the grid function
values of its neighbors, the cells at the very edges of the numerical domain must be updated in a 
special way since---in spherical symmetry---they are lacking one or more neighbors.   We refer 
to such special procedures as \emph{boundary conditions}.  These boundary conditions come in different varieties
depending on where they are used.  
In the next sections we will describe the boundary conditions we use for the metric functions and the 
fluid functions.  

\subsection{Fluid Boundary Conditions}
\label{sec:fluid-bound-cond}

For the outer boundary condition, we use the typical outflow condition 
that simply involves copying the fluid quantities into the ghost region which is essentially
a $1^\mathrm{st}$-order extrapolation.  Since our experience, as well as that of others, indicates that 
this condition is fairly robust and non-reflective, we did not bother to experiment
with more sophisticated conditions.  

The regularity conditions at the origin are, however, more sophisticated.  Since 
the cells on which the fluid fields are defined are not centered on the origin, typical
$O(\Delta r^2)$ regularity conditions are not as well-behaved as those for 
origin-centered cells.  Hence, we have found it helpful to use higher-order, conservative 
interpolation for the fields on the first physical cell.  Since the fluid fields, 
$\bar{\mathbf{q}}_i$, are to be interpreted as cell-averages of some conserved function,
which we will call $\mathbf{Q}(r)$, an interpolation is said to be conservative if the 
integral of the function on a local domain is conserved by the interpolation procedure.
We first assume that the interpolation function $\mathbf{Q}_i(r)$ that is associated with 
a cell $\mathcal{C}_i$  has a polynomial expansion of degree $N-1$:
\beq{
\mathbf{Q}_i (r) = \sum^{N-1}_{n = 0} \mathbf{a}_n 
\left( r - {r}_i \right)^n  
\label{interp-function-def}
}
with $N$ coefficients $\mathbf{a}_n$.  
These coefficients are found by demanding that $\mathbf{Q}_i$ maintains
conservation locally.  That is, a set $\mathcal{S}_{i}$ of $N$ cells are chosen in the neighborhood 
of cell $\mathcal{C}_i$ and requires that $\mathbf{Q}_i$ is 
such that it reproduces the known values $\bar{\mathbf{q}}_k$, where 
$\mathcal{C}_k \in \mathcal{S}_i$.  Specifically, the coefficients 
$\mathbf{a}_n$ are calculated by solving the following set of $N$ equations:
\beqa{
\bar{\mathbf{q}}_k & \ = \ & \frac{1}{V_k} \int_{V_k} \, \mathbf{Q}_i(r)  \ dV \\ 
&\ = \ & \frac{3}{ r_{k+1/2}^3 - r_{k-1/2}^3} \sum^{N-1}_{n = 0} \mathbf{a}_n  
\left[ \int^{r_{k+1/2}}_{r_{k-1/2}}
\left( r - {r}_i \right)^n \ r^2  dr \right] 
\label{ci-ss}
}
for all $\mathcal{C}_k \in \mathcal{S}_i$.
Since this interpolation procedure is used at the origin where local flatness is demanded, then 
we make the assumption that the variation of $\sqrt{{}^{(3)}g\ }$---which should be in the integrand---has
negligible effect and is neglected.  Once (\ref{ci-ss}) is solved for the coefficients $a_n$, then
the interpolation procedure is completed by using this same equation, (\ref{ci-ss}), 
for a cell $\mathcal{C}_j \notin \mathcal{S}_i$ for whose $\qbar_j$ we are interpolating.

From the demand of regularity at the origin, the fields $\rho_\circ, P, D, \tau$
are all even in $r$, at the origin, while $v, S$ are odd in $r$ near $r=0$.   
Thus, $a_n=0$ for odd $n$ in the interpolation function of the even fields, and 
$a_n=0$ for even $n$ in the odd interpolations.  In our case, cell $\mathcal{C}_{N_g+1}$
lies in a uniform domain, $\Omega_a$,  and so the $O(\Delta r^3)$ conservative
interpolation equation can be easily determined:
\begin{itemize}
\item $\underline{\mathrm{\mathbf{For}} \  N = 4, \ j \equiv N_g + 1}$: 
        \begin{list}{}{}
        \item \textbf{Even:} 
                \[ {\bar{\mathbf{q}}}_{_{j}} \ = \  \frac{1}{1627} \left( 
                    3311 \, {\bar{\mathbf{q}}}_{_{j+1}} 
                  - \, 2413 \, {\bar{\mathbf{q}}}_{_{j+2}} 
                  + \, 851  \, {\bar{\mathbf{q}}}_{_{j+3}} 
                  - \, 122  \, {\bar{\mathbf{q}}}_{_{j+4}} \right) \]
        \item \textbf{Odd:} 
                \[ {\bar{\mathbf{q}}}_{_{j}} \ = \  \frac{1}{36883} \left( 
                    35819 \, {\bar{\mathbf{q}}}_{_{j+1}}
                  - \, 16777 \, {\bar{\mathbf{q}}}_{_{j+2}} 
                  + \, 4329  \, {\bar{\mathbf{q}}}_{_{j+3}} 
                  - \, 488   \, {\bar{\mathbf{q}}}_{_{j+4}} \right) \]
        \end{list}
\end{itemize}
Since $\Pi$ and $\Phi$ are combinations of even and odd functions, their regularity conditions are not 
as straightforward.  To determine their behavior at the origin, we first calculate
the interpolated values of $\tau$ and $S$ at  $\mathcal{C}_{N_g+1}$ since the regularity behavior of these two 
functions is known.  Then, $\Pi$ and $\Phi$ are calculated on $\mathcal{C}_{N_g+1}$ by their definitions 
(\ref{Pi-ideal}-\ref{Phi-ideal}) using these interpolated values for $\tau$ and $S$.

\subsection{Geometry Boundary Conditions}
\label{sec:geom-bound-cond}
In solving the Hamiltonian equation (\ref{polar-areal-hamiltonian-const}), we demand that spacetime be 
locally flat at the origin; this implies $a(0,t)=1$.  This condition can always be maintained in a dynamical
evolution, even for cases that lead to black hole formation, since the lapse decays exponentially at the origin 
as a physical 
singularity starts to form.  Hence, the proper time essentially ``freezes'' near the origin
before the singularity can actually arise.  Even though our spacetime foliation 
avoids physical singularities, it is still susceptible to coordinate pathologies
that form near the apparent horizon of the collapsing system because of the metric's
Schwarzschild-like nature. 

The slicing condition (\ref{polar-areal-slicing-condition}) is solved by integrating inward from the 
outer boundary, and we make use of the freedom we have in relabelling constant $t$ surfaces via 
$\alpha \rightarrow k \alpha$, for an arbitrary positive constant, $k$.  This freedom is manifest 
in the slicing condition itself, which is an ODE homogeneous in $\alpha$.  Hence, we use the 
typical parameterization that allows our coordinate time to coincide with proper time at 
$r=\infty$.  Since our grid extends only to a finite $r$, we cannot make this condition hold precisely.  
However, we can employ Birkhoff's theorem, which states that
any compact and spherically symmetric distribution of mass-energy has the same external 
spacetime as the Schwarzschild metric of identical mass, to estimate the correct asymptotic behavior.  If we assume
that all the matter  remains within our grid, then the metric
exterior to the grid should be equivalent to Schwarzschild, and since the Schwarzschild metric 
is asymptotically flat, we can rescale $\alpha$ so that it makes our metric equivalent to 
Schwarzschild at $r_{\max}$.  Specifically, this is done by setting 
\beq{
\alpha(r_{\max}) = \frac{1}{a(r_{\max})}
\label{alpha-bc}
}
This provides the appropriate rescaling to $\alpha$ so that it asymptotes to 1 at $r=\infty$, 
making proper time at space-like infinity coincide with coordinate time.  

\section{Instability at the Sonic Point in the CSS Regime}
\label{sec:instability}


In this section we provide a description of an instability observed to develop 
near the sonic point of near-critical solutions.  The instability made it impossible to 
obtain a consistent bracket about the threshold solution's critical parameter, $p^\star$, 
for $p-p^\star \le 10^{-9}$, when using the approximate Roe solver.  This limited our study 
since we found that we needed to tune quite closely to the threshold solution
in order to calculate an accurate value of the scaling exponent $\gamma$.

The instability manifests itself in different ways, depending on the type of 
cell reconstruction used.  For example when using the conservative variables to 
reconstruct the solution at the cell borders, we found that the conservative variables themselves
remained smooth, but that each primitive function $\mathbf{\bar{w}}$ exhibited
persistent oscillations near the sonic point of order 2-4 grid cells in extent.  On the other hand if we 
reconstructed using the primitive variables, then similar oscillations appeared in the 
conservative variables, while the primitive variables remained smooth.  The oscillations in either
case eventually diverged, leading to super-luminal velocities, negative pressures, and 
erroneous discontinuities.  Also, reconstruction of the cell border states using 
the characteristic variables led to worse stability than the primitive variable reconstruction.  
The so-called \emph{characteristic variables} are 
those variables which embody the fundamental waves of the solution.  The diagonalization of 
the quasi-linear system (\ref{quasilinear-eom}) leads to three independent, or scalar, 
advection equations.  From these equations, the characteristic variables are the advected quantities, while
the eigenvalues of $\mathbf{A}$ are the velocity factors (characteristics speeds) in the scalar advection 
equations.  An example of the instability in an evolution using primitive 
variable reconstruction is shown in Figure~\ref{fig:roe-q1-instability}.  
\vspace{0.2cm}

\centerline{\includegraphics*[bb=0.3in 2.6in 8.0in 9.1in, scale=0.6]{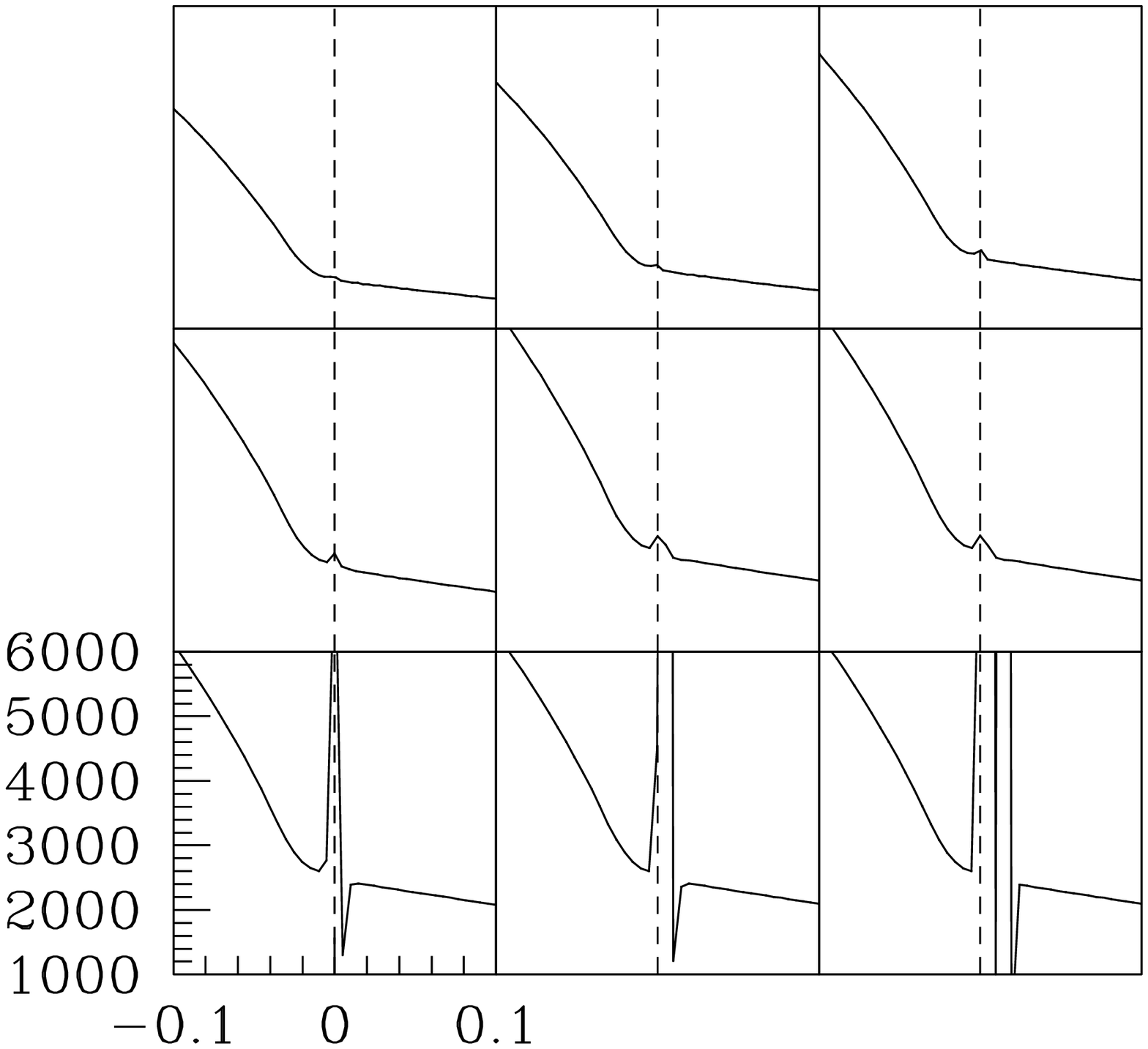}}
\vspace{-0.5cm}
\begin{figure}[h]
\caption[Displayed here is the conservative variable $D(\mathcal{X},\mathcal{T})$ from the most nearly critical evolution 
obtained with the use of  the approximate Roe solver without smoothing.]{Displayed here is the conservative variable 
$D(\mathcal{X},\mathcal{T})$ from the most nearly critical evolution 
obtained with the use of the approximate Roe solver without smoothing. 
$\mathcal{X}$ and $\mathcal{T}$ are the self-similar coordinates
defined in (\ref{X-ss}) and (\ref{T-ss}), respectively.  The dashed line indicates the 
location of the sonic point, which---by definition of $\mathcal{X}$---is always 
at $\mathcal{X}=0$.  No refinement takes place during the period shown here, 
$\Delta r_a\simeq1.55\times10^{-7}$ is the minimum resolution of the spatial coordinates, and the 
Courant factor used was $0.4$.  
From left-to-right and top-to-bottom, the $\mathcal{T}$ values of the frames are
$-10.4109$, $-10.4977$, $-10.5916$, $-10.6938$, $-10.7822$, $-10.7823$, 
$-10.7824$, $-10.7825$, $-10.7826$.
The last five frames are the last 5 times steps before the code crashes, while the 
first four frames are more spread out in $\mathcal{T}$.  Hence, we see that the feature at the 
sonic point exists for a considerable period of time before diverging. 
The initial data used in this solution was a TOV star with central density $\rho_c = 0.05$ 
that is perturbed using profile $U_1$. 
\label{fig:roe-q1-instability}}
\end{figure}

We also found a dependence on the type of slope limiter used to perform the 
cell reconstruction process.  The limiters we tried were the 
\textrm{minmod}, Superbee, and monotonized central-differenced (MC) limiters. 
Typically, the \textrm{minmod} limiter was used since it provided the most diffusion near 
discontinuities and consequently led to more stable evolutions.  The Superbee and MC limiters 
were both found to exhibit slight Gibbs phenomenon in shock tube tests, led to more difficulties 
near the fluid's floor that surrounds the star, and produced more pronounced spurious oscillations 
near the sonic point of near-critical solutions.  Hence, as stated previously, the \textrm{minmod} limiter is 
used throughout the thesis. 

In addition, we ensured that the regridding procedure, as described in 
Section~\ref{sec:refinement-procedure}, was not responsible for the instability.  In order to perform this test we
first evolved a system that was tuned near the critical solution.  We extracted 
$\mathbf{\bar{w}}(r,t)$, $\alpha(r,t)$ and $a(r,t)$ at a specific time $t$ in this evolution, before the 
appearance of instability, and interpolated the functions onto a new grid having more cells near 
the origin.  This allowed for the evolution to continue on a single discrete domain, without the need to regrid.  
We found no significant differences in a comparison of the full evolution with the 
adaptive grid, to the evolution on this new grid.  

Moreover, we have found that the instability does not ``converge away.''  In order to examine the dependence 
of the blow-up on the resolution of the grid in the limit $\Delta r \rightarrow 0$, we tuned the initial data 
towards criticality for three different levels of refinement, where refinement was done locally so that 
$\Delta r_{l}(r) = 2 \Delta r_{l+1}(r) \ \forall \ r$.  As the level of refinement 
increased, the oscillations associated with the instability did not significantly change in magnitude, and 
remained confined to approximately the same number of grid cells.  Also, the evolution eventually 
crashed at the location of the instability in all cases. 
This suggests that the instability may be due to a failure of the numerical methods used. 

In order to understand the source of the instability, we first need to provide a better 
description of the near-critical solution.  
When the initial data has been tuned close to the critical solution at the threshold of black hole
formation, the fluid's evolution becomes increasingly relativistic and its dynamics 
shrink to exponentially smaller scales.  The behavior near the origin is self-similar up to 
the sonic point, $r_s$, where the flow velocity equals the speed of sound, $c_s$.  
If we are to assume that in near-critical solutions the fluid becomes 
ultra-relativistic---e.g. $P \gg \rho_\circ$---in the self-similar regime, then we should 
anticipate that $c_s(r<r_s) \rightarrow 1 $ there.  Also, from previous 
ultra-relativistic studies using
$\Gamma=2$ such as \cite{brady_etal,neilsen-crit}, we 
should expect that $v \rightarrow 1$  for $r > r_s$ as well.  Thus, about the sonic point, 
the characteristic speeds (\ref{ideal-piphi-evalues}) should take the values given 
in Table~\ref{table:ultrarel-char-speeds}.  
\begin{table}[htb]
\begin{center}
\begin{tabular}[htb]{c|c|c}
\hline
Characteristic Speed  &$\lambda(r <  r_s)$ &$\lambda(r > r_s)$  \\
\hline \hline
$\lambda_1$  &$ < 1$   & $\sim 1$ \\
$\lambda_+$  &$\sim 1$   & $\sim 1$ \\
$\lambda_-$  &$\sim -1$   & $\sim 1$ \\
\hline
\end{tabular}
\end{center}
\caption{Asymptotic values of the fluid's characteristic speeds in the ultra-relativistic 
limit. The sonic point is located at $r = r_s$. \label{table:ultrarel-char-speeds}  }
\end{table}

In fact, this is exactly what we find when using the ideal-gas state equation, as seen 
in Figure~\ref{fig:p-rho-ultra-regime} 
and Figure~\ref{fig:crit-characteristics}.  In Figure~\ref{fig:p-rho-ultra-regime}  we
see that $P \gg \rho_\circ$ within the self-similar region, but that $P(r) < \rho_\circ(r)$  for 
$r > r_s$.  Figure~\ref{fig:crit-characteristics} also demonstrates
how well the actual characteristics speeds from the calculation follow the above estimation.  

\clearpage
\centerline{\includegraphics*[bb=0.4in 2.3in 8.1in 9.8in, scale=0.6]{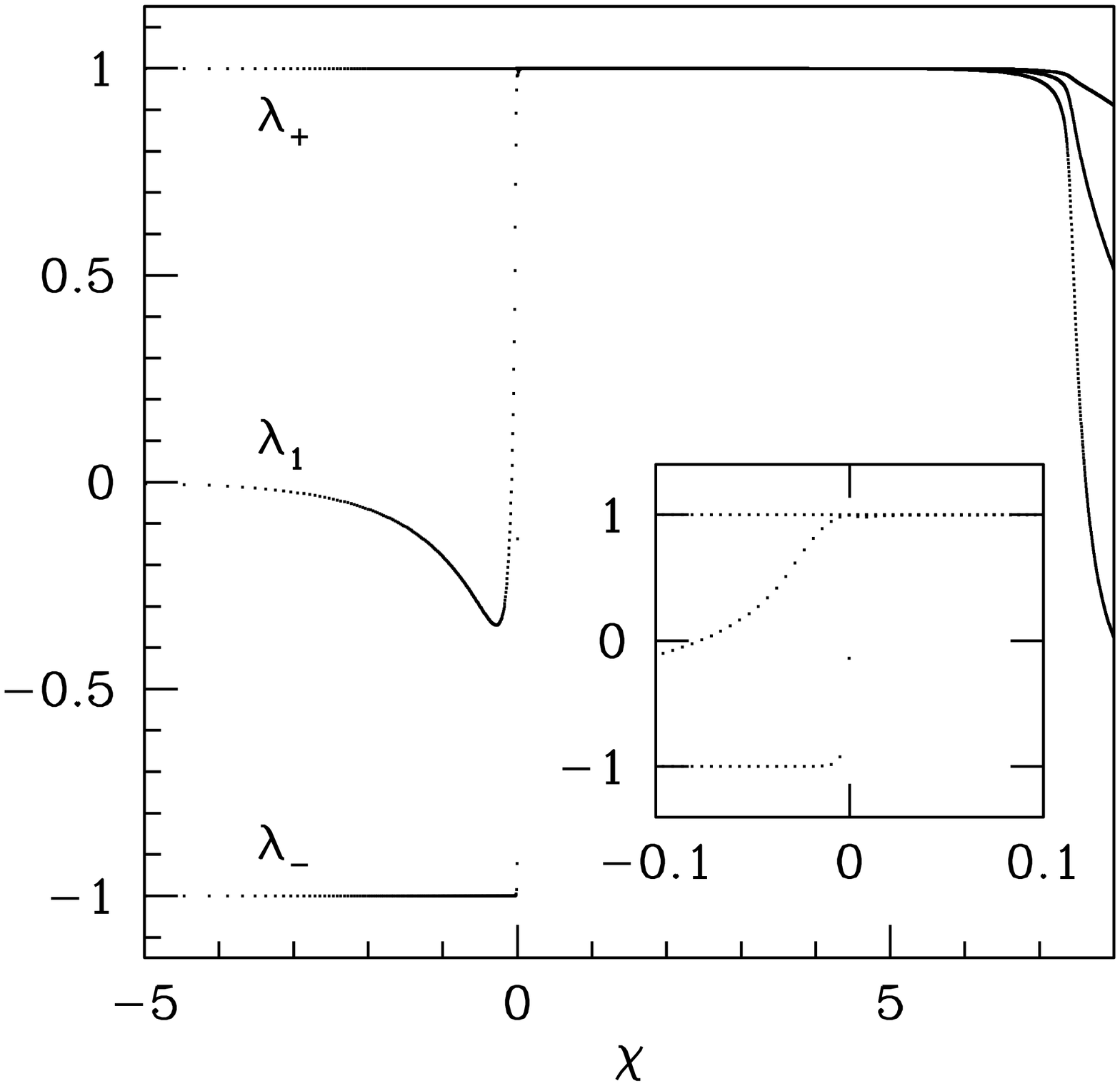}}
\begin{figure}[htb]
\caption[The characteristic speeds of the fluid for the most nearly critical solution obtained with
the approximate Roe solver without smoothing.]{The characteristic speeds of the fluid for the most 
nearly critical solution obtained with the approximate Roe solver without smoothing.  The wave speeds 
are plotted here as functions of the self-similar coordinate $\mathcal{X}$, and are shown at $\mathcal{T}=-10.6938$; 
$\mathcal{X}$ and $\mathcal{T}$ are defined by (\ref{X-ss}),(\ref{T-ss}).  
A closer view of the characteristic speeds near the sonic point is shown as an inset in the lower-right of 
the plot, revealing the severity of the discontinuity in $\lambda_-$ which is discussed in the text.
\label{fig:crit-characteristics}}
\end{figure}

\clearpage 
\centerline{\includegraphics*[bb=0.7in 2.3in 8.1in 9.8in, scale=0.6]{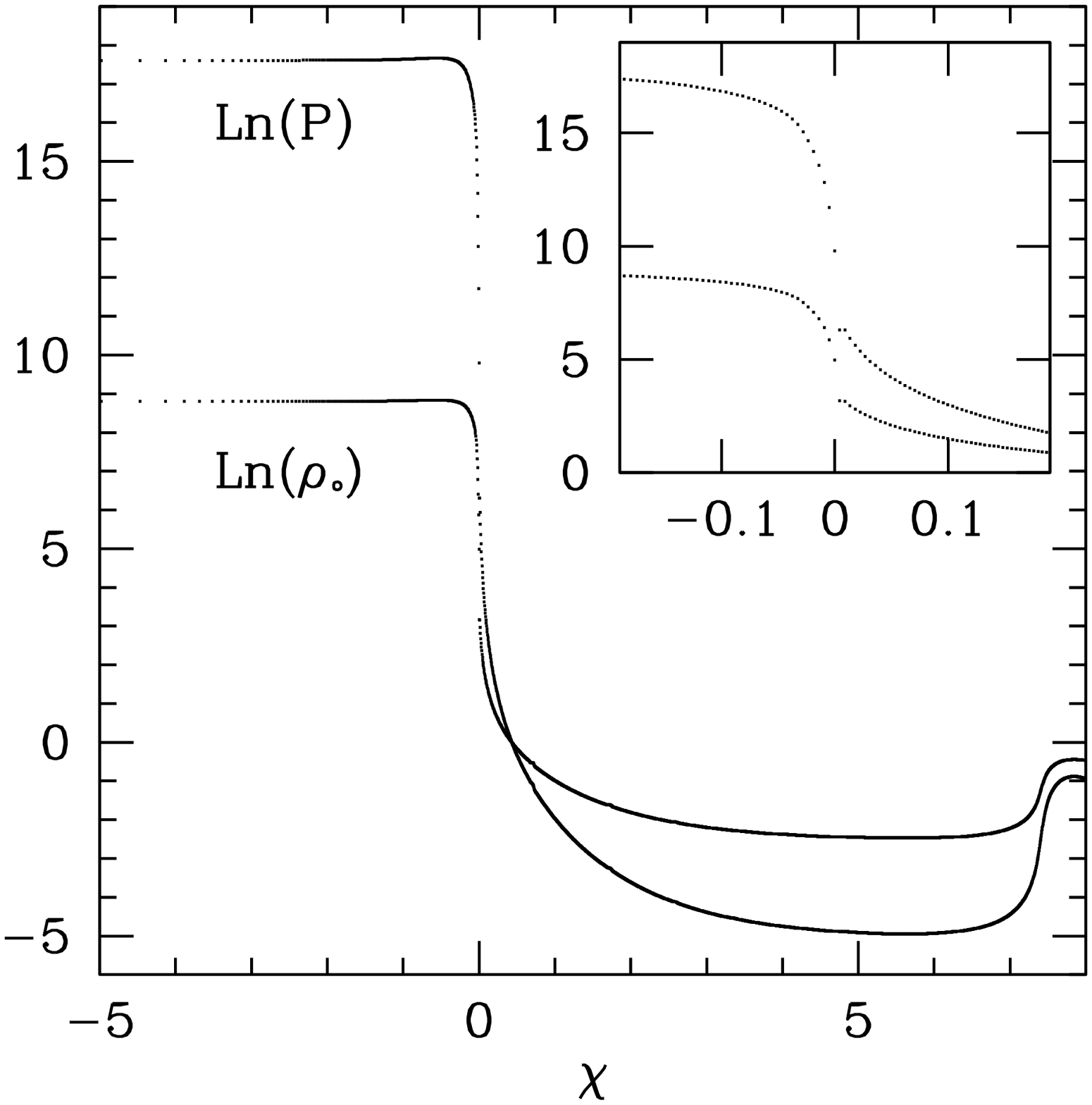}}
\begin{figure}[htb]
\caption[The pressure and rest-mass density of the most nearly critical solution obtained with
the approximate Roe solver without smoothing.]{The pressure and rest-mass density of the most nearly 
critical solution obtained with
the approximate Roe solver without smoothing.  $P$ and $\rho_\circ$ are plotted here versus
the self-similar coordinate $\mathcal{X}$, and are shown at $\mathcal{T}=-10.6938$; $\mathcal{X}$
and $\mathcal{T}$ are defined in (\ref{X-ss}),(\ref{T-ss}).  The fluid is clearly 
shown to be in the ultra-relativistic limit since $P/\rho_\circ \simeq 10^4$ near their maxima.  
However, beyond the sonic point at $\mathcal{X}=0$, this limit no longer holds and $P$ actually 
becomes less than $\rho_\circ$.  
A closer view of the distributions near the sonic point is shown as an inset in the upper-right of 
the plot that more clearly illustrates the formation of an expansion shock as discussed in the text.
\label{fig:p-rho-ultra-regime}}
\end{figure}

Not only do the calculated speeds match those anticipated quite well, but the transition 
from the self-similar, ultra-relativistic regime to the exterior solution is 
quite abrupt; the exterior solution lies at $r>r_s$, is not self-similar, and matches to an asymptotically flat spacetime.
For instance, the discontinuity in $\lambda_-$ is resolved by only a few grid points, 
signifying the presence of a shock which can also be 
seen for $r \sim r_s$ in the plots of $P(r)$ and $\rho_\circ(r)$ shown in Figure~\ref{fig:p-rho-ultra-regime}.  
This shock essentially defines the border of the self-similar region and follows the self-similar portion of 
the solution as it tends toward the origin.  Since $\lambda_-(r<r_s)<0$ and $\lambda_-(r>r_s)>0$,
the discontinuity  represents a point of transonic rarefaction \cite{leveque1}.  Also, the shock appears to 
be an expansion shock, 
which is entropy-violating, since it travels into a region of higher pressure and density. 
LeVeque states in \cite{leveque1} that the Roe solver can lead to entropy-violating shocks 
at transonic rarefactions since the linearization that the Roe solver performs on the EOM leads to 
a Riemann solution having only discontinuities and no rarefaction waves.
He illustrates this point in \cite{leveque2} using a boosted shock tube test that makes the rarefaction transonic.
Other failures of Roe's method that are attributed to its linearization have been 
shown by Quirk \cite{quirk}, and by Donat et al. 
\cite{donat-font} where an unphysical ``carbuncle'' forms in front of a relativistic,
supersonic jet.  

A first attempt to dissipate this apparently unphysical expansion shock involved applying artificial viscosity 
to the region about the sonic point.  Artificial viscosity techniques were first proposed and 
demonstrated by von Neumann and Richtmyer \cite{vonneumann-richtmyer}, and have been 
the traditional method for stably evolving hydrodynamic systems with shocks using finite 
difference techniques.  We followed Wilson's \cite{wilson2} artificial viscosity method and set 
$P \rightarrow P + Q$ in $\mathbf{f}$ alone, where 
\beq{
Q \ = \ c_\mathrm{av} D \, \left(\Delta r \pderiv{v}{r} \right)^2 \quad . \label{av-term}
}
and $c_\mathrm{av}$ is a user-specified parameter. 
Since we observed the instability to worsen as $v \rightarrow 1$, $Q$ became irrelevant as the flow became
more relativistic.  This was because $Q$ did not increase at the same rate as other
terms within $\mathbf{f}$ which contained factors of $W$.  

For our second attempt to circumvent the short-coming of the Roe solver, we performed localized 
smoothing of the conservative variables about the discontinuity in $\lambda_-$ at every 
predictor/corrector step of the fluid update.  Since the matter and geometry do not 
immediately become  self-similar, the smoothing procedure need not be 
used at early times.  Also, those solutions far from criticality do not require smoothing
since they do not enter the ultra-relativistic regime.  Hence, the smoothing is only required
when $p$ is close to $p^\star$, and when the profile of $\lambda_-$ becomes sufficiently discontinuous.  
Specifically, we start to use the smoothing procedure when $p - p^\star \le 10^{-8} - 10^{-9}$, and at
times when $\lambda_-$ begins to be resolved over approximately $10$ or fewer zones.  We can use
the same time, $t_s$, to begin smoothing for all runs since the evolution for $t<t_s$ is almost identical 
for all near-critical values of $p$. 

We also found that the 
instability worsened as the number of points between the origin and the sonic point decreased, 
as occurs in those cases where the solution disperses from the origin instead of forming a black hole.  
To diminish this effect, we performed mesh refinement in such a way as to always 
have an adequate number of points between the origin and the region being smoothed.  
This allowed the 
fluid to disperse even though the discontinuity $\lambda_-$ never reached $r=0$.  We note that the ability 
to follow evolutions through to their dispersal was necessary for our calculation of 
the scaling exponent, since we measured how a solution's global maximum of $T = {T^a}_{a}$ 
scales with $p^\star-p$ and this global maximum usually occurred as the fluid began to disperse. 
In order to resolve the space between the origin and the discontinuity near the sonic point, we performed
grid refinements whenever the discontinuity or ${\max}\left(2m/r\right)$ reached a certain number of grid cells
from the origin.  This allowed us to evolve dispersal cases further in time which, in turn,
granted us the ability to extend our scaling-law calculation further into the critical regime.   
The precise algorithm used to perform the smoothing and extra grid refinement process
is outlined in Table~\ref{table:smoothing-procedure}. 
\begin{table}[htb] 
\centerline{\fbox{\parbox{6in}{
\begin{tabbing}
\hspace*{1cm}\= \hspace{1cm} \= \hspace{1cm} \= \kill	
\textbf{\texttt{If }} $\left(t > t_\mathrm{smooth} \right)$ then \\[0.25cm]
              \> Find the first contiguous set of points, $\{r_\mathrm{sm}\}$, that satisfy\\[0.1cm]
	      \> \T  $-\lambda_-^{\min} \ < \ \lambda_-(r_\mathrm{sm}) \ < \ 
                         \lambda_-^{\min}\ $ 
	       for some constant $\ \lambda_-^min > 0 $. \\[0.25cm]
	      \> After every predictor or corrector step (\ref{predictor})-(\ref{corrector}) 
	            and for all $r_i \in \{r_\mathrm{sm}\}$ do:\\[0.2cm] 
	      \> \>  $\mathbf{\bar{q}}(r_i) 
                         = \frac{1}{2} \left[ \mathbf{\bar{q}}(r_{i-1}) 
                             + \mathbf{\bar{q}}(r_{i+1}) \right]$ \\[0.25cm]
              \> If $\left({\min}\left(\{r_\mathrm{sm}\}\right) \, 
                        < \, r_a / f_\mathrm{reg}\right)$, 
	            then refine grid per Section~\ref{sec:refinement-procedure}. \\[0.25cm]
\textbf{\texttt{End If}}
\end{tabbing}
}}}
\caption{Procedure used to smooth $\mathbf{\bar{q}}$ near the sonic point.  All results in 
the thesis are computed with $\lambda_-^{\min} = 0.95$. 
\label{table:smoothing-procedure}}
\end{table}

The diffusion introduced by the smoothing allowed us to further tune toward the critical
solution, eventually to $p^\star-p \simeq 5 \times 10^{-12}$, which represents a significant
improvement over the use of  Roe's solver alone.  However, we were still unable to calculate
the global maximum of $T$, $T_{\max}$, for the most nearly critical runs even though we could
identify them as being dispersal cases.  For instance, the minimum value of $p^\star-p$ for which 
we could calculate $T_{\max}$ was about $5 \times 10^{-10}$, as illustrated
in Figure~\ref{fig:sub-scaling-alldata}.  This is far smaller, however, than 
we would have been able to achieve without smoothing.

Other, more sophisticated approximate Riemann solvers have been shown to fare better 
than Roe's solver in  certain circumstances.  For example, Donat and
Marquina in \cite{donat-marquina} introduced the so-called Marquina flux formula, which 
attempts to combine Roe's flux with the Lax-Friedrichs flux in an automatic fashion.  
The Lax-Friedrichs aspect
of the method serves as an entropy-fix for the ``Roe'' part of the algorithm and is only
used when a characteristic changes sign across a cell boundary.  A striking difference 
in results obtained from the two methods is given in \cite{donat-font} where it was demonstrated 
how the Marquina method eliminated the aforementioned carbuncle phenomenon seen with Roe's 
solver.  We implemented the Marquina flux in order to see if it would perform better
near transonic rarefactions.  A test of this is shown in Figure~\ref{fig:css-shocktube}, 
where we have evolved a shock tube problem which emulates the fluid state about the sonic point
of near-critical solutions.  The initial conditions used for this test are 
$\left\{\rho_L, v_L, P_L\right\}=\left\{1.0\times 10^3 , -0.3, 1.0\times 10^6\right\}$ and
$\left\{\rho_R, v_R, P_R\right\}=\left\{0.3, 0.9994, 1.0\right\}$; these values are such 
that, initially, ${\lambda_+}_L\simeq0.9995$, ${\lambda_+}_R\simeq0.99998$, 
${\lambda_-}_L\simeq-0.99987$, and ${\lambda_-}_R\simeq0.98296$ which all closely 
follow those in Table~\ref{table:ultrarel-char-speeds}.  
The Roe and Marquina
solutions each used 400 points in the entire grid (only a portion of the grid is shown here)
with $0 \le x \le 1$, and both used a Courant factor of $0.4$.  The exact solution was
obtained from the Riemann solver provided in \cite{marti-muller-living} with $1000$ points, 
using the same range in $x$ and same initial conditions as the Roe and Marquina runs.  
The Marquina method produced
a more diffused solution than the exact solver, but this is expected in any approximate method;
further, this difference is most likely exaggerated by the fact that the exact solution is determined on 
a finer mesh.  In contrast, the Roe solver severely diverges from the exact solution near the 
transonic rarefaction during the first few time steps.  Even though the Roe solution recovers
in the last frame and begins to resemble the Marquina solution in much of the domain, 
a relic feature from the initial divergence still exists and propagates away from the center.
If we were to reverse the evolution of the Roe solution shown here, the sequence would be reminiscent 
of how the instability in $D$ grows near the sonic point
of near-critical solutions (Figure~\ref{fig:roe-q1-instability}).

\centerline{\includegraphics*[bb=0.3in 2.5in 7.4in 9.6in, scale=0.6]{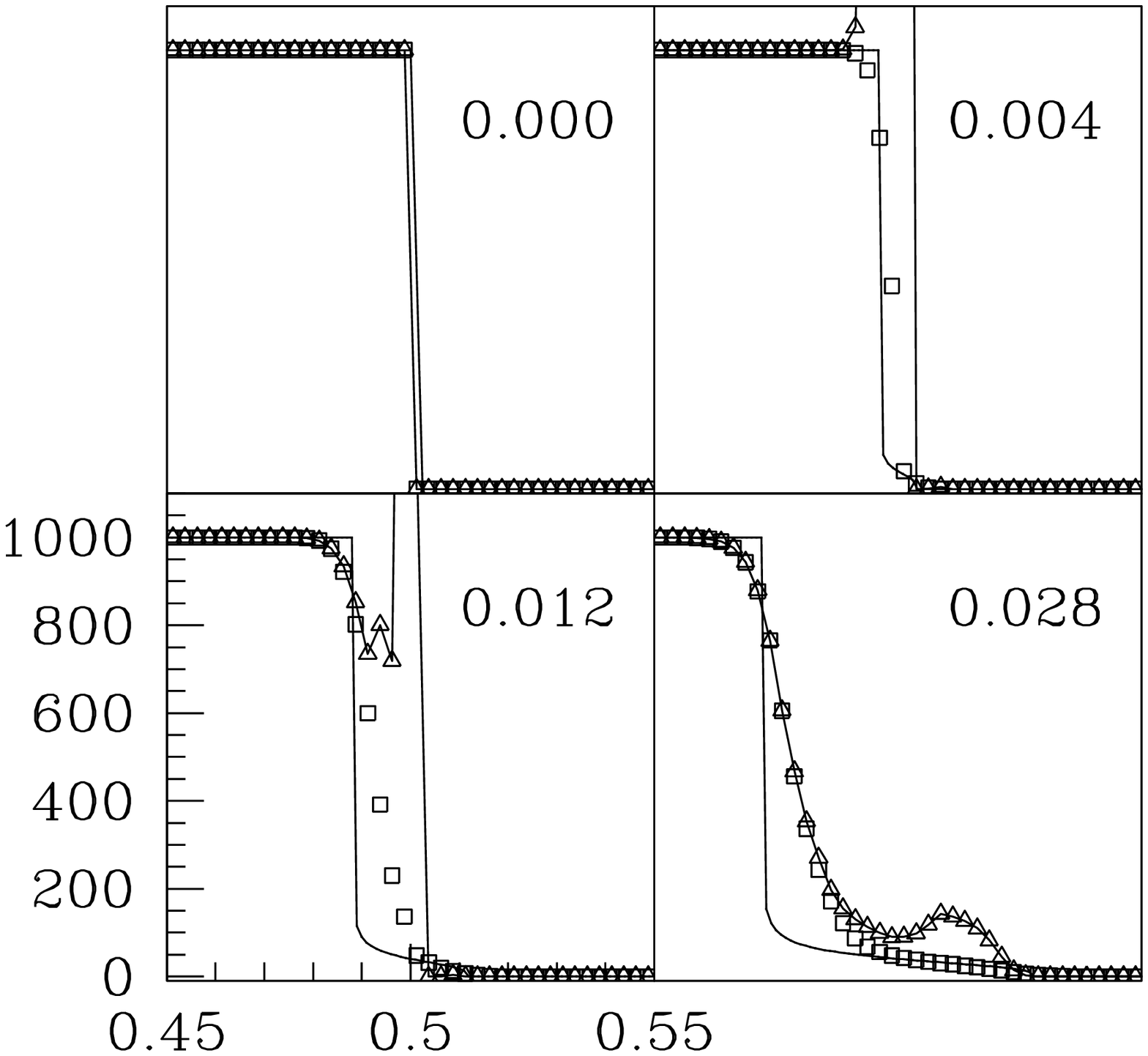}}
\begin{figure}[htb]
\caption[A one-dimensional, slab-symmetric shock tube test that simulates the discontinuity observed 
near the sonic point in solutions near the threshold of black hole formation. ]{A one-dimensional, 
slab-symmetric shock tube test to simulate the discontinuity observed 
near the sonic point of solutions near the threshold of black hole formation.  The rest-mass 
density $\rho_\circ(x,t)$ computed using different Riemann solvers is plotted as a function of the 
Cartesian coordinate $x$ in each frame.  Solution time is shown in the upper-right part
of each frame.  The solid line without points is an exact Riemann solution, the connected 
triangles correspond to the solution
obtained with the approximate Roe solver, and the squares represent the solution from
Marquina's method.  See the text for more details. 
\label{fig:css-shocktube}}
\end{figure}

This shock tube test suggests that the 
origin of the instability in the critical regime may lie in the Roe solver's inability to solve 
this type of Riemann problem.  In order to address this possibility, we implemented the 
Marquina solver in the general relativistic code and 
tuned towards the critical solution.  We were able to tune to 
$p-p^\star \approx 5.0 \times 10^{-11}$, which is approximately a factor $10^2$ closer to $p^\star$ than
we reached with Roe's method without smoothing.  Also, Marquina's method seemed to delay the 
appearance of the observed instability near the sonic point.  However, the use of 
Marquina's flux formula did not completely solve the problem since evolutions using it also 
eventually succumbed to the instability, preventing us from tuning beyond 
$p-p^\star \approx 5.0 \times 10^{-11}$.  Surprisingly, smoothing $\mathbf{\bar{q}}$ about the sonic point 
did not make the Marquina evolutions any more stable; most likely the Marquina flux provided
adequate diffusion on its own.

It is left to future investigation to determine whether or not other Riemann solvers will be able to 
eliminate the instability.  Obvious methods to try are Harten and Hyman's 
entropy-fix \cite{harten-hyman} for Roe's solver, and an improved formula for the flux near
sonic points developed by Roe \cite{roe-1992}.   Harten and Hyman's procedure involves 
estimating the intermediate state in the rarefaction wave which attempts to introduce rarefactions
in the Riemann solution instead of merely discontinuities; a simple description of their 
algorithm is described in \cite{leveque1}.  In contrast, Roe's sonic flux formula
uses the fact that the flux has an extremum at the sonic point in order to derive a better
estimate of the flux there.  As an ultimate test of whether the Riemann solver is the 
cause, an exact Riemann solver can be used at each cell border. 
However, finding the exact solution for each cell at every time step would lead to significantly 
longer run-times, possibly making the process of tuning to the critical solution impractical.  

On the other hand, the ultimate failures of Marquina's method and the 
``smoothed-Roe'' solver may simply 
be due to the overall accuracy of \emph{all} the methods used, and not the result of 
any one part, such as the Riemann solver.  After all, the most nearly critical 
solution is quite relativistic with maximum values of $W \gtrsim 10^6$ just after the sonic point, 
and the pressure obtains a maximum on the order of $10^{13}$ near the origin, relative to $P(0,0) \sim 10^{-2}$
that we typically use.  Some of the error
in these highly-relativistic solutions is undoubtedly due to 
the calculation of $\mathbf{w}$ from $\mathbf{q}$, since this inversion process 
becomes considerably less precise for $W \gtrsim 10^5$ and when $P \gg \rho_\circ$.  

\chapter{Velocity-induced Neutron Star Collapse}
\label{chap:veloc-induc-neutr}

As in many previous works (such as \cite{novak}, \cite{gourg2}, \cite{shap-teuk-1980}), we 
are interested in determining the conditions for black hole formation from unstable
compact stars.  For instance, 
Shapiro and Teukolsky \cite{shap-teuk-1980} asked whether a stable neutron star that has a 
mass below the Chandrasekhar mass 
is able to be driven to collapse by giving it a sufficient amount of in-going kinetic 
energy.  With a mixed Euler-Lagrangian code, they began to answer the question by studying 
stable stars whose density profiles have been ``inflated'' in a self-similar manner such that the 
stars become larger and more massive.  Such configurations were no longer 
equilibrium solutions and had deficits in their central pressures, and  inevitably collapsed 
upon themselves.  By increasing the degree to which 
the equilibrium stars were inflated, they were able to supply more kinetic energy to the system. 
They found, however, that black holes formed only for stars with masses greater than the 
maximum equilibrium mass.  In addition, Shapiro and Teukolsky studied accretion induced collapse, where it 
was again found that collapse to a black hole occurred only when the total mass of the system---in this case 
the mass of the star \emph{and} the mass of the accreting matter---was above the maximum stable mass.  
Both examples seemed to suggest that even perturbed 
stars needed to have masses above the maximum mass in order to produce black holes.
Moreover, they only witnessed three types of outcomes: 1) homologous bounce, wherein
the entire star undergoes a bounce after imploding to maximum compression; 2) non-homologous
bounce where less than $50\%$ of the matter follows a bounce sequence; and 3) direct collapse 
to a black hole.  The survey consisted of 13 different inflated star configurations of 
varying $\Gamma$ and $\rho_c$.  
Also, Baumgarte et al. \cite{baumgarte-etal-1995} using a Lagrangian code based on the 
formulation of Hernandez and Misner \cite{hernandez-misner} qualitatively confirmed these results. 

In order to investigate the question posed by Shapiro and Teukolsky further, Gourgoulhon 
\cite{gourg2} used pseudo-spectral 
methods and realistic, tabulated equations of state to characterize the various ways in 
which a neutron star may collapse when given an \textit{ad hoc}, polynomial velocity profile. 
The particular formulation and methods he used are explained in \cite{gourg1}. 
Such velocity profiles mimic those seen in core collapse simulations as described in 
\cite{may-white-paper},\cite{vanriper1}.  Given a sufficiently large amplitude of the profile, 
Gourgoulhon was able to form black holes from stable stars with masses well below the maximum.  
He also was able to observe bounces off the inner core, but was unable to continue the evolution
significantly past the formation of the shock since spectral techniques typically behave poorly for 
discontinuous solutions.  

To further explore this problem and resolve the shocks more accurately, Novak \cite{novak} used a 
Eulerian code with High-Resolution Shock-Capturing (HRSC) methods.  In addition, he surveyed the 
parameter space
in the black hole-forming regime in much greater detail than previous studies, illuminating  a new scenario in which 
a black hole may form on the same dynamical time-scale as the bounce.  Depending on the 
amplitude of the velocity perturbation, such cases can lead to black holes that have smaller 
masses than their progenitor stars.  This dependence suggests that masses of black holes generated 
by neutron star collapse may not be constrained by the masses of their parents and, consequently, 
could---in principle---allow the black hole mass, $M_{BH}$, to take on a continuum of values. 
In addition, as did the study described in \cite{gourg2}, 
Novak found that the initial star need not be more massive than the maximum mass in order to 
produce black holes.
In fact, he found that for two equations of state---the typical polytropic EOS and a realistic EOS described by
\cite{pons-etal-2000}---arbitrarily small black holes could be made by tuning the initial amplitude of 
the velocity profile about the value  at which black holes are first seen.  Hence, Novak's work
suggests that black holes born from neutron stars are able to have masses in the range 
$0 < M_{BH} \le M_\star$.  

In this section, we present a description of the various dynamic
scenarios seen in perturbed neutron star models, 
as a function of the initial star solution and the magnitude of the initial 
velocity profile.  These results are given to extend and compare with work done 
by Novak \cite{novak} specifically, and others which we will mention along the way. 
We will first provide our description of various phases in the parameter space, giving more 
detail to the regions where no black hole is formed than previous studies have done.  
Then, in the subsequent chapters, we will investigate the critical phenomena observed at 
the threshold of black hole formation.

\section{Parameter Space Survey}
\label{sec:param-space-surv}

Surveying the parameter space of initial possible data sets is essential to the elucidation of new
phenomena in a particular system.  Neutron stars can theoretically take a range of central
densities, and can be driven to instability using a number of mechanisms with varying strength.  
For instance, one can collapse a massless scalar field onto the star, or momentarily change its equation of 
state so that a pressure deficit or surplus arises in the star's interior. 
In this section, we extend work done previously in surveying the parameter space
of initially perturbed neutron star models.  

To drive the neutron star out of equilibrium, it is initially endowed with an in-going 
profile for the coordinate velocity, $U(r,0)$, as described in Section~\ref{sec:init-star-solut}.
We measure the magnitude of this perturbation by the minimum value, $v_{\min}$, of the Eulerian velocity $v$ at 
the initial time.  We find that $v_{\min}$ is uniquely specified by the parameter $U_\mathrm{amp}$ if  
the prescription for generating perturbed TOV stars 
given in Section~\ref{sec:init-star-solut} is followed.  We also note that  $v_{\min}$ is a more physical
quantity than similar parameters---e.g. $U_\mathrm{amp}$---that pertain to the fluid's
gauge-dependent, coordinate velocity.  Consequently, we have created a type of ``phase diagram'' for
the various ways in which perturbed TOV solutions evolve, shown here in Figure~\ref{fig:pspace}.   
Given any combination of the central value of the star's rest-mass density, $\rho_c$, 
and $v_{\min}$, the system will evolve in a fashion specified by the diagram.  In 
Figure~\ref{fig:mass-pspace}, we display the phases in $(M_\star, v_{\min})$ space.  

In order to sample the parameter space, we chose 22 different TOV solutions---specified by 
$\rho_c$---and systematically perturbed each one by varying the parameter $v_{\min}$. 
Approximately 360 $\{\rho_c,v_{\min}\}$ 
sets were run in order to resolve the boundaries to the degree shown here.  
In Figure~\ref{fig:rho-params}, the initial equilibrium solutions used for the parameter space 
survey are displayed along the $M_\star(\rho_c)$ curve for $\Gamma=2$ TOV solutions.  We note that a wide 
spectrum of $\Gamma=2$ stars were chosen, from non-relativistic stars that are relatively large and diffuse, 
to compact and dense relativistic stars.  

\clearpage

\centerline{\includegraphics*[bb=0.6in 2.2in 8in 9.6in, scale=0.6]{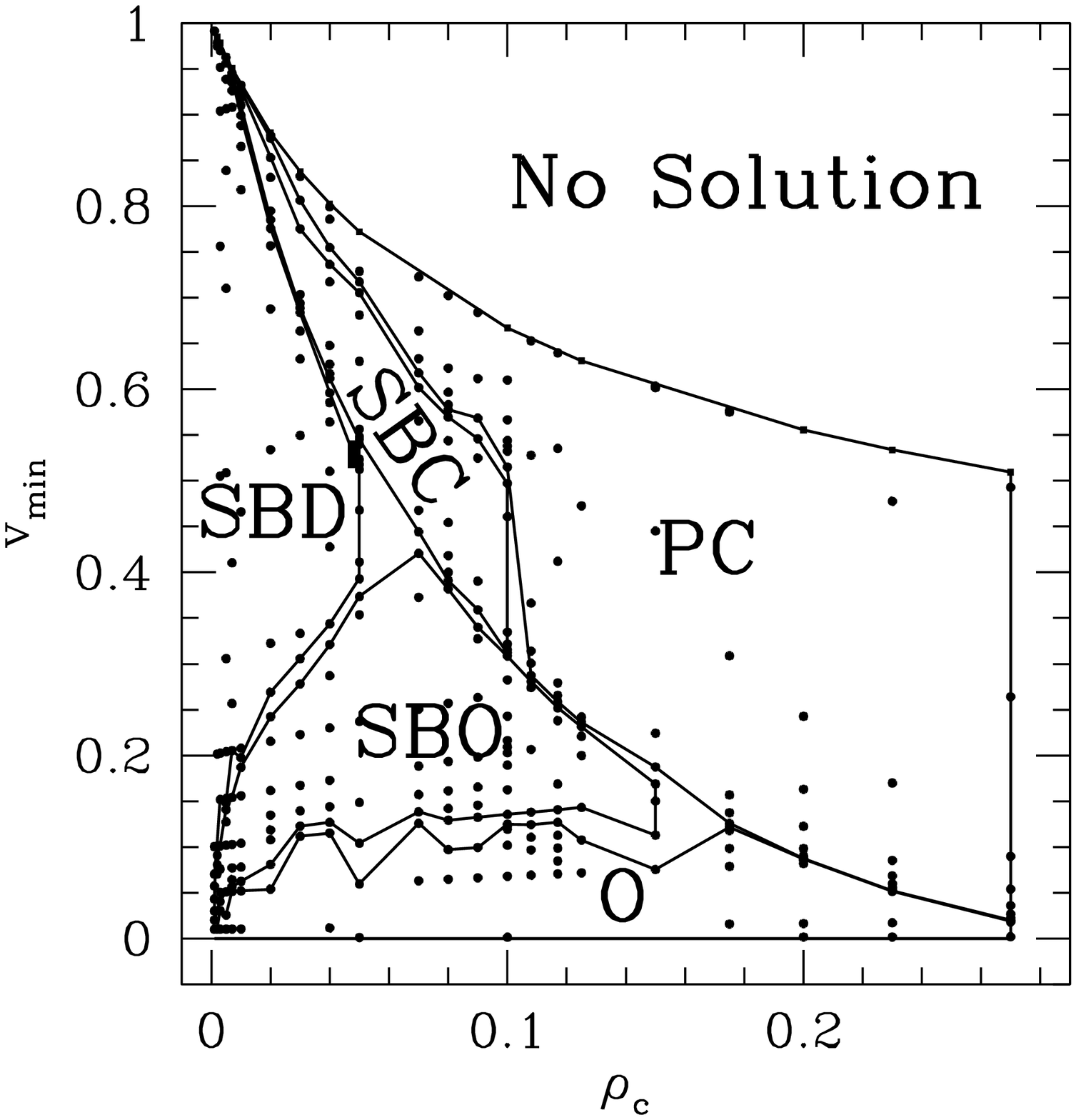}}
\begin{figure}[htb]
\caption[Parameter space surveyed using the initial profile $U_1$ (\ref{v-profile-1}) for the coordinate 
velocity.  The vertical axis is the physical velocity's minimum at the initial time, 
and the horizontal axis is the central density of the TOV solution.]{Parameter 
space surveyed using the initial profile $U_1$ (\ref{v-profile-1}) for the coordinate 
velocity.  The vertical axis is the physical velocity's minimum at the initial time, 
and the horizontal axis is the central density of the TOV solution.
All runs were done using the stiff equation of state $\Gamma=2$; for this EOS, the maximum
mass TOV solution is located at $\rho_c \sim 0.318$.  
The small black rectangular region located at $(\rho_c,v_{\min})\sim (0.05,0.53-0.55)$ represents a set of solutions that 
undergo an SBO-type evolution.  The non-sampled region of parameter space located in the vicinity
$(\rho_c,v_{\min})\sim (0.06,0.45)$ is where the transition from Type~II (smaller $\rho_c$) to Type~I (larger $\rho_c$)
critical behavior takes place; the best estimate for the precise location of this boundary is $\rho_c\approx0.05344$.  
This boundary in critical behavior seems to coincide with the transition from the subcritical
SBD and SBO scenarios.  
\label{fig:pspace}}
\end{figure}

\clearpage

\centerline{\includegraphics*[bb=0.6in 2in 8in 9.6in, scale=0.6]{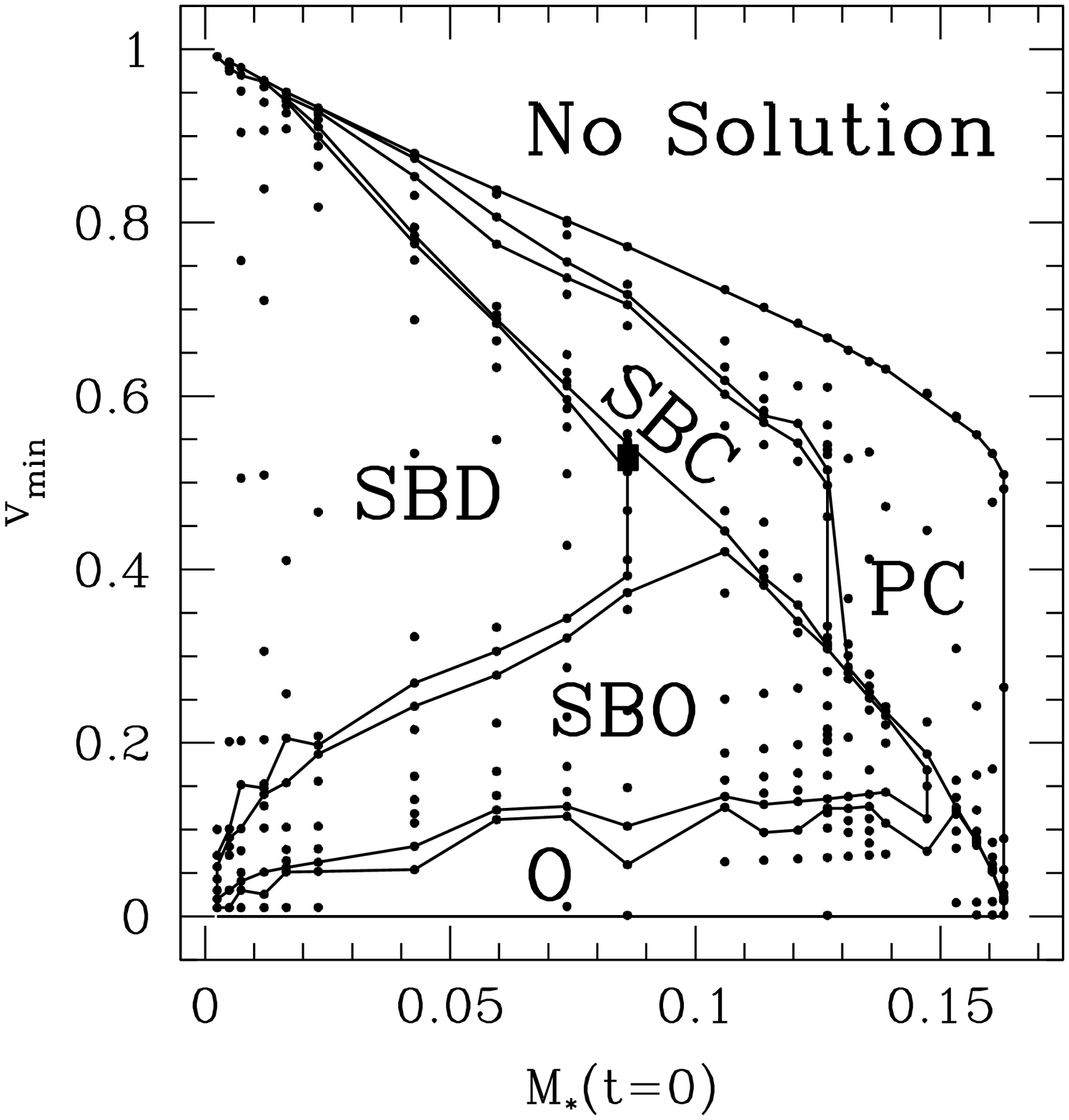}}
\begin{figure}[htb]
\caption[The parameter space in terms of $v_{\min}$ versus initial stellar mass $M_\star$
for the same runs shown in Figure~\ref{fig:pspace}.]{The parameter space in terms of $v_{\min}$ 
versus initial stellar mass $M_\star$
for the same runs shown in Figure~\ref{fig:pspace}.  Note that $M_\star$ here is the gravitational
mass of the \emph{static} star solution used to construct the initial conditions; the gravitational
mass of the star will increase once the velocity profile is ``added'', since this endows the star with non-zero
kinetic energy.  Since $M_\star(\rho_c)$ is monotonic
in the region we sampled (Figure~\ref{fig:rho-params}), this figure is essentially a distortion
of Figure~\ref{fig:pspace}.  The maximum mass TOV solution is located at $\rho_\circ\simeq0.318$ 
and $M_\star \simeq 0.1637$, while the most massive stars shown here are 
TOV solutions with $\rho_c=0.27$ and $M_\star=0.1629$. 
The small black rectangular region located at $(M_\star,v_{\min}) \simeq (0.086,0.53-0.55)$ represents 
a set of solutions that undergo an SBO-type evolution.
The non-sampled region of parameter space located in the vicinity
$(M_\star,v_{\min})\sim (0.095,0.45)$ is where the transition from Type~II (smaller $M_\star$) to Type~I (larger $M_\star$)
critical behavior takes place.  This boundary in critical behavior seems to coincide with the transition from the 
subcritical SBD and SBO scenarios.  
\label{fig:mass-pspace}}
\end{figure}

\centerline{\includegraphics*[bb=0.3in 2.2in 8.2in 9.8in, scale=0.6]{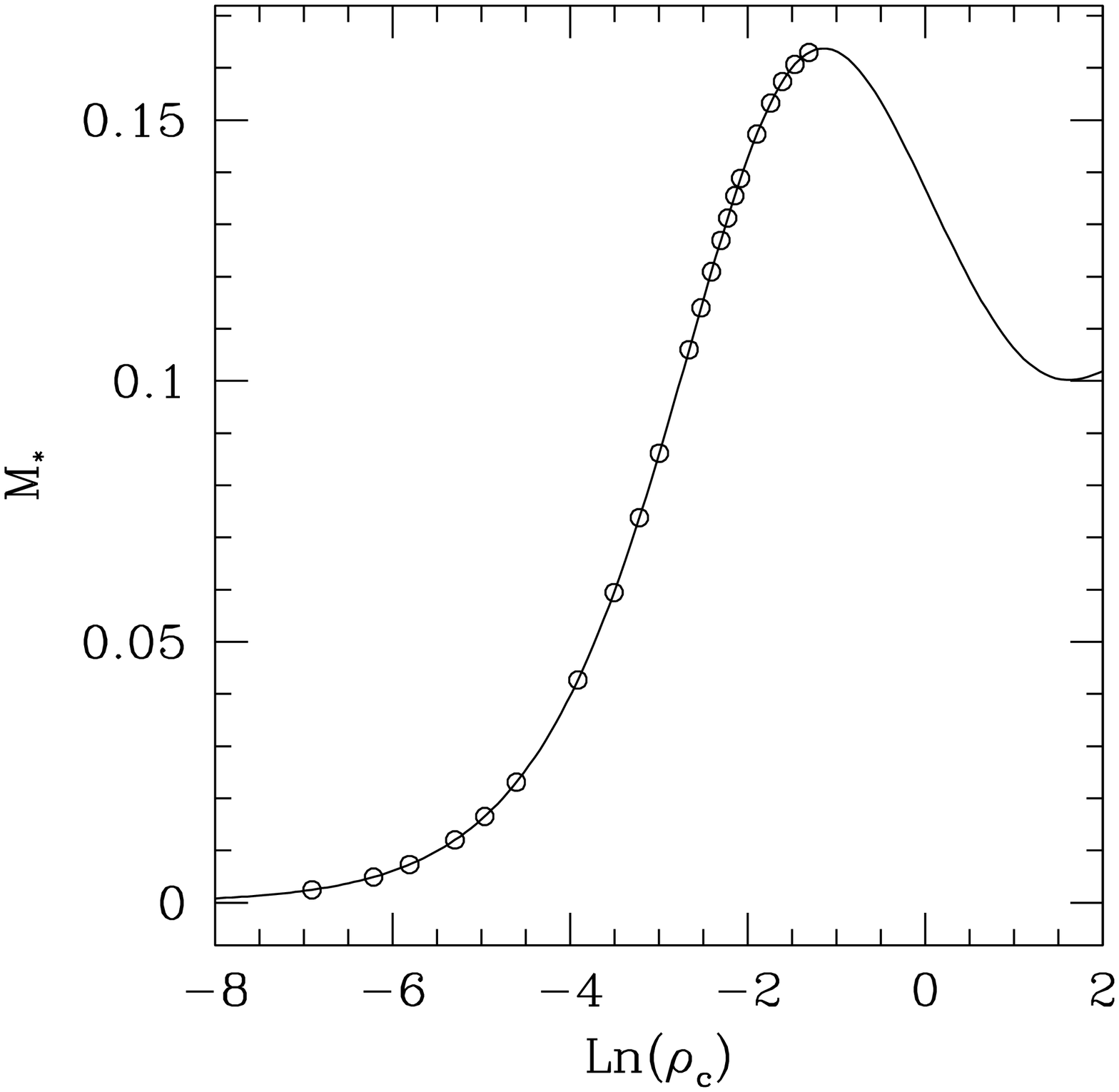}}
\begin{figure}[htb]
\caption{The initial TOV solutions used in the parameter space plots shown in 
Figure~\ref{fig:pspace} and Figure~\ref{fig:mass-pspace} displayed on the $M_\star(\rho_c)$
distribution of equilibrium solutions for $\Gamma=2$ and $K=1$.
\label{fig:rho-params}}
\end{figure}

The types of dynamical scenarios or ``phases'' mentioned in 
Figures~\ref{fig:pspace}~-~\ref{fig:mass-pspace} are described below:
\begin{description}
\item[Prompt Collapse (PC):] For a system of this type, the initial ``perturbation'' is
so strong that the star is driven directly to black hole formation.  The fluid in-falls
homologously---or uniformly---and no significant
amount of matter is ejected before the black hole is formed.  
\item[Shock/Bounce/Collapse (SBC):] In this case, the perturbation is not sufficient to spontaneously
form a black hole, but is still strong enough to eventually drive the star to collapse. 
The outer part of the star in-falls at a faster rate than the interior and eventually bounces 
off the denser core, producing an outgoing shock.  As the shock propagates to larger 
radii, inflow velocities in the vicinity of the shock change to outflow velocities, and 
a portion of the surface material is ejected from the star. 
Meanwhile, the inner portion continues to in-fall and eventually forms a black hole. 
\item[Shock/Bounce/Dispersal (SBD):]  The dynamics involved in an SBD case is quite similar to 
the previous-described SBC scenario, except a black hole never forms.  Instead, the 
star contracts until it reaches some maximum density and pressure at the origin.  The pressure surplus of the 
interior is then great enough to expel the remainder of the star, leaving behind an ever-decreasing amount 
of matter at the origin.  This final 
explosion results in another outgoing shock wave that typically overtakes and engulfs the 
first shock.  
\item[Shock/Bounce/Oscillation (SBO):] As the perturbation is decreased, the rebound of the 
interior no longer results in complete mass expulsion.  Rather, some matter remains after the 
first two shocks propagate outwards and this matter settles into a new equilibrium state by oscillating away any excess 
kinetic energy via the ``shock-heating'' mechanism, wherein shocks created by the oscillations
essentially convert the kinetic energy of the bulk flow into internal energy.  After the 
oscillations dampen away, a ``hot'' star solution remains that is always larger, sparser and
less massive than the original star.  
\item[Oscillation (O):] Finally, if the inward velocity is minimal, then the perturbed
star will undergo oscillations at its fundamental frequency and overtones.  The oscillations 
tend to be insufficient to shock-heat the surface material nor are they strong enough to expel 
significant amounts of matter. 
\end{description}

Differentiating between some of the types of outcomes is difficult.  
To aid in this process, we examined how various quantities varied with time at the star's 
radius, $R_\star(t)$.  Since $R_\star(0)$ is well defined (see Section~\ref{sec:init-star-solut}), 
we can set $R_\star(t)$ to be the radius at which $\rho_\circ(r,t) = \rho_\circ(R_\star(0),0)$ to within
some finite precision.  This served as a fair approximation to tracking the fluid element that 
started at $R_\star(0)$.  
In the future, it would be interesting to see if the results reported here
vary significantly if we set $R_\star(t) = r_L(t)$, where $r_L(t)$ is the world line of 
a Lagrangian observer governed by the characteristic equation, 
\beq{
d r_L / dt = v\left(r_L(t),t\right) \label{lagrangian-world-line} \quad ,
}
with $r_L(0) = R_\star(0)$. 
For instance in \cite{shap-teuk-1980}, (\ref{lagrangian-world-line}) was  numerically integrated 
along with the Einstein equations and the fluid EOM in order to track a set of Lagrangian observers 
starting at different locations.  This procedure more manifestly illustrated the difference between 
stellar collapse evolutions that either have homologous or non-homologous bounces.  
Note, however, that we do not assume that $R_\star(t)$ is that of a Lagrangian observer; in
fact, we sometimes exploit this fact by distinguishing evolution types based upon how the mass, 
$M_\star(t)$, contained within $R_\star(t)$ varies with time, for systems starting from different sets of initial data. 

\clearpage 
\centerline{\includegraphics*[bb=0.3in 2.2in 8.3in 9in, scale=0.6]{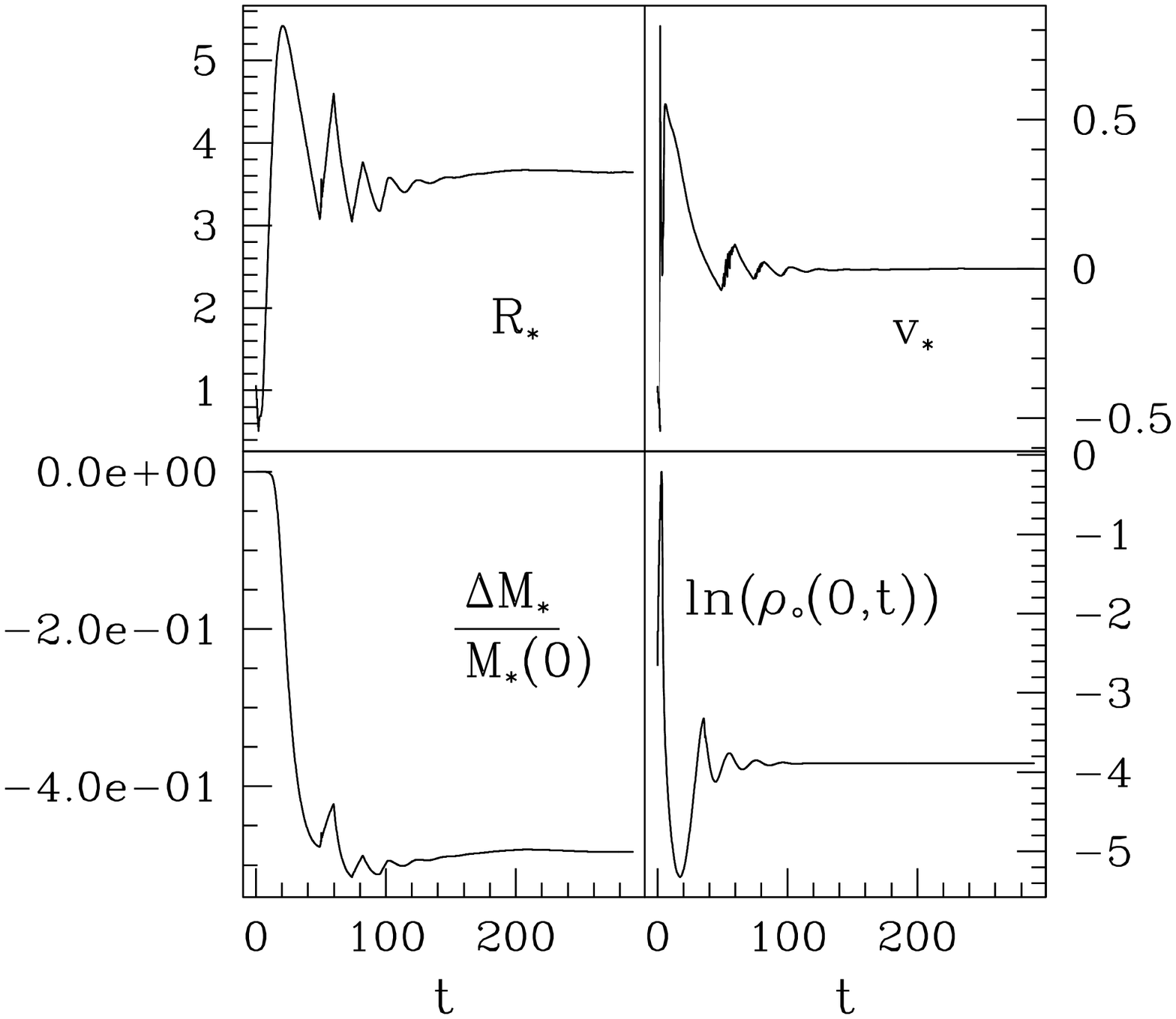}}
\begin{figure}[htb]
\caption[Evolutions of stellar radius ($R_\star$), velocity at $R_\star$ ($v_\star$), relative stellar mass 
deviation from initial time ($\Delta M_\star/M_\star(0)$), and the natural logarithm of the central density 
for a star that is perturbed such that it evolves to a larger, less massive star.]{Evolutions of stellar 
radius ($R_\star$), velocity at $R_\star$ ($v_\star$), relative stellar mass 
deviation from initial time ($\Delta M_\star/M_\star(0)$), and the natural logarithm of the central density 
for a star that is perturbed such that it evolves to a larger, less massive star.  The star 
first undergoes a quick shock and bounce at its edge which seems to play an insignificant role 
in the subsequent evolution.  While the shock propagates out of the star, 
the inner part of the star continues to in-fall and rebounds from the origin, which is responsible for 
ejecting the majority of the matter from the star.  This is shown in the interval of time near $t \approx 3.2$
where the central density obtains its global maximum and decreases, as the star starts to swell 
in size, and as $v_\star$ increases toward its second maximum.  Consecutive, diminishing oscillations 
ensue until the star settles about a state with a smaller central density, larger radius and 
smaller mass than initially.  The defining parameters for this run are 
$\Gamma = 2$, $\rho_\circ(0,0) = 0.02$, $v_{\min}(t=0) = -0.397$, $M_\star(0) = 0.1185$  with profile $U_1$.
\label{fig:sb-oscil}}
\end{figure}

The boundary between SBO and O outcomes may be the most imprecisely determined one. 
This is due to the fact that the shock in SBO cases weakens as the 
perturbation is reduced, making it difficult to tell if a bounce actually happens and whether 
the subsequent oscillations take place about a different star solution.  In addition,
an O system may form a minor shock at first, but still maintain nearly-constant amplitude 
oscillations, indicating the absence of significant shock-heating.  Herein, an O state
is defined as a star which lost less than $1\%$ of its mass over $6$ periods of its
fundamental mode of oscillation.  This choice of cutoff was motivated by two facts:  1) evolutions which 
seemed to be oscillating about the initial solution still lost mass, because the oscillations 
still ejected minute amounts of matter from the star's surface;  2) those evolutions which 
were obviously SBO seemed to eject most of the expelled matter within the first 6 oscillations. 
Using this definition, we estimate the systematic error of the SBO/O boundary to be no larger than 
$0.05$ in the $v_{\min}$ direction.  A more precise definition might be 
to measure the frequency of oscillation of the perturbed star ($\omega(\rho_c,v_{\min})$), 
and then set the SBO/O boundary to be the point at which this frequency equals the fundamental 
frequency associated with the progenitor star ($\omega(\rho_c,0)$).  It is our conjecture that
$\omega(\rho_c,v_{\min}) \rightarrow \omega(\rho_c,0)$ smoothly as 
$v_{\min} \rightarrow 0$. 

\centerline{\includegraphics*[bb=0.4in 2.5in 8.3in 9in, scale=0.6]{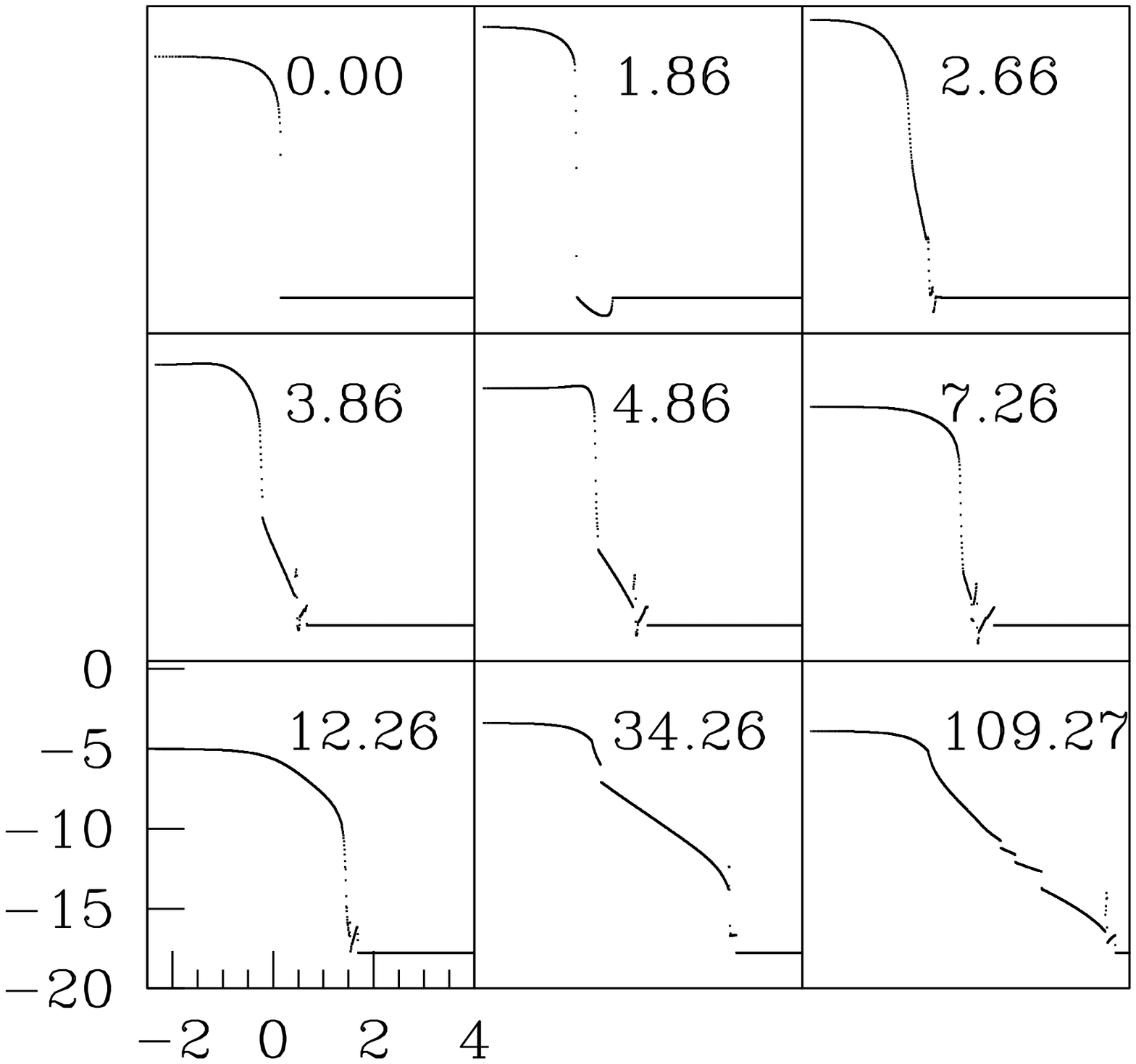}}
\begin{figure}[htb]
\caption[Time sequence of $\ln{\rho_\circ(r,t)}$ versus $\ln(r+0.1)$ for the same 
SBO scenario shown in Figure~\ref{fig:sb-oscil}.]{Time sequence of $\ln{\rho_\circ(r,t)}$ 
versus $\ln(r+0.1)$ for the same SBO scenario shown 
in Figure~\ref{fig:sb-oscil}.  The initial shock is seen at $t=1.86$ as the discontinuity in 
points near the top of the distribution.  The bounce is demonstrated by the increase in density 
at larger radii in the snapshot taken at $t=2.66$.  The rebound from the origin happens between
$t=2.66$ and $t=3.86$, and the shock that results from it can be seen as the discontinuity 
propagating outward at times $t=3.86$ and $t=4.86$.  The shock that heats the exterior of the 
star is the innermost discontinuity visible at $t=34.26$.  
At each time, only every eighth grid point is displayed.
\label{fig:lnrho-sb-oscil}}
\end{figure}

Time evolutions of various quantities pertaining to a perturbed star which epitomizes an SBO state 
are shown in Figure~\ref{fig:sb-oscil}.  In Figures~\ref{fig:lnrho-sb-oscil}~-~\ref{fig:v-sb-oscil} 
we show time sequences of $\ln{\rho_\circ}$, $\ln{\epsilon}$,  and $v$---respectively---for the same 
run.  The initial shock and bounce are clearly seen early on in the time sequences of three 
functions, while the 
subsequent rebounds of the interior are seen later in time.  It can also be clearly seen that the first 
rebound of the core is responsible for most of the ejection of matter, even though the 
initial bounce near the star's surface involves the strongest shock.  This is demonstrated in 
the plots given in Figure~\ref{fig:sb-oscil}.  The apex of the rebound takes place near $t=10$,
when the star reaches extremal size and central density, and when the star begins to lose 
a significant portion of its initial mass---up to $43\%$.  This large change in $M_\star$ signifies
how poorly $R_\star(t)$ follows the path of a Lagrangian observer in this case; however, 
we still feel tracking quantities along this path produces information with which we can consistently 
differentiate outcomes.  The ensuing oscillations after $t \sim 10$ are 
evident in all the quantities shown.  The time-independent character of the 
resultant star is illustrated by the fact that the quantities in Figure~\ref{fig:sb-oscil} asymptote
to constant values, and that $v(r,t) \simeq 0$ within the star at later times, as seen 
in Figure~\ref{fig:v-sb-oscil}.  

\clearpage
\centerline{\includegraphics*[bb=0.4in 2.5in 8.3in 9in, scale=0.6]{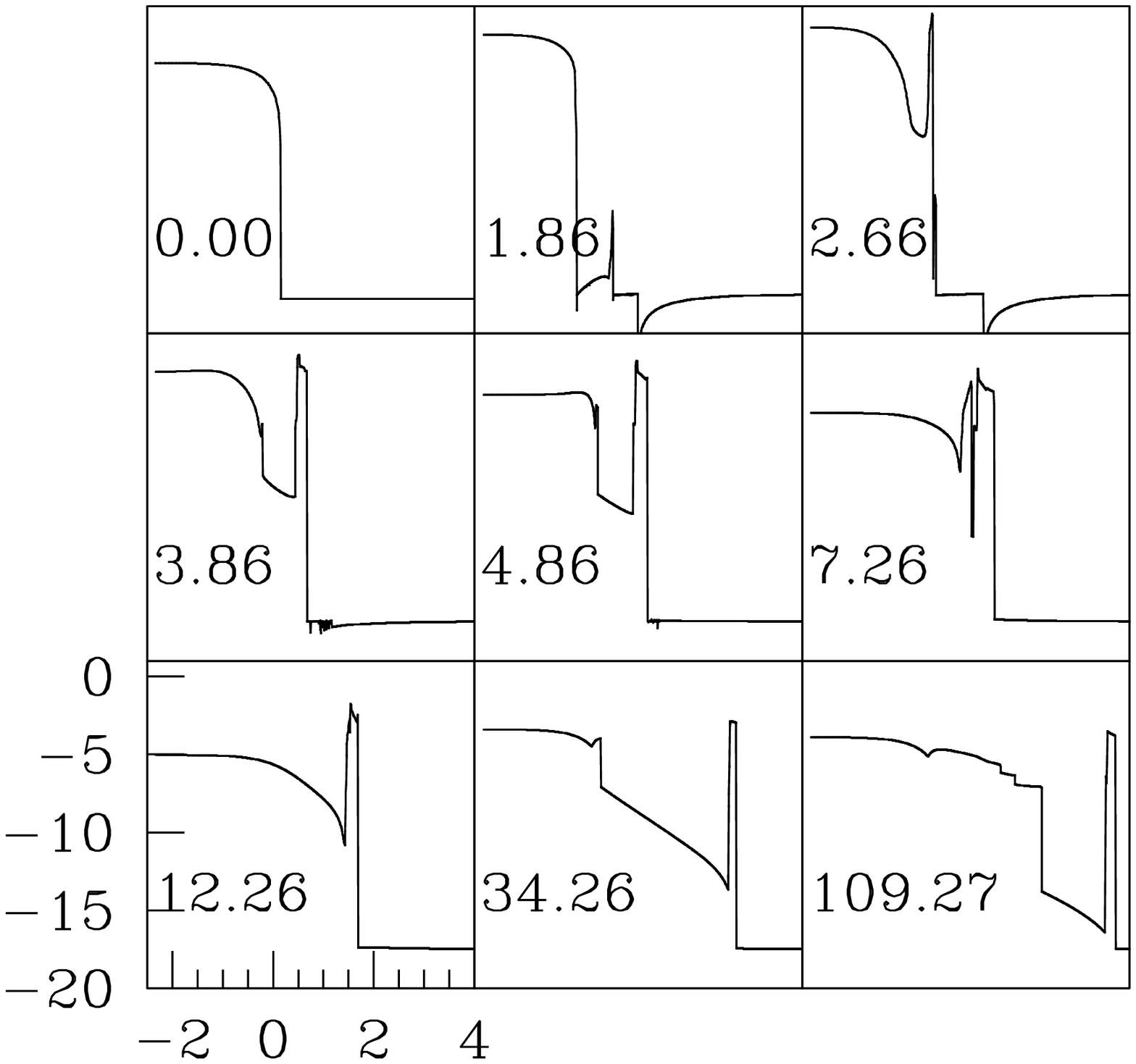}}
\begin{figure}[htb]
\caption[Time sequence of $\ln{\epsilon}$ versus $\ln(r+0.1)$ for the same 
SBO scenario shown in Figure~\ref{fig:sb-oscil}.]{Time sequence of $\ln{\epsilon}$ 
versus $\ln(r+0.1)$ for the same SBO scenario shown 
in Figure~\ref{fig:sb-oscil}.  Here, solution points are connected in order to make certain discontinuities 
more apparent.  The initial shock and bounce occurs near $t=1.86$, but is obfuscated by 
the connecting line.  As the shock wave moves outward, it drastically increases the internal energy 
locally and leaves the material behind it hotter than it was originally. 
The second shock, from the first rebound of the core, can be identified here as the small 
spike at the star's edge at $t=3.86$.  Just after $t=7.26$ do the two regions of high $\epsilon$ 
merge and become a single shock wave.  As the star settles down from the initial rebound, 
subsequent oscillations---whose amplitudes damp rapidly---emit further shocks that 
heat the outer part of the star and leave it in a static, hot state ($t=34.26-109.27$).  
\label{fig:lnepsilon-sb-oscil}}
\end{figure}

Since it is generally impossible to determine whether an arbitrary,  dynamical distribution of
matter is gravitationally bound in general relativity without fully solving Einstein's equations for all 
spacetime occupied 
by the matter, it is sometimes non-trivial to determine the difference between 
SBO and SBD states.  For instance, perturbed stars with smaller $\rho_c$ 
or those on the SBD side near the SBD/SBO boundary often homologously inflate to arbitrary sizes, 
while their maximum densities---still attained at the origin---diminish to magnitudes comparable to the 
floor density.  In contrast more relativistic---and hence denser---stars close 
to the SBC/SBD border tend to disperse completely from the origin in a shell of matter that has
compact support.  In order to ensure that these ``inflated'' stars will not ultimately 
settle into a new equilibrium configuration, we typically let the evolution last until 
the central density of the distribution becomes comparable to the floor density
and increase the size of the grid to accommodate for the expansion.  If, at this time, 
$v(r) > 0$ for all $r$ and $d \rho_\circ(0,t) / dt < 0 $ are still satisfied, then the particular 
case is labelled as a dispersal, or SBD variety. 
An archetypical example of an SBD case involving a relativistic star is shown in 
Figures~\ref{fig:sb-disp}-\ref{fig:v-sb-disp}.

\clearpage

\centerline{\includegraphics*[bb=0.3in 2.5in 8.3in 9in, scale=0.6]{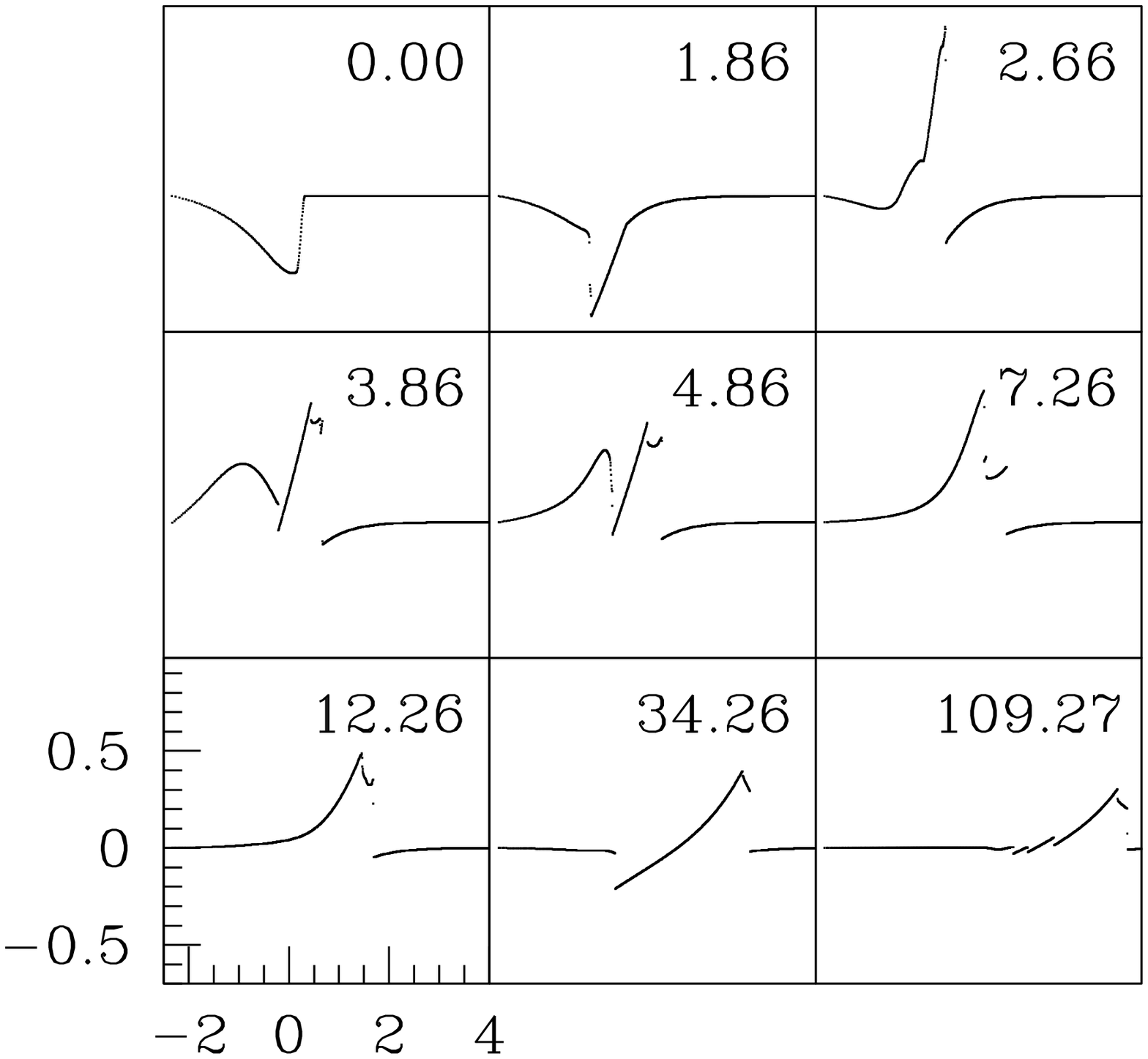}}
\begin{figure}[htb]
\caption[Time sequence of $v(r,t)$ versus $\ln(r+0.1)$ for the same 
SBO scenario shown in Figure~\ref{fig:sb-oscil}. ]{Time sequence  of $v(r,t)$ 
versus $\ln(r+0.1)$ for the same SBO scenario shown in Figure~\ref{fig:sb-oscil}.  The initial shock 
is seen at $t=1.86$, and the bounce is demonstrated by the shock's outward propagation, visible
in successive frames.  The rebound from the origin happens between
$t=2.66$ and $t=3.86$, and the shock that results from it can be seen as the innermost discontinuity 
propagating outward at times $t=3.86$ and $t=4.86$.  The shock from the first rebound travels 
faster than the bounce shock and overtakes it just before $t=12.26$, at which time only one shock is 
observed.  The shock that heats the exterior of the 
star is visible as the innermost discontinuity in points at $t=34.26$.  
At each time, only every eighth grid point is displayed.
\label{fig:v-sb-oscil}}
\end{figure}

\clearpage

\centerline{\includegraphics*[bb=0.3in 2.2in 8.3in 9.8in, scale=0.6]{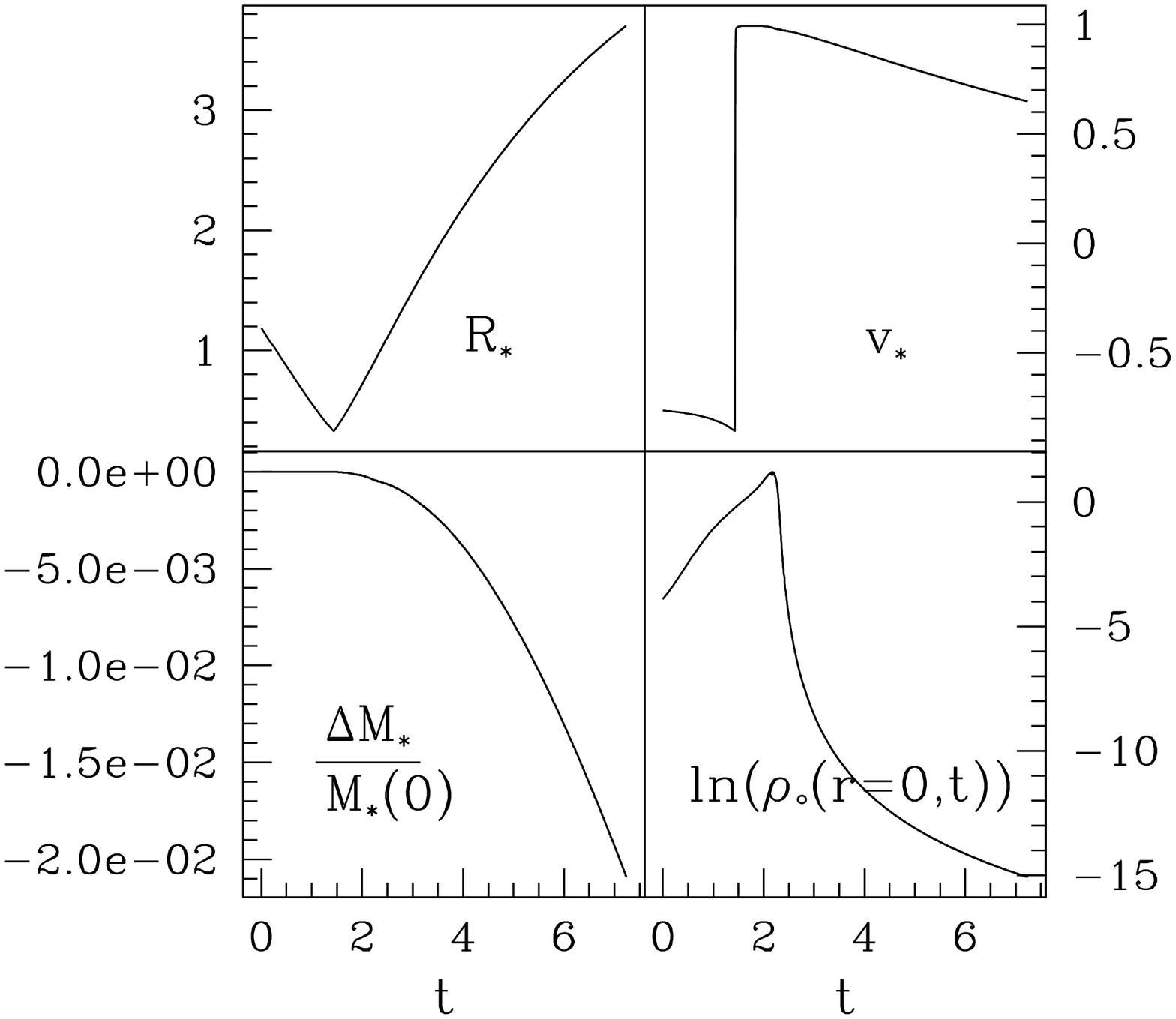}}
\begin{figure}[htb]
\caption[Evolutions of stellar radius ($R_\star$), velocity at $R_\star$ ($v_\star$), relative stellar mass 
deviation from initial time ($\Delta M_\star/M_\star(0)$), and the natural logarithm of the central density 
for a star that is perturbed such that it also undergoes a shock and bounce before rebounding from 
the origin.]{Evolutions of stellar radius ($R_\star$), velocity at $R_\star$ ($v_\star$), relative stellar mass 
deviation from initial time ($\Delta M_\star/M_\star(0)$), and the natural logarithm of the central density 
for a star that is perturbed such that it also undergoes a shock and bounce before rebounding from 
the origin.  The rebound causes the star's matter to eventually disperse away from the origin and, 
most likely, become gravitationally unbound.  At the end of this particular run, the bulk of the matter 
had propagated beyond $r=27$, which is more than $14$ times the original stellar radius, $R_\star=1.1885$.  
The defining parameters for this run are 
$\Gamma = 2$, $\rho_\circ(0,0) = 0.02$, $M_\star(0) = 0.0726$, and $v_{\min}(t=0) = -0.766$ with profile $U_1$.
\label{fig:sb-disp}}
\end{figure}

The small rectangle near the upper-right corner of the SBD region in 
Figures~\ref{fig:pspace}-~\ref{fig:mass-pspace} represent 3 runs with $\rho_c=0.05$
that exhibited SBO behavior.  It remains
to be seen whether or not these cases are dominated by numerical artifacts---that is, the 
remnant star may converge away as $\Delta r \rightarrow 0$---or, if they instead 
represent the sparsest instances of SBD type evolutions along the black hole threshold line. 
If they are real solutions, then each section of the parameter space diagram may not be as 
homogeneous as illustrated here.  Interestingly, these 3 runs are near the region where the 
black hole threshold behavior changes from being of Type~II to Type~I ($\rho_c\approx0.05344$).  
\vspace{0.5cm}

\centerline{\includegraphics*[bb=0.4in 2.5in 7.3in 9in, scale=0.6]{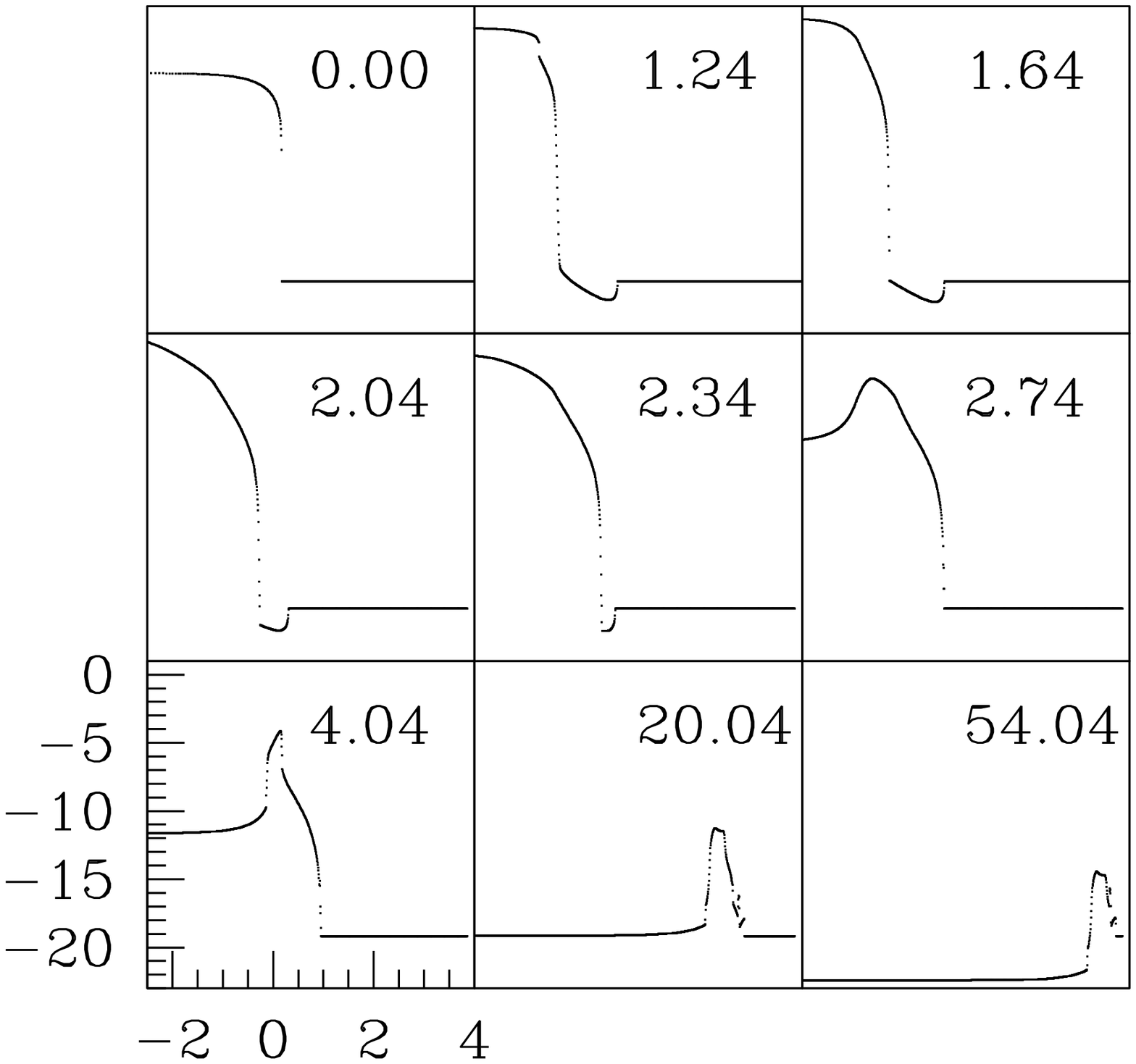}}
\begin{figure}[htb]
\caption[Time sequence of $\ln{\rho_\circ(r,t)}$ versus $\ln(r)$ for the same 
shock/bounce/dispersal scenario shown in Figure~\ref{fig:sb-disp}.]{Time sequence of 
$\ln{\rho_\circ(r,t)}$ versus $\ln(r)$ for the same 
shock/bounce/dispersal scenario shown in Figure~\ref{fig:sb-disp}.  The bulk of the stellar matter 
is seen leaving the numerical domain in a compact distribution. At $t=54.04$, $\rho_\circ$ has fallen well below the 
floor's density in the vicinity of the origin.
At each time, every eighth grid point is displayed. 
\label{fig:lnrho-sb-disp}}
\end{figure}

\centerline{\includegraphics*[bb=0.3in 2.5in 7.3in 9in, scale=0.6]{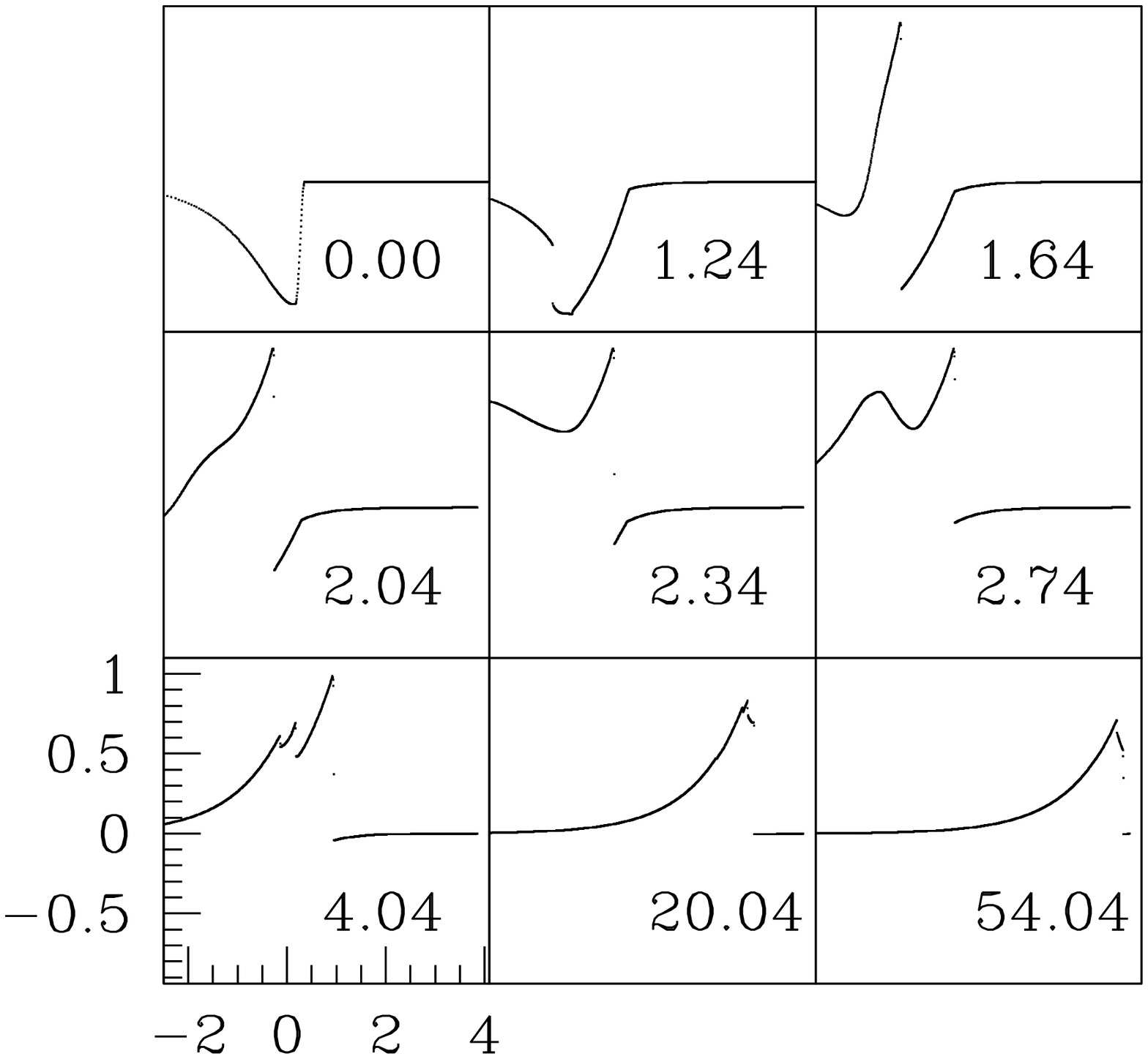}}
\begin{figure}[htb]
\caption[Time sequence $v(r,t)$ versus $\ln(r)$ for the same 
shock/bounce/dispersal scenario shown in Figure~\ref{fig:sb-disp}.]{Time sequence of $v(r,t)$ 
versus $\ln(r)$ for the same 
shock/bounce/dispersal scenario shown in Figure~\ref{fig:sb-disp}.  The shock from the initial bounce is 
first seen at $t=1.64$.  The rebound, responsible for ejecting the majority of the stellar matter, 
forms a shock that is first visible here at $t=4.04$ as the discontinuity closest to the 
origin.  By $t=20.04$, the two shocks have merged into a single shock. 
At each time, every eighth grid point is displayed. 
\label{fig:v-sb-disp}}
\end{figure}

In coordinate systems such as the one we use (\ref{metric}), 
initial data sets that lead to black hole formation are typically characterized by a late-time coordinate 
pathology---$a(r,t)$ diverges---in the vicinity of the radius, $R_\mathrm{BH}$, where an apparent horizon 
would first appear.  Also, the 
lapse, $\alpha(r,t)$ tends to zero for $r < R_\mathrm{BH}$, giving the appearance
that the dynamics of the fluid is ``frozen out.''
In addition, the velocity of the flow typically tends to $-c$ for $r\simeq R_\mathrm{BH}$, 
indicating that matter is trapped within this region.  
In Figure~\ref{fig:prompt-collapse}, the accumulation of matter onto the core is illustrated 
by the behavior of $R_\star(t)$ and $\rho_\circ(0,t)$, while $v_\star(t)$ reveals the asymptotic behavior of $v(r,t)$ 
close to the incipient trapped surface.  This star seems to undergo a homologous free-fall, 
$\Delta M_\star(t)$ varies only on the order of its numerical error and the other quantities 
are monotonic over the course of the collapse.  

Since our choice of coordinates (\ref{metric}) precludes a black hole from forming in finite time, 
we need a fairly rigorous prescription for \emph{predicting} when they would form.
Empirically, we have found that those systems which attain 
${\max}(2m/r) > 0.7$ will asymptote to a state that resembles a black hole in 
our coordinates---where $a$ diverges and $\alpha$ shrinks to an exponentially small magnitude
at the origin.  These all provide strong evidence that the simulated spacetime contains a black hole.
If all goes well, we label any spacetime that reaches ${\max}(2m/r) > 0.995$ 
a ``black hole''.
Since such spacetimes involve extremely steep gradients, it is often difficult to stably integrate
the equations of motion until this threshold is achieved.  Consequently we assume that any evolution, 
which crashes and satisfies ${\max}(2m/r) > 0.7$, will eventually give rise to a black hole.  
Otherwise, the system is assumed to be one without a black hole and is either of the type O, SBO or SBD. 

\clearpage 

\centerline{\includegraphics*[bb=0.3in 2.1in 8.3in 9.8in, scale=0.6]{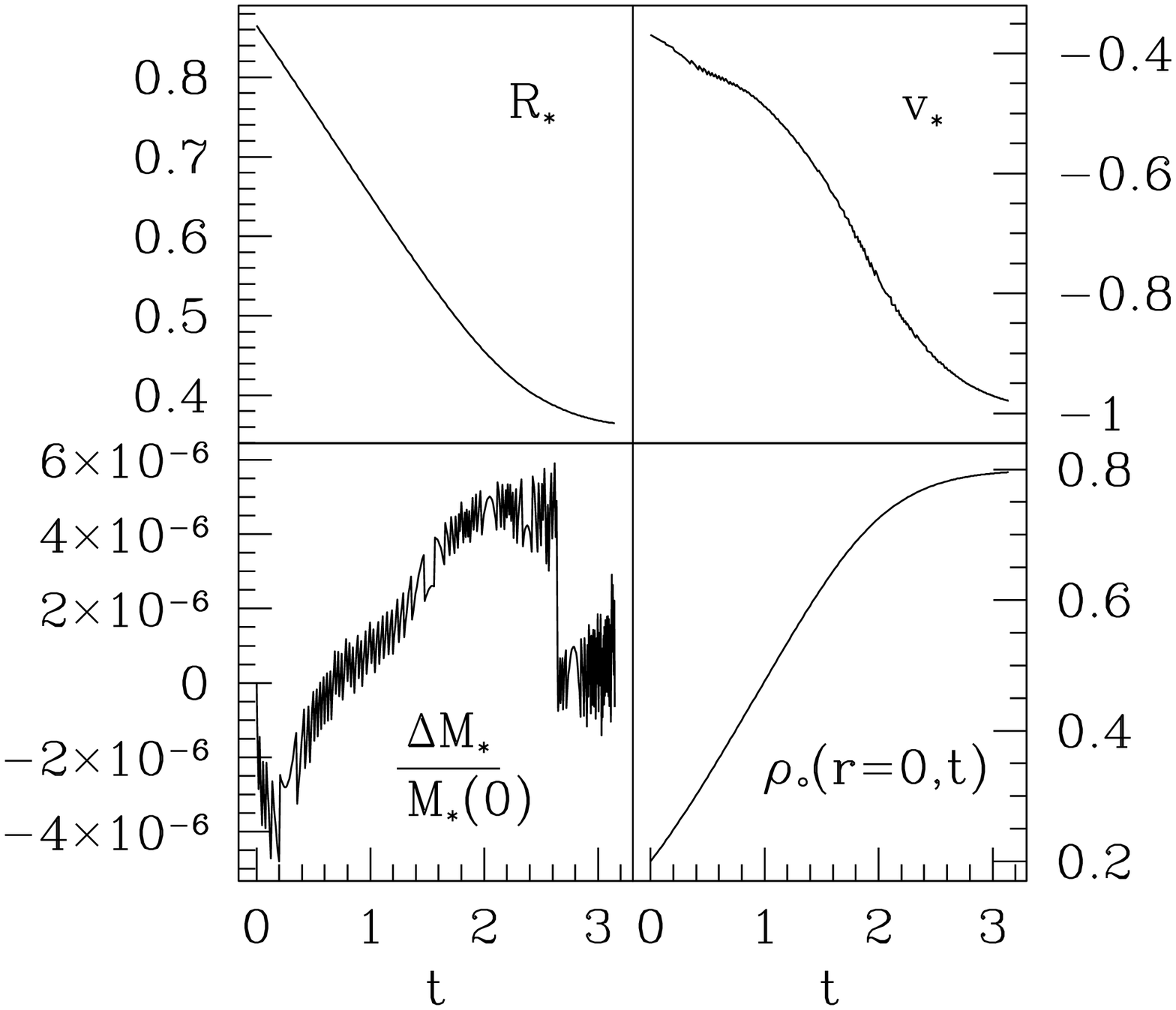}}
\begin{figure}[htb]
\caption[Evolutions of stellar radius ($R_\star$), velocity at $R_\star$ ($v_\star$), relative 
stellar mass deviation from initial time ($\Delta M_\star/M_\star(0)$), and central density for a star 
that is perturbed such that it undergoes prompt collapse to a black hole. ]{Evolutions of stellar 
radius ($R_\star$), velocity at $R_\star$ ($v_\star$), relative 
stellar mass deviation from initial time ($\Delta M_\star/M_\star(0)$), and central density for a star 
that is perturbed such that it undergoes prompt collapse to a black hole.  The maximum of value 
of $2m(r)/r$ observed for this run is $0.98$ attained at a time immediately before the run crashed.  The high-frequency 
oscillations observed
in $R_\star$, $v_\star$, and $\Delta M_\star$ are the result of $R_\star$ being restricted to a discrete domain, 
i.e. the stellar radius may jump back-and-forth between two 
adjacent grid points that have different values of $v$ and $r$.  The lower-frequency variation
in $\Delta M_\star/M_\star(0)$, however is most likely due to truncation errors and small amounts of accretion of the
atmosphere due to the fluid floor.
\label{fig:prompt-collapse}}
\end{figure}

\centerline{\includegraphics*[bb=0.3in 2.2in 8.3in 9.8in, scale=0.6]{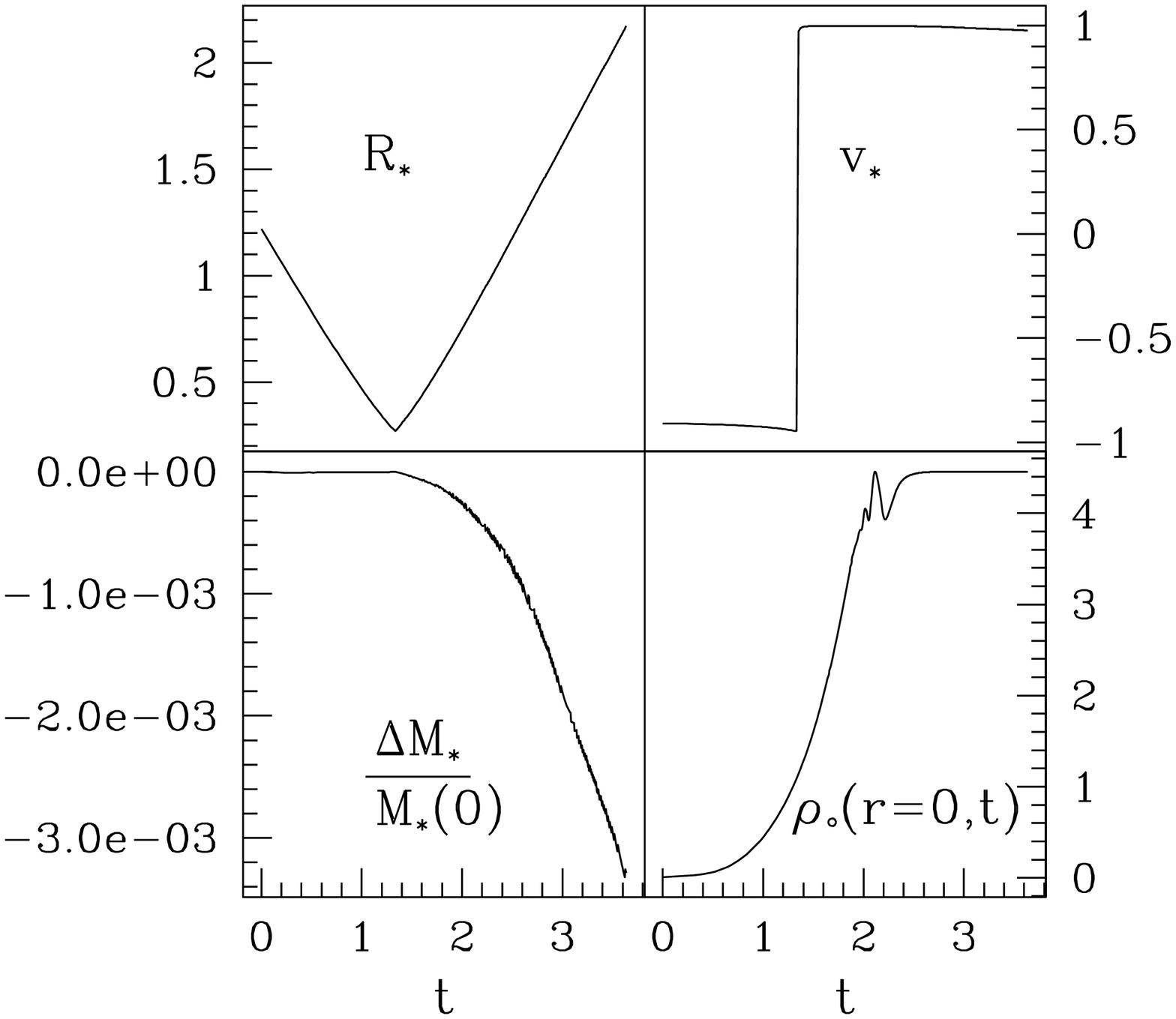}}
\begin{figure}[htb]
\caption[Evolutions of stellar radius ($R_\star$), velocity at $R_\star$ ($v_\star$), stellar mass deviation 
from initial time ($\Delta M_\star$), and central density for a star that is perturbed such that it 
undergoes a shock and bounce before forming a black hole.]{Evolutions of stellar radius ($R_\star$), 
velocity at $R_\star$ ($v_\star$), stellar mass deviation 
from initial time ($\Delta M_\star$), and central density for a star that is perturbed such that it 
undergoes a shock and bounce before forming a black hole.  In this particular case, the matter
at the stellar radius has near-luminal velocity and appears to be escaping from the gravitational field of the black hole.
The perturbed star has an initial mass of $0.062$ and forms a black hole with 
a mass of $0.037$.  Even though the perturbed star forms a black hole that is $40\%$ 
less massive than its initial state, only a negligible amount of matter escapes beyond $r=R_\star$ 
because $R_\star$ travels outward with the rebounding matter.  It is hard to say from our numerical scheme
how much of the rebounding matter actually escapes the gravitational bounds of the black hole. 
For this run, the global maximum 
of $2m/r$ calculated is $0.995$, and the global minimum of $\alpha$ attained is approximately 
$8.9\times 10^{-10}$.  The defining parameters here are 
$\Gamma = 2$, $\rho_\circ(0,0) = 0.01$, $M_\star(0)=0.062$, and $v_{\min}(t=0) = -0.91$ with profile $U_1$.  
\label{fig:sbc-escape}}
\end{figure}

A dynamical scenario is said to be of the type SBC if a black hole forms, a shock/bounce occurs, 
and $\left(\Delta M_\star(t) / M_\star(0)\right)$ decreases over the entire course of the evolution by 
an amount greater than $10$ times the numerical error in calculating $\left(\Delta M_\star(t) / M_\star(0)\right)$. 
By tracking how $M_\star(t)$ evolves, we wish to examine whether the perturbation can force
the star to expel a significant portion of its mass before collapsing to a 
black hole, and also to estimate the prevalence of these cases in the parameter space diagram.  
Since some of the matter is ejected from the gravitational field  of the black hole, these systems
produce black holes with masses smaller than their progenitor stars. 
The behavior of various quantities at $R_\star(t)$ are shown for a typical SBC system in 
Figure~\ref{fig:sbc-escape}.  Not surprisingly, we see that the shock/bounce abruptly alters
the flow's direction at $R_\star(t)$, while the central rest-mass density increases.  Also, 
we see that $M_\star(t)$ decreases by only a small amount over the lifetime of the evolution.  Indeed, 
$R_\star(t)$ seems to approximate a Lagrangian world line quite well, and so very little mass 
fluxes through the corresponding shell.  However, even though $R_\star(t)$ may closely follow paths of constant $m$,
we consistently see a decrease in $M_\star$ in all SBC cases.  Hence, we believe this 
is a valid way of differentiating them from PC cases.  The ejection of the matter 
is particularly evident in the time sequence of $\ln \rho_\circ$ given in Figure~\ref{fig:lnrho-sbc-escape}, 
whereas the shock formation and subsequent out-moving flow due to the bounce 
is illustrated by $v(r,t)$ in Figure~\ref{fig:v-sbc-escape}.  

\centerline{\includegraphics*[bb=0.4in 2.5in 7.3in 9in, scale=0.6]{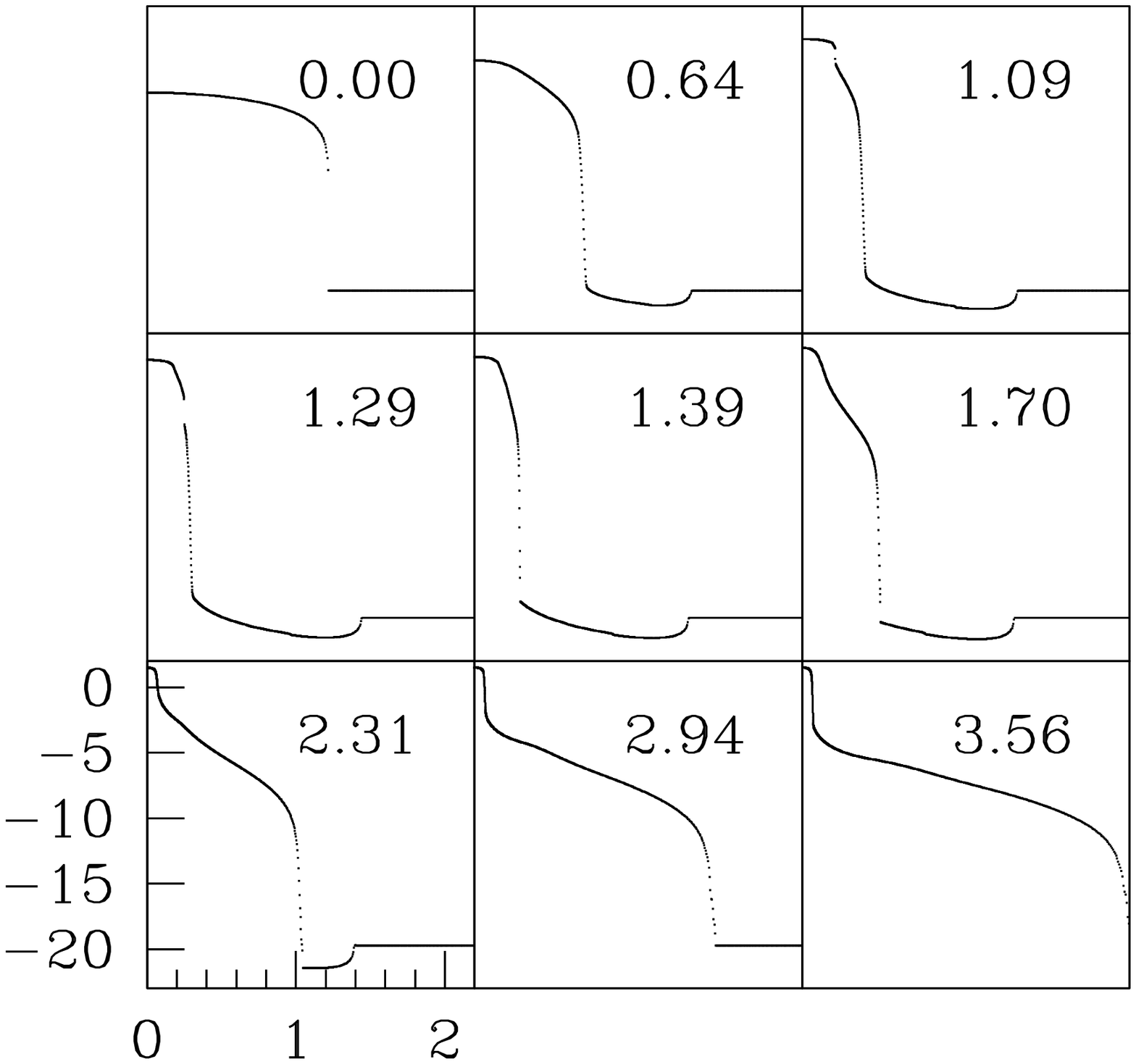}}
\begin{figure}[htb]
\caption[Time sequence of $\ln{\rho_\circ(r,t)}$ for the same shock/bounce/collapse scenario shown 
in Figure~\ref{fig:sbc-escape}.]{Time sequence of $\ln{\rho_\circ(r,t)}$ for the same shock/bounce/collapse scenario shown 
in Figure~\ref{fig:sbc-escape}.  The shock from the bounce is first seen at $t=1.09$ as the discontinuity 
near the origin, and leaves the domain at a time just before $t=3.56$.  The compact 
distribution near the origin seen at later times illustrates the extent of the forming black hole.  
In each frame of this figure, every fifth grid point is displayed.
\label{fig:lnrho-sbc-escape}}
\end{figure}

The distinction between SBC and PC states is somewhat arbitrary, 
however, because we are unable to measure the eventual steady-state mass 
of the nascent black holes, due to restrictions imposed by our current code's coordinate system.
Further, some SBC states seem such that most of the star's matter is still 
trapped even after the shock and bounce, as illustrated in the time evolutions of 
Figure~\ref{fig:sbc-bound}.  This example demonstrates that not all SBC scenarios
result in black holes that are less massive than their progenitors.  
Nonetheless, the method we use provides a consistent way for approximating the location 
of the boundary between those stars that collapse to black holes entirely and those 
which possibly expel matter before forming a black hole. 

\centerline{\includegraphics*[bb=0.3in 2.5in 7.3in 9in, scale=0.6]{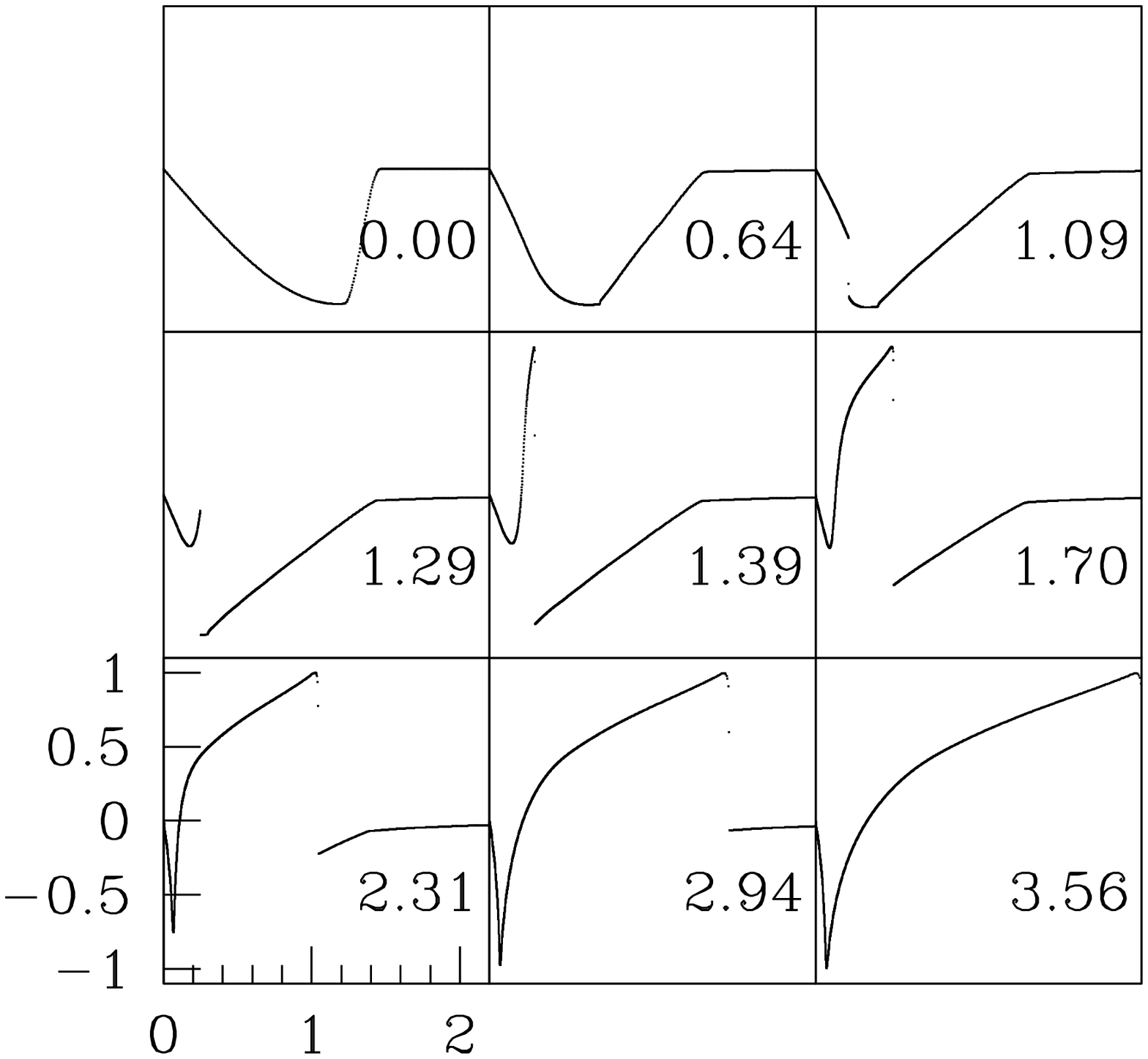}}
\begin{figure}[htb]
\caption[Time sequence of $v(r,t)$ for the same shock/bounce/collapse scenario shown 
in Figures~\ref{fig:sbc-escape}~-~\ref{fig:lnrho-sbc-escape}.  ]{Time sequence of $v(r,t)$ 
for the same shock/bounce/collapse scenario shown 
in Figures~\ref{fig:sbc-escape}~-~\ref{fig:lnrho-sbc-escape}.   The shock from the 
bounce is first seen at $t=1.09$ as the discontinuity 
near the origin, and leaves the domain at a time just before $t=3.56$.  
Instead of rebounding, the matter in the interior of the star collapses to a black hole, whose approximate 
size is represented at $t=3.56$ by the distance between the origin and the spike in $v(r)$.  
In each frame of this figure, every fifth grid point is displayed.
\label{fig:v-sbc-escape}}
\end{figure}

The phase boundaries---with the possible exception of that 
between the SBO/O states---appear to be quite smooth.  This uniformity lends itself to 
global characterizations, such as a comparison of the dynamical scenarios possible between 
non-relativistic stars (low $\rho_c$) and relativistic stars (high $\rho_c$).  
Only less relativistic---or compact---stars can undergo a complete 
explosion, one which completely disperses the star's matter to infinity.  Also, less compact stars  require 
significantly larger perturbations to develop into black holes.  Both of these aspects of 
non-relativistic stars are intuitive since, as they are the sparsest, they generate 
spacetimes with less curvature.  More compact stars induce greater spacetime 
curvature, and so are more difficult---and apparently impossible in some cases---to completely disperse
from the origin.  

For less relativistic stars, it is natural to justify the existence of the transition between
SBD to SBO scenarios.  If we follow evolutions of a particular star---say one with $\rho_c=0.03$---
for various $v_{\min}$, we see that the initial velocity perturbation results in dispersal of 
more and more of the stellar material as $v_{\min}$ increases.  The central 
densities and masses of the resultant SBO stars decrease as the SBO/SBD boundary is reached, 
implying that the transition is continuous.  For instance, if $\rho_c^f$ and $M_\star^f$ are
the final central density and mass, respectively, of the product star, then we should see that 
$\rho_c^f,M_\star^f \rightarrow 0$ as 
$v_{\min} \rightarrow v_{\min}^{\star-}(\rho_c)$, where $v_{\min}^\star(\rho_c)$ is the 
threshold value of $v_{\min}$ that separates the SBO and SBD states.
From our coarse tuning of $v_{\min} \rightarrow v_{\min}^\star(\rho_c)$ 
for various $\rho_c$, we have found that this seems to be the case.  For instance, after 
tuning $v_{\min} \rightarrow v_{\min}^\star(0.03)$ to an 
approximate precision of $10\%$, $\rho_c^f \simeq 0.0045$---which is about an $85\%$ decrease
in central density.  Alternatively, we can think of this transition in terms of the compactness 
of the star solution varying while $v_{\min}$ is held constant.  That is, if we choose a 
specific $v_{\min}$ and start perturbing stars with larger $\rho_c$ with this velocity profile, 
we see that---as
the stars become less compact---the velocity distribution is able to expel more and more
matter from the central core.  In turn, smaller and smaller stars will form for a given $v_{\min}$
as $\rho_c \rightarrow \rho_c^{\star+}(v_{\min})$, where $\rho_c^{\star}(v_{\min})$ is the value of $\rho_c$ at
the SBO/SBD boundary for a given value of $v_{\min}$.  It would be interesting to calculate
the scaling behavior of $M_\star^f$ as a function of $\rho_c - \rho_c^\star(v_{\min})$ 
or $v_{\min}^\star(\rho_c)-v_{\min}$.   An accurate calculation of this
scaling law would require many runs in this regime, which---as mentioned previously---is 
one of the more computational intensive regimes; as the central density decreases, the radius of 
the star would increase.  Hence, in this limit, we would be required to evolve 
a wide range range of scales in order to resolve the initial dynamics of the compact progenitor 
star through to the new star settling to equilibrium.  Such calculations might require a
full-fledged adaptive mesh refinement (AMR) code to be able to efficiently treat the large range 
of length scales. 

From the results of our survey, we have seen that it is not possible to drive some of the less compact stars 
to black hole formation, regardless of the size of the initial velocity perturbation.
Black holes arise through SBC dynamical scenarios for $\rho_c\ge0.007$, and homologous 
collapse to a black hole (PC) only occurs for stars with $\rho_c\ge0.01$.  Since we 
observe Type~II critical phenomena for $0.01\le\rho_c\lesssim 0.05343$ (see 
Chapter~\ref{chap:type-ii-critical} for more details), we surmise that arbitrarily small
black holes can form for this entire range of TOV solutions.  For $\rho_c\gtrsim0.05344$, we find 
that the threshold solutions are Type~I solutions, suggesting the smallest black holes that can evolve
from such stars have finite masses.  
The Type~I behavior seen in perturbed stars will be discussed in Section~\ref{chap:type-i-critical}.

\clearpage

\centerline{\includegraphics*[bb=0.3in 2.2in 8.3in 9.8in, scale=0.6]{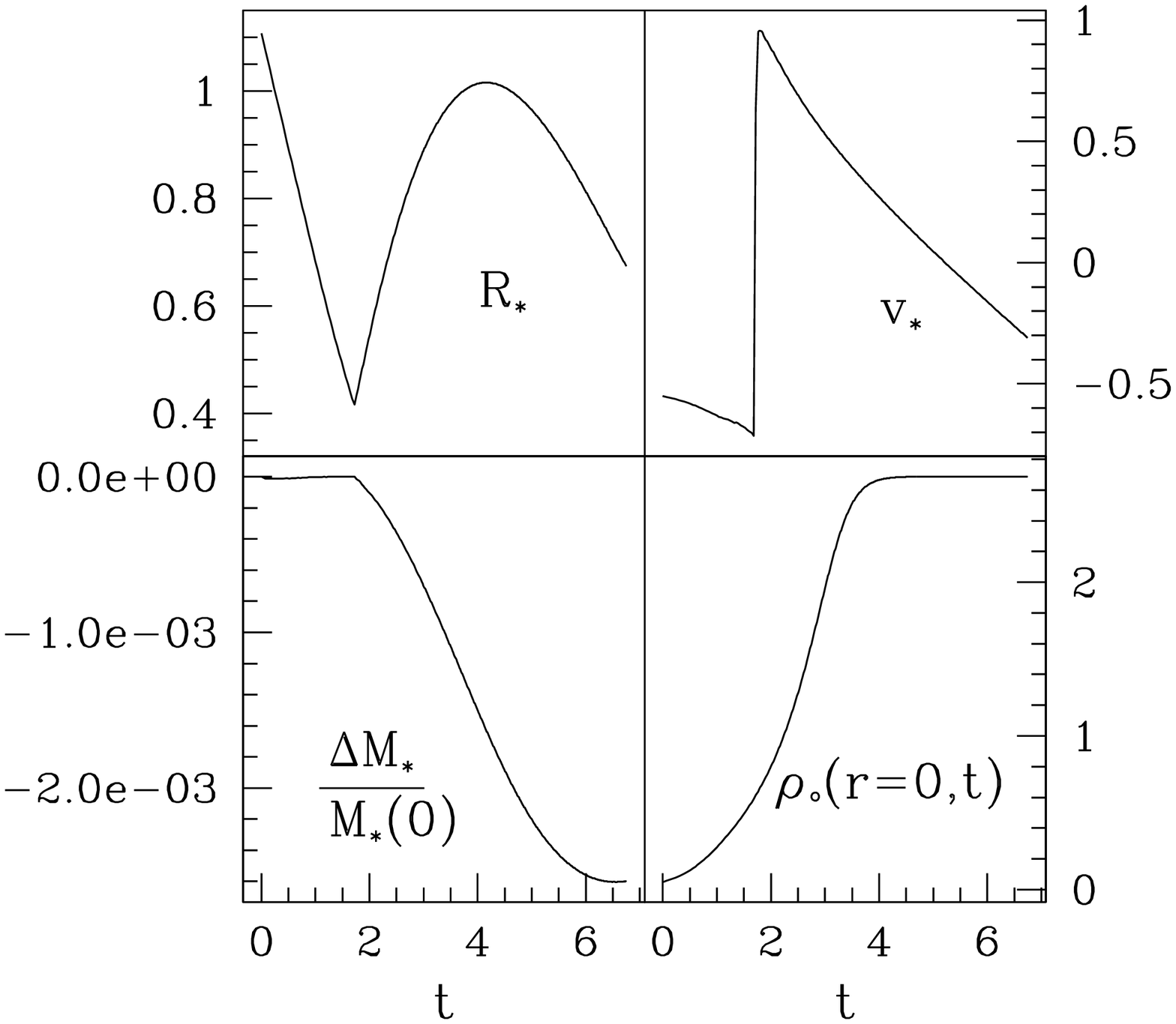}}
\begin{figure}[htb]
\caption[Evolutions of stellar radius ($R_\star$), velocity at $R_\star$ ($v_\star$), relative stellar mass 
deviation from initial time ($\Delta M_\star/M_\star(0)$), and central density for a star that is perturbed 
such that it also undergoes a shock and bounce before forming a black hole, but where matter at the 
star's surface seems to be bound to the black hole.]{Evolutions of stellar radius ($R_\star$), velocity 
at $R_\star$ ($v_\star$), relative stellar mass 
deviation from initial time ($\Delta M_\star/M_\star(0)$), and central density for a star that is perturbed 
such that it also undergoes a shock and bounce before forming a black hole, but where matter at the 
star's surface seems to be bound to the black hole.  Initially, the surface matter
begins to recoil until it finally succumbs to the curvature of the 
forming black hole and begins to descend to smaller radii.  The fact that $R_\star$ decreases 
and $v_\star$ becomes in-going after the bounce suggests that the outer parts of the star do,
indeed, accrete onto the collapsing interior.  Another indication that the star is not shedding
matter is the fact that $\Delta M_\star/M_\star(0)$ stops decreasing towards the end of 
the run.  The evolution was stopped when the maximum value of $2m/r$ obtained a value of $0.995$, 
at which point the mass of black hole was calculated to be about $0.1080$ and the minimum of $\alpha$ 
was $1.0\times 10^{-8}$.  The defining parameters for this run are 
$\Gamma = 2$, $\rho_\circ(0,0) = 0.05$, and $v_{min}(t=0) = -0.556$ with profile $U_1$; and, 
$M_\star(0) = 0.1092$.  
\label{fig:sbc-bound}}
\end{figure}

In order to compare our results to Novak's, we need to transform our scale to his.  
However, it is unclear what scale Novak; for example, he states masses in terms of solar masses, but 
fails to specify the units of $K$ and simply says ``$K=0.1$'', which possibly suggests that he is 
using geometrized units in that case.  We will attempt to compare our values to his by determining the $K$ 
that makes the mass of our  last stable TOV solution (i.e. the solution with the maximum mass) 
to correspond to his value of $3.16 M_\odot$.  
We will place a ``hat'' over all quantities that we state in his units. 
Using the methods described in Appendix~\ref{app:unit-conversion}, we find that this
factor of $K$ is 
$\hat{K}_\mathrm{Novak}=5.42\times 10^5 \mathrm{cm}^{5} \mathrm{g}^{-1} \mathrm{s}^{-2} $.
Let $M_1$ be the mass of the least massive star that can form a black hole through any scenario, and $M_2$ be 
the mass of the least massive star that we observe to undergo prompt collapse to a black hole.   
In our units, we find $M_1 \simeq 0.01656$ at $\rho_c=0.007$, and $M_2 \simeq 0.02309$ at $\rho_c=0.01$. 
Using $\hat{K}_\mathrm{Novak}$ to convert our masses to his yields a value 
for the least massive star that forms any type of black hole to be 
$\hat{M}_1=0.320 M_\odot$, opposed to his value of 
$\hat{M}^\mathrm{Novak}_1=1.155 M_\odot$.  
Moreover, the least massive star to admit prompt collapse evolutions that we 
see is $\hat{M}_2=0.446 M_\odot$, contrary to his value of  
$\hat{M}^\mathrm{Novak}_2\approx2.3 M_\odot$.  Note that $\hat{M}^\mathrm{Novak}_2$ is estimated from Figure~5 
of \cite{novak}, where a velocity profile equivalent to our $U_2$ (\ref{v-profile-2}) 
profile is used.  Since we have only performed the parameter space survey for $U_1$ (\ref{v-profile-1})
we cannot say what we would get for $M_\mathrm{min2}$ when using $U_2$.  However, 
Novak performed a comparison between these two profiles and found that his estimates for 
$M_\mathrm{min1}$ deviated by about $1\%$ between the two.  Hence, we believe it is adequate 
to quote his result for $U_2$ in order to compare to our result for $U_1$. 

The difference in masses is also obvious in our respective phase diagrams from the parameter space surveys, where 
the point along the $\rho_c$ axis---or $n_B$ in Novak's case---at which black holes are 
attainable occurs for noticeably more compact stars in Novak's case.  
Since Novak uses $K=0.1$ and since $\rho_c$ scales as $K$, then we may compare our values to his by 
transforming his multiplying $0.1$ to his unit-less density, $n_B$.
Another significant distinction we see in our phase space plot is an absence of 
SBC states for larger $\rho_c$.  Novak seems to observe such scenarios all the way to the 
turnover point ($\rho_c=0.318$), whereas we find that they no longer happen for $\rho_c\ge0.108$. 

Additionally, we believe our study is the first to extensively
cover the subcritical region of neutron star collapse.  While the method by which
the neutron stars are perturbed may not be the most physically relevant prescription, we 
are able to see all the collapse scenarios found in  previous  works. 
Much of the previous research focussed on compact stars near the turnover point or studied some other 
region exclusively (e.g. \cite{vanriper2}, \cite{vanriper-arnett}, \cite{romero}, 
\cite{font-etal2}, \cite{siebel-font-pap-2001}),
while others individually observed much of the phenomenon without thoroughly scrutinizing the 
boundaries between the phases (\cite{shap-teuk-1980}, \cite{novak}, \cite{gourg2}).  

In addition to the overall picture the parameter space survey illustrates, it sheds light on the 
critical behavior observed at the threshold of black hole formation. 
That is, we see that the phase boundary separating SBD and SBO cases on the subcritical side of the diagram 
seems to be correlated with the transition from Type~II to Type~I critical behavior on the 
line separating initial data sets that do form and do not form black hole spacetimes.   
The Type~II threshold is at the boundary between the SBD and SBC scenarios, while the Type~I 
threshold occurs along the line that separates SBO and O cases from black hole-forming cases.  We have 
been able to determine that $\rho_c\approx0.05344$ is the approximate point at which the 
transition from Type~II to Type~I behavior occurs.  For Type~II minimally subcritical solutions in this regime, 
the matter disperses from the origin, but it is difficult to say  if it escapes to infinity since 
our grid refinement procedure is incapable of coarsening the domain.  Consequently, the time step is too 
small to allow for longtime evolutions of these dispersal cases, and so we are unable to guarantee that 
they do indeed disperse to infinity.  In addition, the self-similar portion one of these marginally subcritical solutions 
entails only a small amount of the original star's matter, the remainder of which 
could, in principle, collapse into a black hole at a time after the inner self-similar component departs from the
origin.  Hence, it is difficult to determine, at this point with the current code, the ultimate fate of these dispersal
scenarios.  


\chapter{Type~II Critical Phenomenon}
\label{chap:type-ii-critical}

In Section~\ref{sec:crit-phen-gener}, we described the important role that perfect fluid 
studies played in today's picture of general relativistic critical phenomena.  Most of these
investigations, however, have involved ultra-relativistic fluids (please see 
Section~\ref{sec:ultra-relat-fluid} for the description of an ultra-relativistic perfect fluid)
that are explicitly scale-free. 
The reason for the predominance of this type of fluid is due to the fact that  Cahill and Taub \cite{cahill-taub}
showed that only those perfect fluids which follow state equations of the form (\ref{ultra-eos})---e.g.
the so-called ultra-relativistic EOS---can give rise to spacetimes that admit a homothetic symmetry. 
Hence, it is not completely unreasonable to expect that Type~II, CSS critical solutions would 
only appear in such fluids, or at least in fluids that admit an ultra-relativistic limit.  
To study this conjecture, Neilsen and Choptuik \cite{neilsen-crit} 
considered the evolution of a typical perfect fluid (see Section~\ref{sec:relat-perf-fluids})  with 
the $\Gamma$-law EOS (\ref{ideal-eos}) that includes the rest-mass density.  They argued that Type~II
critical collapse scenarios are 
typically kinetic energy dominated and entail large central pressures in order to setup the tenuous 
balance between the matter dispersing 
from the origin or collapsing to a black hole.  Therefore, they thought that if one would be able 
to give the fluid sufficient kinetic energy, then it would naturally enter 
into an ultra-relativistic phase.  Specifically, if the fluid undergoes a collapse
such that $\epsilon \rightarrow \infty$ dynamically, then  $\rho_\circ$ will effectively become 
negligible in the equations of motion and the system would be able to follow a scale-free---hence 
self-similar---evolution.  To see if their hypothesis was correct, they collapsed a compact distribution 
of perfect fluid, whose EOS was $P=0.4\rho_\circ \epsilon$ ($\Gamma=1.4$), and were able to tune toward 
the threshold solution. 
The critical solutions they obtained by solving the full set of PDE's (\ref{conservationeq}) closely 
matched the precisely self-similar solutions, which they calculated by assuming that a model governed by 
the ultra-relativistic EOM had an exact homothetic symmetry.  Further, they found that the scaling exponent, $\gamma$, 
defined by (\ref{mass-scaling}) matched that of the ultra-relativistic critical solution for $\Gamma=1.4$. 
Since the ultra-relativistic fluid exhibited Type~II phenomena 
for all considered values of the adiabatic index in the range $0 < \Gamma \le 1$, then 
the results of \cite{neilsen-crit} suggested that the ultra-relativistic solution for given $\Gamma$
should be the same as that for an ideal-gas perfect fluid for the same $\Gamma$.  

This hypothesis is not without precedence, since several models have been found to exhibit 
DSS or CSS collapse, even when explicit length scales are present.  For instance, 
Choptuik \cite{choptuik-1994} found Type~II behavior in the Einstein-Klein-Gordon (EKG) model---that 
is a massive scalar field governed by (\ref{generalscalareom}) with
$V(\phi)=\frac{1}{2}m^2\phi^2$---even though it has an explicit length 
scale set by $1/m$.  His heuristic argument was that the potential term is 
naturally bounded since $\phi$ itself is bounded in the critical regime, 
but that the kinetic term---$\Box \phi$---diverges in the critical limit.
Hence, the kinetic term overwhelms the potential term and essentially makes the critical evolution scale-free.

Systems with an explicit scale dependence can also exhibit Type~I behavior as well as  
Type~II behavior.  The boundary separating  the two types has been studied extensively in 
the SU(2) Einstein-Yang-Mills (EYM) model \cite{choptuik-chmaj-bizon,choptuik-hirshmann-marsa} 
and the aforementioned EKG model \cite{brady-chambers-goncalves}.  
In the latter case it was found that when 
the length-scale, $\lambda$, which characterizes the ``spatial extent'' of 
the 2-parameter family of initial data used was small compared to the scale set by the scalar field, 
Type~II behavior was observed.  The transition from Type~II to Type~I behavior was calculated for 
different families and was found to occur when $\lambda m \approx 1$. 

The one study by \cite{neilsen-crit} that showed Type~II behavior in a perfect fluid 
with an ideal gas EOS remained unverified until Novak \cite{novak} announced 
results on  neutron star models driven to black hole formation.  In order to determine the possible 
range in the masses of nascent black holes formed from stellar collapse, he performed a parameter space survey
using the 1-parameter family of TOV solutions with $\Gamma=2$, and varied the amplitude of the 
initial coordinate velocity 
profile (see Chapter~\ref{chap:veloc-induc-neutr} for further details on the 
survey performed in \cite{novak}).  The Type~II behavior observed was quantified by fitting to the 
typical black hole mass scaling relation (\ref{mass-scaling}), where Novak used the initial 
velocity amplitude $U_\mathrm{amp}$ as the tuning parameter $p$.  A significant aspect of his 
study is that Novak was able to observe such a scaling behavior even with a realistic equation of 
state formulated by Pons et al. \cite{pons-etal-2000}.  This was somewhat surprising since 
Type~II phenomena was never expected to be observed in such realistic models \cite{gundlach-rev2}. 
However, this is not entirely surprising so long as the model (EOS) admits an ultra-relativistic limit.

Even though Novak observed Type~II behavior, he did not find the same scaling exponent as 
had been observed by Neilsen and Choptuik for the $\Gamma=2$ ultra-relativistic fluid.  In addition, 
he claimed that $\gamma$ was 
a function of central density $\rho_c$, which parameterizes the initial star solution, as well as 
the EOS.  He observed that the fit to (\ref{mass-scaling}) worsened as $\rho_c$ increased
to that of the maximum mass solution, and that it eventually broke down completely.  Specifically, 
he found for (\ref{ideal-eos}):
\beq{
\gamma_{_{N1}} \ \simeq \  0.52 \quad , \label{novak-exponent1}
}
and when using the realistic EOS 
\beq{
\gamma_{_{N2}} \ \simeq \ 0.71   \quad . \label{novak-exponent2}
}
These values are significantly different from the values most recently calculated 
with the $\Gamma=2$ ultra-relativistic fluid in \cite{brady_etal} using a variety of 
methods:
\beq{
\gamma_{_B} \ \simeq \ 0.95 \pm 0.1  \quad , 
\label{brady-exponent}
}
where we have taken the average of the three independent values \cite{brady_etal} calculated,
and the uncertainty here is the standard deviation from the set.  This uncertainty, however, does not
include the systematic errors inherent in the distinct calculations.

However, Novak admitted that his code was not designed to simulate the formation of very small black holes, and apparently
was only able to tune to a precision of $p-p^\star\simeq10^{-3}$.  Hence, we wish to investigate
the Type~II behavior in this particular system in order to investigate his claims and to obtain 
a better measurement of the scaling exponent.  Along the way, we  also provide consistency checks in order
to ensure that the critical solutions obtained are, in fact, genuine and not the result of our approximate 
numerical procedure.

If not otherwise stated, the results in the following sections use $U_1$ (\ref{v-profile-1}) 
for the initial velocity profile, $\Gamma=2$ perfect fluids (\ref{ideal-eos}), 
and $K=1$ for the factor in the polytropic EOS (\ref{polytropic-eos}) that is used at $t=0$.
Also, the tuning parameter $p$ used is the initial amplitude of the in-going 
velocity amplitude, $U_\mathrm{amp}$ (\ref{v-profile-1}). 

\section{The Type~II Critical Solution}
\label{sec:type-ii-solution}
In this section, we study the Type~II, CSS critical solution found at the black hole-forming threshold
of the parameter space described in Chapter~\ref{chap:veloc-induc-neutr}.  
As mentioned there, the region exhibiting Type~II collapse consisted of the least relativistic stars, e.g. 
the sparsest, that we could drive to 
collapse. 
We were able to form black holes from stars with an initial rest-mass central density 
greater than $\rho_c^{\min} = 0.007$, but have closely tuned towards critical solutions
for only a handful of these initial states.  In Table~\ref{table:type-ii-stars}, we list the stars 
in which Type~II behavior was actually observed, and quantify how close to the critical value 
we were able to tune.
The instability described in Section~\ref{sec:instability} limited the tuning in all instances.  
\begin{table}[htb]
\begin{center}
\begin{tabular}[htb]{|c|c|c|c|}
\hline
$\rho_c$   &${\min}(M_\mathrm{BH})/M_\star$ &${\min}\left|p-p^\star\right|/p$ &$p^\star$\\
\hline 
\hline
$0.01$  &$1\times 10^{-6}$  &$2\times 10^{-9}$   &$0.88942207$\\
$0.02$  &$6\times 10^{-7}$  &$1\times 10^{-9}$   &$0.74611650$\\
$0.03$  &$3\times 10^{-7}$  &$5\times 10^{-10}$  &$0.633712118$\\
$0.04$  &$6\times 10^{-8}$  &$2\times 10^{-11}$  &$0.543143513$\\
$0.05$  &$2\times 10^{-8}$  &$6\times 10^{-12}$  &$0.46875367383$\\
\hline
\end{tabular}
\end{center}
\caption[The star solutions in which we observed Type~II behavior, and the minimum black hole masses
we were able to form from them.]{The star solutions in which we observed Type~II behavior, and the 
minimum black hole masses we were able to form from them.  The first column lists the
stars' initial central rest-mass densities which parameterize the star solutions used.  
We denote the mass of the smallest black hole found for a given $\rho_c$ by 
${\min}(M_\mathrm{BH})$, $M_\star=M_\star(\rho_c)$ is the mass of the initial star solution, 
and ${\min}\left|p-p^\star\right|/p$ is the relative precision reached in $p^\star$ per star. 
The final columns lists the critical parameters obtained
\label{table:type-ii-stars} }
\end{table}

From Table~\ref{table:type-ii-stars} it is clear that the instability's effect on 
our ability to find the critical parameter increases with decreasing $\rho_c$.  This is most likely 
due to the fact that sparser stars require greater in-going velocities in order to collapse, giving rise to 
more relativistic and, consequently, less stable evolutions.  We note, however, that our results represent 
great improvement over the precision obtained in \cite{novak}; the smallest 
black hole attained in that study was ${\min}(M_\mathrm{BH})/M_\star \sim 10^{-2}$. 
The success of our code is most likely due to our use of adaptive/variable mesh procedures and the great 
lengths we went to combat the sonic point instability.  

Unless otherwise stated and for the remainder of the section we  focus 
on behavior seen in the $\rho_c=0.05$ star.

\clearpage

\centerline{\includegraphics*[bb=0.7in 2.3in 8.1in 9.8in, scale=0.5]{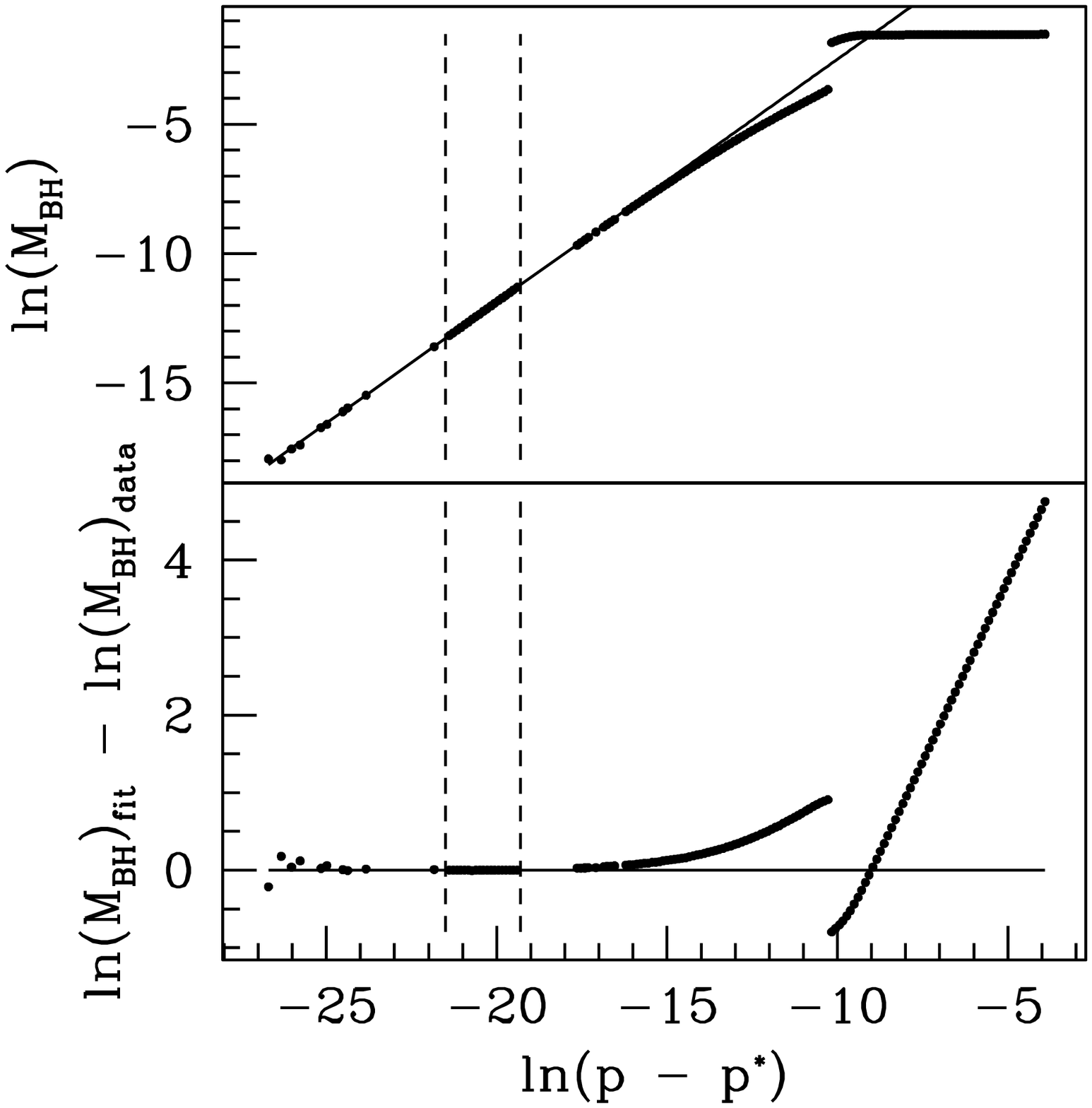}}
\begin{figure}[htb]
\caption[Plot of the scaling behavior for supercritical solutions, e.g. those that form black holes,
for solutions far from and near the critical solution.]{Plot of the scaling 
behavior for supercritical solutions, e.g. those that form black holes,
for solutions far from and near the critical solution. 
The top plot illustrates how the points from a series of supercritical runs follow the scaling 
law for the black hole mass (\ref{mass-scaling}), while the bottom plot 
shows how the data deviate from our best fit to this scaling law.  The two dotted 
lines delineate the data used in making the best fit; this data is plotted separately in
Figure~\ref{fig:super-scaling-bestfit}.  Black holes were assumed to have formed when 
${\max}(2m/r) \ge 0.995$.  The gaps between some of the points represent those 
runs that crashed before ${\max}(2m/r)$ reached this value.  Smoothing was used for 
$\ln{\left(p - p^\star\right)} < -19.3$, which is also where we start our fit. 
These runs used $\rho_c = 0.05$, $U = U_1$ and an initial grid defined by
$\{N_a, N_b, N_c, l\} = \{300, 500, 20, 0\}$.  
\label{fig:super-scaling-alldata}}
\end{figure}

\centerline{\includegraphics*[bb=0.3in 2.3in 8.1in 9.8in, scale=0.5]{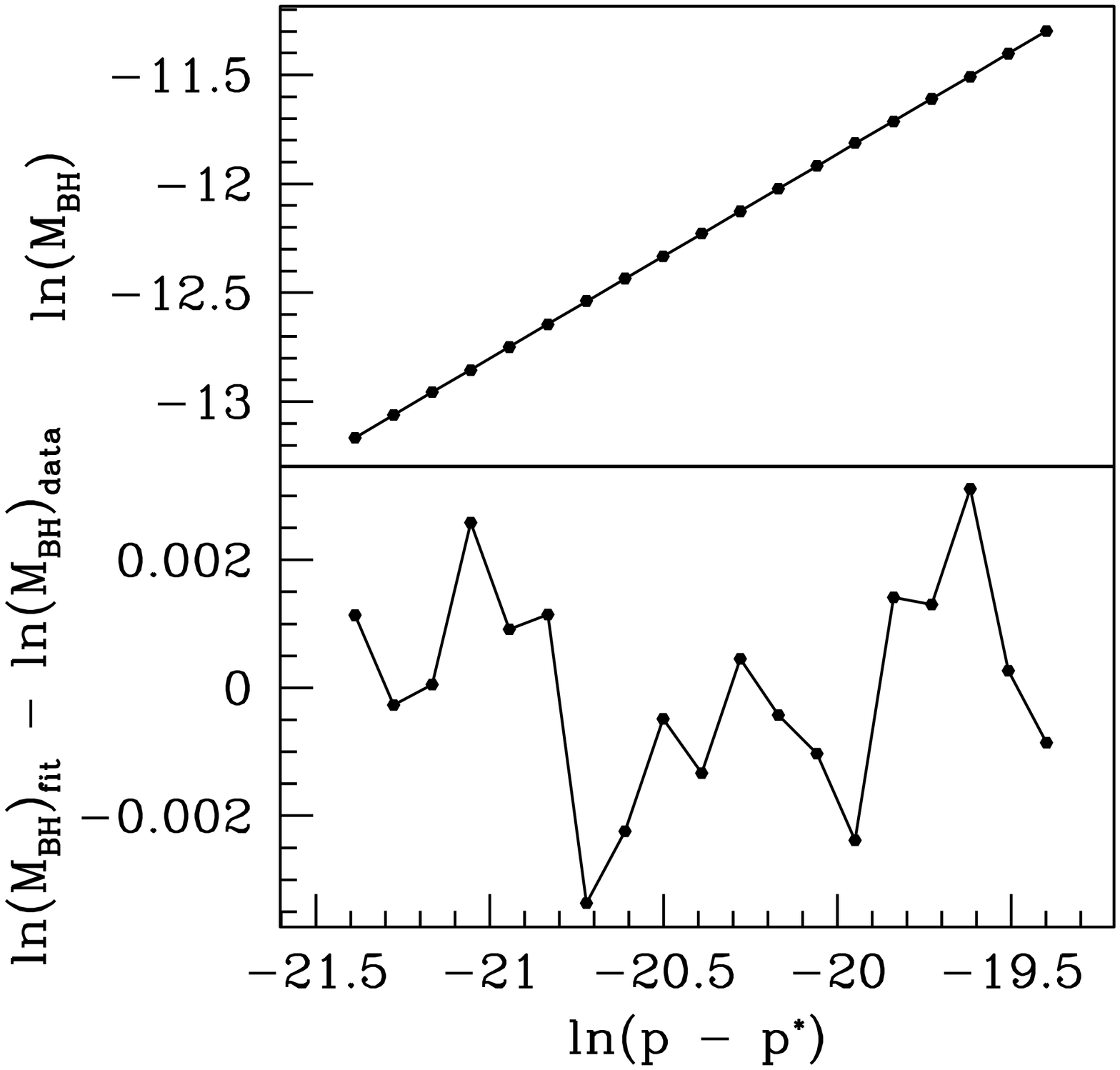}}
\begin{figure}[htb]
\caption{The best-fit for the scaling behavior of black hole masses near the critical regime. 
The top plot shows calculated masses and the fitting line, while the bottom plot shows the 
deviation between the two.  The scaling exponent for this fit, which is just the slope of 
the line, was found to be $\gamma=0.94$.  
\label{fig:super-scaling-bestfit}}
\end{figure}

To demonstrate the scaling behavior of $M_\mathrm{BH}$ as $p$ tends toward $p^\star$, we show in 
Figure~\ref{fig:super-scaling-alldata}  a plot of  $\ln\left(M_\mathrm{BH}\right)$ versus 
$\ln\left(p-p^\star\right)$ for a wide range of supercritical solutions.  The slope of the trend 
determines the scaling exponent, $\gamma$.  We will compare our values for $\gamma$ later in this
section to those from previous studies.  From the figure, we can clearly
see that the scaling law provides a good fit only in the limit $p\rightarrow p^\star$ as expected \cite{koike-etal-1995}. 
The jump seen at larger $\left(p-p^\star\right)$ represents the point at which the fluid is able to 
enter a dynamical phase
where the center part of the star has enough kinetic energy to dominate the length scale 
set by $\rho_\circ$.  The fluid in this regime are then able to follow a CSS-type evolution.

In addition, Figure~\ref{fig:super-scaling-alldata} is meant to illustrate the code's problem 
with handling the formation of the apparent horizon in the critical regime.  The data in the plot 
was made by a script that ran the simulation for a distribution of $p$ values evenly spaced 
in $\ln\left(p-p^\star\right)$.  Hence, the plot is \emph{supposed} to have points evenly spaced 
along the horizontal axis.  Any gaps represent where the code crashed before it could 
determine that an apparent horizon was about to form.  The flow velocity 
becomes discontinuous and nearly luminal as a black hole forms which seems to exacerbate the instability mentioned 
in Section~\ref{sec:instability}, and results in the evolution halting before ${\max}(2m/r)$
surpasses its threshold.  However, for a set of parameter values, indicated by the dashed lines, 
we were able to find a good fit to a scaling law.  The fit, and the data's deviation from it,
is shown in Figure~\ref{fig:super-scaling-bestfit}.  The data seems to follow the scaling law 
quite nicely, as indicated by the small, apparently random deviation from the fit.  From the slope, we 
calculate a scaling exponent of $\gamma=0.938$, which agrees well with previous studies 
for $\Gamma=2$ \cite{brady_etal,neilsen-crit}, and disagrees significantly with that calculated
for the same system in \cite{novak}. 

\centerline{\includegraphics*[bb=0.3in 2.1in 8.0in 9.8in, scale=0.5]{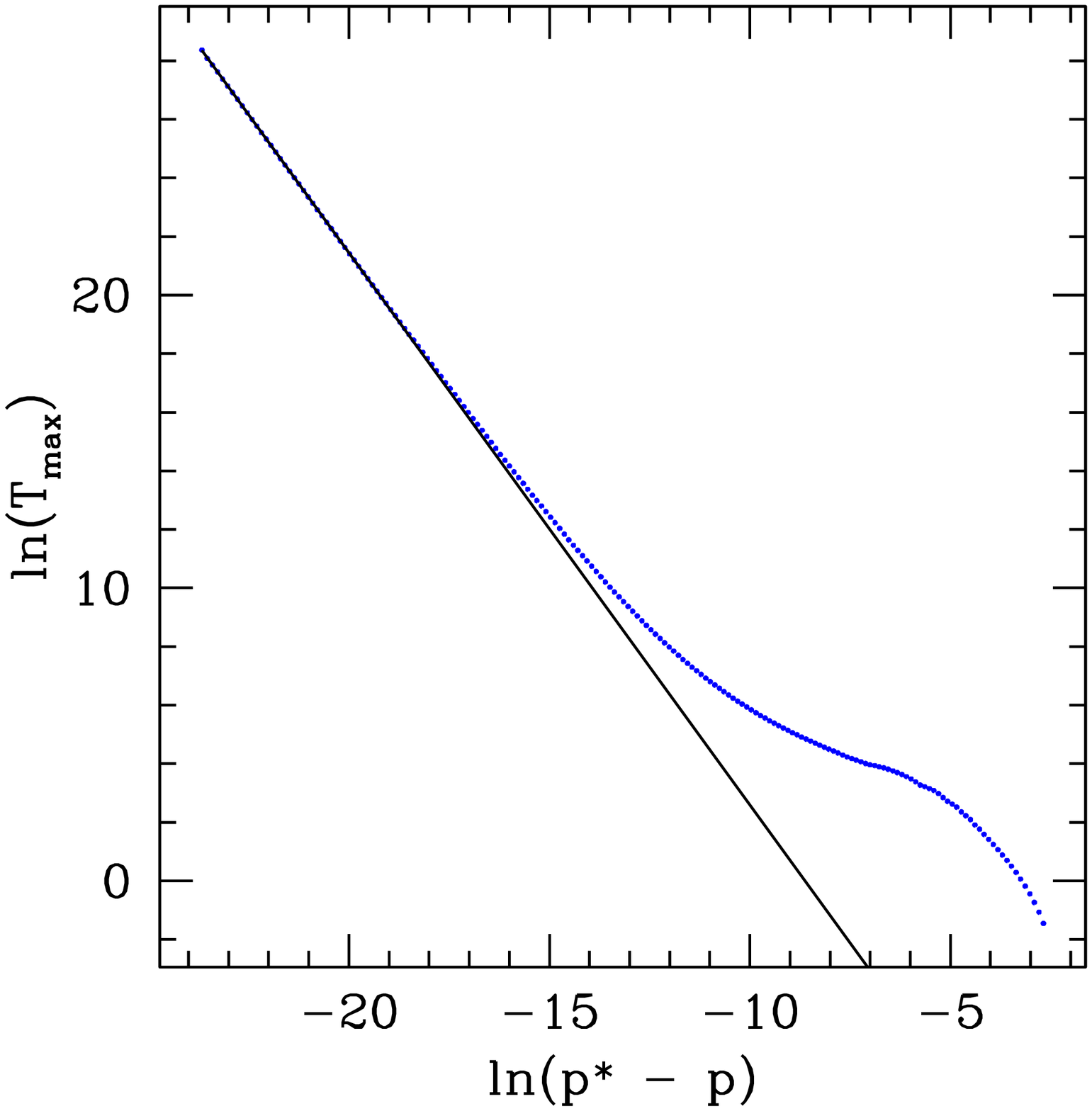}}
\begin{figure}[htb]
\caption[This is a plot of the scaling behavior in $T_{\max}$ for subcritical 
solutions, e.g. those not forming black holes.  Points far from and near the critical solution 
are shown in order to illustrate how the solutions behave in a more ultra-relativistic
manner---and hence tend toward a straight line in this plot---as the solutions tend towards criticality.]{
This is a plot of the scaling behavior in $T_{\max}$ for subcritical 
solutions, e.g. those not forming black holes.  Points far from and near the critical solution 
are shown in order to illustrate how the solutions behave in a more ultra-relativistic
manner---and hence tend toward a straight line in this plot---as the solutions tend towards criticality.  
The line shown here is the best-fit for the expected scaling law (\ref{stress-scaling}) when
using only the solutions closest to criticality; for a better view of those points involved
in the fit, please see the fit called ``Original'' in Figure~\ref{fig:scaling-difffloors}. 
These runs used $\rho_c = 0.05$, $U = U_1$ and an initial grid defined by
$\{N_a, N_b, N_c, l\} = \{300, 500, 20, 0\}$.  
\label{fig:sub-scaling-alldata}}
\end{figure}

\centerline{\includegraphics*[bb=0.8in 2.1in 7.8in 9in, scale=0.5]{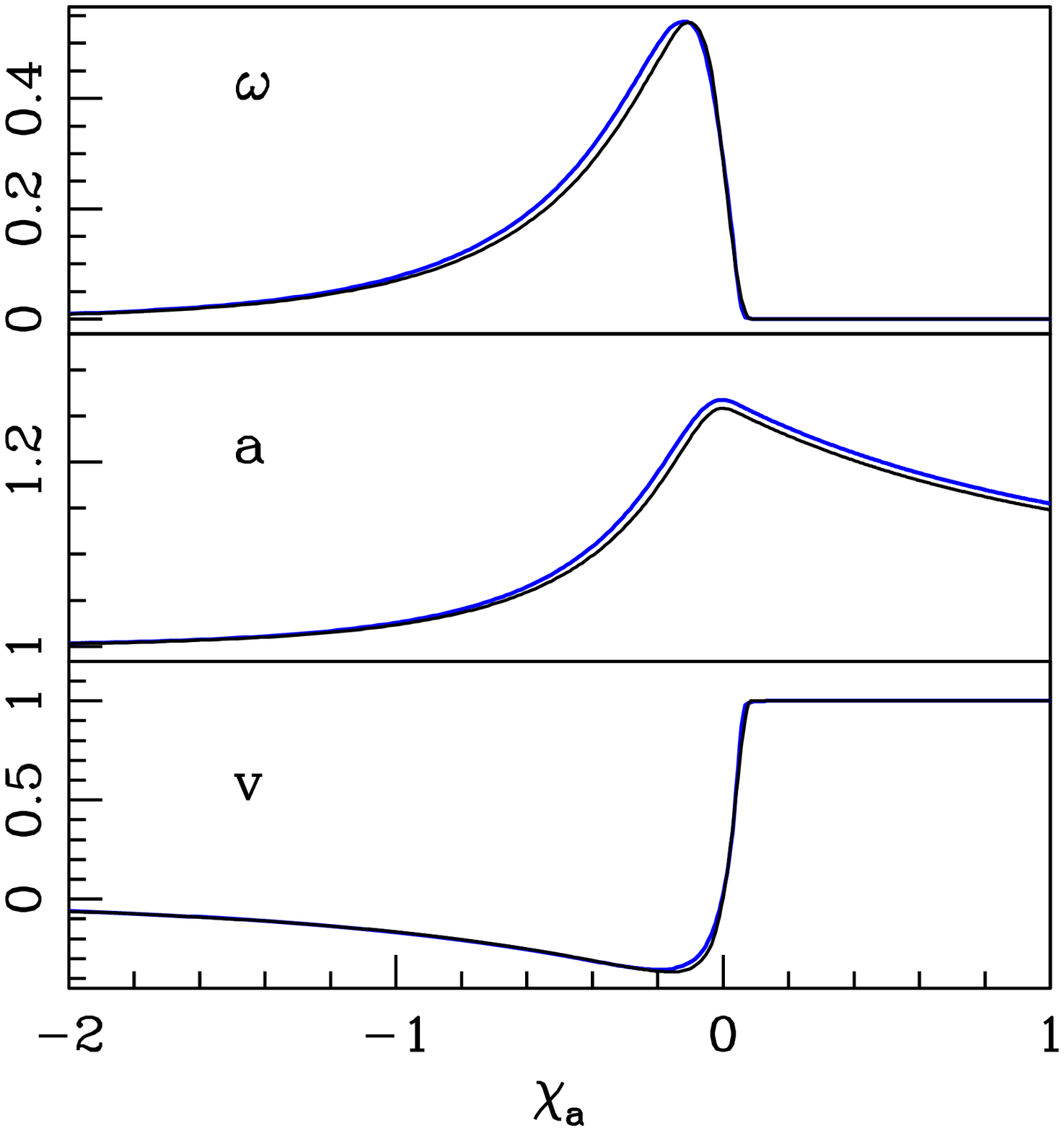}}
\begin{figure}[htb]
\caption[This is a plot of three scale-free quantities from near-critical 
evolutions in self-similar coordinates for the ideal-gas system (blue line) and the ultra-relativistic 
system (black line).]{This is a plot of three scale-free quantities of near-critical 
solutions in self-similar coordinates for the ideal-gas system (blue line) and the ultra-relativistic 
system (black line).  We can see they are quite
similar, but have noticeable discrepancies.  The deviation of the two could be 
due to the smoothing operation performed throughout the ideal-gas evolution. 
\label{fig:ideal-ultra-critsoln}}
\end{figure}

To obtain another measure of the scaling exponent, we also calculated how the global maximum of 
the stress tensor's trace, $T_{\max}$,
scales as $p \rightarrow p^\star$ from the subcritical side (\ref{stress-scaling}).  With this 
additional measurement we can get an estimate of the systematic error in our results.  Also, the code was more 
successful at calculating $T_{\max}$ than 
$M_\mathrm{BH}$ in the  $p\rightarrow p^\star$ limit.  The scaling behavior for this quantity can be seen in  
Figure~\ref{fig:sub-scaling-alldata} where $\ln T_{\max}$ is plotted versus $\ln\left(p^\star-p\right)$. 
The solutions far from criticality seem to smoothly asymptote toward the critical regime.  
The line shown in this plot only uses those points in the critical regime that provide the best 
linear fit; a closer view of the points used in the fit are shown, for instance, in 
Figure~\ref{fig:scaling-difffloors}.  
Since the slope of the line now represents $-2\gamma$ (\ref{stress-scaling}), we find 
from this fit that $\gamma=0.94$, which is most likely within systematic error from 
our value found with the scaling of $M_\mathrm{BH}(p)$.

Although our calculated scaling exponents match well to results previously obtained for the 
ultra-relativistic fluid with $\Gamma=2$, 
this does not necessarily say how well the ideal-gas critical solutions compare to the ultra-relativistic 
ones in detail.  To obtain the ultra-relativistic critical solutions, we let an adjustable
distribution of ultra-relativistic fluid free-fall and implode at the origin; specifically, the initial 
data is set so that $\tau$ is a Gaussian distribution and $S=0$ for the ultra-relativistic fluid, and 
the amplitude of the Gaussian is used as the tuning parameter.  The scale-free functions 
from the critical solutions of the velocity-induced neutron star system and the 
ultra-relativistic system are shown in Figure~\ref{fig:ideal-ultra-critsoln}.  Here, $a$ is the metric
function and $v$ is the Eulerian velocity of the fluid.  The function $\omega$ is another 
scale-free function determined from metric and fluid quantities:
\beq{
\omega \ \equiv \ 4 \pi r^2 a^2 \rho 
\label{omega-css}
}
In order to make the comparison between the two solutions, the grid functions were transformed 
into the self-similar coordinates $\mathcal{T}$ (\ref{T-ss}) and $\mathcal{X}_a$ (\ref{X-ss2}) 
using the solutions' respective values of $r_a(\mathcal{T})$ and accumulation times, which are the times
at which their critical solutions are estimated to converge upon the origin.  We found the sonic point we 
calculated for the ideal-gas fluid 
did not follow a continuous world line and was thus a bad point of reference for making the self-similar
coordinate transformation.  The discontinuous sonic point trajectory was probably caused
by the smoothing procedure (Table~\ref{table:smoothing-procedure}), since the smoothing process deforms the fluid
quantities and, hence, can lead to errors in determining when $v$ and $c_s$ intersect.  
On the other hand the spatial maximum of $a$ followed a smooth path, so we used this point to rescale the
coordinates of the ideal-gas fluid's evolution.  Either $\mathcal{X}_a$ and $\mathcal{X}$ 
are---in principle---adequate coordinates to follow the solution's self-similar scaling, since they are both
calculated using lengths scale inherent to the anticipated self-similar solution.

Our results indicate that the ideal-gas system does asymptote to the ultra-relativistic self-similar solution 
in the critical limit.
While the ultra-relativistic fluid  enters a self-similar phase shortly after the initial time, 
the ideal-gas solution seems to tend toward the critical solution and then eventually diverge away from 
it.  The agreement between the ideal-gas and ultra-relativistic solutions improves as $p\rightarrow p^\star$, 
as expected, and Figure~\ref{fig:ideal-ultra-critsoln} shows profiles at a time when the difference 
between the solutions was minimized. The $l_2$-norms of the deviations between the three scale-free functions
are shown in Figure~\ref{fig:ideal-ultra-deviation}; it can be easily gleaned from this figure that the 
minimum of the average deviations occurs at  approximately $\mathcal{T}=-13.1$, which the time at which 
we  have displayed the profiles in Figure~\ref{fig:ideal-ultra-critsoln}.  The $l_n$-norm of a 
discretized function, $f_i$, is defined by 
\beq{
l_n-\mathrm{norm} f_i \ \equiv \ \left|\left| f_i \right|\right|_2 \ = \ \left( \sum_i f_i^2 \right)^{1/n}
\label{ln-norm}
}
Also, Figure~\ref{fig:ideal-ultra-deviation}
graphically illustrates how the ideal-gas solution exponentially---in $\mathrm{T}$---asymptotes 
to the ultra-relativistic critical solution 
at early times.  The deviations for the three functions seem to have the same qualitative trend, indicating 
that metric \emph{and} fluid quantities asymptote to their ultra-relativistic counterparts.

\clearpage 
\centerline{\includegraphics*[bb=0.5in 2.2in 8.1in 9.8in, scale=0.5]{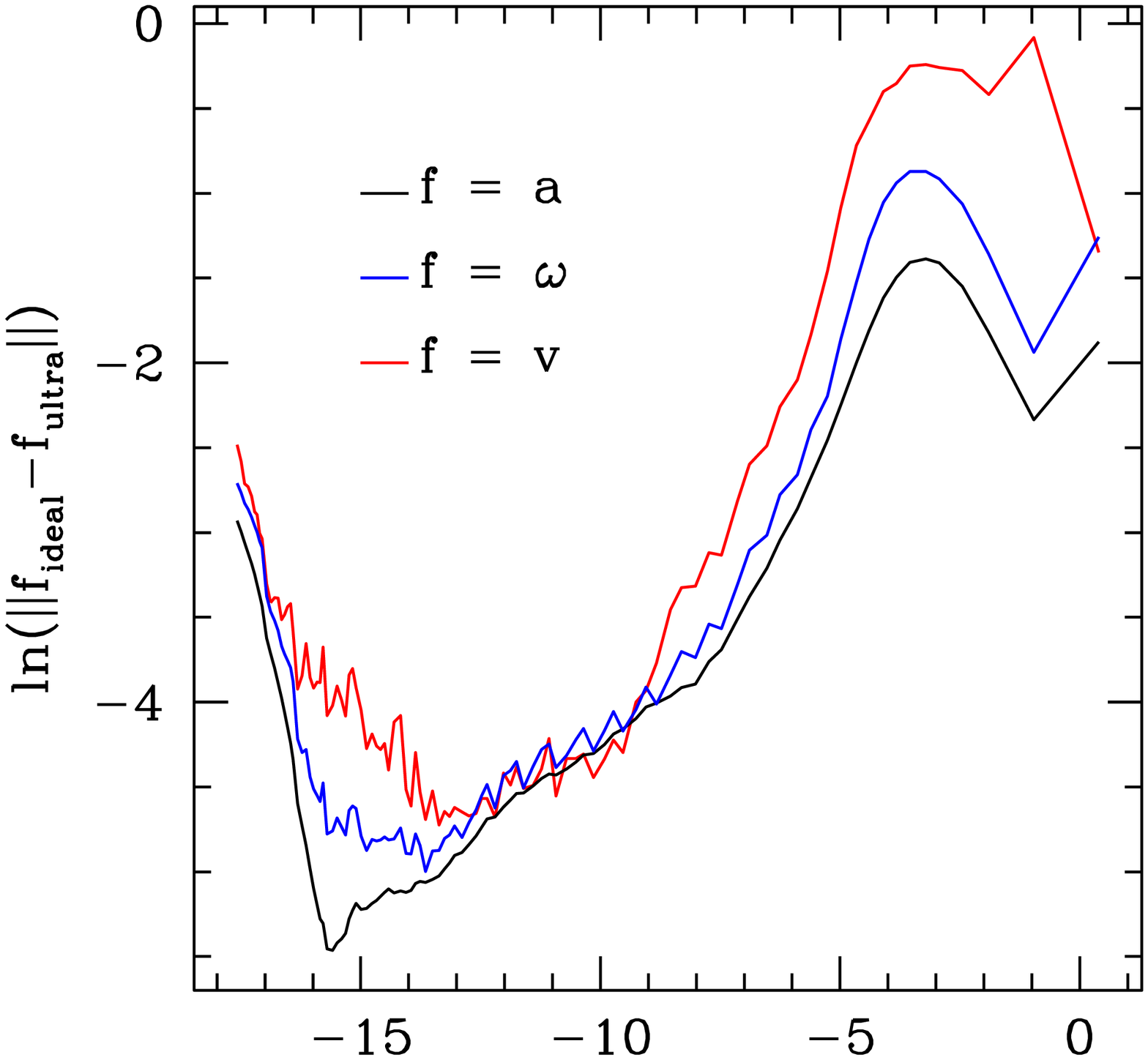}}
\begin{figure}[htb]
\caption[The deviation over time of those quantities displayed in 
Figure~\ref{fig:ideal-ultra-critsoln}.]{The deviation over time of those quantities displayed 
in Figure~\ref{fig:ideal-ultra-critsoln}.
Here, $||f||$ denotes the $l_2$-norm of the function $f$.  The $l_2$-norm is taken of these differences
at every time satisfying $\mathcal{X}_a < 2$ (\ref{X-ss2}), and its logarithm is plotted 
versus $\mathcal{T}$, a self-similar coordinate defined by (\ref{T-ss}).  Note that physical time flows in the opposite 
direction than $\mathcal{T}$, or $\mathcal{T}\rightarrow-\infty$ as the solution approaches the 
accumulation time.  As the evolution proceeds from the initial time, the two solutions asymptote 
toward each other.  After $\mathcal{T} \approx -13$, the deviation between the two solutions increases 
as the ideal-gas near-critical solution departs from the asymptotic critical solution and 
eventually disperses from the origin.  
\label{fig:ideal-ultra-deviation}}
\end{figure}

\centerline{\includegraphics*[bb=0.3in 2.2in 8.1in 9.8in, scale=0.6]{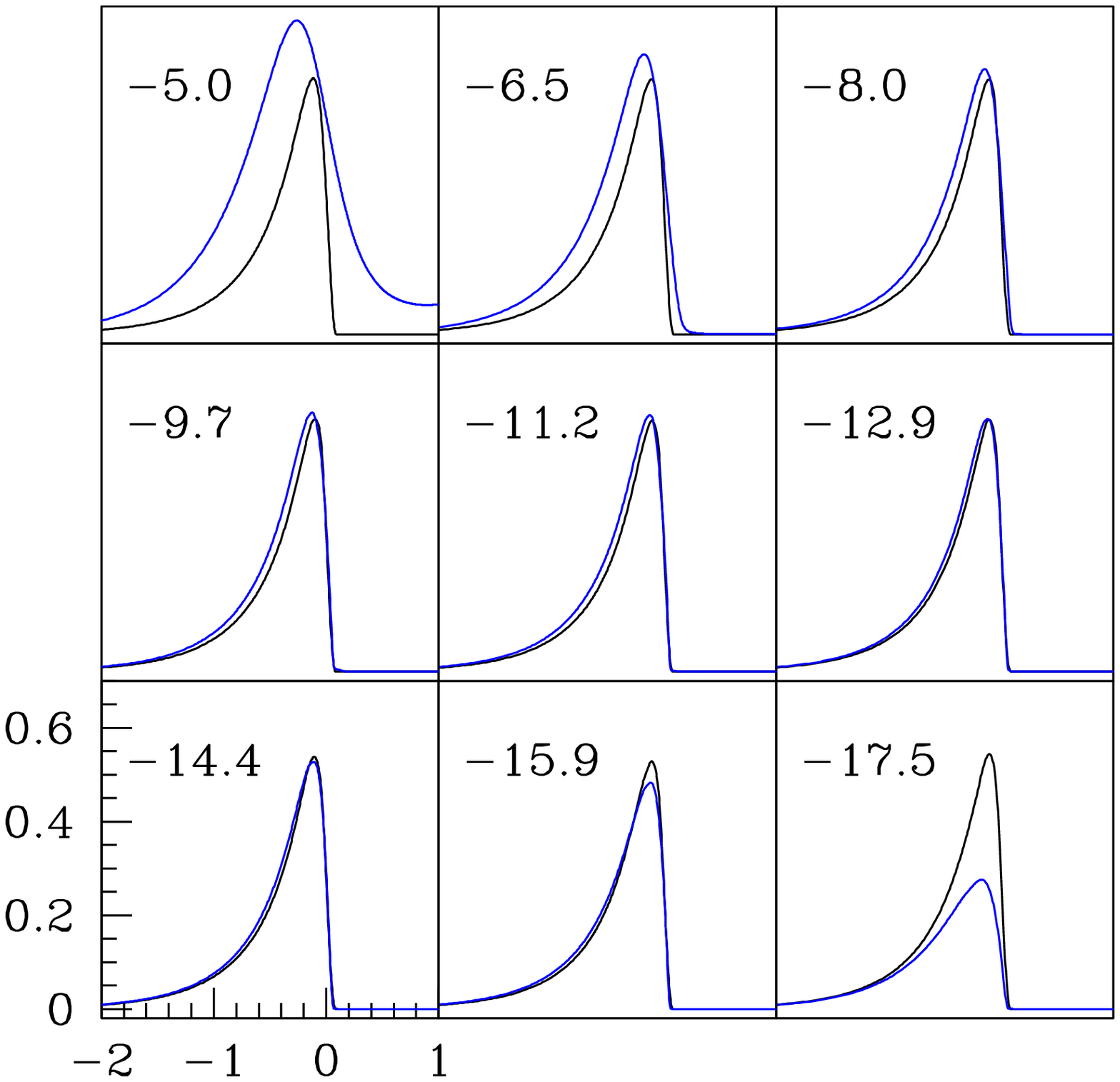}}
\begin{figure}[htb]
\caption[A time sequence of $\omega$ for the most nearly critical solution obtained 
with the ideal-gas EOS (blue) and $\omega$ for the most nearly critical ultra-relativistic solution (black).  
The solid line is $\omega$ of the most nearly critical 
ultra-relativistic solution. ]{A time sequence of $\omega$ for the most nearly critical solution obtained 
with the ideal-gas EOS (blue) and $\omega$ for the most nearly critical ultra-relativistic solution (black).  
The solid line is $\omega$ of the most nearly critical ultra-relativistic solution. Both functions have been 
transformed into self-similar coordinates, based upon 
their respective accumulation times and respective positions of their first maxima of $a(r)$.  
The number in the upper-left corner of a frame is the value of the self-similar 
time-like coordinate $\mathcal{T}$ (\ref{T-ss}) at 
which each frame's functions are displayed.  Note that the ultra-relativistic $\omega$ is varying 
slightly frame-to-frame contrary to appearances.  The ideal-gas solution requires more time to 
form the self-similar solution since the length scale set by $\rho_\circ$ only becomes insignificant
in the ultra-relativistic limit, $P/\rho_\circ \gg 1$.  
\label{fig:crit-omega-evolution}}
\end{figure}

This exponential approach of the ideal-gas solution to the self-similar solution may be better 
seen in the time sequences of $\omega$ from the ideal-gas and ultra-relativistic fluids, shown in 
Figure~\ref{fig:crit-omega-evolution}.  In the series of snapshots, $\omega_\mathrm{ultra}$ 
has already entered its self-similar form, while $\omega_\mathrm{ideal}$ takes significantly 
longer to enter an analogous form and only remains there for approximately $3$ or $4$ of the $9$ 
frames.  

\subsection{Universality and Consistency}
\label{sec:universality-consistency}
As in any scientific endeavor, it is vital that the methods used in obtaining physical results---albeit 
from simulation in this case---be rigorously tested and that the results be repeatable using similar, but 
different, means.  We present calculations in this section to justify that our methods are sound and 
that the results are not artifacts of the computational techniques used. 
These tests also provide a measure of the systematic
error in our calculation of $\gamma$.  In order to verify previous claims that critical solutions 
in perfect fluids of the same adiabatic index $\Gamma$ reside in the same universality class, we
will also measure $\gamma$ for different initial conditions while keeping $\Gamma$ constant. 
When making the comparisons, the methods, parameters, and initial data used to make
Figures~\ref{fig:super-scaling-alldata}-~\ref{fig:sub-scaling-alldata} will be referred to as 
the ``original'' configuration.  A tabulation of the values of $\gamma$ and $p^\star$ calculated 
from the different simulation configurations is given in Table~\ref{table:gammas}. 

\begin{figure}[htb]
\centerline{\includegraphics*[bb=0.3in 2.3in 8.1in 9.8in, scale=0.5]{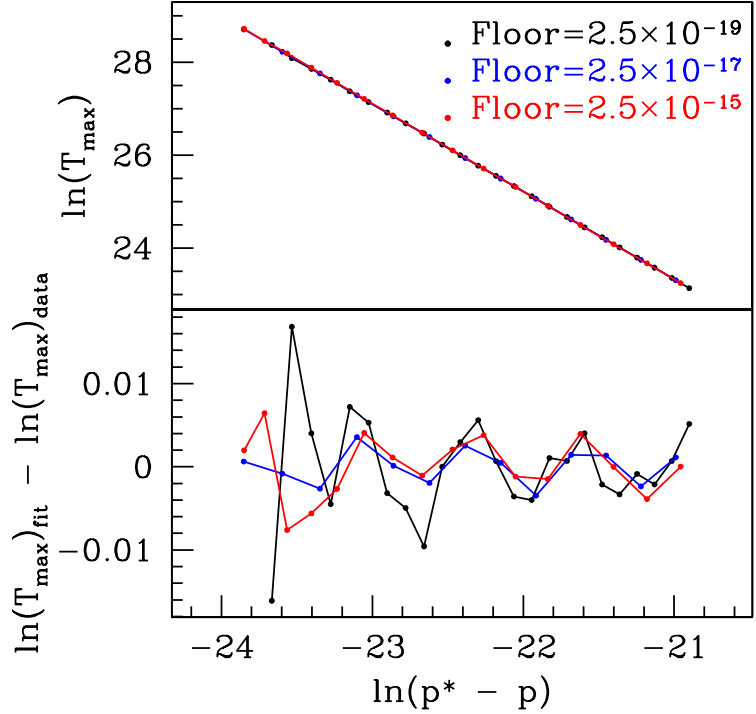}}
\caption[The scaling behavior in $T_{\max}$ near the critical 
solution for runs using different values of $P_\mathrm{floor}$ and $\delta$.  ]{The scaling behavior 
in $T_{\max}$ near the critical 
solution for runs using different values of $P_\mathrm{floor}$ and $\delta$.  
The black line connects data from the original configuration, while the blue and red data points 
are from runs using the significantly larger floor values listed in the upper-right of the plot.  
The scaling exponents $\gamma$ for these runs are listed in Table~\ref{table:gammas}
\label{fig:scaling-difffloors}}
\end{figure}

The effect on the scaling behavior due to the fluid's floor (Section~\ref{sec:floor}) will be 
estimated first.  
Since the floor is employed merely to alleviate our numerical scheme's inability to treat the fluid  dynamics 
at the relative scale of numerical round-off and represents nothing physical, it is crucial to 
verify that any results obtained with such methods are independent of the magnitude of the floor.
To test this, we replicated the ``original'' results shown in 
Figures~\ref{fig:super-scaling-alldata}-~\ref{fig:sub-scaling-alldata} using different values 
of the floor while keeping all other parameters fixed.  Both $P_\mathrm{floor}$ and $\delta$ 
were increased by the same factor to keep their relative magnitudes the same.  
The scaling behavior obtained  using these different floor values is illustrated 
in Figure~\ref{fig:scaling-difffloors}.  The blue and red lines correspond to floor values that 
are factors of $10^2$ and $10^4$, respectively, larger than the original configuration, which itself
used $\delta=2.5\times 10^{-19}$ and $P_\mathrm{floor}=10^3\delta$.  The minimal influence of the floor 
on solutions in the critical regime is clearly seen by the fact that all the points follow 
nearly the same best-fit line.  In fact, Table~\ref{table:gammas} indicates that all estimated values of 
$\gamma$ agree to within $\simeq0.5\%$ and that all estimates of $p^\star$ 
coincide to within $4\times10^{-4}\%$.  The deviations of the calculated sets 
$\{\ln\left(T_{\max}\right),\ln\left(p^\star-p\right)\}$ from their respective best-fit lines 
for the different floor values even follow the same functional form, suggesting that the 
observed ``periodic'' deviations from linearity are not due to the floor.  
The fact that the scaling behavior is hardly affected by differences in the floor at these 
magnitudes is not too surprising since the component of the fluid that undergoes self-similar 
collapse is never rarefied enough to trigger the floor's use.  For instance, at a time when the central part of 
the star begins to resemble an ultra-relativistic critical solution, the maximum 
values of $\{D,\Pi,\Phi\}$ are, respectively, $\{\sim10^2, \sim10^3, \sim10^3 \}$---far above 
the typical floors used.  Only for $r \gtrsim R_\star$ will the floor be activated, and dynamics 
in this region cannot affect the interior solution once the self-similar collapse 
initiates due to the characteristic structure of near-critical solutions as described in 
Figure~\ref{table:ultrarel-char-speeds}. 
From this figure we see that all the waves of the fluid are traveling outward once it escapes 
from the self-similarity region.  Therefore, we see that the presence of an artificial definition 
of the fluid's vacuum state has a negligible effect on the observed scaling behavior.  

\begin{figure}[htb]
\centerline{\includegraphics*[bb=0.3in 2.3in 8.1in 9.8in, scale=0.5]{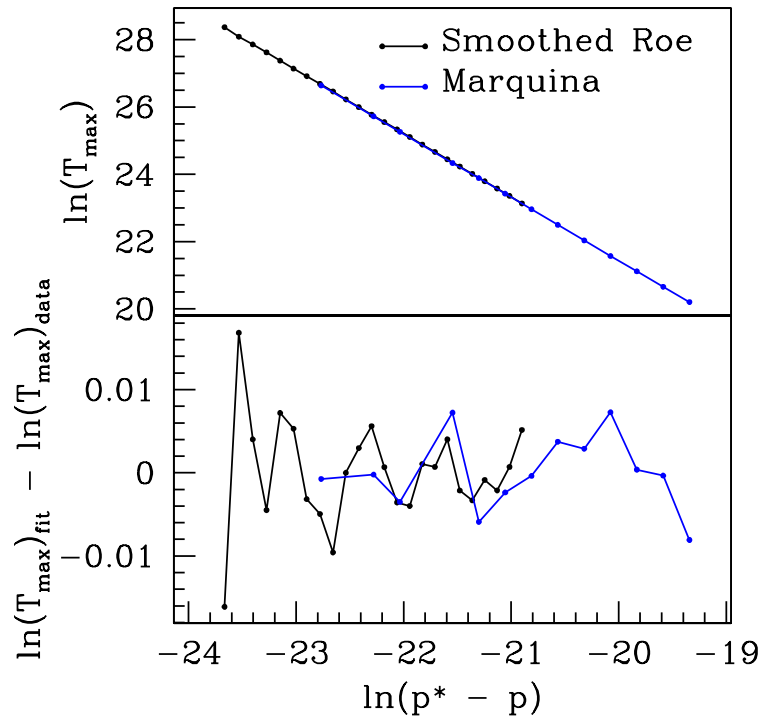}}
\caption[A comparison of the scaling behavior in $T_{\max}$ near the critical 
solution obtained with two different Riemann solvers. ]{A comparison of the scaling behavior 
in $T_{\max}$ near the critical 
solution obtained with two different Riemann solvers.  The ``Smoothed Roe'' line corresponds
to runs made with the approximate Roe solver with a smoothing procedure outlined in 
Table~\ref{table:smoothing-procedure}; this distribution is also called ``Original'' or ``$\mathrm{level}=0$''
in other figures.  The other line was generated using the Marquina method, with all other parameters fixed.
The scaling exponents, $\gamma$, for these runs are listed in Table~\ref{table:gammas}
\label{fig:scaling-diffsolvers}}
\end{figure}

Now we discuss the effect of the Riemann solver used on the scaling behavior.
As mentioned in Section~\ref{sec:instability}, an instability, which is apparently numerical in origin,
forms at the sonic point of those solutions that have been tuned near the threshold of black hole 
formation.  We found that the Marquina Riemann solver performs better than the approximate
Roe solver \emph{without} smoothing, so we wish to find out if it leads to the same $\gamma$ obtained
using the Roe solver with smoothing enabled.  From Figure~\ref{fig:scaling-diffsolvers}, we see that the scaling 
behavior of $T_{\max}$ from the two methods is remarkably close.  Even though the Roe method 
with smoothing allows us to determine $\ln\left(T_{\max}\right)$ for smaller 
values of $\ln\left(p^\star-p\right)$, the deviations from the best-fit  of the 
two data sets are of the same order of magnitude for common values of $\ln\left(p^\star-p\right)$.   
From Table~\ref{table:gammas}, we see that the respective values of $\gamma$ agree to within $0.3\%$ 
and that values of $p^\star$ agree to within $10^{-3}\%$.  These differences are quite small---comparable 
to those found as a result of varying the floor.  Hence, we may conclude that the choice in Riemann solvers has 
little, if any, effect on the computed scaling behavior, indicating that the smoothed approximate Roe solver 
is adequate for our purposes.  It remains to be seen if, in fact, the instability is affected when using 
other Riemann solver, to see if the instability is not just an artifact of these two solvers.  

\centerline{\includegraphics*[bb=0.3in 2.3in 8.2in 9.8in, scale=0.6]{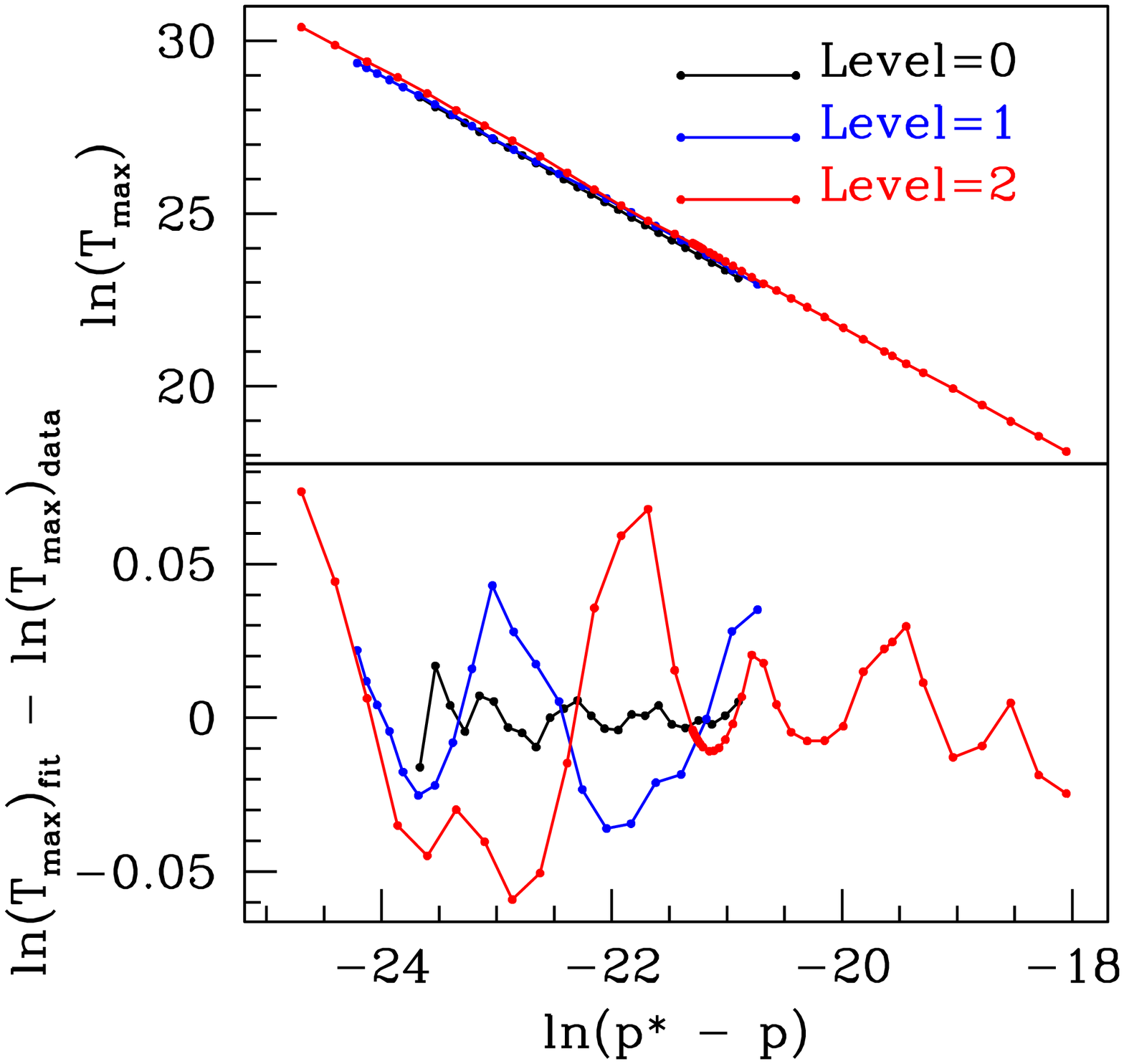}}
\begin{figure}[htb]
\caption[The scaling behavior in $T_{\max}$ near the critical 
solution for runs using different ``levels'' of resolution.]{The scaling 
behavior in $T_{\max}$ near the critical 
solution for runs using different levels of resolution.  The runs were made with 
$\rho_c = 0.05$, $U = U_1$, and the black line was generated from runs using the original configuration.  
The $\mathrm{level}=1,2$ runs, respectively, used computational grids that were locally $2$ times 
and $4$ times as refined.  
The scaling exponents, $\gamma$, for these runs are listed in Table~\ref{table:gammas}
\label{fig:scaling-difflevels}}
\end{figure}

When using finite difference methods, it is vital to verify that the order to 
which the derivatives are approximated by difference operators is the same as the 
order of the solution error.  For example, our HRSC scheme should be $O(\Delta r^2)$
accurate in smooth region and $O(\Delta r)$ near shocks, so we should expect this scaling 
behavior of the error as $\Delta r$ is varied.  First, we wish to see if our estimate for $\gamma$
converges as the grid is refined.  Figure~\ref{fig:scaling-difflevels} shows a plot of 
$\ln\left(T_{\max}\right)$ versus $\ln\left(p^\star-p\right)$  for the original configuration (black)
along with others computed at higher resolutions (blue and red).  Please see Section~\ref{sec:refinement-procedure}
for a description on how the nonuniform mesh is refined.  We first see that the three distributions 
follow lines shifted by a constant amount of approximately the same slope, while the deviation of the best-fits
seems to increase slightly with resolution.  Also, we can see that an increase in resolution 
allowed us to follow the collapse through to dispersal for solutions closer to the critical threshold,
allowing for the scaling law to be sampled at smaller $\ln\left(p^\star-p\right)$. 
Even though the deviations from the best-fits for $l=1,2$ are
quite small compared to the typical size of  $\ln\left(T_{\max}\right)$, it is a little worrisome
that they are larger than those from the lowest resolution runs.  However, this
increase can likely be attributed to the sonic point instability and the smoothing procedure used to 
damp it.  In particular, the ``hump'' of the instability sharpens with increasing resolution spanning a 
roughly constant number of grid cells (see Section~\ref{sec:instability} for more details).  
Consequently, the instability's impact on the solution may also increase  with 
decreasing $O(\Delta r)$, since the discretized difference operators will---in turn---lead
to larger estimates for spatial derivatives.  In addition, the smoothing operation is always 
performed using nearest-neighbors, so the smoothing radius physically shrinks with resolution,
diminishing the impact of the smoothing.

\centerline{\includegraphics*[bb=0.5in 2.1in 7.5in 9in, scale=0.6]{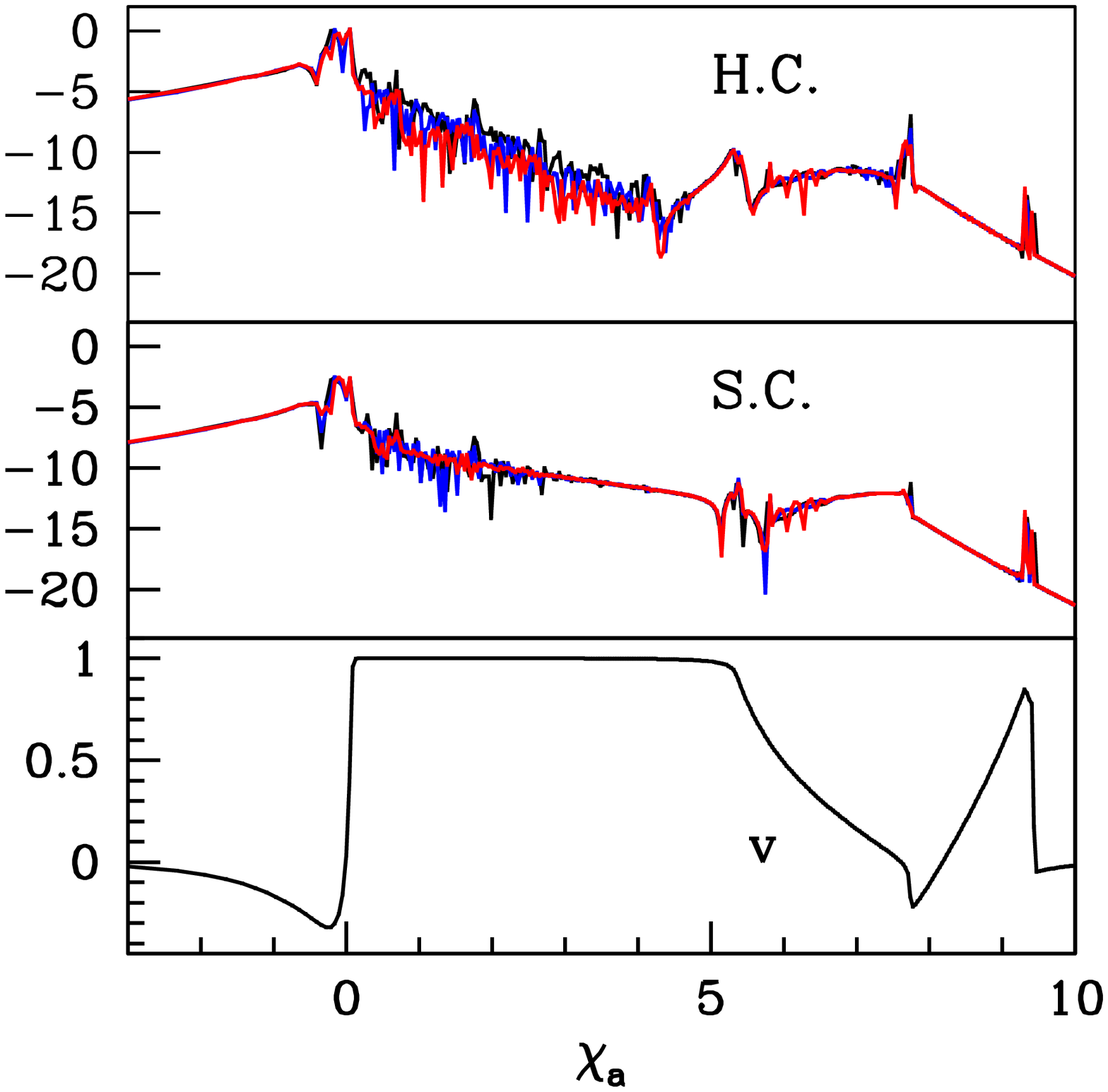}}
\begin{figure}[htb]
\caption[The logarithm of scaled, independent 
residuals of the Hamiltonian constraint (\ref{hamconstraint}) 
and slicing condition (\ref{slicingcondition}) for three levels of resolutions
calculated from solutions in the self-similar regime.]{The logarithm of scaled, independent 
residuals of the Hamiltonian constraint (\ref{hamconstraint}) 
and slicing condition (\ref{slicingcondition}) for three levels of resolutions 
calculated from solutions in the self-similar regime.  The blue (red) lines are from 
a run which used $2$ ($4$) times the local spatial and temporal resolutions of the original run, which represented 
by the black lines; 
the red residual was scaled by a factor of $16$ and the blue by $4$ in order to make 
the $O(\Delta r^2)$ convergence of the solution more apparent.
Each distribution is from a solution that has been tuned to  $\ln \left(p^\star-p\right) \simeq -19$
with respect to each resolution's value of $p^\star$. 
Every tenth grid point of each level's distribution is displayed here.  The physical velocity of the fluid for 
the $l=0$ run is shown in the bottom frame in order to facilitate comparison of features in the truncation error
to those in the solution.  
\label{fig:crit-resids}}
\end{figure}

In order to verify that the code is converging in the self-similar regime, we computed the 
independent residuals of the Hamiltonian constraint (\ref{hamconstraint}) and slicing condition
(\ref{slicingcondition}) for the three levels of resolution (Figure~\ref{fig:crit-resids}).  
The independent residuals used for the metric equations are given in Section~\ref{sec:numer-solut-metr}. 
Each residual was first scaled such that they would all coincide assuming $O(\Delta r^2)$
convergence; that is, the $l=2$ residuals were scaled by a factor of $16$ and the $l=1$ residuals
were scaled by a factor of $4$.  The overlap of residuals seen in the figure indicates $O(\Delta r^2)$ 
convergence.   Note that smoothing procedure has not been used to calculate the solutions shown here. 
We see that the scaled residuals follow similar magnitudes of smooth form in all regions 
except those which have been processed by shocks, namely $\mathcal{X}_a=[0,4.5],\simeq7.8,\simeq9.4$.
Because the self-similar solutions are converging at the expected rate, we surmise that the 
variations observed in $\gamma$ for the three resolutions does not indicate a problem with convergence, but
demonstrates the effect of truncation error on the scaling behavior.  With only three levels of 
resolution, it is hard to make definite claims as to whether $\gamma$ is or is not converging to a 
particular value.  
Even so, the standard deviation of $\gamma$ determined from the three evolutions 
is about $1\%$ of their mean, suggesting that the observed variation in $\gamma$ values is not significant.
In fact, the variation of $\gamma$ as a function of resolution is comparable to that 
found with the simpler ultra-relativistic perfect fluid \cite{neilsen-crit}.  In the convergence 
test performed, their values of $\gamma=0.9989,0.9837,0.9600$ were obtained for $l=0,1,2$, 
which suggests a relative standard deviation of $2\%$.

\centerline{\includegraphics*[bb=0.3in 2.3in 8.1in 9.8in, scale=0.6]{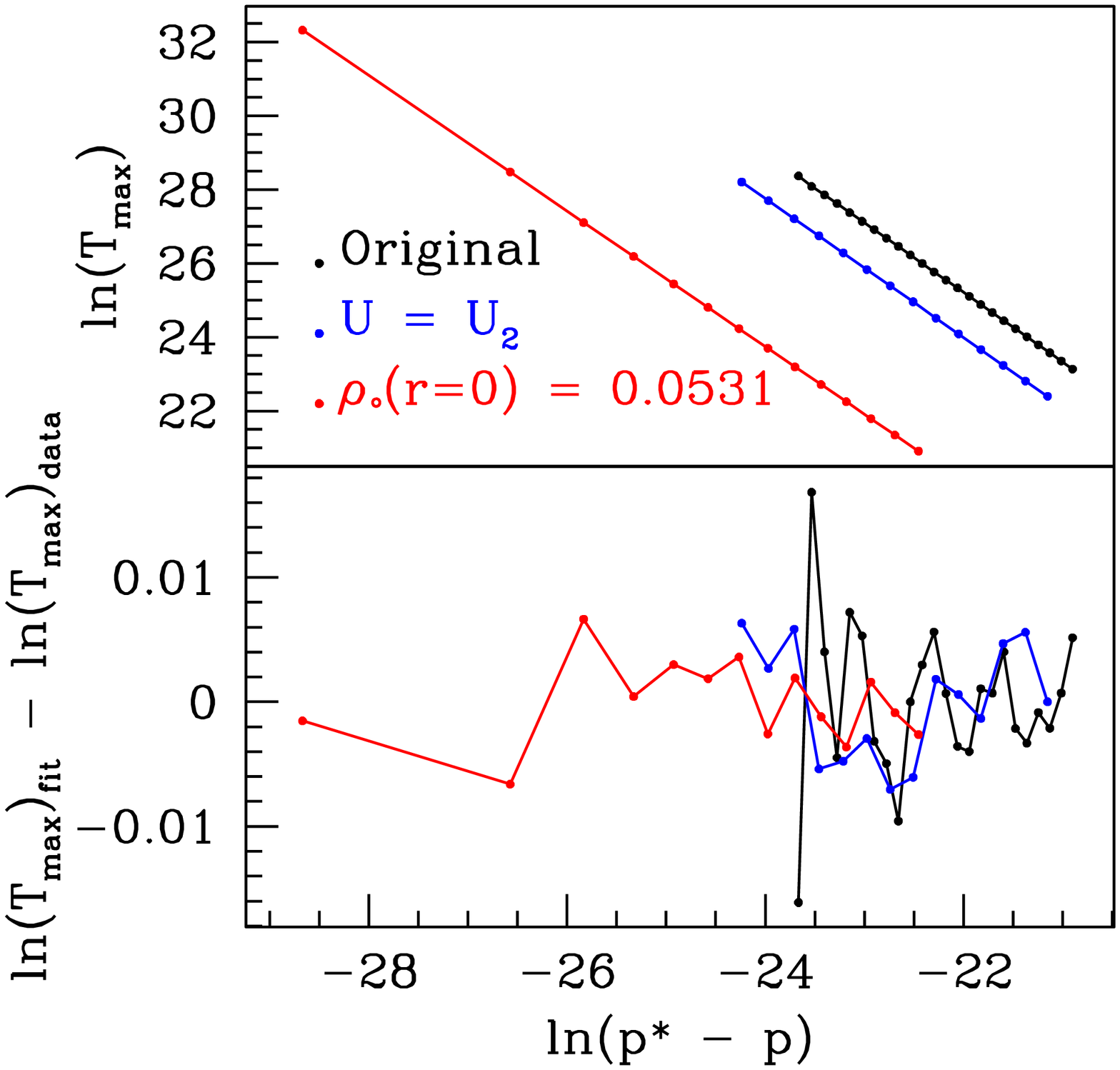}}
\begin{figure}[htb]
\caption[The scaling behavior in $T_{\max}$ near the critical 
solution for several ``families'' of initial data. ]{The scaling 
behavior in $T_{\max}$ near the critical 
solution for several ``families'' of initial data.  The ``Original'' line was made from runs with 
$\rho_c = 0.05$, $U = U_1$, and whose initial grid was made with the following parameters 
$\{N_a, N_b, N_c, \mathrm{level}\} = \{300, 500, 20, 0\}$.   The blue line shows the scaling behavior 
for runs that used a different  initial velocity profile, $U = U_2$.  And, the red line was made from  runs 
with a different TOV solution, defined by $\rho_c = 0.531$.  The 
scaling exponents, $\gamma$, for these runs are listed in Table~\ref{table:gammas}
\label{fig:scaling-difftypes}}
\end{figure}

The final comparison we discuss entails varying the physical initial conditions of the system 
to investigate the universality of the critical phenomena computed with the ideal-gas 
EOS.  The primary constituents of our model are the initial
star solution and the form of the perturbation with which we drive the star to collapse.  Hence, 
we chose to perform sets of runs to measure the scaling law using 1) a different initial star 
solution and 2) a different functional form of the initial velocity profile.  The scaling behavior 
of $\ln\left(T_{\max}\right)$ versus $\ln\left(p^\star-p\right)$ for these different
configurations are compared to the results from the original configuration in 
Figure~\ref{fig:scaling-difftypes}.  For the distribution found with a star specified by 
the central density $\rho_c=0.0531$ (red points), we kept everything else the same as that used in 
the original configuration except for the initial star solution.  The blue points show data from runs that 
used $U_2$ (\ref{v-profile-2}) for the initial profile of the coordinate velocity.   Naturally, we see that the three 
distributions are shifted from each other since each set evolved from significantly distinct 
profiles of mass-energy which obviously sets the scale for $T_{\max}$ at specific values of $p$.
However, we are interested in the slopes of the curves which determine $\gamma$ for the particular systems. 

From the values listed in Table~\ref{table:gammas}, we see that $\gamma$ varies more significantly 
with the particular star solution used,  than with the form of the velocity profile.  In fact, 
we were able to tune closer to the critical solution with the more compact star, possibly because 
it required a smaller perturbation to enter the self-similar phase so that the global maximum of 
the Lorentz factor, $W$, was smaller for the same relative point in the tuning process or same 
$\ln\left(p^\star-p\right)$.  We actually observe that the global maximum of $W$ for the 
most nearly critical solutions in both cases was $\sim10^6$ even though the $\rho_c=0.0531$ 
solution was tuned significantly closer to criticality.  Nonetheless, the different star's scaling exponent 
agrees with the original's to an accuracy of $2\%$.  

The change in the function used for the initial velocity profile hardly affected the computed value 
of $\gamma$.  The deviation in $\gamma$ found for the two profiles is surprisingly small: $0.04\%$.
Thus, we find that the initial form profile has very little to do with the observed scaling exponent. 
This suggests that other methods of perturbation would also yield close to the same value.
These three different families of initial data imply that universality of critical solutions
is maintained for perfect fluids of given $\Gamma$ that follow the ideal-gas EOS.  It would be 
interesting to see whether these results are maintained with even more realistic equations of state.

\subsection{Final Determination of $\gamma$}
\label{sec:final-determ-gamma}

Using the calculated values of $\gamma$ from the various methods, floor sizes and
grid resolutions, we are able to provide an estimate of the systematic error inherent in our 
numerical model.  Further, by assuming that the universality is strictly true, we can even use the 
variation for the different families used in this estimation.   Taking the average and calculating the standard deviation
from these values for the ideal-gas EOS given in Table~\ref{table:gammas}, we find that our 
value of the exponent is 
\beq{
\gamma \ = \ 0.94 \pm 0.01   \quad . 
\label{final-gamma}
}
This is in agreement with $\gamma$ from the black hole mass scaling fit 
Figure~\ref{fig:super-scaling-bestfit}. 

\begin{table}[htb]
\begin{center}
\begin{tabular}[htb]{|c|c|c|c|c|c|c|}
\hline
Method &$\rho_c$   &$\delta$ &$l$  &$U$  &$\gamma$ &$p^\star$  \\
\hline \hline
Roe &0.05   &$2.5\times10^{-19}$  &0 &$U_1$ &0.94    &0.4687536738   \\ 
Roe &0.05   &$2.5\times10^{-17}$  &0 &$U_1$ &0.94    &0.4687535028   \\ 
Roe &0.05   &$2.5\times10^{-15}$  &0 &$U_1$ &0.95  &0.4687516089   \\ 
Roe &0.05   &$2.5\times10^{-19}$  &1 &$U_1$ &0.92     &0.4682903094    \\ 
Roe &0.05   &$2.5\times10^{-19}$  &2 &$U_1$ &0.93    &0.4682461196    \\ 
Roe &0.05   &$2.5\times10^{-19}$  &0 &$U_2$ &0.94 &0.4299031509    \\
Roe &0.0531 &$2.5\times10^{-19}$  &0 &$U_1$ &0.92   &0.44820474298   \\ 
\hline
Marquina  &0.05   &0  &0 &$U_1$ &0.94    &0.46876822118    \\ 
\hline
Ultra-rel. &&&&&0.97 &\\
\hline
\end{tabular}
\end{center}
\caption[The scaling exponents $\gamma$ and critical parameters $p^\star$ determined from
fits to the expected scaling behavior in $T_{\max}$.]{The scaling exponents $\gamma$ 
and critical parameters $p^\star$ from fits to the expected scaling behavior in $T_{\max}$.  
The runs labelled ``Roe'' use the approximate Roe solver with smoothing, the ``Marquina'' run used 
the Marquina flux formula instead, and the ``Ultra-rel.'' scaling exponent was computed from our results
involving the collapse of Gaussian profiles of ultra-relativistic fluid.  
\label{table:gammas} }
\end{table}

In addition, we can compare our final estimate of $\gamma$ to values previously found for the ultra-relativistic
fluid.  As already mentioned, Neilsen and Choptuik \cite{neilsen-crit} measured $\gamma$ at three
different refinement levels, and quoted a value
\beq{
\gamma_{NC} \ \lesssim  \ 0.96 \quad . \label{gamma-neilsen-choptuik}
}
Instead of solving the full set of PDE's, $\gamma$ can also be found by solving the 
eigenvalue problem that results from performing $1^\mathrm{st}$-order perturbation theory about the CSS solution.  
This was done in two ways by \cite{brady_etal}: using the common shooting  method, and solving the 
linear system directly after differencing the equations to $2^\mathrm{nd}$-order.   The scaling exponents
calculated were, respectively, $\gamma=0.9386\pm0.0005$ and $\gamma=0.95\pm0.01$.  

We find our value (\ref{final-gamma}) to agree well with those found by \cite{brady_etal}, 
and agree with $\gamma_{NC}$ to within the uncertainty quoted by Neilsen and Choptuik.  The discrepancy between 
the value from the ideal-gas equations and that determined from the ultra-relativistic PDE's 
is also seen when \emph{we} solve the ultra-relativistic equations.  Our ultra-relativistic value, 
$\gamma=0.97$, agrees well with those calculated by Neilsen and Choptuik, but deviates 
by 3 standard deviations from our ideal-gas calculations.  It somewhat interesting, yet possibly 
coincidental, that our results from the ideal-gas system of equations lead to estimates of $\gamma$ 
that agree with the perturbation calculations better than those values found from the 
ultra-relativistic PDE calculations.  

Hence, some of the claims made by Novak \cite{novak} have been found to be significantly inaccurate for the 
ideal-gas EOS with $\Gamma=2$.  It remains to be seen whether the scaling behavior we have observed is also seen
with more 
realistic state equations such as the one Novak used \cite{pons-etal-2000}.  Since accurate measurements 
of $\gamma$ have only been found for equations of state with constant adiabatic indexes $\Gamma$, 
and since $\gamma$ seems to only depend on $\Gamma$ for perfect fluids, it remains to be seen 
what the scaling behavior---if any---will be like for realistic state equations that 
do not guarantee that $\Gamma$ be constant throughout the fluid.  

\section{Type~II Phenomena with Scalar Field Perturbation}
\label{sec:t2-scalar-perturb}

It is important to mention that we had been studying perturbed neutron stars before 
\cite{novak} was published.  Instead of using an initial velocity, however, a minimally-coupled, 
massless scalar field was used to perturb the star purely through their mutual gravitational interaction. 
That is, the energy of the scalar field leads to spacetime curvature to which the star responds, and 
vice versa.  In order to search for critical phenomena, we tuned the magnitude of the initial
Gaussian shell of scalar field about the threshold of black hole formation.  Type~I behavior was 
studied extensively with this multi-matter system, and is described in 
Chapter~\ref{chap:type-i-critical}. 
Surprisingly, we were unsuccessful in driving the star's matter to CSS collapse with the 
scalar field perturbation.  Those stars which did not follow Type~I behavior were sparser and 
less massive, requiring a larger excitation to collapse.  The scalar field profile needed
to collapse the star was sufficiently strong that it exhibited Type~II behavior itself instead
of merely perturbing the star.  That is, when the scalar field profile was tuned about the critical 
point, we found that the near-threshold solution was the scalar field DSS solution found in the first 
critical phenomena study \cite{choptuik-1993}.  

\centerline{\includegraphics*[bb=0.7in 2in 8.1in 9.8in, scale=0.6]{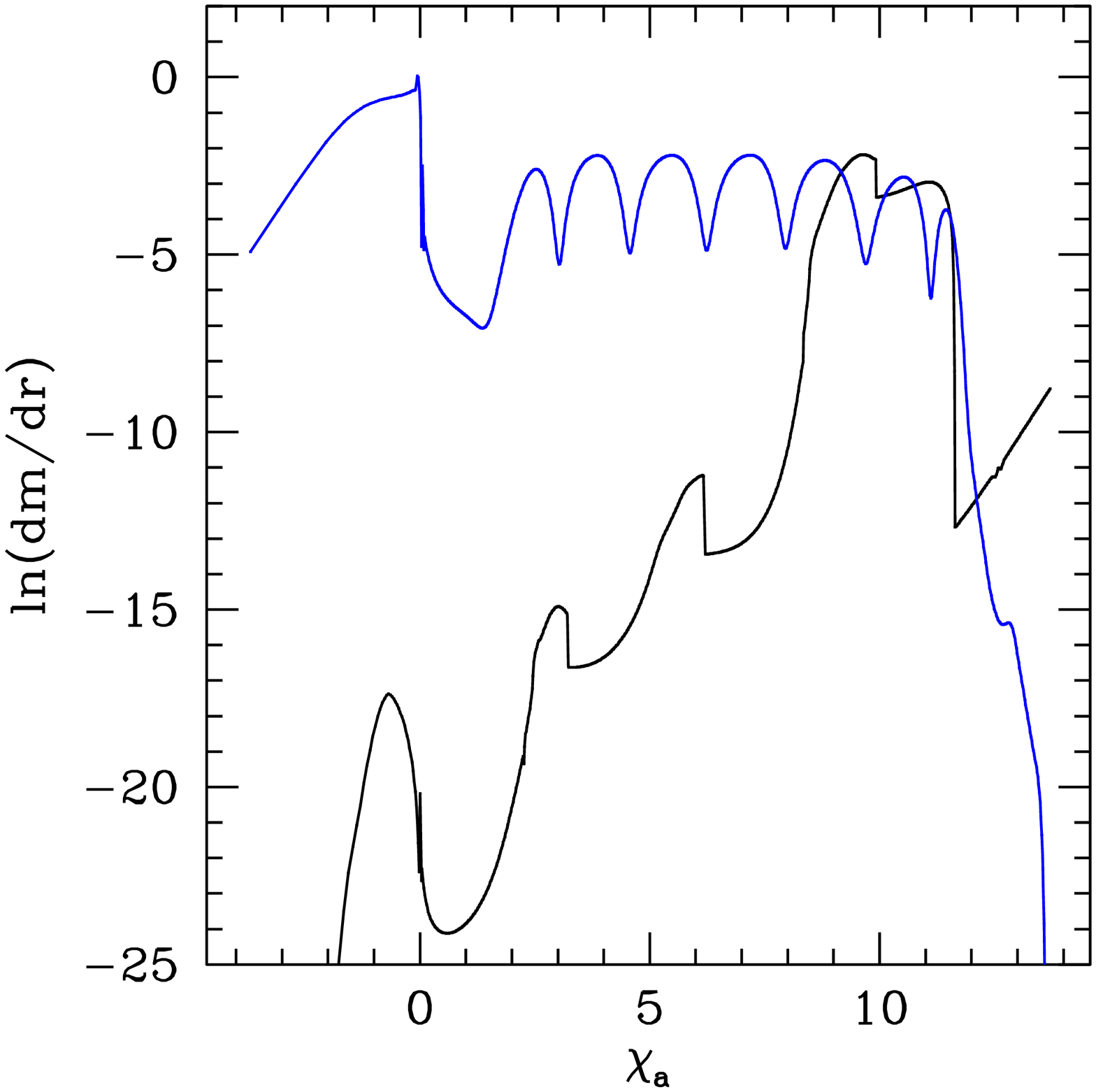}}
\begin{figure}[htb]
\caption[A snapshot of the separate contributions to the energy density from the massless scalar 
field and from the fluid for Type~II collapse involving a coupling of the two fields.]{A snapshot 
of the separate contributions to the energy density from the massless scalar 
field and from the fluid for Type~II collapse involving a coupling of the two fields.
This particular frame shows the configuration just prior to black hole formation for the most nearly critical 
solution.  The scalar field contribution is shown in blue 
while the fluid contribution is shown in black.  The two are plotted against a self-similar coordinate 
$\mathcal{X}_a$ which tracks the maximum of the metric function $a$ (\ref{X-ss2}).
The star shown here corresponds to $\rho_c=0.02$ and $\Gamma=2$.  Every fifth grid point is shown here
for each distribution. 
\label{fig:dmdr-scalar-fluid}}
\end{figure}

For example, Figure~\ref{fig:dmdr-scalar-fluid} shows the scalar field and fluid contributions to 
$dm/dr$ (\ref{dmdr}) for a Type~II collapse with the massless scalar field and star.  
The scalar field and fluid contributions are, respectively, 
\beq{
\frac{d}{dr} m_{\,_{\mathsmallest{\mathrm{scalar}}}} 
\ = \ 4 \pi r^2 \varrho_{\,_{\mathsmallest{\mathrm{scalar}}}} \quad , \quad 
\frac{d}{dr} m_{\,_{\mathsmallest{\mathrm{fluid}}}}
\ = \ 4 \pi r^2 \varrho_{\,_{\mathsmallest{\mathrm{fluid}}}} 
\label{dmdr-scalar-and-fluid}
}
where $\varrho_{\,_{\mathsmallest{\mathrm{scalar}}}}$ and $\varrho_{\,_{\mathsmallest{\mathrm{fluid}}}}$
are given in (\ref{adm-rho-scalar}) and (\ref{adm-rho-fluid}).  The periodic echoing of the scalar
field's DSS collapse can be clearly seen in the oscillations of 
$d m_{\,_{\mathsmallest{\mathrm{scalar}}}}/dr$
for $\mathcal{X}_a>0$.  The presence of the oscillations in this late-time snapshot
illustrates how the  non-self-similar part of the fluid ``freezes out'', or evolves at an exponentially 
slower rate than the interior part of the solution; in this way, the spatial profile of the distributions
serve as a sort of historical record of the collapse.  Also, it appears that the fluid reacts to the 
echoing of the spacetime, indicated by the periodic discontinuities in
$d m_{\,_{\mathsmallest{\mathrm{fluid}}}}/dr$ that occur at $\mathcal{X}_a=3,6,10$.  
Especially interesting, though, is that the echoing of the scalar field contribution 
occurs twice as frequently as the 
fluid's.  From the evolution of the fluid velocity $v(r,t)$ and the discontinuities in this figure, 
we see that shocks seem to develop at every other echo.  In addition, the disparate magnitudes of 
$d m_{\,_{\mathsmallest{\mathrm{scalar}}}}/dr$ and 
$d m_{\,_{\mathsmallest{\mathrm{fluid}}}}/dr$ 
demonstrate how irrelevant the fluid is in this region of spacetime.  
We may conclude, then, that the fluid was a passive element 
as the scalar field---and the spacetime it dominated---collapsed in a discretely self-similar fashion.

The next chapter contains further discussion regarding the dynamics of a massless
scalar field coupled to neutron star models through their gravitational interaction. 


\chapter{Type~I Critical Phenomena}
\label{chap:type-i-critical}

Compared to Type~II phenomena in general relativity, Type~I behavior is far simpler to study 
in many respects and involves systems that are not quite as exotic as the Type~II variety. 
Instead of the critical solutions
having self-similar symmetry, Type~I critical solutions have always been found to exhibit 
continuous (static) or discrete (oscillatory) symmetry with respect to time.  In this chapter, 
the first thorough study of Type~I behavior of perfect fluid solutions is presented.  Other 
Type~I studies have involved a great variety of other field theories.  For example, the first model in 
which Type~I behavior 
was explored was the self-gravitating SU(2) gauge field, or Einstein-Yang-Mills (EYM) system, 
\cite{choptuik-chmaj-bizon}.   In this work, Choptuik et al. 
found that the threshold solution of certain EYM systems is the static $n=1$---where $n$ parameterizes
the number of zero-crossings of the Yang-Mills field---Bartnik-McKinnon solution \cite{bartnik-mckinnon} 
which has one unstable mode.  Next, Brady et al. \cite{brady-chambers-goncalves} 
were the first to discover Type~I collapse involving an oscillating critical solution in their 
study of a real massive scalar field coupled to gravity.  The critical solutions 
they found belong to the class of oscillating solitonic solutions constructed by Seidel and 
Suen \cite{seidel-suen-1991}.  In these studies, the two ``fixed points'' in phase space 
involve either a black hole  or  flat  space (vacuum).  However, in the 
Einstein-Skyrmion model, whose Type~I behavior was first examined by Bizon and 
Chmaj \cite{bizon-chmaj-1998}, the non-black hole fixed point is one containing a stable, static 
Skyrmion solution.  After approximating the unstable static solution for some time, the 
near-critical Skyrmion field would either form a  black hole or expand to a stable, equilibrium solution. 

Possibly the most similar study to ours was done by Hawley and Choptuik \cite{hawley-choptuik-2000}. 
They investigated perturbed stable boson star solutions, which are massive complex scalar field solutions whose 
only time-dependence is a phase that varies linearly with time.  In order to perturb the stable 
boson stars, they collapsed a spherical pulse of massless scalar field onto it from a distance far outside
the star's radius, to ensure that the two distributions were initially distinct. 
As the pulse collapses through the origin, the two energy distributions interact 
solely through the gravitational field.  The 
increase in curvature within the star from the massless field was 
observed to be enough to drive the boson stars inward, resulting in either black hole formation
or a sequence of large 
oscillations. Note that in the original paper by Hawley and Choptuik \cite{hawley-choptuik-2000}, they did 
not find that the subcritical fixed point involved a periodic spacetime, but assumed that the 
stars would disperse to spatial infinity.  Upon evolving subcritical evolutions longer, Lai \cite{lai}
found that they were, in fact, bound and oscillatory systems.  Later, Hawley \cite{hawley-pc} confirmed
these results. 
During the non-trivial gravitational interaction of the massless scalar field and the boson star, 
a transfer of mass-energy from the massless scalar field to the complex scalar field was observed, increasing
the gravitating mass of the boson star.  Type~I critical solutions were found by varying 
the magnitude of the initial pulse of massless scalar field, and it was shown that the 
critical solutions were unstable boson star solutions with masses somewhat larger than their stable progenitors.
Boson stars are similar to 
their hydrostatic analogues in that their stable solutions are on the branch below the maximum of 
$M_\star(\phi(0))$ graph, while the unstable solutions are on the other 
side (see Section~\ref{sec:init-star-solut} for a discussion regarding the hydrostatic star solutions).
Finally, as with any Type~I phenomena investigation,
Hawley and Choptuik found that the lifetime of a solution tuned close to the threshold scales as a power-law of the 
deviation of the tuning parameter, $p$, from the critical value, $p^\star$:
\beq{
T_0(p) \propto - \sigma \ln \left| p - p^\star \right|  \label{type-i-scaling}
}
They verified that the scaling exponent, $\sigma$, for a given 
critical solution is the inverse of the real part of the Lyapunov exponent, $\omega_{Ly}$, 
for the corresponding unstable boson star solution.  They calculated $\omega_{Ly}$ for several 
cases using the ODE's resulting from linear perturbation theory about the unstable solutions.  
Since boson stars model many of the characteristics of TOV solutions, it was 
conjectured that the critical behavior they discovered would carry over to their fluid analogues.  
We will see shortly that in many respects it does.  

Before proceeding to  the presentation of results, we would like to first mention that 
the threshold between hydrostatic solutions and black hole formation has been studied in a variety
of ways in the past.  For instance, the first time-dependent numerical simulations 
of a fully-coupled general relativistic system involved the collapse of adiabatic perfect fluid spheres 
of constant density and were performed by May and White in 1966 
\cite{may-white-paper} (see \cite{may-white-book} for a more thorough 
explanation of the methods used by May and White and see the work by Misner and Sharp \cite{misner-sharp} for 
the origin of the formulation they used).
About five years later, Wilson \cite{wilson1} studied the core collapse supernova problem using 
an approximate method for the neutrino transport in full, spherically-symmetric general relativity. 
Van Riper in 1979 \cite{vanriper2} studied the purely hydrodynamic collapse of iron core models of 
different masses in order to determine the maximum mass for the resultant neutron star.  
Interestingly enough, he tuned this critical mass to within $0.005\%$, but the ``critical'' 
or threshold solution he found never reached densities above the Oppenheimer-Volkoff limit, 
above which the TOV solutions become unstable.  

Recently, Siebel et al. 
\cite{siebel-font-pap-2001} sought to measure the maximum neutron star mass allowed by  the 
presence of a perturbing pulse of minimally-coupled, massless scalar field.  A general relativistic hydrodynamic 
code using a characteristic formulation was used to investigate the spherically symmetric system.
Instead of varying the massless scalar field, however, 
they studied five star solutions of assorted central densities that straddled 
the threshold of black hole formation.  
They found that the perturbation either led to a black hole or to oscillations of the star about its initial 
configuration.   Further, in order to test their new 3-dimensional general relativistic fluid 
code, Font et al. \cite{font-etal2} dynamically calculated the fundamental and harmonic mode frequencies
of spherical TOV solutions that were perturbed only by their initial truncation error. 
In this fashion, they were able to observe the transition of a TOV solution on the \emph{unstable} branch to the 
stable branch by evolving the unstable solution with only a truncation error level perturbation.  The expansion 
of the unstable star initially overshoots the stable solution, resulting in a series of oscillations.

\section{Model Description}
\label{sec:model-description}
As others have done \cite{hawley-choptuik-2000,siebel-font-pap-2001}, we chose to use a minimally-coupled,
massless Klein-Gordon (EMKG) field to perturb our star solutions dynamically.  The EMKG field is 
advantageous 
for several reasons. First, the fact that the two matter models 
are both minimally-coupled to gravity with no explicit interaction between the two ensures that any 
resulting dynamics from the perturbation is entirely due to their gravitational interaction.
Second, the EOM of the EMKG field are straightforward to solve numerically and provide little 
overhead to the hydrodynamic simulation.  Third, since gravitational waves cannot exist in spherical symmetry 
and the EMKG field only couples to the fluid through gravity, it can serve as a plausible first approximation 
to gravitational radiation acting on these spherical stars.  

We will continue to approximate neutron stars
as polytropic solutions of the TOV equations with $\Gamma=2$; and the factor in the polytropic
EOS (\ref{polytropic-eos}) will still be set to $K=1$ to keep the system unit-less.   Since all stellar 
radii $R_\star$ satisfy $R_\star<1.3$ for such solutions, we will---by default---position the initial scalar field 
pulse at $r=5$.  This has been found to be well outside any star's extent and so ensures that 
the two models are not initially interacting and thus represent two independent distributions of energy at $t=0$.

\section{The Critical Solutions} 
\label{sec:critical-solutions}

\begin{figure}[htb]
\centerline{\includegraphics*[bb=0.2in 2in 8.6in 9.9in, scale=0.34]{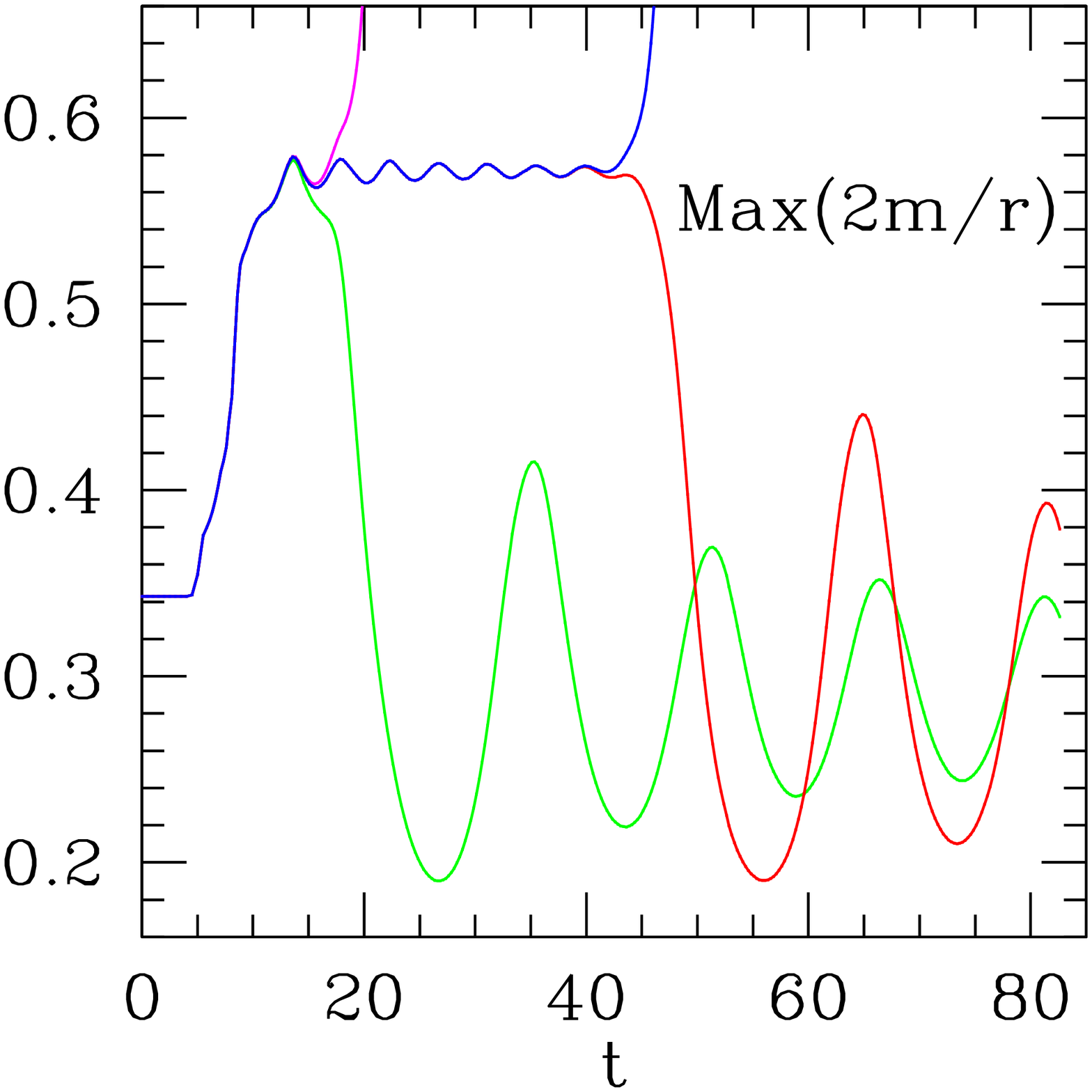}
\includegraphics*[bb=0.2in 2in 8.1in 9.9in, scale=0.34]{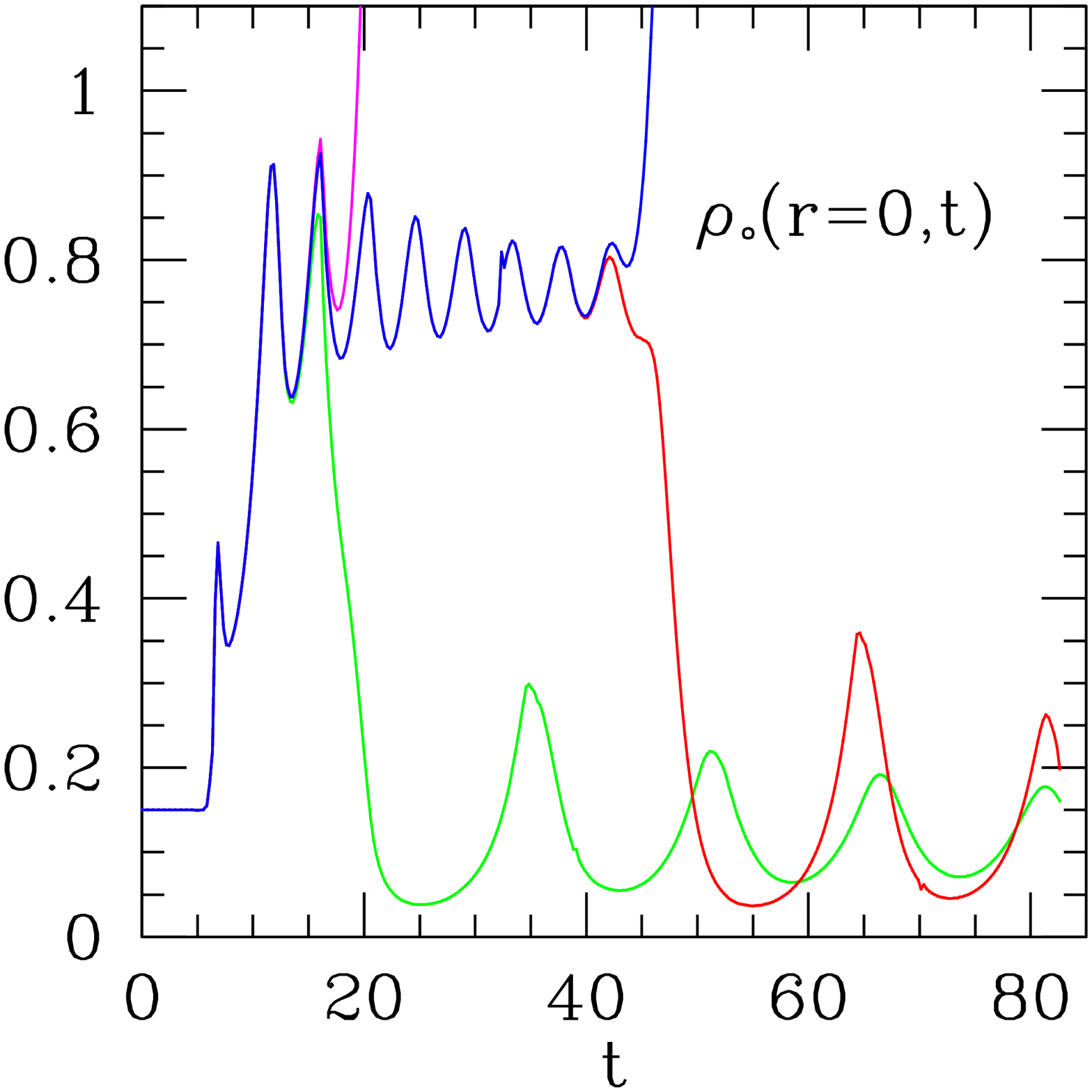}}
\caption[Evolutions of ${\max}(2m/r)$ and $\rho_\circ(r\!\!=\!\!0,t)$ from $4$ solutions near 
the black hole threshold for a star parameterized by $\Gamma=2$, $\rho_c=0.15$.]{Evolutions of ${\max}(2m/r)$ 
and $\rho_\circ(r\!\!=\!\!0,t)$ from $4$ solutions near 
the threshold of a star parameterized by $\Gamma=2$, $\rho_c=0.15$. The purple (green) line is
from a solution far from the threshold on the supercritical (subcritical) side.  The 
blue (red) line pertains to a supercritical (subcritical) solution whose parameter has been tuned 
to within machine precision of the critical value.
\label{fig:tuning-maxtmr-rhoc}}
\end{figure}

The evolution of the star towards the critical solution and the critical solutions themselves will 
be described in this section.  As the scalar field pulse travels into the star, the star undergoes a
compression phase whereby the exterior implodes at a faster rate than the interior.  This is 
reminiscent of the velocity-induced shock-bounce scenarios described in 
Chapter~\ref{chap:veloc-induc-neutr}.  If the perturbation is weak, then the star will undergo 
oscillations with its fundamental frequency after the scalar field disperses 
through the origin and finally escapes to null infinity (higher harmonics are also excited). 
On the other hand, when the initial scalar 
shell of  sufficiently large amplitude, the star can be driven to prompt collapse, trapping some of the scalar field 
along with the entire star in a black hole.  Somewhere in between, the scalar field can compactify 
the star to a nearly static state that resembles an unstable TOV solution of slightly increased mass.  
The length of time the perturbed star emulates the unstable solution, which we will call the 
\emph{lifetime}, increases as the initial pulse's amplitude is adjusted closer to the critical value, $p^\star$.  
It is expected from this scaling behavior that a perfectly constructed scalar 
field pulse with $p=p^\star$ would perturb the star in such a way that it would resemble the unstable solution 
forever.  This putative infinitely long-lived state is referred to as the critical solution of the progenitor star. 

\begin{figure}[htb]
\centerline{\includegraphics*[bb=0.2in 2in 8.6in 9.9in, scale=0.34]{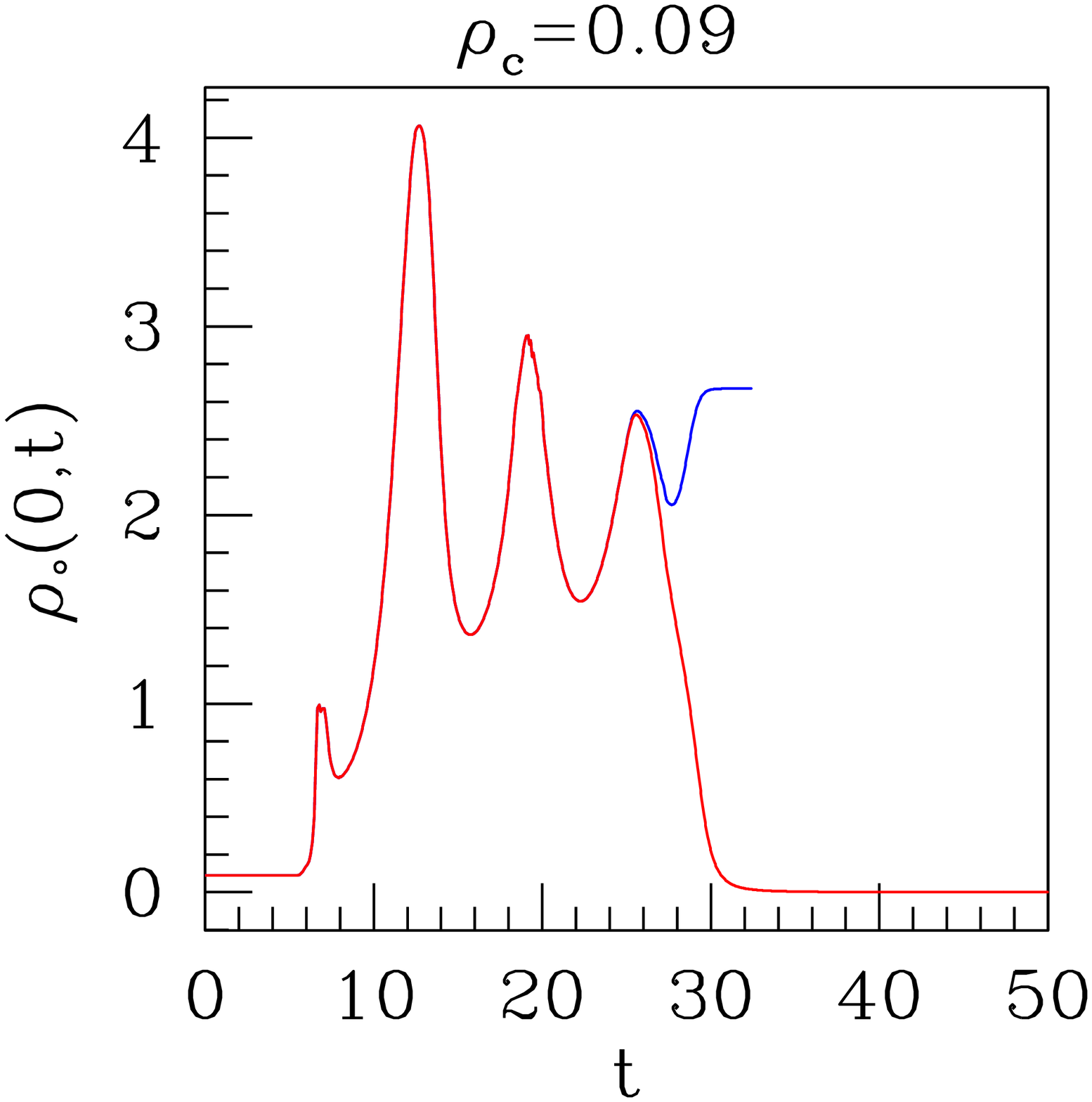}
\includegraphics*[bb=0.2in 2in 8.1in 9.9in, scale=0.34]{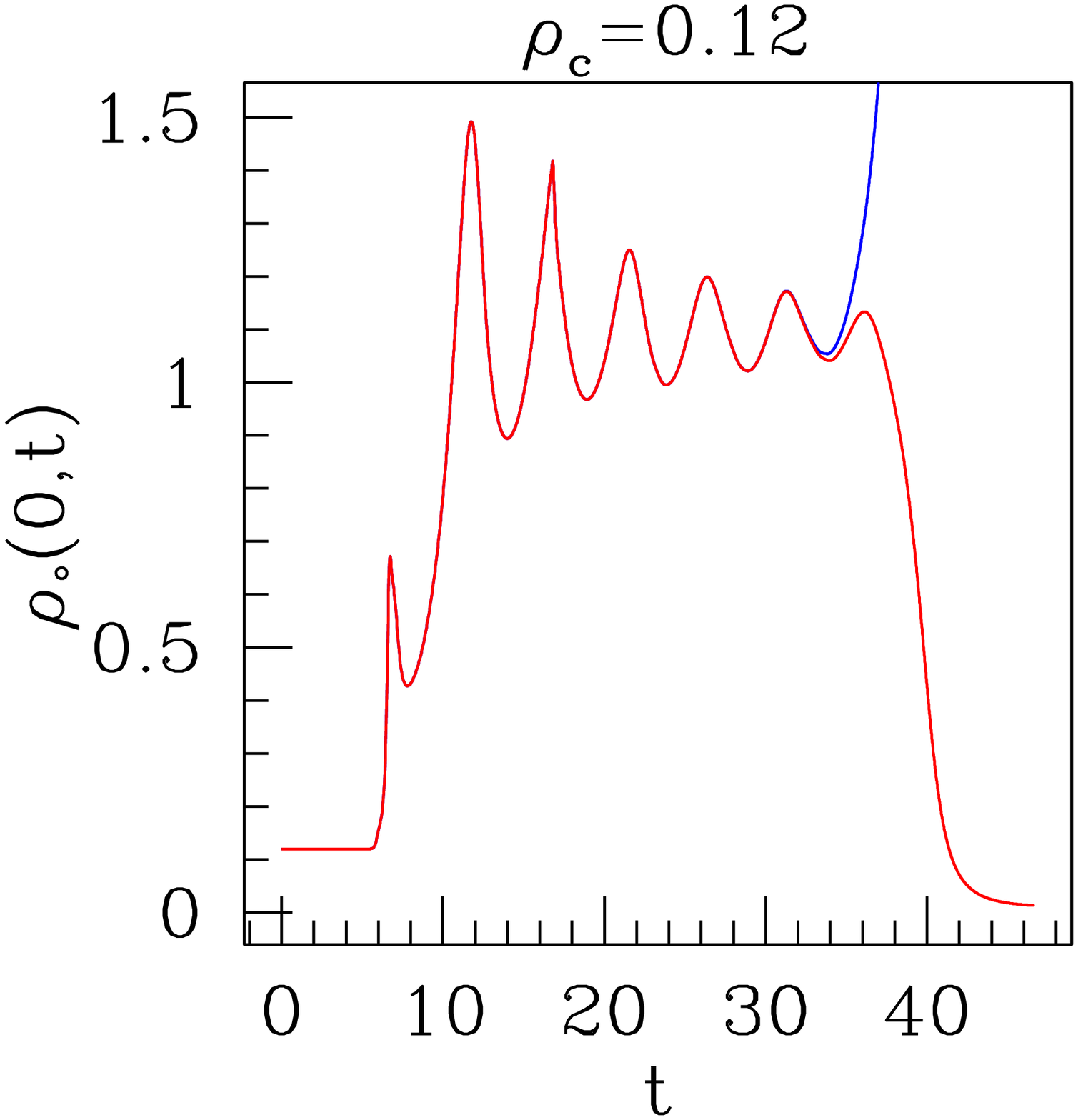}}
\caption[Sample evolutions of the central rest-mass density for supercritical (blue) and subcritical 
(red) solutions from progenitor stars  with $\rho_c=0.09$ and $\rho_c=0.12$.]{Sample evolutions of the 
central rest-mass density for supercritical (blue) and subcritical 
(red) solutions from progenitor stars  with $\rho_c=0.09$ and $\rho_c=0.12$. The solutions have been 
tuned to within machine precision of criticality  in each case.  Note that for $\rho_c=0.09$, $\rho_\circ(0,t)$
for the supercritical calculation tends to a constant value since the 
collapse of the lapse has effectively frozen the star's evolution near the origin.  Also, even though 
it may seem from the figures that the subcritical solutions for both stars evacuate the origin, both 
inflate to larger, sparser star solutions along the lines of the shock-bounce-oscillate (SBO) 
scenario described in Section~\ref{sec:param-space-surv}. 
\label{fig:rho-samples-0.09-0.12}}
\end{figure}

\begin{figure}[htb]
\centerline{\includegraphics*[bb=0.2in 2in 8.6in 9.9in, scale=0.34]{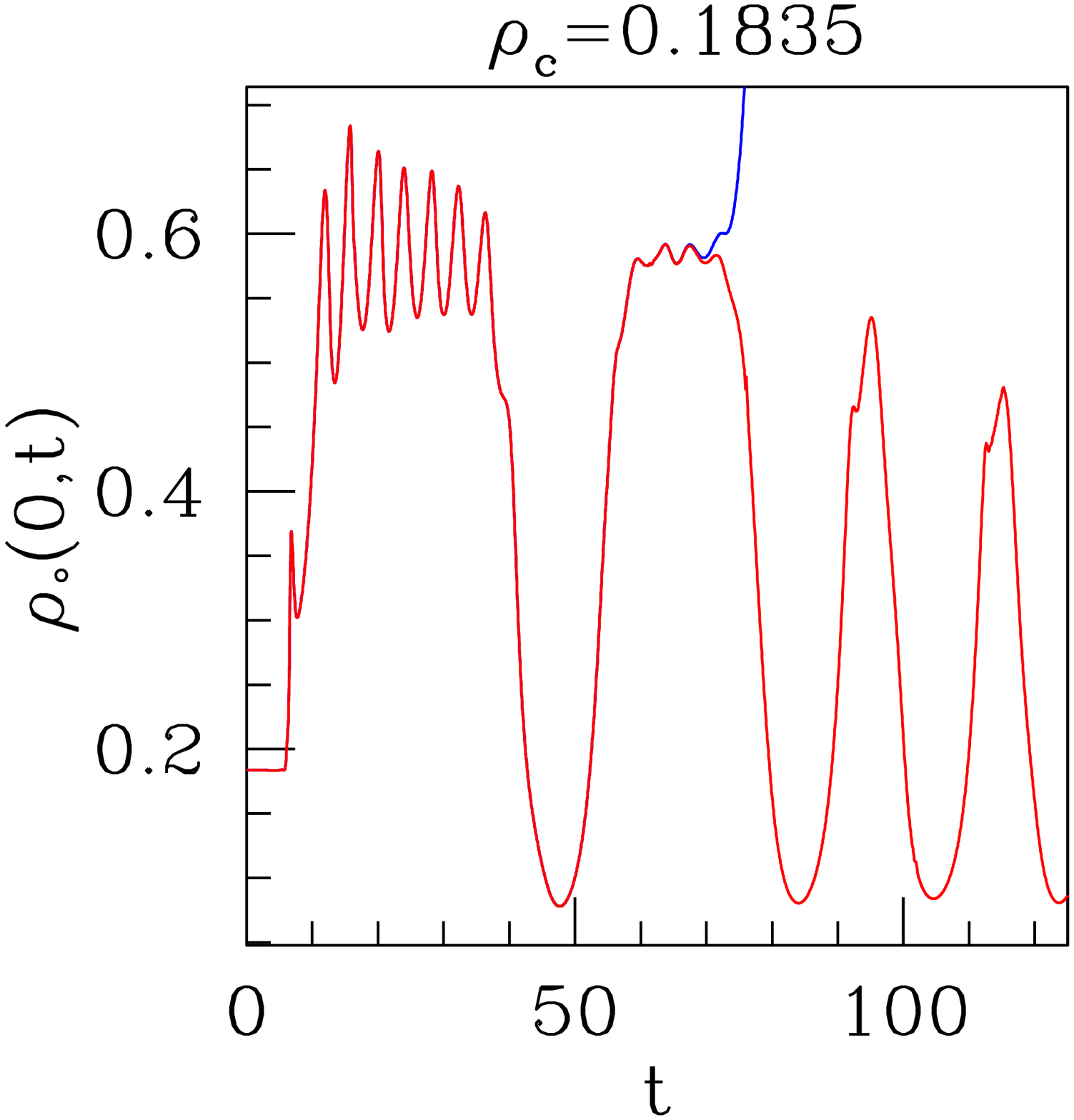}
\includegraphics*[bb=0.2in 2in 8.1in 10in, scale=0.34]{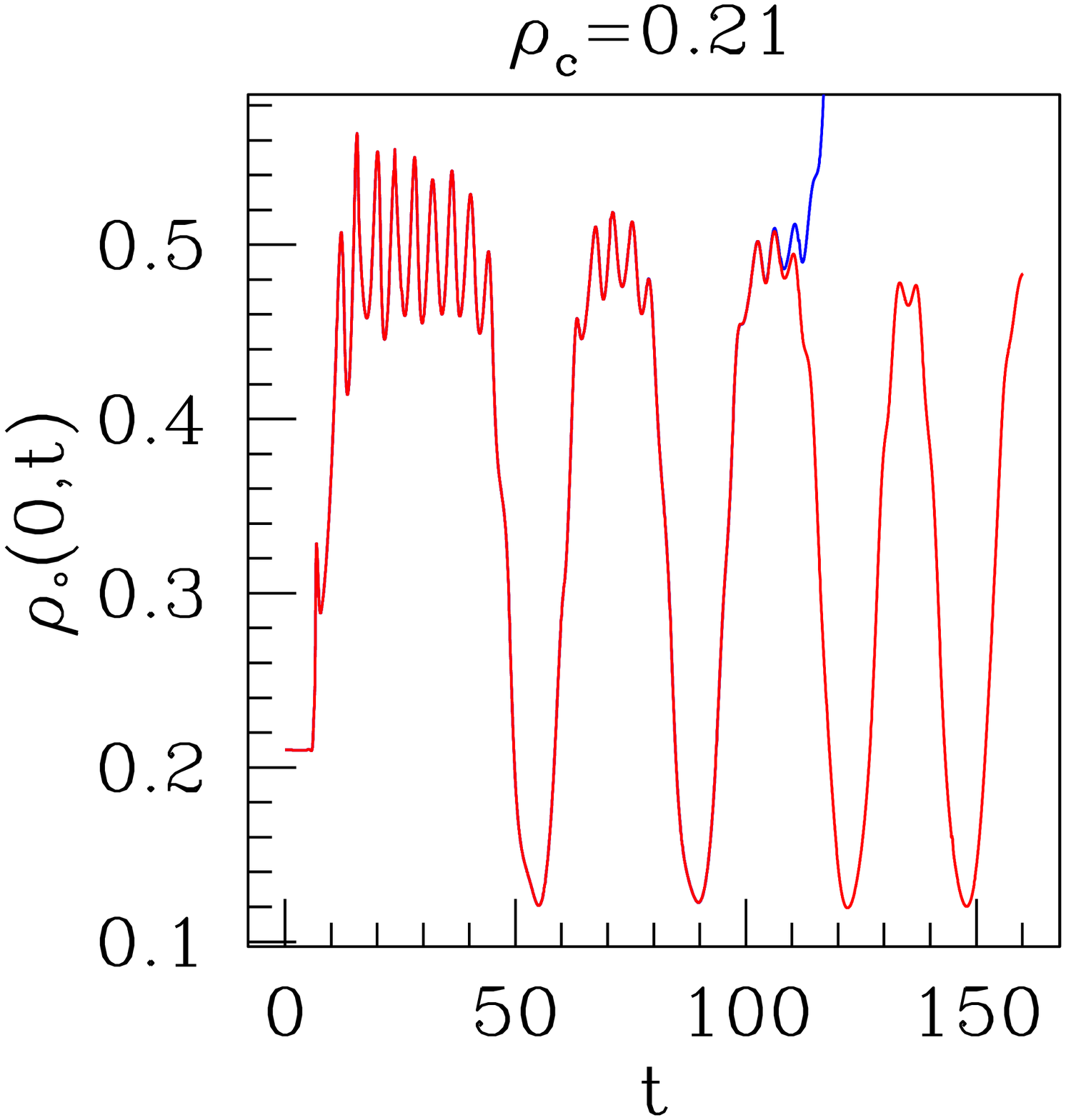}}
\caption[Further examples of the central density variation over time for the most nearly 
critical solutions from two stars, $\rho_c=0.1835$ and $\rho_c=0.21$.]{
Further examples of the central density variation over time for the most nearly 
critical solutions from two stars, $\rho_c=0.1835$ and $\rho_c=0.21$; again, the subcritical solutions are 
plotted in red, while the supercritical solutions are plotted in blue.  The $\rho_c=0.1835$ star is the star 
with the smallest initial central density whose nearest-to-critical solution exhibits a momentary departure from the 
unstable, equilibrium solution; this is indicated by the break between the two ``plateaus'' in the graph.  
This behavior is seen for most stars above $\rho_c=0.1835$, as exemplified by the other star's
solutions.  The $\rho_c=0.21$ star is the sparsest initial solution found whose most nearly critical 
solutions have two departures or three plateaus. 
\label{fig:rho-samples-0.1835-0.21}}
\end{figure}

Examples of solutions near and far from the critical solution are illustrated in 
Figure~\ref{fig:tuning-maxtmr-rhoc} for a star with $\rho_c=0.14$.  Here we show the 
evolution of the spatial maximum of $2m/r$, 
${\max}(2m/r)$, and the central density of the star for a series of solutions.  The 
quantity $2m/r$ is, effectively, a measure of the degree of compactification;
the global maximum that $2m/r$ can attain for the static TOV solutions studied herein is approximately $0.61$, and 
$2m/r\rightarrow 1$ when a black hole would form.  The purple lines clearly show 
that the supercritical systems far from the threshold quickly collapse to black holes, represented
here by the divergence of the central density and compactification factor. 
On the opposite side of the spectrum, we see from the periodic nature of the green ${\max}(2m/r)$ and $\rho(0,t)$
distributions that subcritical solutions undergo a series of 
oscillations.  The blue and red lines, respectively, illustrate the long lifetimes of 
marginally supercritical and subcritical solutions.  The plateau shown 
in the plots represents the period of time during which the marginally subcritical and supercritical 
solutions resemble the critical solution.  We will see shortly that this critical 
solution is actually a star-like configuration oscillating about an unstable TOV solution.  

\begin{figure}[htb]
\centerline{\includegraphics*[bb=0.2in 2in 8.6in 9.9in, scale=0.34]{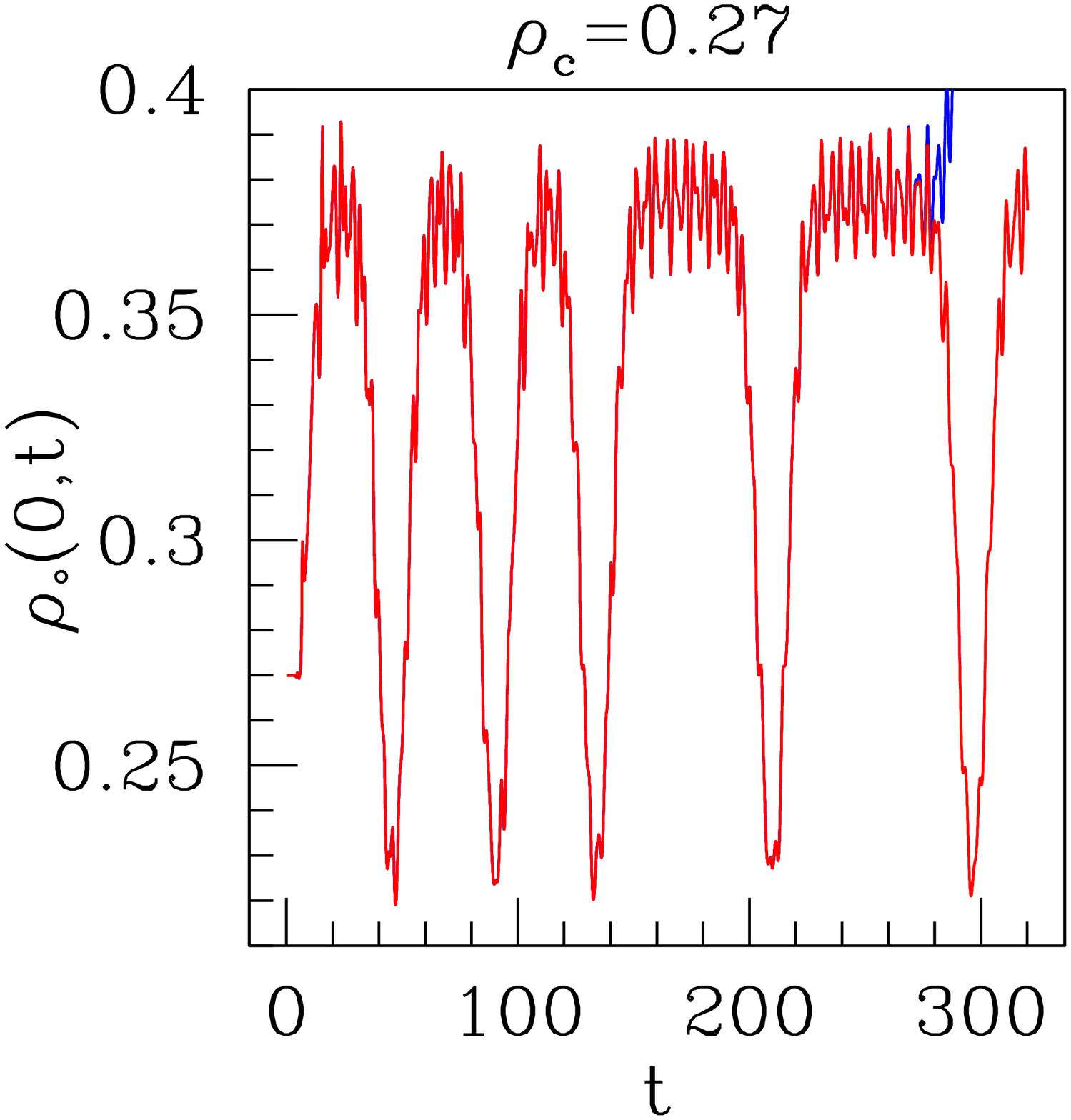}
\includegraphics*[bb=0.2in 2in 8.1in 10in, scale=0.34]{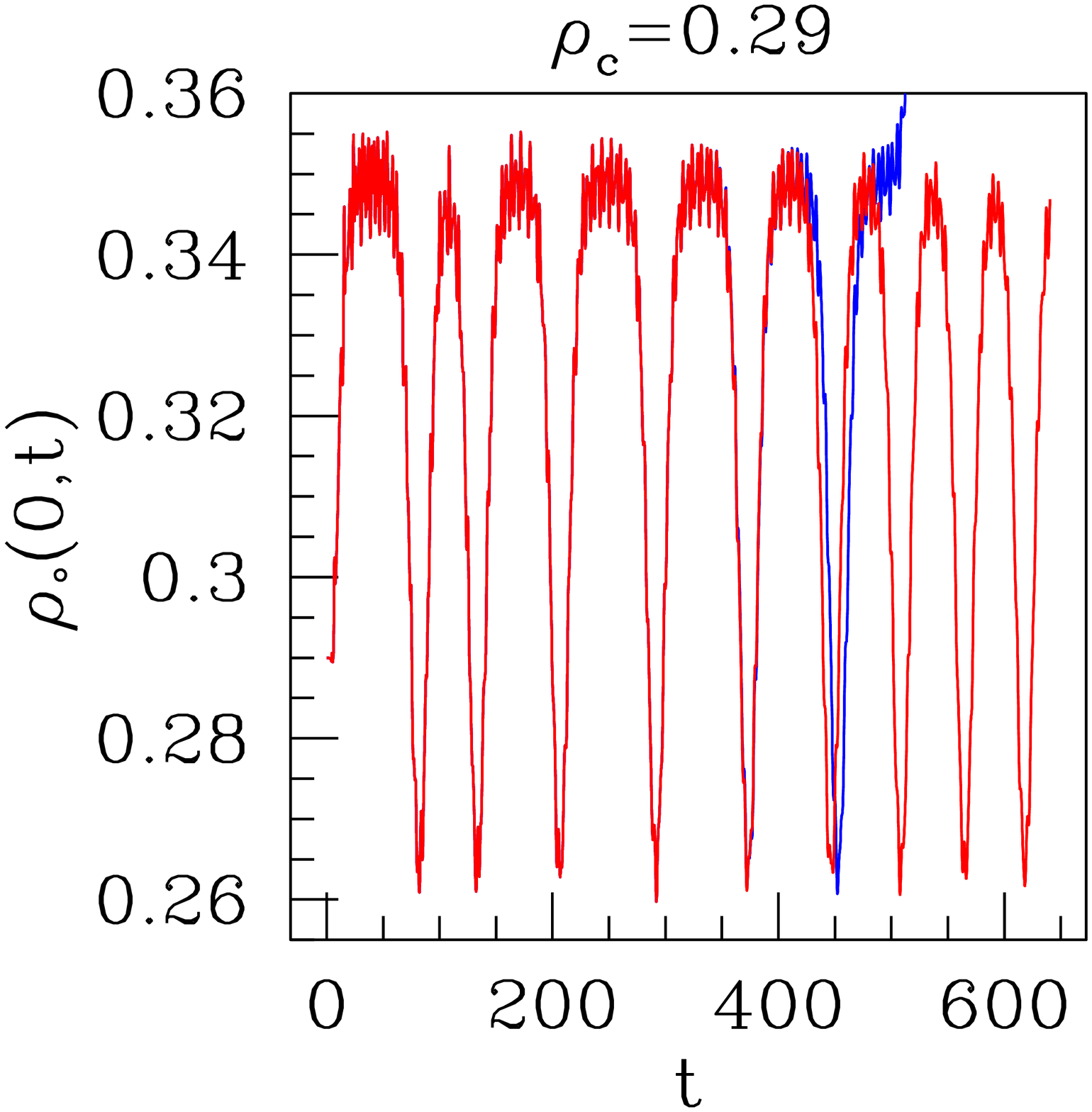}}
\caption[The central density evolutions from solutions tuned within machine precision 
for progenitor stars $\rho_c=0.27$ and $\rho_c=0.29$.  These stars are close to the maximum mass 
equilibrium solution, $\rho_c=0.318$. ]{The central density evolutions from solutions tuned within machine precision 
for progenitor stars $\rho_c=0.27$ and $\rho_c=0.29$.  These stars are close to the maximum mass 
equilibrium solution, $\rho_c=0.318$. 
The supercritical solutions are plotted in blue, and the subcritical solutions are shown in red. 
The nearest-to-critical solutions from the $\rho_c=0.27$ star shows four departures, while those
from the $\rho_c=0.29$ star shows five.  The supercritical solution from the $\rho_c=0.29$ initial
star undergoes a curious sequence not seen in many cases---after it deviates from the subcritical 
solution---instead of collapsing to a black hole from the unstable, equilibrium configuration, it 
departs from it one last time only to return again, and \emph{then} collapses.  
\label{fig:rho-samples-0.27-0.29}}
\end{figure}

\clearpage
\centerline{\includegraphics*[bb=0.3in 2.3in 7.8in 9.5in, scale=0.65]{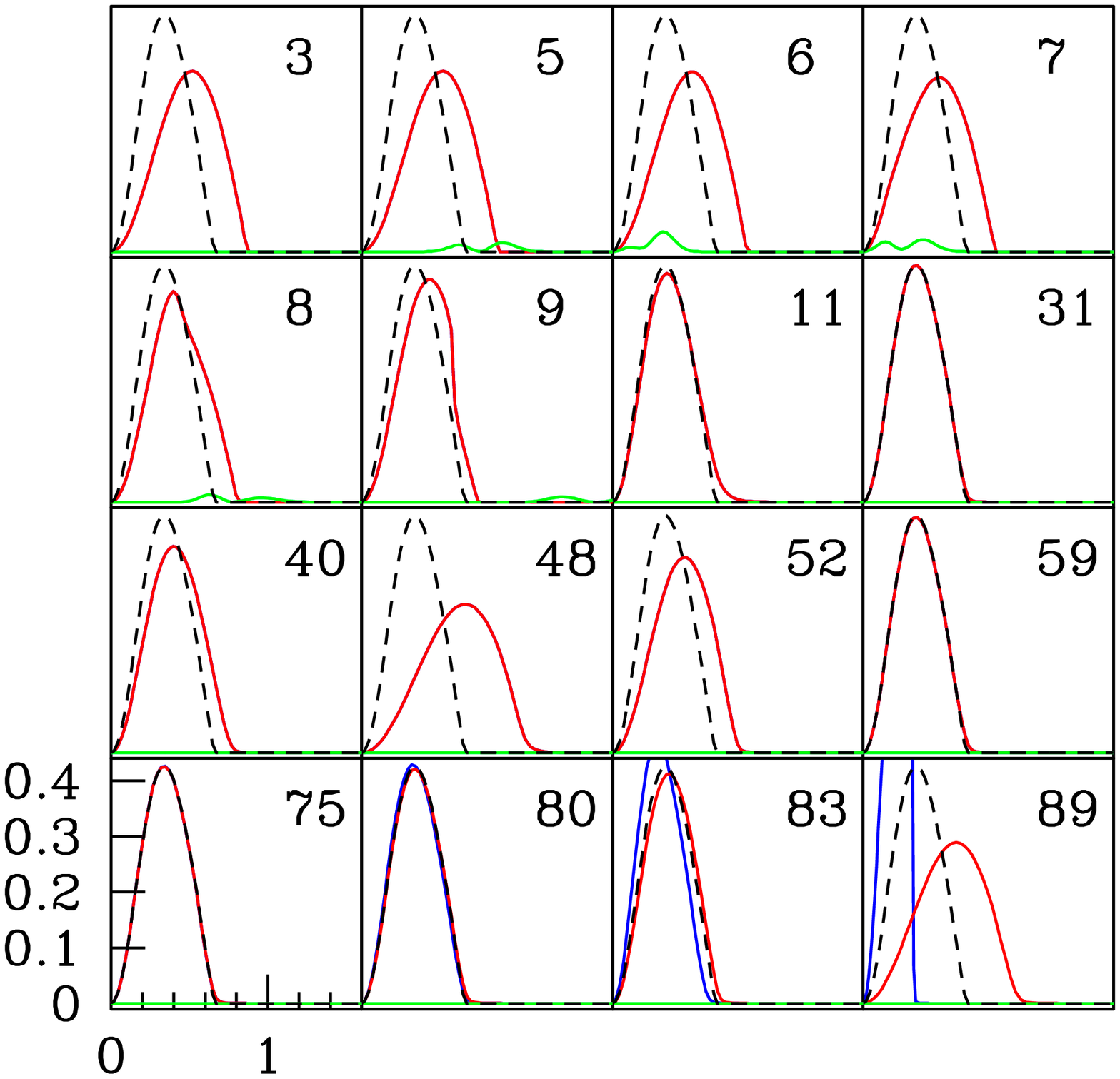}}
\begin{figure}[htb]
\caption[Time series of fluid and scalar field contributions to $dm/dr$ for the most nearly
critical solutions corresponding to the $\rho_c=0.197$ star.]{Time series of fluid and scalar 
field contributions to $dm/dr$ for the most nearly critical solutions corresponding to the $\rho_c=0.197$ star.
The supercritical (subcritical) fluid contribution is colored blue (red), and the scalar field 
contribution for the supercritical (subcritical) solution is green (cyan); the dotted black line 
is $dm/dr$ of the unstable, equilibrium solution that most closely approximates our
critical solution.  The elapsed time of each frame is shown in the upper-right corner. 
Since the differences between
the supercritical and subcritical scalar field perturbations is on the order of machine precision, the 
subcritical scalar field contribution is completely obscured by the supercritical one.  
Indeed, the supercritical and subcritical fluid contributions are nearly identical until $t=80$, when
the two solutions begin to diverge from the critical solution.  In every frame, 
only every tenth grid point is displayed for each distribution.
\label{fig:dmdr-evol-0.197}}
\end{figure}

Instead of dispersing to spatial infinity as do the solitonic oscillon stars of 
\cite{brady-chambers-goncalves}, the marginally-subcritical 
TOV stars ultimately settle into bound states.  Depending on the magnitude of $p^\star$ for a particular 
progenitor star, the final star solution will either be larger and sparser than the original 
(large $p^\star$), or it will oscillate indefinitely about the original solution. In reality, the star will radiate away 
the kinetic energy of the oscillation via some viscous mechanism.  In our model, however, the only 
dissipation is the trivial amount from the numerical scheme, and that from the star shock-heating its
atmosphere---transferring the kinetic energy of the bulk flow into internal energy. 
If the subcritical star settles to a sparser solution, it will 
do this through a series of violent, highly-damped oscillations similar to the SBO scenarios of 
velocity-perturbed stars described in Section~\ref{sec:param-space-surv}.  Examples of such 
subcritical SBO solutions are depicted in 
Figures~\ref{fig:tuning-maxtmr-rhoc}-~\ref{fig:rho-samples-0.09-0.12}.  The damped oscillations are best
illustrated in the marginally subcritical solutions shown in Figure~\ref{fig:tuning-maxtmr-rhoc}, since 
the oscillations of the subcritical solution of $\rho_c=0.09$ occur at an imperceptible scale 
and those of $\rho_c=0.12$ occur at later times than are shown in Figure~\ref{fig:rho-samples-0.09-0.12}.  

For these less relativistic and sparser stars, the perturbation required to generate near-critical evolution
is quite large and, consequently, is such that it drives the star 
to significantly \emph{overshoot} the unstable TOV solution, setting it to ring about the unstable 
solution instead.  This meta-stable ringing decreases with decreasing $p^\star(\rho_c)$, or increasing
$\rho_c$.  For instance, the critical solution of the $\rho_c=0.09$ star seems to correspond to 
an unstable TOV star with central density $\rho_c^\star\simeq2$ that oscillates such that $0<\rho_\circ(0,t)<4$.
The increase in central density---from the initial stable star to the unstable star solution---represents 
an increase by a factor of $22$.
This is to be contrasted with the critical solution for the $\rho_c=0.29$ star which has a central density 
$\rho_c^\star\simeq0.35$---an increase by a factor of $1.2$; this critical solution oscillates 
such that $0.32<\rho_\circ(0,t)<0.38$.  This trend will be discussed further in the next section.  

In addition to smaller oscillations about the meta-stable states for denser initial stars, we see 
from Figures~\ref{fig:rho-samples-0.1835-0.21}-~\ref{fig:rho-samples-0.27-0.29} that near-critical evolutions
can momentarily depart from their meta-stable states.  The departures
are illustrated by a break in the plateaus of the $\rho_\circ(0,t)$ distributions.  As $\rho_c$
increases and gets closer to the turnover point, which is located at $\rho_c=0.318$, we see that the 
number of distinct plateaus increases.  The $\rho_c=0.1835$ solution is the smallest initial central 
density where two plateaus are observed, and $\rho_c=0.21$ is the first one where three are seen.  
For higher densities we see an ever-increasing number of plateaus, most likely because the difference 
between the progenitor solution and its corresponding critical solution is diminishing.

As we can see in the time sequence of the scalar field and fluid contributions to $dm/dr$ 
in Figure~\ref{fig:dmdr-evol-0.197}, the marginally subcritical and supercritical stars leave the 
unstable TOV star configuration only to return to it after one oscillation about the progenitor solution. 
The unstable star was found by taking the time average of $\rho_\circ(0,t)$ for the most nearly critical
solutions as described in detail below.  The shock from the outer layers of the star reacting first to 
the increase in curvature is first seen at $t=9$ of this figure, and the time referred to in this figure 
coincides with proper time at spatial infinity.

\centerline{\includegraphics*[bb=0.3in 2in 8.2in 9.7in, scale=0.6]{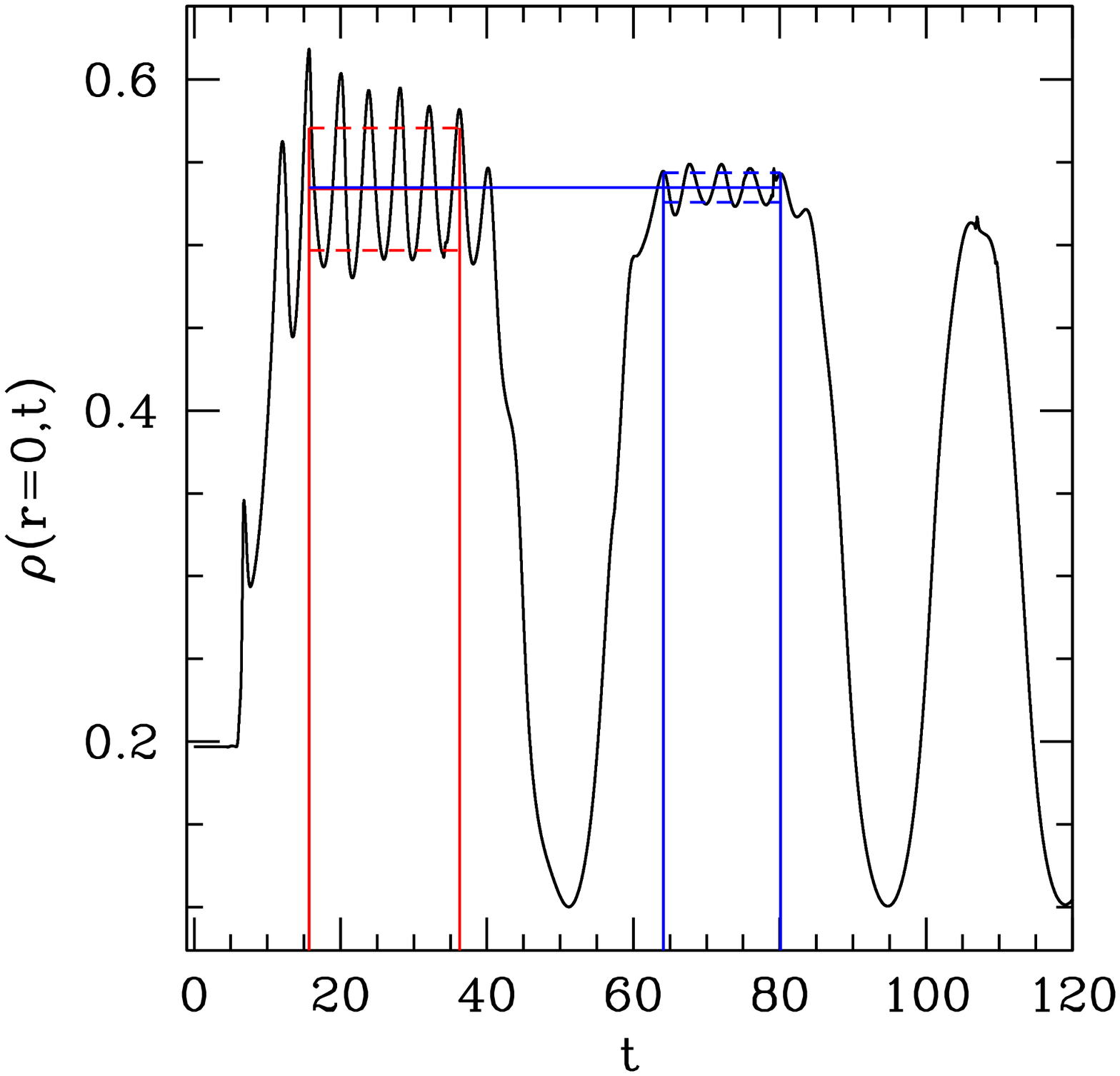}}
\begin{figure}[htb]
\caption[Illustration of the fitting procedure used to determine the central density of the 
critical solution. ]{Illustration of the fitting procedure used to determine the central density of the 
critical solution, $\rho_c^\star$.  The progenitor star corresponds to a star with $\rho_c=0.197$.  The critical 
solution shown here exhibits two plateaus, and  we calculate $\rho_c^\star$ from both plateaus.  
The time-spans used to calculate both averages are determined by the first and last peaks that 
seem to represent complete sets of oscillations for the unstable star.  These periods of time
are shown here by the solid, vertical lines.  For instance,
the last peaks on each plateau are significantly smaller than the other plateaus' peaks suggesting 
that star has already begun its departure from the unstable solution. 
\label{fig:rhoc-fit}}
\end{figure}

Making a quantitative comparison of the critical solution to an unstable star is not easy since the 
critical solution is not exactly static.  If we make the assumption that the oscillation is 
sinusoidal, we can take a time-average of the solutions when the critical solution most resembles
an unstable star.  Figure~\ref{fig:rhoc-fit} graphically depicts how we go about this for 
for the $\rho_c=0.197$ critical solution as an example.  We first start with the subcritical solution
which is tuned closest to the critical solution.  The periods at which the solution best 
approximates the unstable solution are determined by qualitatively judging where the sequences of 
quasi-normal oscillations begin and end for the unstable star.  For instance, in this figure we can 
clearly see that that the 
``first'' peak---located at $t\simeq12$---of the first plateau does not ``belong'' to the 
sequence of quasi-normal oscillations since it is distinctly smaller than the ``second'' peak 
of this plateau.  Thus, we start the time-average from this second peak and stop before the 
last peak since it, too, seems out-of-character with this particular sequence of oscillations.  
The central density, $\rho_c^\star$, of the unstable 
star solution corresponding to the critical solution is then calculated by taking the time-average 
of $\rho_\circ(0,t)$ over a given period.  This is repeated for other plateaus if present, so 
Figure~\ref{fig:rhoc-fit} would yield---for instance---two estimates of $\rho_c^\star$. 
For each system with multiple plateaus studied here, we have found the plateau averages all agree
with each other to within their standard deviation.  Hence, we feel that this is a consistent method 
for identifying the unstable star associated with a critical solution.
The standard deviations of $\rho_\circ(0,t)$ for $\rho_c=0.197$ about its calculated $\rho_c^\star$ 
are represented by the red and blue sets of dashed lines, whereas 
the average for each plateau is given by a solid, horizontal line.   

\centerline{\includegraphics*[bb=0.4in 2in 8.5in 9.9in, scale=0.35]{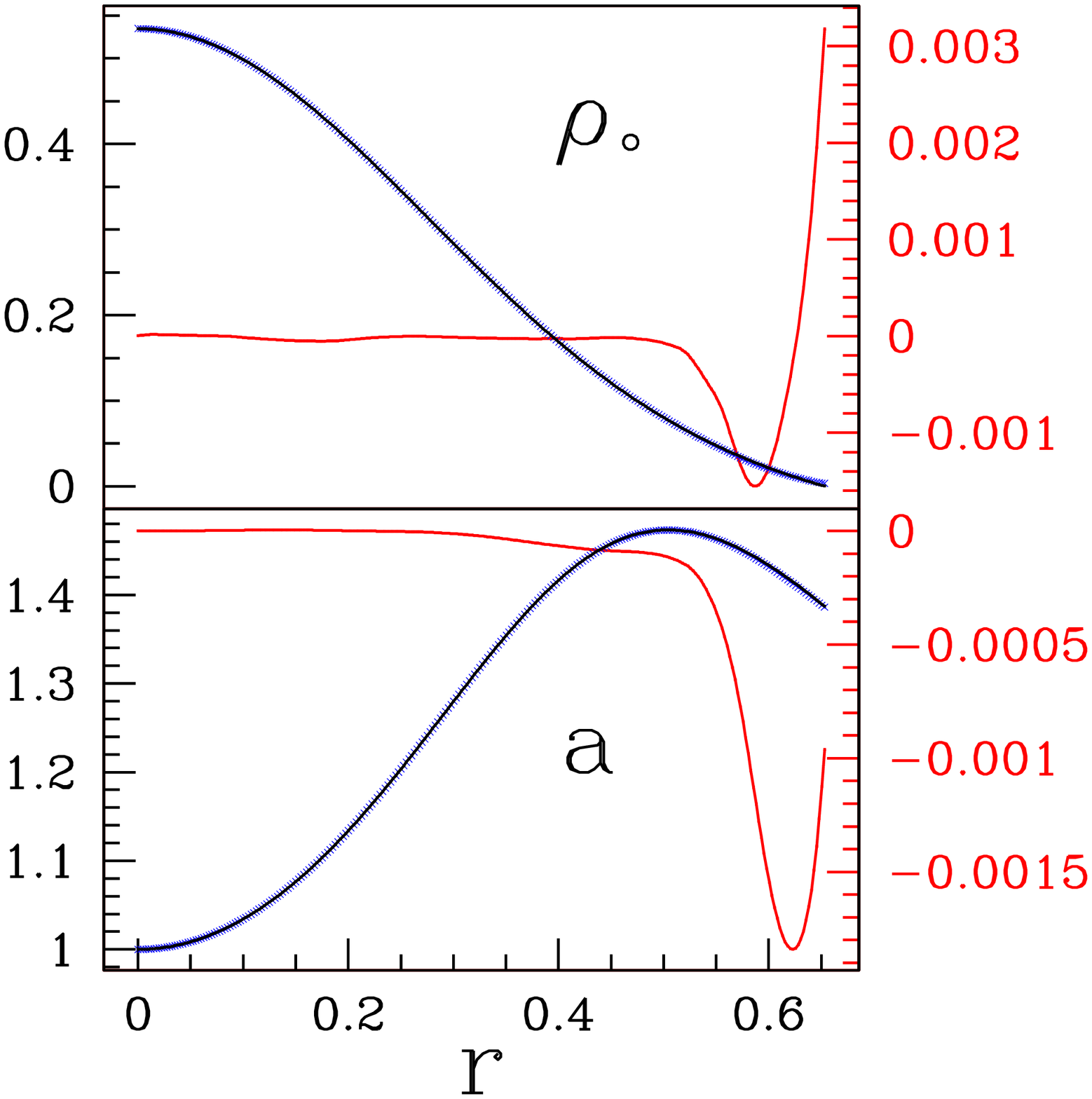}
\includegraphics*[bb=0.3in 2in 8.3in 9.9in, scale=0.35]{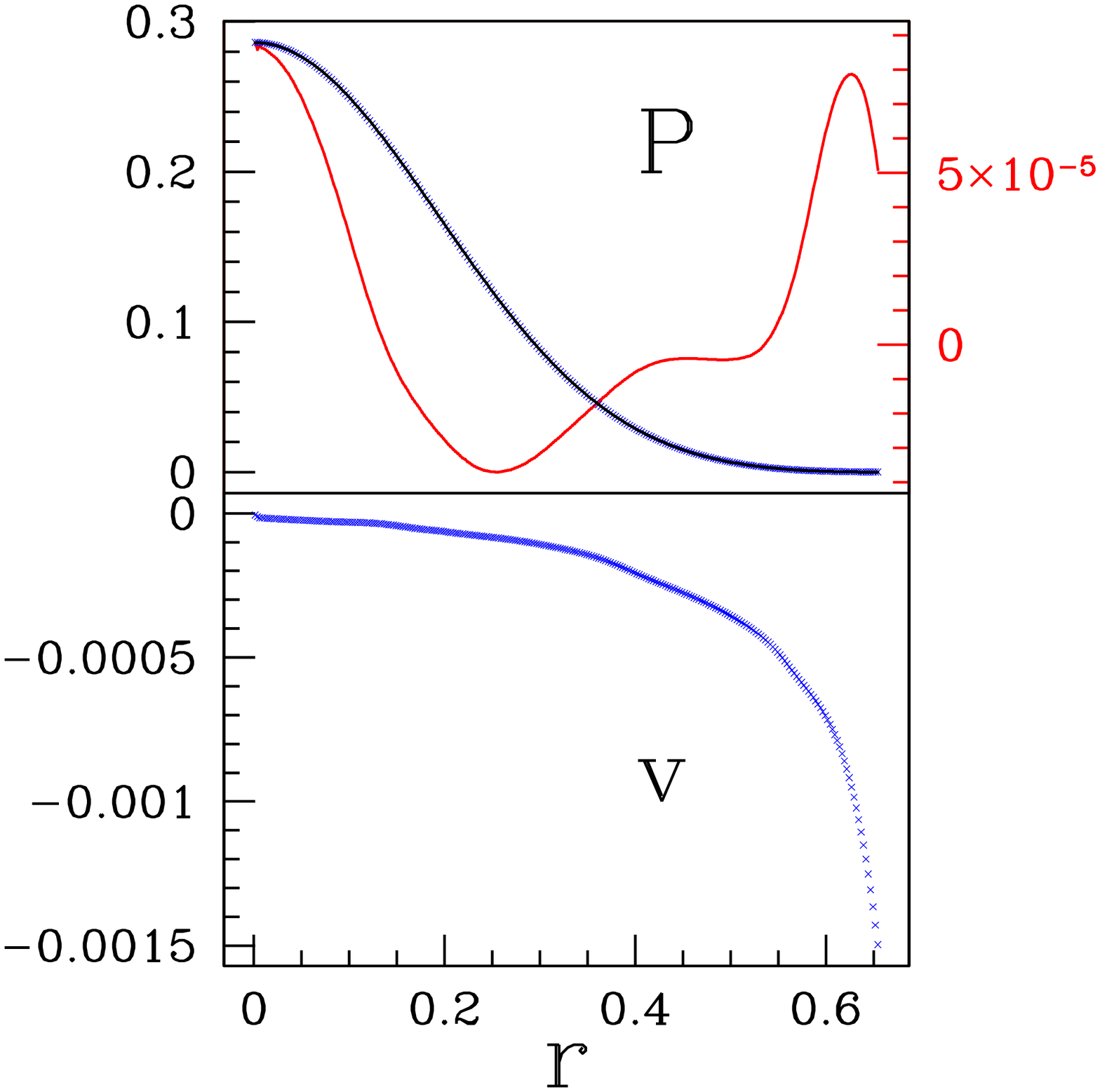}}
\begin{figure}[htb]
\caption[The time-average (blue crosses) of a marginally subcritical solution compared to the unstable 
TOV solution (black line) it best approximates.]{The time-average (blue crosses) of a marginally 
subcritical solution compared to the unstable 
TOV solution (black line) it best approximates.  The time-average was performed while the solution dwelled 
on the second plateau, shown in Figure~\ref{fig:rhoc-fit}.  The unstable star was calculated by 
numerically solving the TOV equations using $\rho_c^\star$ for as the solution's central density.  
The distributions shown in red, whose ranges are given on the right-hand sides of the plots, are 
the point-wise differences between the other two functions plotted.  The solutions and their deviations are only shown 
here within the stellar radius, $R_\star$.
\label{fig:type-i-critsol-comparison}}
\end{figure}

After identifying a perturbed star's associated unstable solution, its shape with 
the solution it oscillates about during a plateau.  To perform this comparison for $\rho_c=0.197$, 
we first took the time-average of the perturbed star during the second plateau.  This 
time-average serves as an approximation to the static solution about which the critical solution
varies, assuming that the deviations are sinusoidal in nature.  The time-averaged solution can 
then be compared to the TOV solution with central density $\rho_c^\star$.  The results of this comparison
for the critical solution of the $\rho_c=0.197$ perturbed star are shown in 
Figure~\ref{fig:type-i-critsol-comparison}.  Metric and fluid functions from the time-average (black)
and the estimated unstable TOV solution (blue) are shown together along with their differences (red). 
This figure clearly shows that, during ``plateau epochs'', the critical solution 
closely approximates 
that of an unstable TOV solution of the same central density.  The relative deviation between the two 
solutions increases near the radius of the star, $R_\star$, which is not surprising since the fluid's time-averaged
velocity is greatest there.  Also, near $R_\star$ the star is most likely interacting with the atmosphere in a 
non-trivial way, which could alter its form near its surface.  In fact, a similar discrepancy 
was observed in the critical boson star solutions in \cite{hawley-choptuik-2000}; they found that 
the critical solutions had a longer ``tail'' than their corresponding static solutions.  Still, 
the differences we see here are encouraging, and suggest strongly that the critical solutions 
we obtain are perturbed stellar solutions from the unstable branch.

\section{Mass Transfer and the Transition to the Unstable Branch}
\label{sec:mass-transf-trans}

Not only does the perturbing scalar field momentarily increase the spacetime curvature near the origin 
as it implodes through the star, the gravitational interaction of the two matter fields involves
an exchange of mass from the 
scalar field to the star.  A hint of this was shown in Figure~\ref{fig:dmdr-evol-0.197} by the 
difference in heights of $dm_\mathrm{scalar}/dr$ before and after the interaction.  
In Figure~\ref{fig:mass-transfer-0.197-0.09}, we provide a more explicit illustration of the 
mass exchange for two marginally subcritical solutions of stars with $\rho_c=0.197$ and $\rho_c=0.09$.
The figure shows the mass contributions for each matter component, as well as, the total gravitating mass.  
$M_\mathrm{total}$
is calculated via (\ref{massaspect}), while $M_\mathrm{fluid}$ ($M_\mathrm{scalar}$) is found 
by integrating $d m_\mathrm{fluid}/dr$
($d m_\mathrm{scalar}/dr$) from the origin to the outer boundary. 
For each case, the non-trivial gravitational interaction of the fluid and scalar field 
can be recognized by the sudden change in their integrated masses, which occurs near $t=7$ in each plot.  
The perturbation for the $\rho_c=0.197$ star is small and does not transfer a significant portion of 
its mass to the star, whereas the perturbation required to drive the $\rho_c=0.09$ 
star to its marginally subcritical state transfers more than an third of its mass to the star 
before leaving the bounds of the star.  This dramatic interaction drives the star to oscillate
wildly about its unstable counterpart---as seen in Figure~\ref{fig:rho-samples-0.09-0.12}---and 
it eventually expels a great deal of its mass as it departs from this highly energetic, yet unstable, 
bound state.  The slow leaking of the ejected matter from the grid is clearly 
seen in Figure~\ref{fig:mass-transfer-0.197-0.09} as 
the long tail of $M_\mathrm{fluid}/M_\mathrm{total}$ that starts well after the scalar field leaves 
the grid.  
\vspace{0.5cm}

\centerline{\includegraphics*[bb=0.4in 2in 7.9in 8.4in, scale=0.6]{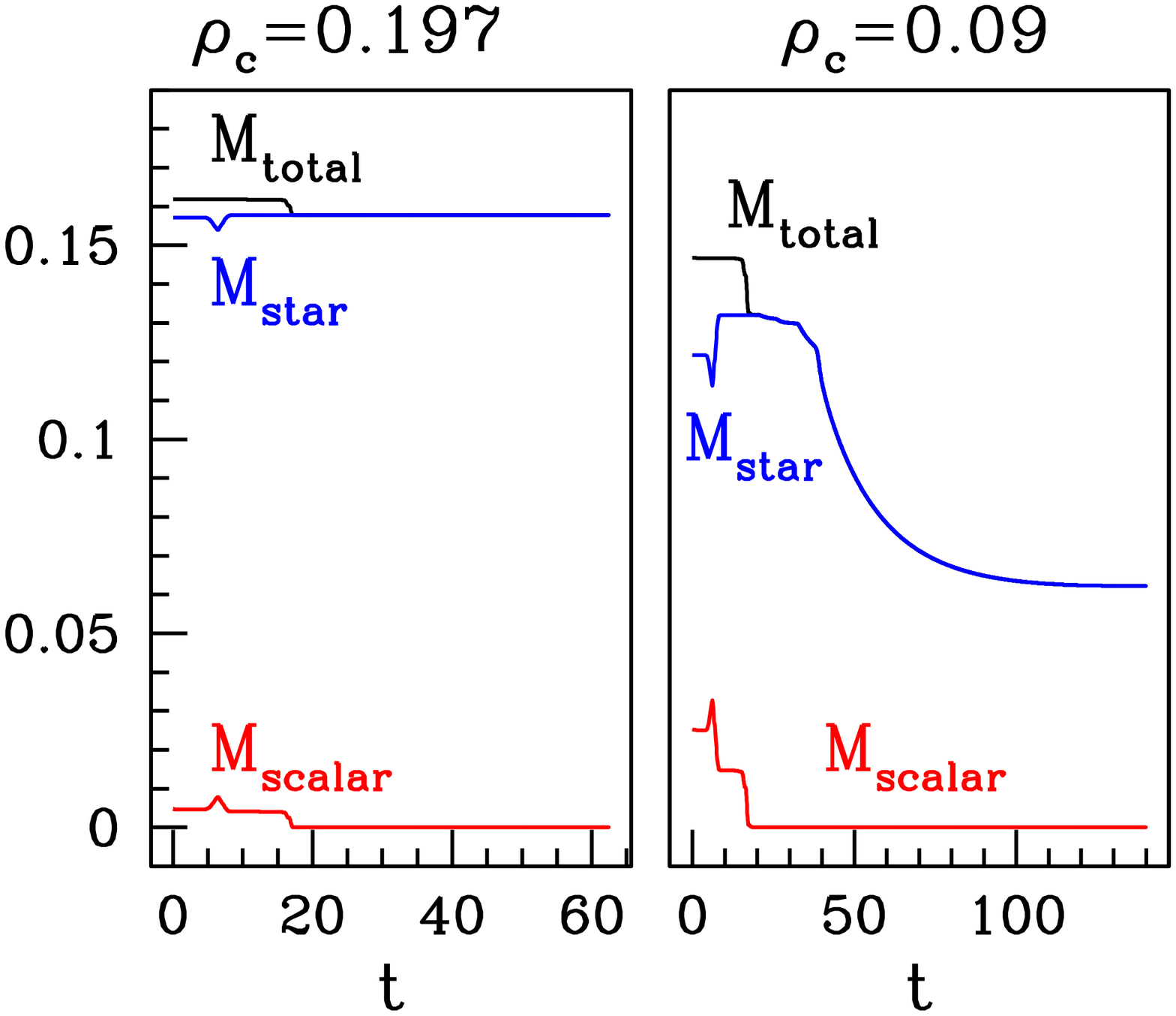}}
\begin{figure}[htb]
\caption[The integrated masses of the matter fields as a function of time for marginally subcritical
solutions and progenitor stars with $\rho_c=0.197$ and $\rho_c=0.09$. ]{The integrated masses of the 
matter fields as a function of time for marginally subcritical
solutions and progenitor stars with $\rho_c=0.197$ and $\rho_c=0.09$. 
The decrease in $M_\mathrm{total}$ at the same time as 
$M_\mathrm{scalar}$ vanishes signifies the scalar field leaving the numerical grid; from the time
it leaves, $M_\mathrm{total}=M_\mathrm{fluid}$.
\label{fig:mass-transfer-0.197-0.09}}
\end{figure}

\centerline{\includegraphics*[bb=0.2in 2in 8.5in 9.9in, scale=0.6]{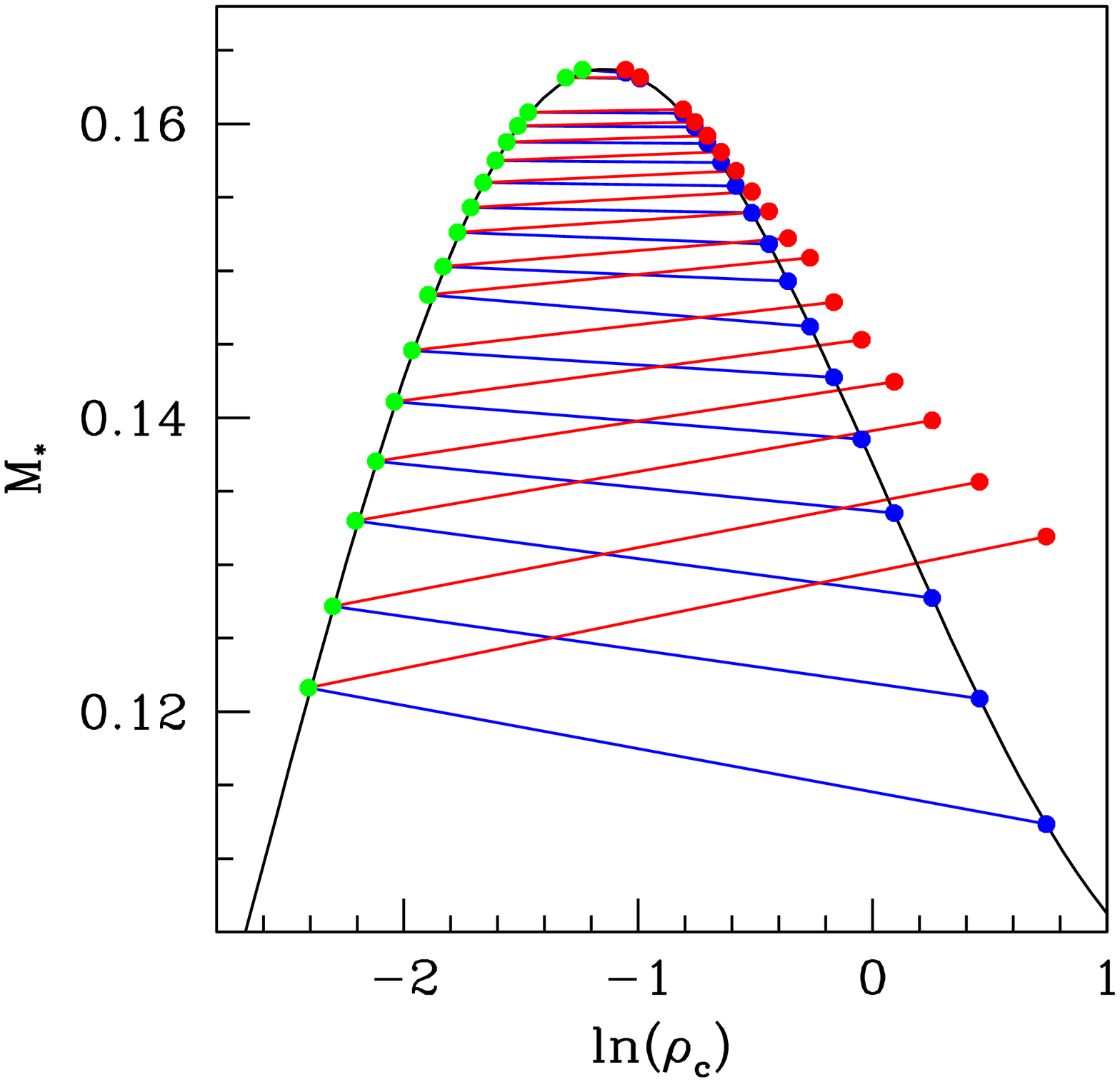}}
\begin{figure}[htb]
\caption[Mass versus the log of the central density for equilibrium solutions (solid black line),
a few of the initial data sets used (green dots), and the critical solutions obtained from 
these initial data sets (blue and red dots).]{Mass versus the log of the central density for 
equilibrium solutions (solid black line),
a few of the initial data sets used (green dots), and the critical solutions obtained from 
this initial data sets (blue and red dots).  The central densities of the critical solutions
are obtained by taking a time average of the central density when the star is in resembling the 
critical solution.  The blue dots refer to equilibrium solutions with central densities that match
those of  the critical solutions, while we have used the mass of all the fluid in the numerical 
domain in determining the locations of the red and green dots.  The red and blue lines 
match the initial solutions to their critical solutions. 
\label{fig:mass-crit-sols}}
\end{figure}

To examine how the amount of mass exchange  varies for different critical solutions and to see 
where exactly critical solutions fall on the $M_\star$ versus. $\rho_c$ graph of equilibrium solutions, we 
construct Figure~\ref{fig:mass-crit-sols}.  The initial star solutions are indicated here on the left
side---the stable branch----while their critical solutions are shown on the right near the unstable
branch.  The central densities of the red and blue dots use the values of $\rho_c^\star$ calculated by fitting 
$\rho_\circ(0,t)$ during periods when the star emulates the critical solution, as described in the
previous section.  Only the masses of the blue and red dots vary; masses of the blue solutions are 
those corresponding to the unstable TOV solutions with central density equal to $\rho_c^\star$, and 
the masses of the solutions represented  by the red dots are $M_\mathrm{fluid}$ calculated after the 
scalar field has left the numerical domain.  The amount of mass transferred to a particular star from the scalar 
field is represented here by the mass difference of the red and blue dots corresponding 
to the same $\rho_c$.  We can see that the total fluid mass is always larger than its initial mass, 
whereas the mass of the critical solution's associated unstable star solution is always \emph{smaller} 
than its stable progenitor.  In addition, as the turnover is approached, both of these deviations diminish 
until, at turnover, the final mass of the fluid distribution corresponds to its initial mass \emph{and} the 
mass of the unstable TOV solution.  

The fact that the unstable TOV solution is always smaller than the progenitor may be explainable in a number of ways.  
First, the oscillations of the critical solution about the unstable star configuration may not 
be sinusoidal, thereby leading to central density estimate that is possibly larger than it should be. 
A larger central density would then lead to a mass estimate that is less than it should be, since 
$dM_\star/d\rho_c<0$ on the unstable branch.  Second, it was seen in 
Figures~\ref{fig:rho-samples-0.09-0.12}-~\ref{fig:rho-samples-0.27-0.29} that the oscillations 
of the critical solutions decrease with increasing $\rho_c$.  The decrease in the amount of 
energy in these kinetic modes seems to be correlated with the decrease in the exchanged mass.  
It is most certainly the case that a large portion of exchanged mass goes into perturbing the unstable star 
solution.

\section{Type~I Scaling Behavior}
\label{sec:type-i-scaling}

As the initial pulse of scalar field is adjusted toward $p^\star$, the lifetime of the meta-stable, 
near-critical configuration increases.  To quantify the scaling for a given 
initial star solution, the subcritical solution closest to the critical one is first determined.  
This is done by tuning the amplitude of the scalar field pulse, $p$, until consecutive bisections 
yield a change in $p$ smaller than machine precision.  Let $p^\mathrm{lo}$ be the value of $p$
that yields the subcritical solution that most closely approximates the critical solution. 
For each $p$, a unique solution is 
produced that resembles this marginally subcritical solution for different lengths of time, 
determined by how close $p$ is from $p^\star$.  Assuming that the $p^\mathrm{lo}$ solution 
resembles the critical solution longer than any other from our code, the lifetime, $T_0(p)$, is then
determined from the proper time measured at the origin that elapses until 
\beq{
{\max}(2m/r)[T_0(p),p] - {\max}(2m/r)[T_0(p),p^\mathrm{lo}] \ > \  
0.01 {\max}(2m/r)[T_0,p^\mathrm{lo}]
\label{lifetime-criterion}
}
where ${\max}(2m/r)[T_0',p]$ is the value of ${\max}(2m/r)$ at central proper time
$T_0'$ for the solution specified by $p$.  This lifetimes are then plotted against the natural
logarithm of the deviation of $p$ from $p^\star$ to find the scaling exponent from the 
expected trend (\ref{type-i-scaling}).  An example of a linear fit to such values is given 
in Figure~\ref{fig:lifetime-fit}.  Since supercritical solutions resemble the critical solution 
as well as subcritical solutions, then both kinds can be used when determining the scaling 
exponent $\sigma$.  The exponent is then found to be the negative of slope of the line.  The deviation 
of the code-generated data from the best-fit has an obvious modulation, most likely due to the 
periodic nature of the critical solution.  

\centerline{\includegraphics*[bb=0.3in 2in 8.5in 9.9in, scale=0.6]{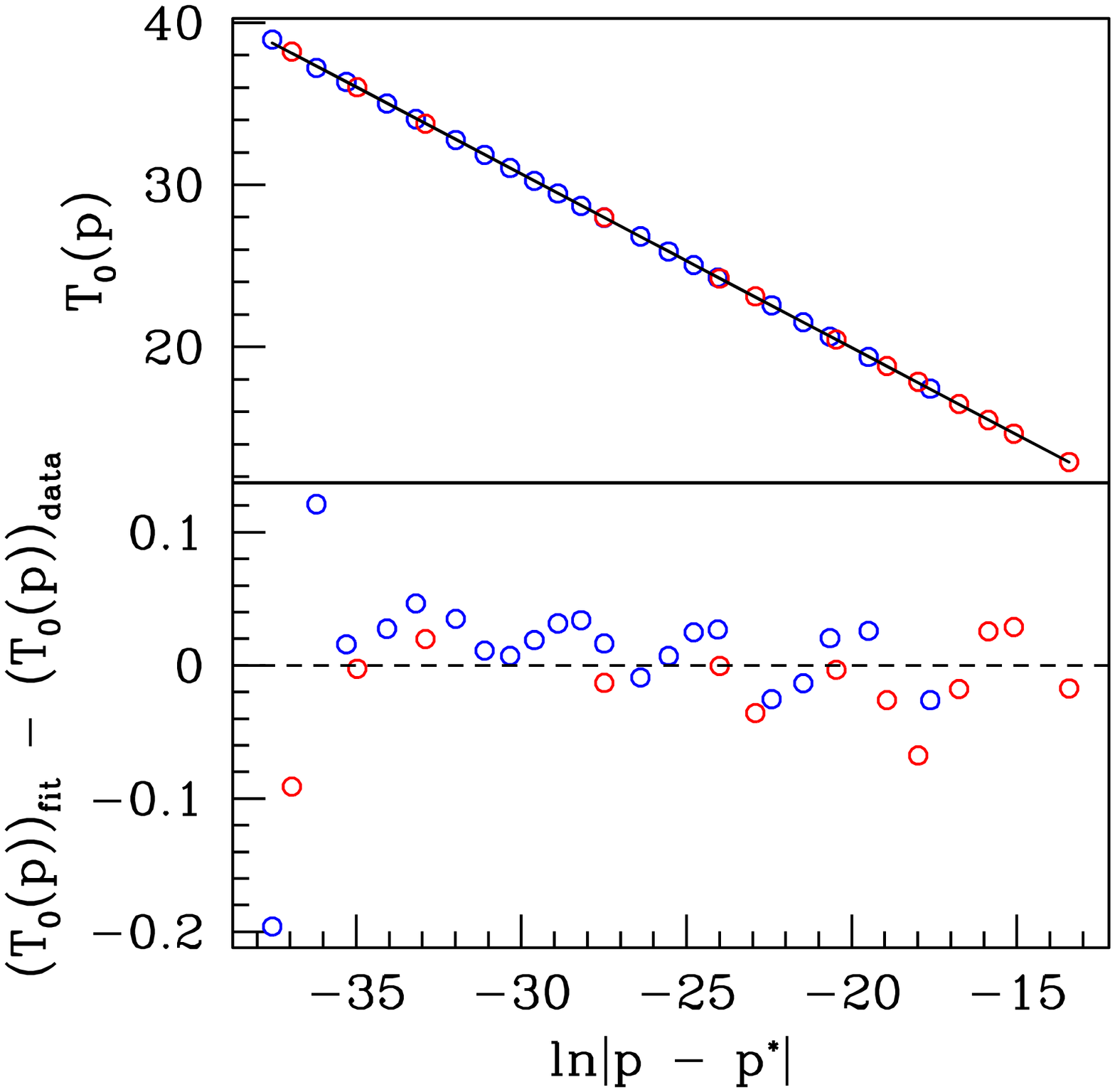}}
\begin{figure}[htb]
\caption[The lifetimes, $T_0(p)$, for solutions near the critical solution for a star with $\rho_c=0.14$.]{The 
lifetimes, $T_0(p)$, for solutions near the critical solution for a $\rho_c=0.14$ star.  
The scaling exponent, $\sigma$, is found from the negative of the slope of the best linear
fit to  the points.  The fact that both supercritical and subcritical solutions can be used
for calculating $T_0(p)$ is illustrated here by our inclusion of both sets of points: 
the blue points show data from supercritical solutions and the red points come from subcritical calculations.  
The lifetimes here are actually those as measured at spatial infinity; see the text for further information. 
\label{fig:lifetime-fit}}
\end{figure}

In practice, the lifetime is determined using the 
proper time elapsed at spatial infinity, $T_\infty$, instead of that measured at the origin.  
In order to get the correct scaling exponent, which would correspond to $1/\omega_{Ly}$ of the 
unstable mode, $\sigma_\infty$ must be rescaled.  Since $T_\infty$ is the same as our coordinate time,
$t$, then 
\beq{
dT_0(t) = \alpha(0,t) \, dt   \quad . \label{propertime-interval}
}
In order to estimate the rescaling factor, we assume that $\alpha(0,t)$ does not vary much when the 
solution is in the near-critical regime, so that 
\beq{
\alpha(0,t) \approx \alpha^\star(0) \label{propertime-approx}
}
where $\alpha^\star$ is the central value of the lapse of  the unstable TOV solution that corresponds
to the critical solution.  The corrected value of $\sigma$ is then calculated using:
\beq{
\sigma = \alpha^\star \sigma_\infty  \quad . \label{sigma-scaling}
}

We have performed fits for $\sigma_\infty$ and then rescaled them using the above procedure to obtain 
an estimate of $\sigma$
for $55$ different initial TOV stars.  The variation of $\sigma$ with $\rho_c^\star$ 
is shown in Figure~\ref{fig:type-i-scaling}.  We find that $\sigma(\rho_c^\star)$
fits surprisingly well by the linear relationship
\beq{
\sigma = 5.93 \rho_c^\star - 1.475  \quad . \label{sigma-rhoc-fit}
}

In order to verify that the calculated $\sigma$ values are, indeed, equal to $1/\omega_{Ly}$, we
would need to calculate the fundamental modes of the unstable star solutions.  To the extent of
my knowledge and others \cite{lindblom,sterg}, this has not been done before for the particular 
EOS used, and we leave this for future work.  

\centerline{\includegraphics*[bb=0.3in 2in 8.5in 9.9in, scale=0.6]{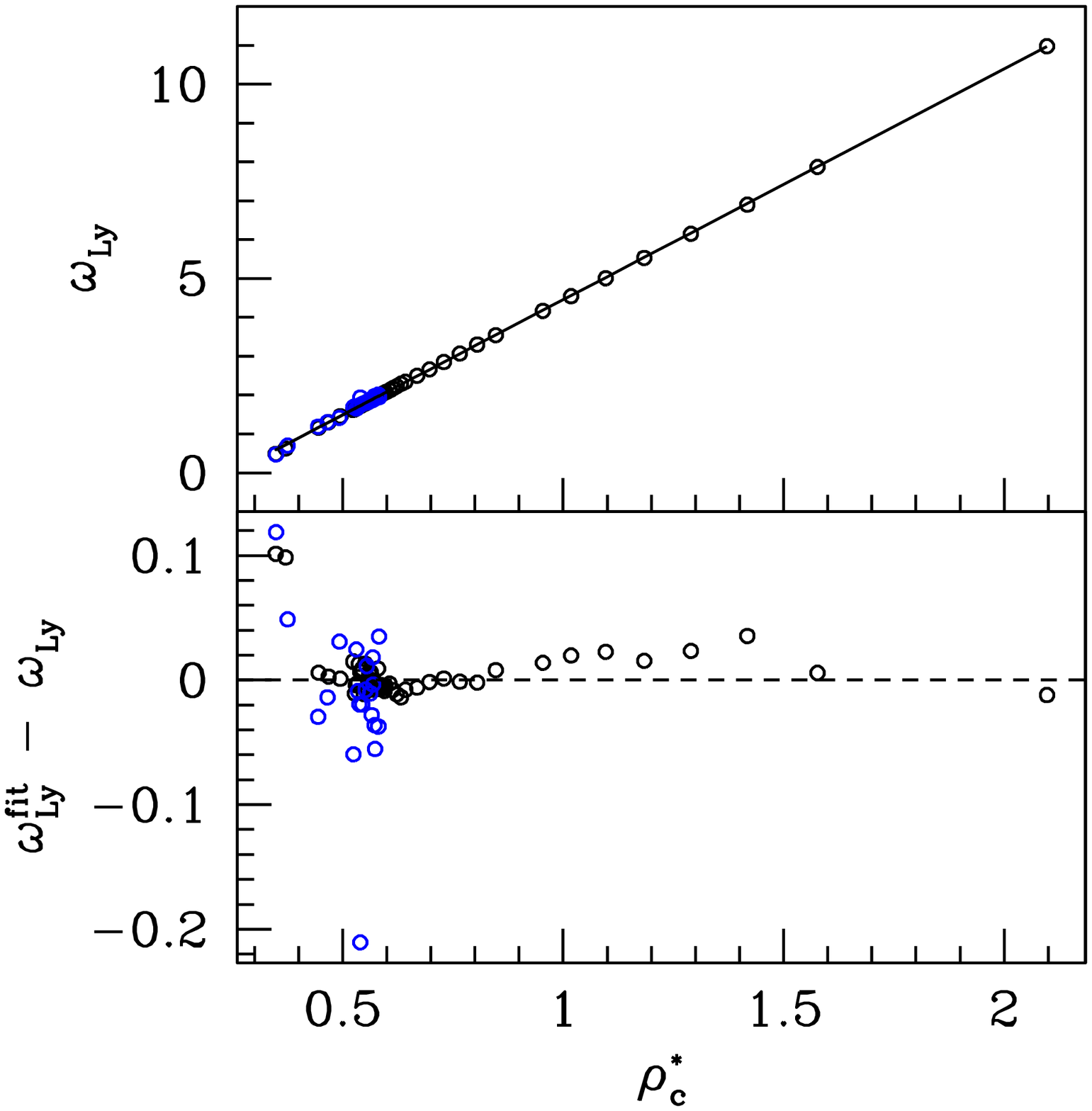}}
\begin{figure}[htb]
\caption[The real part of the estimated Lyapunov exponent for a given star solution 
parameterized by $\rho_c^\star$. ]{The real part of the estimated Lyapunov exponent for a given star solution 
parameterized by $\rho_c^\star$.  The black dots were calculated from data from the first ``plateau'', 
while the blue dots from the second.  ${\max}(2m/r)$ was used to calculate the $\omega_{Ly}$
shown here.  \label{fig:type-i-scaling}}
\end{figure}

\section{The Plateaus}
\label{sec:plateaus}

In order to gather a better understanding of what causes the critical solutions to temporarily 
depart from the unstable branch, we performed a series of bisection searches for different 
values of various control parameters of our numerical simulations.  
For instance, to see if the presence of the departures is affected by changes in the 
floor, we tuned to the critical solution for three different sets of values for 
$\{\delta,P_\mathrm{floor}\}$.  The most marginally subcritical solutions from these 
searches are shown in Figure~\ref{fig:rhoc-floor}.  In addition, the effect of the outer boundary's
location, $r_{\max}$ is seen in Figure~\ref{fig:rhoc-rmax}.   To see if the 
time at which the pulse collides with the star has any effect, the initial position of the pulse, 
$R_\phi$ was varied; the results from this particular analysis are shown in Figure~\ref{fig:rhoc-r0pp}. 

In general, we see all these aspects to have significant and non-trivial effect on the 
critical solution's departure from the unstable solution. But, all the different marginally-subcritical 
solutions finally depart from the unstable solution at approximately the same time.  

Whether because of its magnitude or extent, 
the solution's departure seems to be affected by the floor.  Increasing the size of the floor seems
to hasten the initial departure; even though they represent only two points of reference, 
the similarity of the solutions with the two highest floor values may suggest that the floor's effect
``converges'' to one behavior as its size increases. 

On the other hand, changes in the size of the computational domain and $R_\phi$ 
seem to have no \emph{consistent} effect on the first departure time.  

The exact cause of these departures remains unsolved, and is left for future examination.  

\clearpage

\centerline{\includegraphics*[bb=0.3in 2in 8.2in 9.8in, scale=0.6]{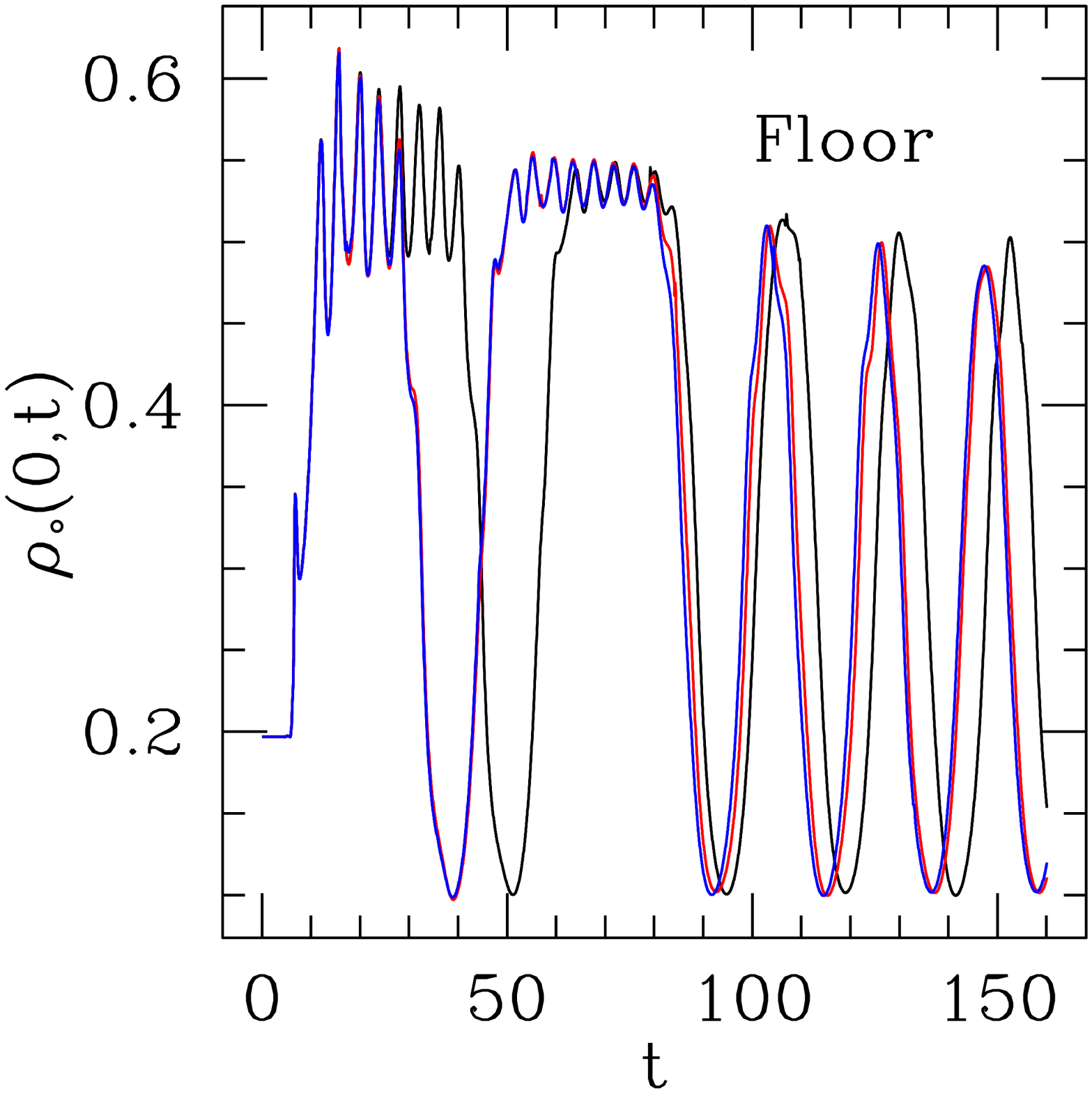}}
\begin{figure}[htb]
\caption[Comparisons of $\rho_\circ(0,t)$ for marginally subcritical solutions 
obtained when using varying values of the fluid's floor. ]{Comparisons of $\rho_\circ(0,t)$ for 
the marginally subcritical solutions 
obtained when using varying values of the fluid's floor.  The original, reference solution is 
shown in black and used $P_\mathrm{floor}=3.8809\times10^{-15}$, $\delta=3.8809\times10^{-18}$.  
The red and blue lines are from critical searches that used floor values $10$ and $100$ times 
greater, respectively, than those of the original solution.  Variations can be seen between 
each floor size, even though this difference is smallest between solutions with the larger floor 
quantities.  The star's initial central density was $0.197$ for all the runs shown here. 
\label{fig:rhoc-floor}}
\end{figure}

\centerline{\includegraphics*[bb=0.3in 2in 8.2in 9.8in, scale=0.6]{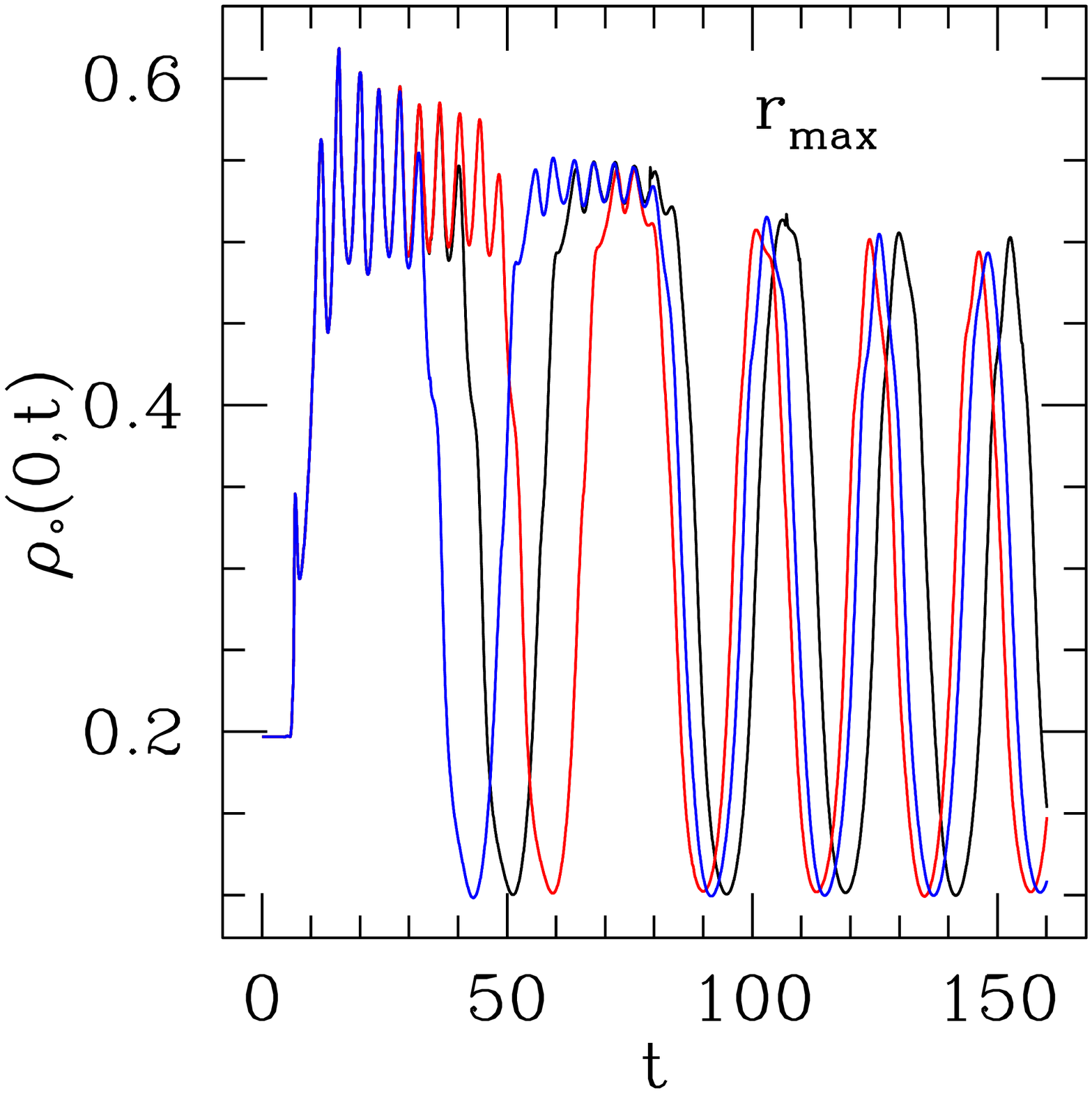}}
\begin{figure}[htb]
\caption[The central density as a function of time of the most nearly critical, subcritical solutions
obtained with physical domains of various sizes.]{The central density as a function of time of the most 
nearly critical, subcritical solutions obtained with physical domains of various sizes.  The 
red (blue) sequence used a domain twice (thrice) as large as that of the original configuration,
which is shown here by the black line.  The star's initial central density was $0.197$ for 
all the runs shown here. 
\label{fig:rhoc-rmax}}
\end{figure}

\centerline{\includegraphics*[bb=0.3in 2in 8.2in 9.8in, scale=0.6]{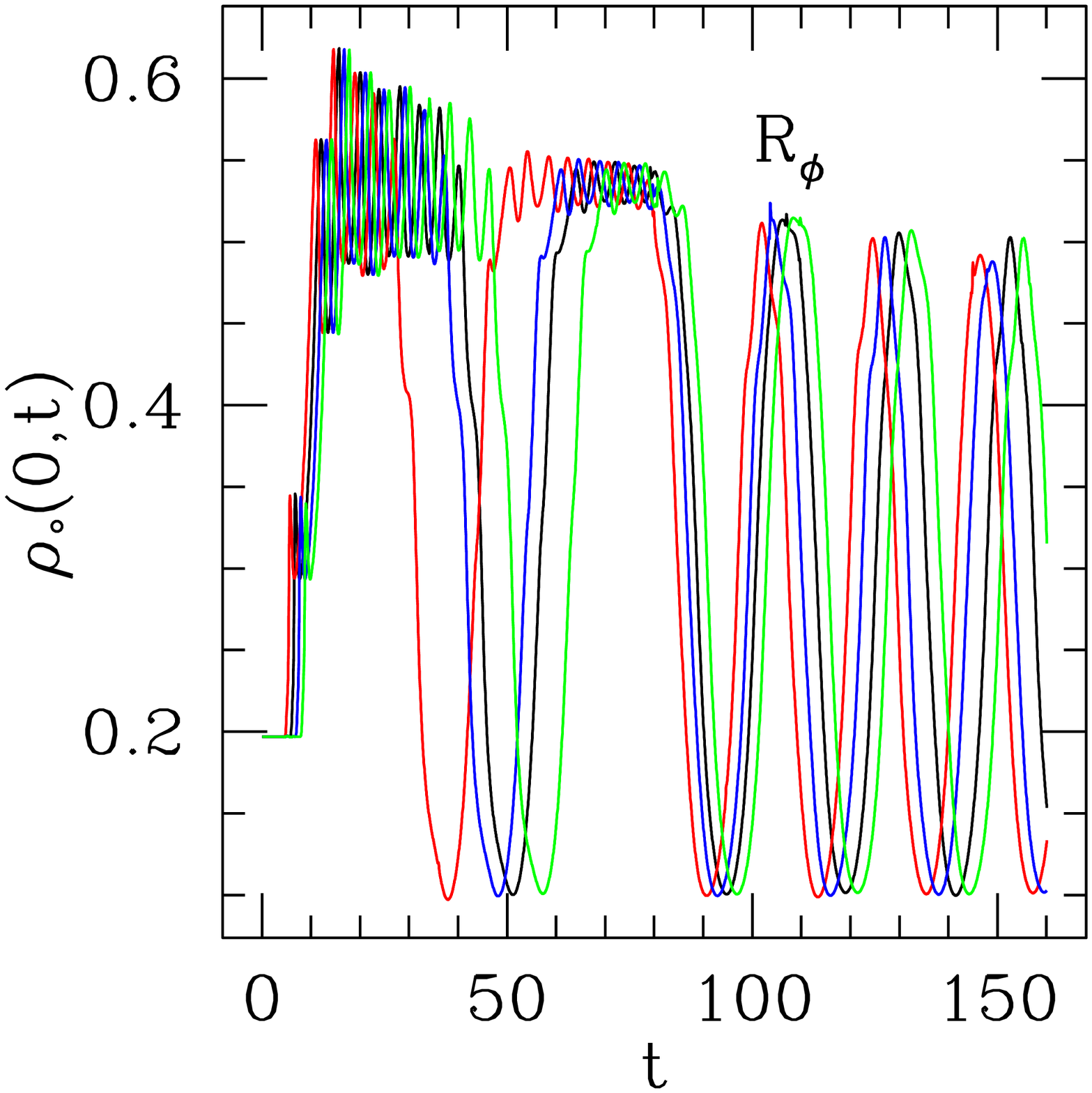}}
\begin{figure}[htb]
\caption[The central density as a function of time of the most nearly critical, subcritical solutions 
obtained by using different initial locations of the initial scalar field distribution, $R_\phi$.]{The 
central density as a function of time of the most nearly critical, subcritical solutions 
obtained by using different initial locations of the initial scalar field distribution, $R_\phi$. 
Specifically, the scalar field at $t=0$ takes the form of a Gaussian distribution, and the 
position of the center of this Gaussian is unique for each color shown here.  In the units 
used for these runs, the radius of the progenitor star was $r=0.87$, while the initial positions 
of the scalar field pulses were located at $r=4$ (red), $r=5$ (black), $r=6$ (blue), $r=7$ (green).  
The star's initial central density was $0.197$ for all the runs shown here. 
\label{fig:rhoc-r0pp}}
\end{figure}

\chapter{Conclusion}
\label{chap:concl-future-work}

In this work, we simulated spherically-symmetric relativistic perfect fluid flow in the strong-field regime of 
general relatively.  Specifically, a perfect fluid that admits a length scale, for example one that 
follows a relativistic ideal gas law,  was used to investigate the dynamics of 
compact, stellar objects.  These  stars were modeled as neutron stars by using a stiff equation 
of state, approximating the behavior of some realistic state equations.  These models were then used
to study the dynamics of neutrons so far out of equilibrium that they driven to gravitational  collapse.

Since the behavior in neutrons stars driven catastrophically to collapse entails 
highly-relativistic fluid motion and strong, nonlinear effects from the fluid-gravitational 
interaction, a numerical treatment is challenging.  To achieve stable evolutions in near-luminal flows, 
the primitive variable solver required improvements.  In addition, an unusual instability was found to 
develop near the threshold of black hole formation, which required the use of 
new computational methods. 

The star models served as initial data for a parameter survey, in which we drove the stars to 
collapse using either an initial velocity profile or a pulse of massless scalar field.   Both types of 
critical phenomena were observed using each of the two mechanisms.  The parameter space survey 
provided a description of the boundary between Type~I  and Type~II behavior, and illustrated the 
wide range of dynamical scenarios involved in stellar collapse.  We found that the non-black hole 
end states of solutions near the threshold of black hole seemed to be correlated to the 
type of critical behavior observed.  For instance, Type~I behavior seemed to always entail 
subcritical end states that were bound and star-like.  Type~II behavior, 
on the other hand, was observed to coincide with dispersal end states.

To refute recent claims that driven neutron stars lead to Type~II critical behavior with characteristics 
at odds with 
previous ultra-relativistic fluid studies, we performed accurate calculations of the 
scaling exponent for such scenarios.  Using different stars and  velocity profiles, 
and by varying other aspects of the numerical model, we found that our observed scaling behavior
was insensitive to approximations made in the numerical solution and was universal with respect to 
different families of initial data.  We found that the scaling exponent and critical solution 
agreed remarkably well with their ultra-relativistic counterparts.  Type~II behavior with a 
neutron star and a scalar field was also studied.  Since the scalar field pulse
required to drive the star to collapse was so strong, the scalar field was found to dominate the 
critical behavior.  Hence, for this scenario, Type~II scaling behavior of the perfect fluid was not 
seen.  

Since meta-stable, star-like states of perfect fluid systems have been known for decades, many 
anticipated  the Type~I behavior observed here.  However, this thesis describes the first in depth analysis
of Type~I phenomena associated with hydrostatic solutions.  The Type~I critical solutions were 
found to coincide with perturbed unstable hydrostatic solutions which were typically more massive than their
progenitor stars.  Also, the Lyapunov exponents of the critical solutions were measured, and were
found to follow a linear relationship as a function of the time-averaged central densities of their associated
critical solutions.  

In the future, we hope to address a great number of topics that expand on this work.  
First, the Lyapunov exponents of the Type~I critical solutions need to be calculated in order 
to verify that they match the measured scaling exponents.  Second, the 
supercritical section of parameter space demands further exploration, in order to investigate 
how much matter can realistically be ejected from shock/bounce/collapse scenarios.  In addition, the 
ability to follow spacetimes after the formation of an apparent horizon would allow us to study the 
possible simultaneous overlap of Type~I and Type~II behavior.  Ultimately, it is our goal to 
expand the model a great deal, making the matter description more realistic and eliminating 
symmetry.  As a first step, we wish to develop Adaptive Mesh Refinement procedures 
for conservative systems that will be  required to study critical phenomena of stellar 
objects in axial-symmetry \cite{choptuik-etal-2003}.  Also, we wish to 
examine how Type~II behavior changes in the context of realistic equations of state.  For example, 
realistic equations of state effectively make the adiabatic index of the fluid a function of the fluid's
density and temperature, and, to date, critical behavior in perfect fluids has only been described for 
fluids with constant adiabatic index.   

The numerical simulation of relativistic perfect fluids on the brink of gravitational 
collapse is a difficult, yet rewarding, endeavor.  The wide range of phenomena that result from 
relativistic fluids admitting a length scale still requires a great deal of future study.  
This thesis has advanced our ability to faithfully model such systems, 
and it has furthered our understanding of black hole formation in fluids.

\appendix
\index{Appendix@\emph{Appendix}}%

\chapter{Conversion of Units and Scale}
\label{app:unit-conversion}

When theoretical calculations are made in the theory of general relativity, it is customary 
to use ``geometrized units'' in which $G = c = 1$ (see Appendix~E of 
\cite{wald} for a comprehensive discussion on the conversion to and from 
geometrized units, only a few key ideas will be mentioned here). 
In such units, scales or dimensions of 
mass ($M$) and time ($T$) are transformed into scales of length ($L$) only, by multiplying 
by appropriate factors of $G$ and $c$.  For instance, by the mass and time 
scale dependence of $G$ and $c$, one can easily derive that a quantity $\mathcal{Q}$
that scales like $L^lM^mT^t$, can be converted into geometrized units by 
multiplication of $c^t \left(G/c^2\right)^m$.  After the conversion to 
geometrized units, $\mathcal{Q}$  scales as $L^{l+m+t}$.  

Since the equations governing the ultra-relativistic fluid are all invariant
under changes in the fundamental length scale $L$, such fluids naturally follow 
self-similar behavior \cite{cahill-taub}.  The inclusion of $\rho_\circ$ in the 
system eliminates this intrinsic scale-invariance via the equation of state.  For example, 
when using the polytropic equation of state, $P = K \rho_\circ^\Gamma$, 
the constant $K$ has dimensions $L^{2\left(\Gamma - 1\right)}$ in geometrized units and 
$L^{3\Gamma-1} M^{1-\Gamma} T^{-2}$ in arbitrary units. 
Hence, one may set the fundamental length-scale of the system
by choosing a value for $K$ \cite{cook-shap-teuk-1992},\cite{cook-shap-teuk-1994}.
Since all physical quantities are expressible in dimensions of $L$ in geometrized 
units, the quantities of static \emph{and} dynamic systems 
which use one set $\{K,\Gamma\}$ should be exactly the same as 
those using another set $\{\hat{K},\hat{\Gamma}\}$,
modulo a rescaling of each quantity by the factor  
\beq{
\left(\hat{L}/L\right)^n = 
\left({\hat{K}}^{1/2\left(\hat{\Gamma}-1\right)} / {K}^{1/2\left(\Gamma-1\right)}\right)^n
\label{rescaling-factor}
}
where $n$ depends how the particular quantity scales with length.
Thus, setting $K=1$ makes the 
system dimensionless, and this is the approach used in the thesis.  This choice 
makes clearer the comparison of two solutions having different values of $K$ and $\Gamma$.

In order to transform from our dimensionless system to one with dimensions, one must 
first set the scale by fixing $K$.  Let $\hat{X}$ be a quantity that has 
dimensions of $L^lM^mT^t$, and $X$ be the corresponding dimensionless quantity.  In order to 
transform $X$ into $\hat{X}$, one may use the following equation
\beq{
\hat{X} = K^x c^y G^z X 
\label{unit-transform}
}
where 
\beq{
x = \frac{l + m + t }{ 2 \left( \Gamma - 1 \right)}  
\quad , \quad 
y = \frac{ \left( \Gamma - 2 \right) l + \left( 3 \Gamma - 4 \right) m  - t }{\Gamma - 1}
\quad , \quad 
z = - \frac{ l + 3 m + t }{2}
\label{unit-conv-exponents}
}

When presenting results of TOV solutions using polytropic state equations, it is 
customary to choose $K$ in such a way that the maximum stable mass for 
the given polytrope corresponds to that of the Chandrasekhar mass, $1.4 M_\odot$. 
As an example, a mass $\hat{M}(K)$ expressed in units can be 
calculated from the 
dimensionless $M(0)$ via the above formula (since $\hat{M}$ has dimensions of only mass, then
$l=0, m=1, t=0$):
\beq{
\hat{M}(K)  =  K^{1/2\left(\Gamma-1\right)} c^3 c^{-1/\left(\Gamma-1\right)} G^{-3/2} 
          M\left(K=1\right) \    \label{mass-units}
}

Since the TOV solutions for $\Gamma=2$ and $K=1$ yield a maximum stable mass of 
$0.164$, then the $K$ that would make $\hat{M}(K) = 1.4 M_\odot$ would be approximately 
$10^5 \mathrm{cm}^{5} \mathrm{g}^{-1} \mathrm{s}^{-2}$, in cgs units.  
The radius of this maximum mass star is $0.768$ with $K=1$, and is 
about $9.4\,\mathrm{km}$ with $K=10^5 \mathrm{cm}^{5} \mathrm{g}^{-1} \mathrm{s}^{-2}$.

\begin{vita} 
Scott Charles Noble, son of Barbara Sullenger and Charles  Noble,  was born at 2:16 PM on 
August 25, 1975, in Royal Oak, Michigan.  After growing up in Lansing, MI, Scott moved to Okemos, 
MI, where he began to show an interest in science.  Along the way, he became a 
Semi-Finalist in the Westinghouse Science Talent Search of 1993 for his study on the angular correlations
of nuclear fragmentation in heavy ion collisions, which was completed the previous summer at the 
National Superconducting Cyclotron Laboratory located on the campus of Michigan State University (MSU).  
He graduated from Okemos High School in June of 1993, and worked a few weeks as a lab assistant 
in a Nematology lab at MSU later that summer.  Then, Scott spent four years at 
Caltech, and graduated with a BS in Physics on Friday, June 13, 1997. 
While there he accomplished research in high-energy physics and quantum computation.  
In August of 1997, he entered the graduate program in physics at the University of Texas in Austin, 
and began working in numerical relativity under the supervision of Matthew Choptuik.  After 
graduation, Scott will become a postdoctoral fellow at the University of Illinois at Urbana-Champaign.
\end{vita}              

\end{document}